\documentclass[10pt]{revtex4}

\usepackage{import}
\usepackage{color}
\usepackage{graphicx}
\usepackage{dcolumn}
\usepackage{lineno,hyperref}
\usepackage{amsmath, amsfonts}
\usepackage{amssymb}
\usepackage[caption=false]{subfig}
\usepackage{lipsum}
\usepackage{booktabs}
\usepackage{titlesec }
\usepackage{bm}
\usepackage{verbatim}
\usepackage{epsfig}
\usepackage{tikz}        
\usepackage{etoolbox}
\usepackage{lineno}
\usepackage[caption=false]{subfig} 
\usepackage{mwe}
\usepackage{subfig}

\begin{document}
\bibliographystyle{revtex}

\title[Short Title]{Properties of nuclear pastas}

\author{Jorge A. L\'opez}

\affiliation{Department of Physics, University of Texas at El
Paso, El Paso, Texas 79968, U.S.A.}

\author{Claudio O. Dorso}

\affiliation{Departamento de F\'isica, FCEN, Universidad de Buenos
Aires, N\'u\~nez, Argentina}

\author{Guillermo Frank}

\affiliation{Unidad de Investigaci\'on y Desarrollo de las Ingenier\'ias, Universidad Tecnol\'ogica Nacional, Facultad Regional Buenos Aires, Buenos Aires, Argentina}

\date{\today}

\begin{abstract}

In this Review we study the nuclear pastas as they are expected to be formed in 
neutron star cores.  We start with a study of the pastas formed in nuclear 
matter (composed of protons and neutrons), we follow with the role of the 
electron gas on the formation of pastas, and we then investigate the pastas in 
neutron star matter (nuclear matter embedded in an electron gas). \\

Nuclear matter (NM) at intermediate temperatures ($1$ MeV $\lesssim T \lesssim 
15$ MeV), at saturation and sub-saturation densities, and with proton content 
ranging from 30\% to 50\% was found to have liquid, gaseous and liquid-gas mixed 
phases. The isospin-dependent phase diagram was obtained along with the critical 
points, and the symmetry energy was calculated and compared to experimental data 
and other theories.  At low temperatures ($T \lesssim 1$ MeV) NM produces 
crystal-like structures around saturation densities, and pasta-like structures 
at sub-saturation densities. Properties of the pasta structures were studied 
with cluster-recognition algorithms, caloric curve, the radial distribution 
function, the Lindemann coefficient, Kolmogorov statistics, Minkowski 
functionals; the symmetry energy of the pasta showed a connection with its 
morphology. \\

Neutron star matter (NSM) is nuclear matter embedded in an electron gas. The 
electron gas is included in the calculation by the inclusion of an screened 
Coulomb potential. To connect the NM pastas with those in neutron star matter 
(NSM), the role the strength and screening length of the Coulomb interaction 
have on the formation of the pastas in NM was investigated.  Past was found to 
exist even without the presence of the electron gas, but the effect of the 
Coulomb interaction is to form more defined pasta structures, among other 
effects. Likewise, it was determined that there is a minimal screening length 
for the developed structures to be independent of the cell size. \\

Neutron star matter was found to have similar phases as NM, phase transitions, 
symmetry energy, structure function and thermal conductivity. Like in NM, pasta 
forms at around $T\approx 1.5$ MeV, and liquid-to-solid phase changes were 
detected at $T\approx 0.5$ MeV.  The structure function and the symmetry energy 
were also found to depend on the pasta structures. \\

\end{abstract}

\maketitle

\newpage
\tableofcontents

\newpage

\section{Introduction}\label{intro}

Neutron stars are the final product of the death of a massive star.  Stars die 
when the internal thermonuclear fusion can no longer balance gravitational 
compression, and a supernova shock ejects most of the mass of the star leaving 
behind a dense core.  If the star is big enough as to generate a remnant core 
with a mass larger than the Chandrasekhar limit, the gravity will be so strong 
that the electron degeneracy pressure will not be able to support the system, 
and the star will compress down to nuclear density. Since during the collapse a 
big part of electrons and protons turn into neutrons and escaping neutrinos 
through electron capture, the produced core tends to have an excess of neutrons 
over protons, thus justifying the name of a neutron star. A neutron star is a 
neutral object composed of neutron, protons and electrons, with the charge of 
the protons balanced by the one of electrons. \\

Neutron stars range in mass between 1 and 3 solar masses, with a radius is of 
the order of about 10 km, and the average density of a degenerate Fermi gas 
composed of nucleons, $10^{15}$ $g/cm^3$ or about 20 times that of normal 
nuclei. Energy considerations
indicate~\cite{20,23} that the cores of neutron stars have a 1-km crust where 
the $\beta$ decayed-produced neutrons form neutron-rich nuclear matter immersed 
in a sea of electrons. The density of neutron star crusts goes from normal 
nuclear density ($\approx 3\times10^{14}$ $g/cm^3$, or $\approx \rho_0=0.16$ 
nucleons/fm${}^3$) at a depth of $\approx 1 \ km$ to the neutron drip density 
($\approx 4\times10^{11}$ $g/cm^3$) at $\approx 1/2 \ km$, to, finally, an even 
lighter mix of neutron-rich nuclei also embedded in a sea of electrons with 
densities decreasing down to practically zero in the neutron star envelope. \\

It is known that nuclear systems at low temperatures and sub-saturation 
densities can exist in non-uniform structures, known as ``pastas'', and it is 
suspected that the crusts of neutron stars may conform to such 
configurations~\cite{20,23,Hashimoto,Koonin,24,25,Maruyama,27,28}. Knowing the 
structures attained by neutron crusts is an important factor in the 
understanding of star-quakes, pulsar frequencies and neutron star evolution. 
Indeed since the cooling of non-pulsar neutron stars is due mostly to neutrino 
emission from the core, the interaction between neutrinos and the crust 
structure is relevant to their thermal evolution~\cite{dorso2017}. The study of 
such structures is the main goal of this review.\\

\subsection{The pasta}

The pasta structures of nuclear systems have been studied using different 
models~\cite{20,23,Hashimoto,Koonin,24,25,Maruyama,27,28}. Such structures 
appear to form due to the interplay between attractive-repulsive nuclear and 
Coulomb forces. As the density, temperature and proton fraction vary, the 
structure changes from a uniform phase at the core to configurations with voids 
filled with small clusters ``gnocchi-like'', to ``lasagna-like'' layers of 
nuclear matter and gas, to ``spaghetti-like'' rods of matter embedded in a 
nuclear gas, to ``gnocchi'' clumps, to a practically uniformly dissolved gaseous 
phase~\cite{23}.\\

To attain these structures, nucleons should reach states of a free energy 
minimum. In the original studies of the 1980s the energy minima were determined 
using static methods such as the liquid drop model~\cite{20,Hashimoto}, mean 
field theories~\cite{Page} and Thomas-Fermi models~\cite{Koonin}. These methods, 
however, usually work only at zero temperature, locate only the ``traditional'' 
global minima and tend to miss the ``non-traditional'' local minima of energy 
barriers; to gain access to the complete set of structures (traditional and 
non-traditional) temperature-dependent dynamical models are needed.\\

Dynamical studies, such as quantum molecular 
dynamics~\cite{Maruyama,27,horo_lambda,gw-2002,schuetrumpf,fattoyev}, predict 
the formation of the pasta phases by dynamical means, but tend to be limited to 
states of global energy minima.  To properly obtain the pasta structures, it is 
best to use models that can achieve phase changes, clusterization and identify 
local minima by cooling, such as classical molecular dynamics 
models~\cite{dorso2014,Horo2004,dor12, dor12A}.\\

In this review we will use the classical molecular dynamics model (CMD) to study 
the pasta in nuclear matter and in neutron star matter; systems which will be 
defined in the next Section. To avoid technical distractions, the description of 
the CMD model is relegated to Appendix~\ref{cmd}.\\

\subsection{The pasta in nuclear matter and in neutron star matter}

The neutron star crust is composed of protons, neutrons and an embedding gas of 
electrons.  To study such system we systematically divide the review in three 
parts. First, a study of nuclear matter (NM), e.g. systems composed solely of 
protons and neutrons; second, an investigation of the role electrons have on the 
properties of NM; and third, a study of neutron star matter (NSM), i.e. systems 
of protons and neutrons embedded in an electron gas. \\

Nuclear matter exhibits fascinating complex phenomena at subsaturation densities 
and warm and cold temperatures. At densities below the saturation density, 
$\rho_0 = 0.16\,\mathrm{fm}^{-3}$, and temperatures, say, between 1 MeV and 5 
MeV, nuclear systems exist in liquid and gaseous phases, as well as in a mixture 
between the two. At lower temperatures, crystals and structures resembling the 
so-called nuclear pastas appear with different morphologies depending on the 
temperature, density and isospin content and, furthermore, nucleons inside such 
structures can undergo phase transitions. \\

This review studies nuclear matter first. At intermediate temperatures we review 
the bulk properties, phases and phase transitions of NM, all with varying 
percentages of isospin content to allow the study of the symmetry energy. At 
lower energies, the formation of the pasta is investigated along with its 
properties, phases, transitions, and symmetry energy.\\

This is followed by a review of the role the electron gas has on the pasta of 
nuclear matter. As the electron gas is introduced in the CMD calculation by 
means of a screened Coulomb potential, the study focuses on the effects the 
strength and range of the interaction have on the morphology of the pastas.\\

Finally, the properties of the pastas in neutron star matter, i.e. systems with 
protons, neutrons and electrons, are reviewed with special attention to their 
shapes, phases, phase transitions, and symmetry energy.\\

In understanding the pastas both in NM and in NSM we also study how the energy 
varies with respect to the isospin content, i.e. the symmetry energy. It is in 
this type of studies that one can appreciate the value of CMD, as to perform 
such a study at intermediate and low temperatures and sub-saturation densities, 
a model capable of exhibiting clustering phenomena and phase changes is needed. 
\\

Similarly, as mentioned before, the cooling of neutron stars is related to pasta 
structure of the crust as the neutrino opacity is enhanced by coherent 
scattering. Here, we also study the very-long range order of the pasta phases, 
focusing on its influence on the opacity of the crust to low-momentum neutrinos. 
\\

In this review the properties of nuclear matter will be studied in 
Section~\ref{Nuclear-Matter}, the role of the electron gas on the NM will be 
investigated in Section~\ref{Electron-gas} and, finally, the neutron star matter 
will be studied in Section~\ref{nsm}.  Analytical tools used, such as the 
classical molecular dynamics method used, algorithms for the recognition of 
clusters, the radial correlation functions, Lindemann coefficient, Kolmogorov 
statistics, Minkowski functionals, and the procedure used to calculate the 
symmetry energy are presented in the various appendices.\\

\newpage


\section{Nuclear matter}\label{Nuclear-Matter}

In this section we study properties of nuclear matter. NM is composed of protons 
and neutrons in varying proportions. The properties of NM vary depending on the 
temperature and density of the system, as well as on its isospin content. At 
intermediate temperatures, that is between $1$ MeV $\lesssim T \lesssim 
T_{critical}\approx15$ MeV, NM can exist in liquid form, gaseous form, and in a 
mixture between these phases. At lower temperatures crystal structures develop 
around normal saturation nuclear density, $\rho_0 = 0.16$ fm$^{-3}$, and 
non-homogeneous structures (pasta-like) at sub-saturation densities. We will 
study NM at intermediate temperatures in Section~\ref{intT}, and at lower 
temperatures in Section~\ref{nm-LT}. \\

For the purposes of the present investigation, nuclear matter is taken as 
protons and neutrons interacting only through the Pandharipande potentials, i.e. 
the proton-proton interaction is ignored (since the Pandharipande potentials 
yield the correct scattering cross sections, it can be though of as containing 
the proton-proton Coulomb interaction effectively). Later, in 
Section~\ref{Electron-gas}, NM will be embedded in an electron gas and the 
proton-electron interaction will be taken explicitly through an screened Coulomb 
Potential; this will effectively constitute neutron star matter as studied in 
Section~\ref{nsm}.

\subsection{Nuclear matter at intermediate temperatures}\label{intT}

Here we study systems with different values of isospin content, density and 
temperature to obtain bulk properties such as the energy per nucleon, pressure, 
equilibrium density, compressibility, symmetry energy, as well as phases and 
phase transitions. An overview of the bulk properties of nuclear matter at 
intermediate temperatures is presented in Section~\ref{bulk}, followed by a 
discussion of the existence of phases in Section~\ref{phasesIT}, the phase 
diagram in Section~\ref{phasediag}, and an estimation of the nuclear symmetry 
energy in Section~\ref{nseIT}; these findings are summarized in 
Section~\ref{nm-s-IT}.\\

\begin{figure}  
\begin{center}
\includegraphics[width=3.4in]{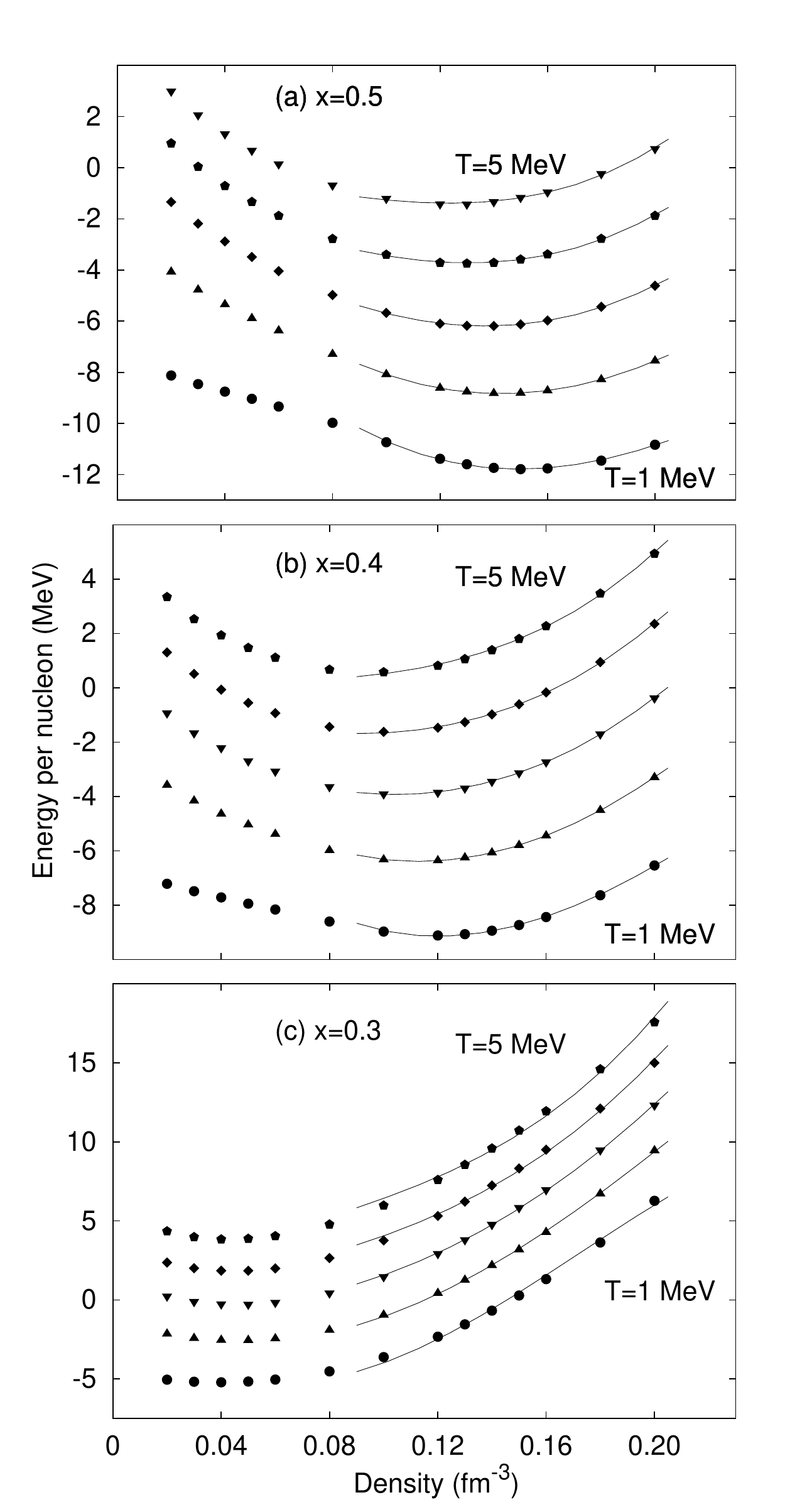}
\end{center}
\caption{Energy per nucleon as a function of the density for three
different isospin contents. In each case the curves correspond to
temperatures ranging from T = 1 MeV (lower curve) to 5 MeV (upper
curve) with the intermediate curves corresponding to 2, 3 and 4
MeV. The lines indicate the fits used in Section~\ref{nse} to
estimate the symmetry energy.} \label{e-d}
\end{figure}




\begin{figure*}[!htbp]
\centering
\subfloat[Pressure\label{pressure}]{
\includegraphics[width=0.33\columnwidth]
{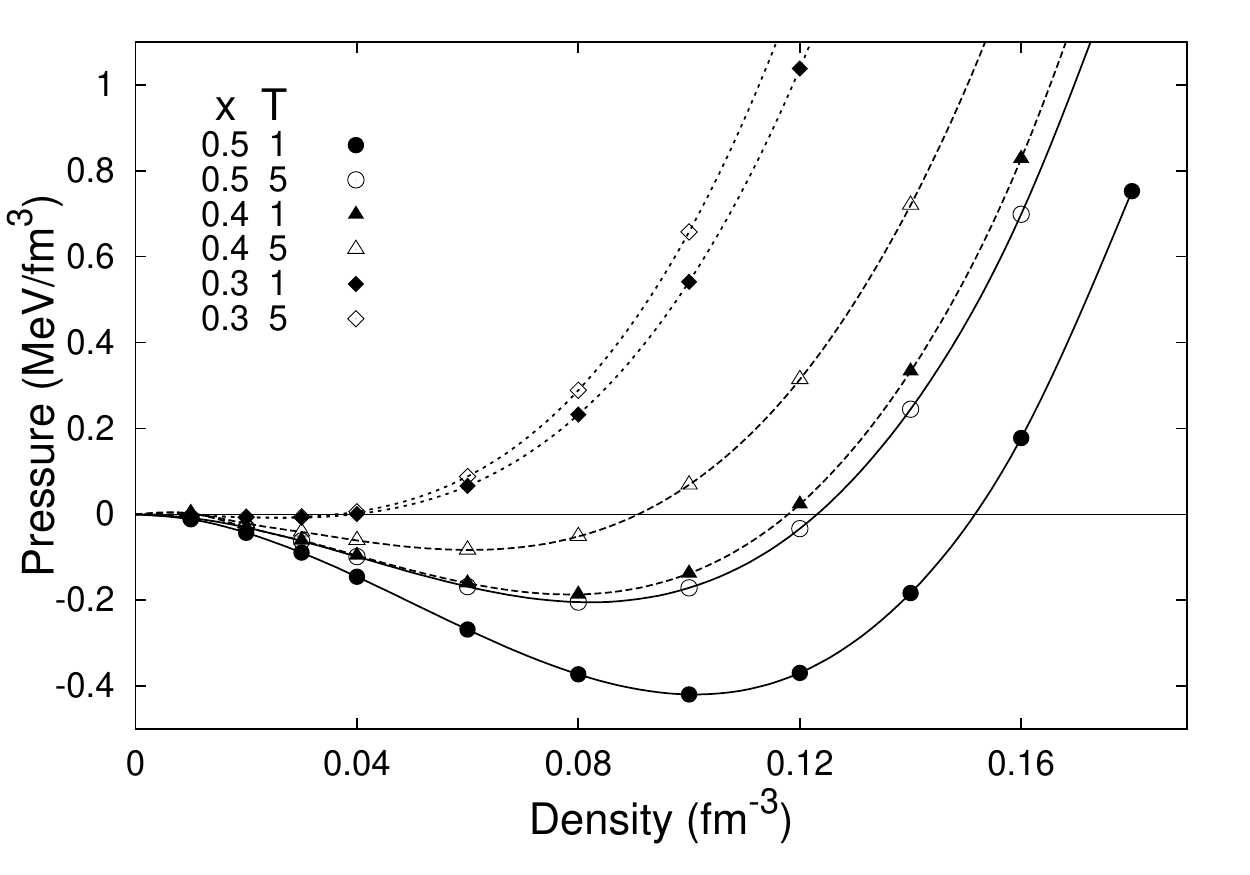}
}
\subfloat[Saturation density\label{Min-dens-vs-T}]{
\includegraphics[width=0.33\columnwidth]
{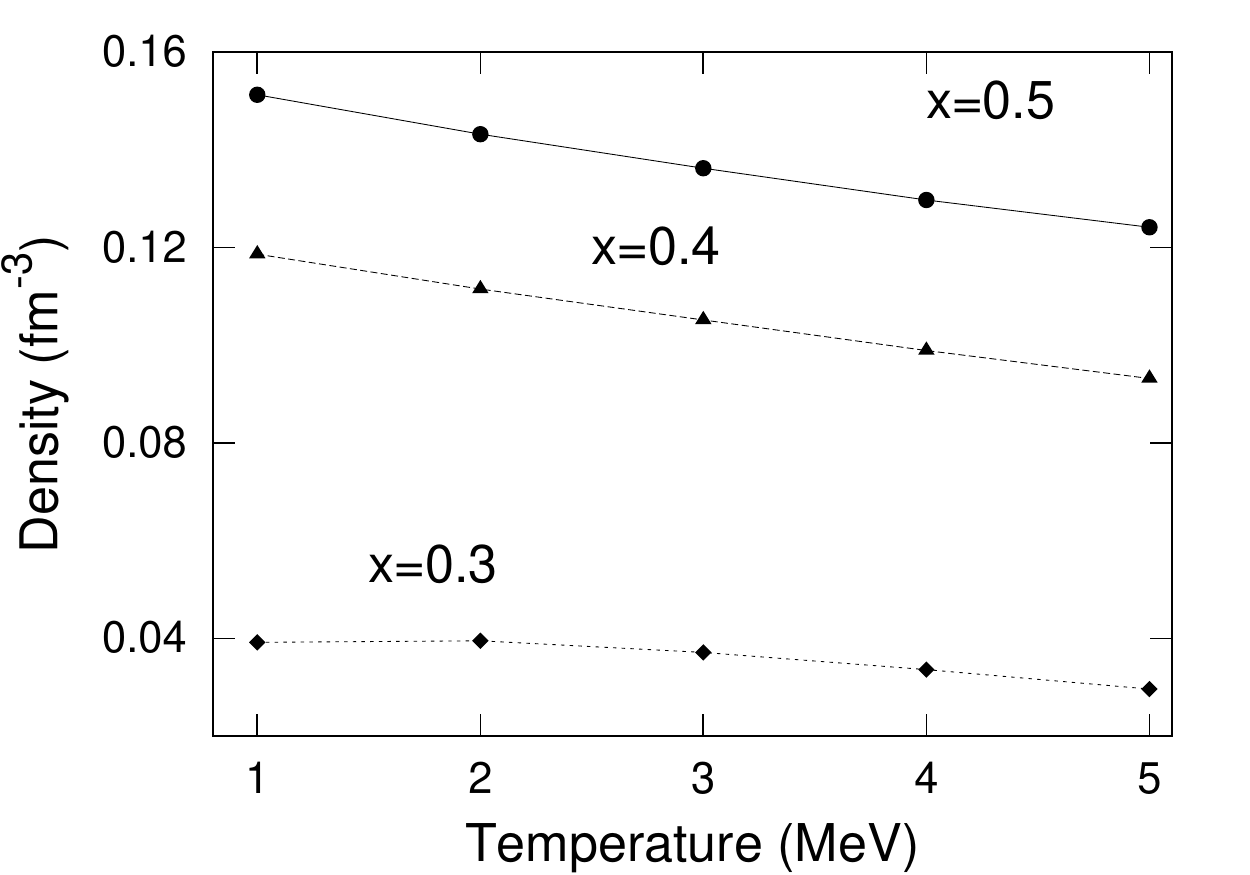}
}
\subfloat[Compressibility\label{CompVsTemp}]{
\includegraphics[width=0.33\columnwidth]
{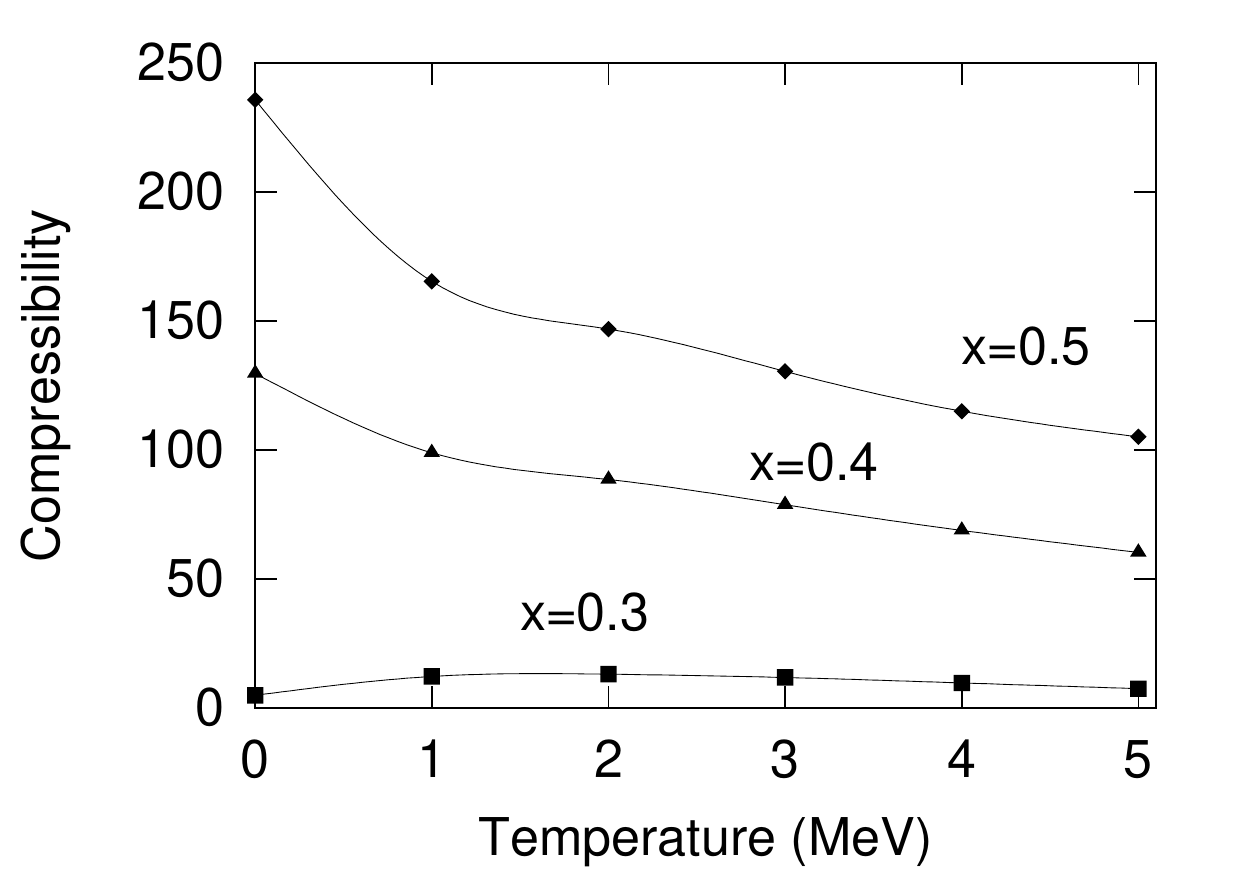}
}
\caption{(a) Pressure versus density, (b) saturation density versus temperature, 
and (c) compressibility versus temperature, all for $x=0.3$, 0.4 and 0.5. The 
value of the compressibility at $T\approx 0$ MeV is from the study presented in 
Section~\ref{nmltsd}.}
\end{figure*}

\subsubsection*{Bulk properties of nuclear matter at intermediate 
temperatures}\label{bulk}

The studied systems were created with the molecular dynamics code described in 
Appendix~\ref{cmd}, implemented according to the details described 
in~\ref{CMD-IT}. In summary, a total of 2,000 nucleons were placed in cubic 
cells of the appropriate size to have densities $\rho=$0.02, 0.03, ... 0.2 
fm${}^{-3}$, and with periodic boundary conditions to simulate an infinite 
system. The ratio of protons to neutrons was fixed to values $x=Z/A=0.3$, 0.4 
and 0.5, where $Z$ is the number of protons, and $A$ is the total number of 
nucleons. The temperatures of the systems studied are $T = 1$, 2, 3, 4, and 5 
MeV. \\

The energy per nucleon (kinetic plus potential) is shown in Figure~\ref{e-d} as 
a function of the density at various temperatures.  The curves show 
characteristic ``$\cup$'' shapes around their saturation density (minimum of the 
$\cup$); these shapes resemble the predictions of Skyrme-Hartree-Fock and 
relativistic mean-field calculations for zero temperature~\cite{Tanihata}.  At 
lower temperatures all three cases presented in Figures~\ref{e-d} have regions 
of negative binding energy which correspond to bound (liquid) matter; at high 
temperatures the systems become unbound. At lower densities the curves depart 
from the $\cup$ shapes showing a phase transitions from a uniform medium around 
saturation densities to non-homogeneous media at sub-saturation densities.\\

\begin{figure}  
\begin{center}
\includegraphics[width=3.4in]{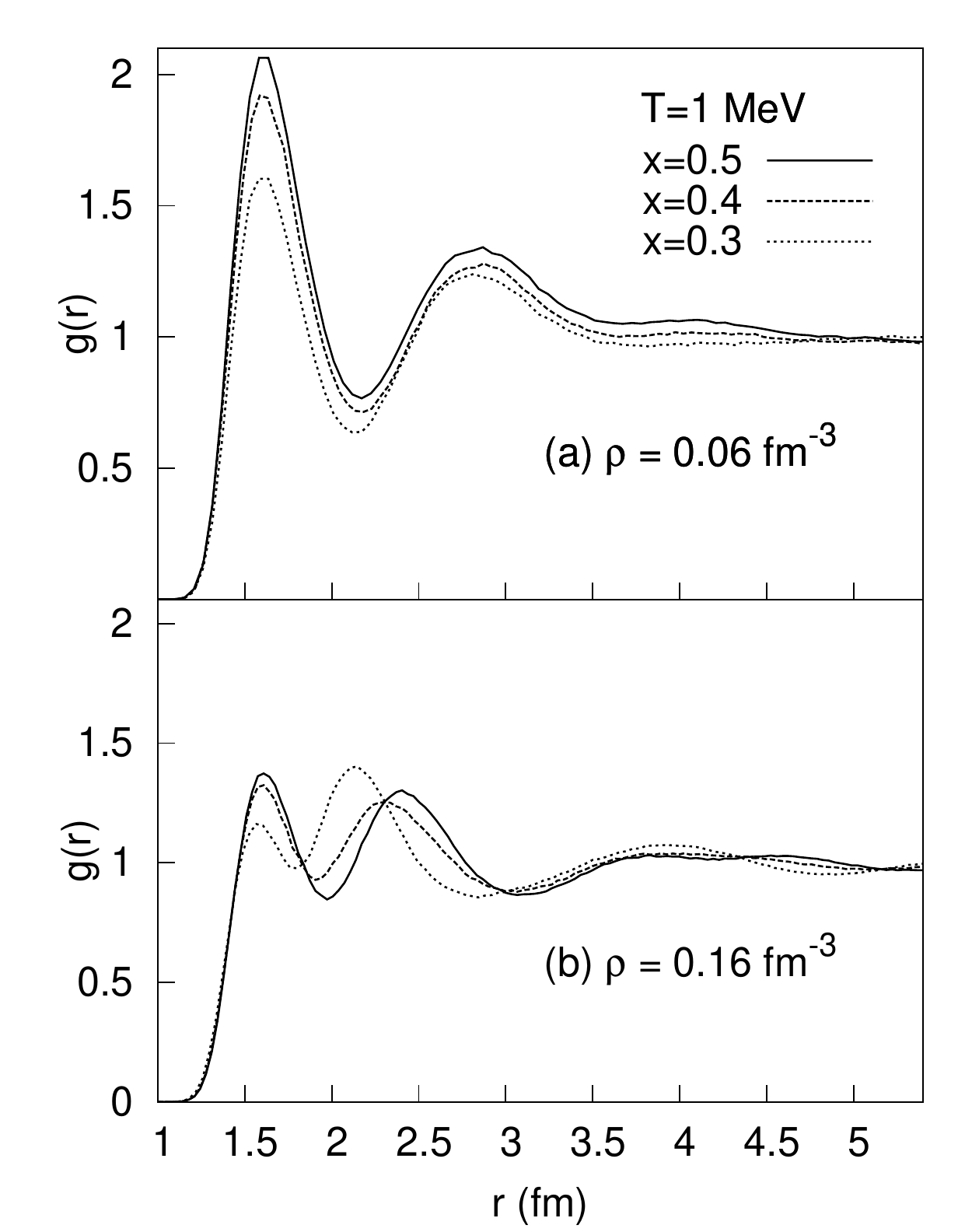}
\end{center}
\caption{Radial distribution functions of systems with $x=0.3$,
0.4 and 0.5 at $T = 1$ MeV and densities $\rho=0.06 \ fm^{-3}$ in
panel (a) and $\rho=0.16 \ fm^{-3}$ in panel (b).} \label{rdf}
\end{figure}

Figure~\ref{pressure} shows the pressure-density curves for $T=1$ and 5 MeV, and 
$x=0.3$, 0.4 and 0.5 and, corroborating the previous findings, the equilibrium 
densities (zero pressure points) correspond to the minima of the energy-density 
curves. It must be remarked that at low densities the CMD pressure is not 
correct as the system should approximate a free nucleon gas, which is a quantum 
system (see Appendix~\ref{App-Max}). The corresponding saturation densities can 
be estimated through a fit of the minima of $E(T,\rho)$; 
Figure~\ref{Min-dens-vs-T} shows the saturation density as a function of the 
temperature. At near zero temperatures $\rho_0$ varies from $0.16 \ fm^{-3}$ at 
$x=0.5$, to a low $0.04 \ fm^{-3}$ at $x=0.3$. The compressibility at saturation 
density, obtained from the energy through $K(T,\rho)=9\rho^2\left[\partial^2 
E/\partial\rho^2 \right]_{\rho_0}$, is presented in Figure~\ref{CompVsTemp}.  
Nuclear matter becomes a very soft fluid for $x=0.3$ at all temperatures, and it 
varies drastically with $x$, and reduces by $\approx 30\%$, as $T$ increases 
from 1 to 5 MeV for $x=0.4$ and 0.5. This softening with excitation energy is 
consistent with IQMD simulations of $^{197}Au+^{197}Au$ at 600 
MeV/A~\cite{kumar13} and with BUU calculations of similar reactions at 1 
GeV/A~\cite{Dan02}. \\

\begin{figure}
\begin{center}
\includegraphics[width=3.4in]{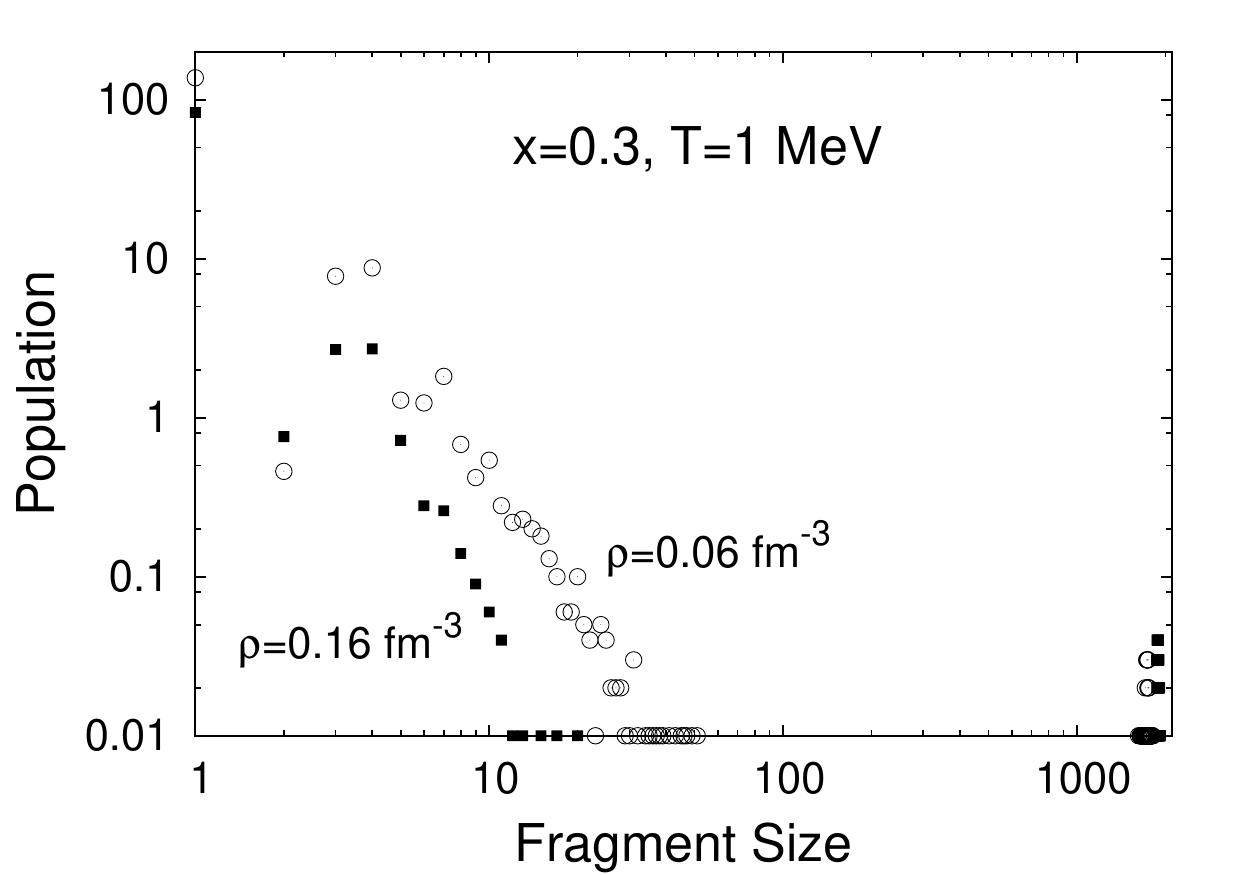}
\end{center}
\caption{Mass distribution obtained with $x=0.3$, $T = 1$ MeV at
$\rho=0.06$ and $0.16 \ fm^{-3}$.} \label{MassDist}
\end{figure}

\subsubsection*{Phases of nuclear matter at intermediate 
temperatures}\label{phasesIT}

The existence of phases in nuclear matter can be seen from Figure~\ref{e-d}.  In 
the $T = 1$ MeV curve the transition from a smooth $\cup$ shape to an extraneous 
curve happens at $\rho \approx 0.10 \ fm^{-3}$ for $x=0.5$, and at $\rho \approx 
0.08 \ fm^{-3}$ for $x=0.4$, but it is not as pronounced for the $x=0.3$ case.  
The smooth $\cup$ shape corresponds to a uniform crystal-like phase  at low 
temperatures and liquid-like phase at higher temperatures, and the lower-density 
part corresponds to non-homogeneous crystal-gas mixture at low temperatures or 
liquid-gas mixture at higher temperatures. As we will see in 
Section~\ref{phasediag}, systems with $x < 0.3$ would appear to be unable to 
reach such a liquid-gas mixture region and would always stay in a liquid-like 
continuous medium down to very low densities. \\

To further investigate these phases, we examine the $T = 1$ MeV systems through 
the radial distribution function and their mass distribution. As explained in 
Appendix~\ref{RadialDist}, and more specifically in Section~\ref{g-for-pasta}, 
the radial distribution function $g(\mathbf{r})$ signals the presence of 
neighbors and their average distance, which is different in liquid and gaseous 
states.  Figure~\ref{rdf} shows $g(\mathbf{r})$ for systems at $T=1$ MeV at a 
liquid density ($\rho=0.16 \ fm^{-3}$, bottom panel) and at a liquid-gas mixture 
density ($\rho=0.06 \ fm^{-3}$, top panel).  The strengths of the 
nearest-neighbor peaks show that at low densities nucleons tend to be more 
correlated than at higher densities indicating that at $\rho=0.06 \ fm^{-3}$ the 
main contribution at short distances is from nucleons in droplets, while at 
$\rho=0.16 \ fm^{-3}$ the larger nucleon mobility reduces such correlation. This 
is also observed in the second-neighbor peaks which appear at the same distance 
for all values of $x$ at $\rho=0.06 \ fm^{-3}$ but not at $\rho=0.16 \ fm^{-3}$ 
indicating again a reduced mobility of nucleons in the droplets.  The growth of 
the second-neighbors peak over the first peak for the case of $x=0.3$, is due to 
the large number of nn repulsive interactions which exceed the smaller number of 
np attractive interactions at such large discrepancy between the number of 
neutrons (70\%) and protons (30\%).\\

Further information about the phases found before can be obtained by the mass 
distributions attained by the systems in the liquid phase and in the liquid-gas 
mixture; the method used to recognize cluster is outlined in 
Section~\ref{cluster}. Looking at the $x=0.3$ cases, Figure~\ref{MassDist} shows 
the mass distribution in the central cell (i.e. not continuing the fragments 
into neighboring cells) at $\rho=0.06 \  fm^{-3}$ and $T = 1$ MeV in the 
liquid-gas mixture at $\rho=0.16 \ fm^{-3}$ compared to that in the liquid-gas 
mixture at $\rho=0.06 \  fm^{-3}$. Figure~\ref{MassDist} shows that in the 
mixture zone ($\rho=0.06 \ fm^{-3}$) there are more intermediate-mass droplets 
than in the liquid phase (at $\rho = 0.16 \ fm^{-3}$). The relatively large 
number of clusters with $A>1000$ in the liquid phase, indicates that the system 
remains in a continuous medium; notice that such large cluster necessarily must 
have low populations for obvious reasons.\\

The results obtained so far indicate that liquid and gaseous phases exists in 
nuclear matter with isospin asymmetries in the range from $x = 0.3$ to 0.5. 
Transitions between these phases appear to be are alive and well in these cases 
of high isospin asymmetry.  A complementary observation is that these systems 
appear become unbound at temperatures
as low as T = 2 MeV for $x=0.3$.  It is known that pure neutron matter us 
unbound, i.e. it exists only in gaseous phase~\cite{myers95}; in the next 
section we determine the boundaries of these phases.\\

\subsubsection*{Phase diagram}\label{phasediag}

The phase diagram of nuclear matter has been studied for isospin symmetric 
matter in~\cite{siemens,lopezlibro,EntropyCalCur}, and evaluated for isosospin 
asymmetric matter in~\cite{muller}, but it was obtained dynamically with full 
inclusion of the
isospin degree of freedom the first time in~\cite{lopez2020}. The phase diagram 
of nuclear matter can be obtained from the pressure-density isotherms of 
Section~\ref{bulk}, c.f. Figure~\ref{pressure}, by means of the technique known 
as the Maxwell construction, described in Appendix~\ref{App-Max}.\\

\begin{figure}  
\begin{center}
\includegraphics[width=3.4in]{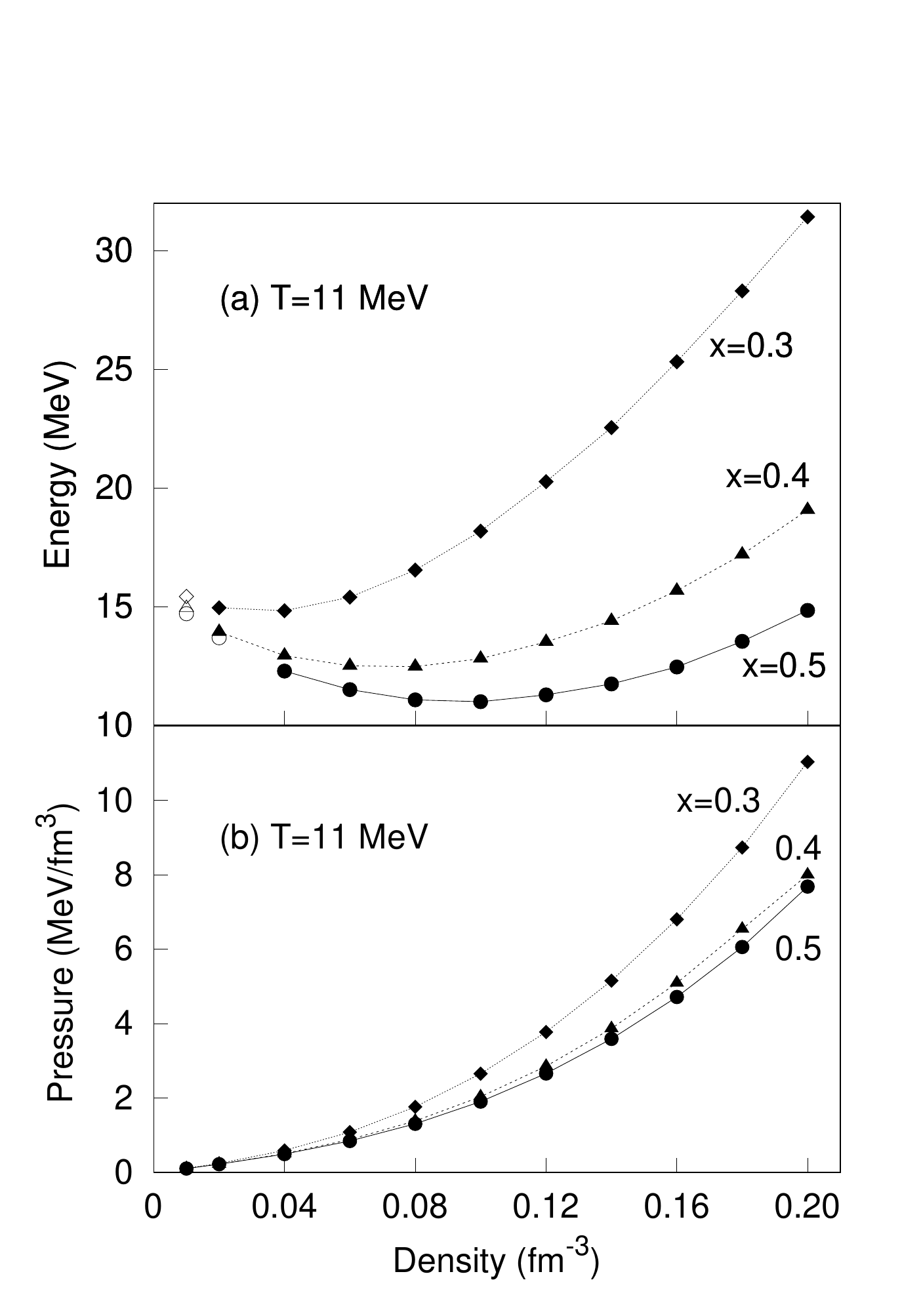}
\end{center}
\caption{Energy per nucleon (a) and pressure (b) versus density for nuclear 
matter systems with $x = 0.3$, 0.4 and 0.5 at $T=11$ MeV.}
\label{Fig-2}
\end{figure}

Pressure-density isotherms, obtained by CMD simulations of systems with isospin 
content of $x = 0.3$, 0.35, 0.4, 0.45 and 0.5, at temperatures varying between 
$T=1, 2, \cdots 15$ MeV, and densities $\rho = 0.01, 0.02, \cdots \, 0.2$ 
fm$^{-3}$, were used in the Maxwell construction. Figure~\ref{Fig-2} shows the 
$T = 11$ MeV isotherms for $x = 0.3$, 0.4 and 0.5.\\

\begin{figure}  
\begin{center}
\includegraphics[width=6.2in]{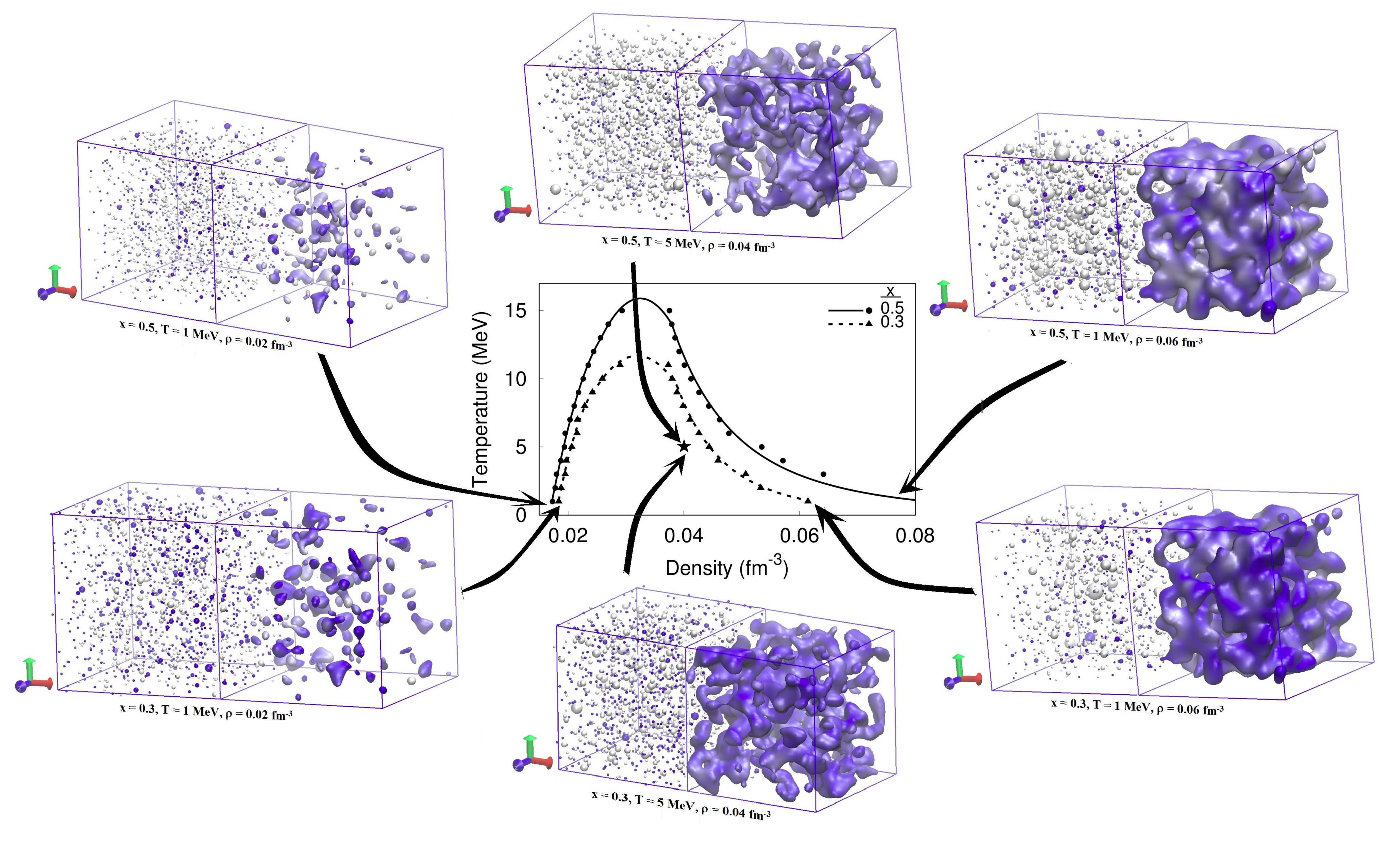}
\end{center}
\caption{\label{Fig-6}Coexistence region for $x = 0.5$ and $x = 0.3$. The points 
indicate the limits of the liquid-gas coexistence region obtained through the 
Maxwell constructions. The continuous and dashed lines are simple approximations 
to indicate liquid-gas mixture regions. The boxes show characteristic 
configurations for six cases (see text).}
\end{figure}

As explained in Appendix~\ref{App-Max}, the application of the Maxwell 
construction to each pressure isotherm yields the value of the density 
corresponding to the boundary between the liquid-gas mixture and the gaseous 
phase, $\rho_{Gas}$ and that of the limit of the mixture and the liquid phase, 
$\rho_{Liq}$ for such isotherm. When plotted together in a density-temperature 
plane, all such points outline the liquid-gas coexistence region; 
Figure~\ref{Fig-6} shows the points obtained from the Maxwell construction for 
the cases $x = 0.5$ and $x = 0.3$. The area delineated by such points indicates 
the liquid-gas coexistence region; the area to the left and to the right of the 
coexistence region correspond, respectively, to gaseous and liquid phases; the 
continuous and dashed lines are simple approximations to indicate the boundaries 
of the liquid-gas mixture regions for the two values of $x$. The maximum of each 
curve are the respective critical points for each value of $x$.\\

Also shown in Figure~\ref{Fig-6} are six characteristic configuration of protons 
and neutrons for two cases on the boundary of the gaseous phase: $T = 1$ MeV and 
$\rho$ = 0.02 fm$^{-3}$ for $x = 0.3$ and 0.5 (left two frames), two cases in 
the liquid-gas mixed regions: $T = 5$ MeV and $\rho$ = 0.04 fm$^{-3}$ for $x = 
0.3$ and 0.5 (central top and bottom frames), and two more cases on the boundary 
of the liquid phase: $T = 1$ MeV and $\rho$ = 0.06 fm$^{-3}$ for $x = 0.3$ and 
0.5 (right two frames). Each configuration shows the protons with dark spheres 
and neutrons with white ones. For clarity the boxes are shown approximately of 
the same size, but in reality they are drawn at different scales. The varying 
sizes of the spheres are due to the perspective effect of the $Visual$ 
$Molecular$ $Dynamics$ (VMD) software used, and are not intended to represent 
clusters. \\

The right part of each of the six configurations shown in Figure~\ref{Fig-6} is 
an attempt to quantify the connectivity of the phases. The surfaces shown have 
similar particle densities obtained from a volumetric Gaussian density map 
computed from particles in a grid.  In all of these cases the isosurfaces were 
calculated with VMD's $QuickSurf$~\cite{QS} application with radius scale of 
1.0, density isovalue of 2.5 and grid spacing of 0.5, and clearly show an 
evolution from low connectivity in the gaseous phase, to mid-connectivity in the 
liquid-gas mixed phase, to high connectivity in the liquid phase. \\

\begin{figure}  
\begin{center}
\includegraphics[width=5.2in]{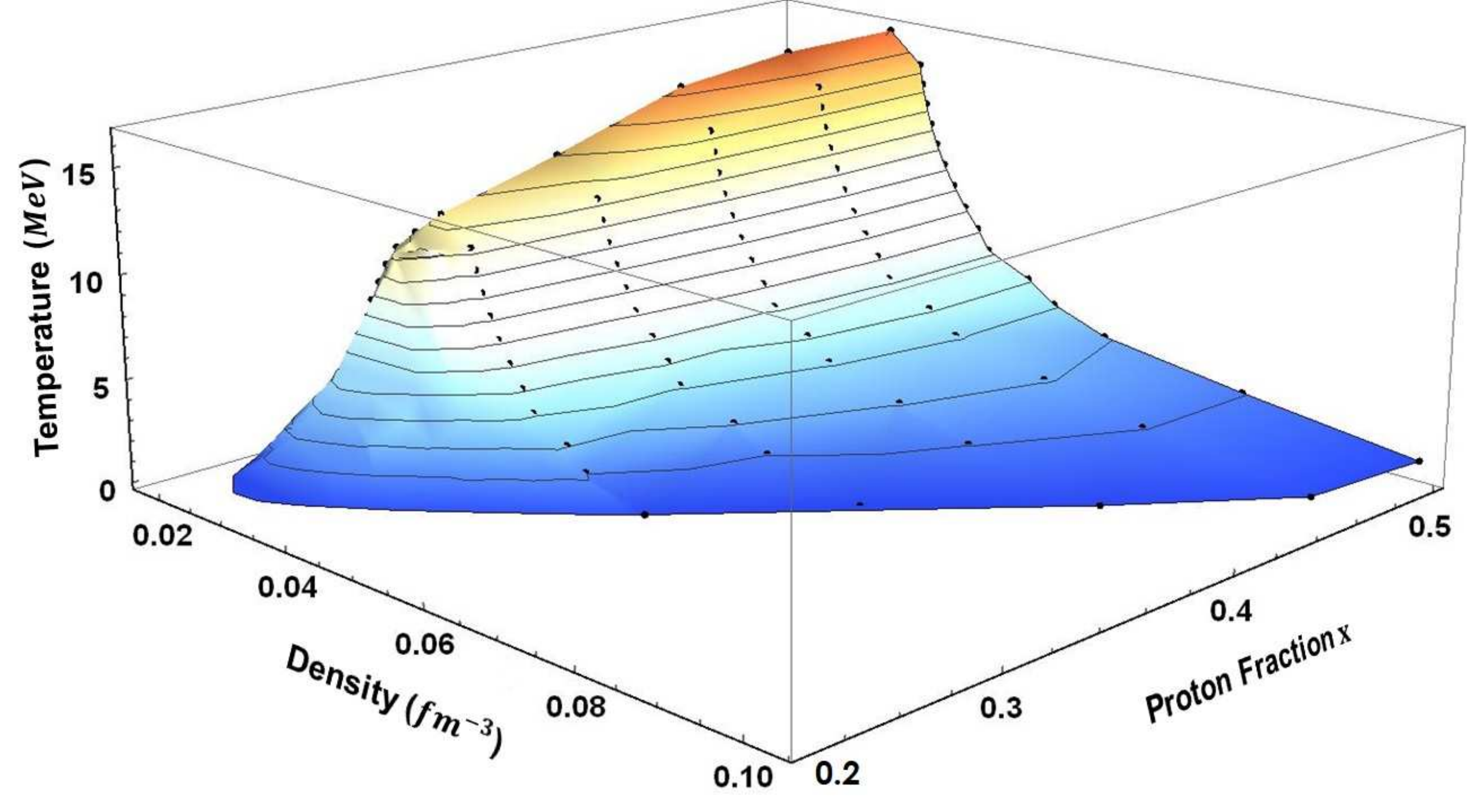}
\end{center}
\caption{\label{Fig-7}Density-$x$-temperature phase diagram. The liquid-gas 
coexistence region is the volume under the surface. The points are the 
density-temperature points obtained from the Maxwell construction for the 
various values of $x$. The critical points now form a {\it critical ridge}}
\end{figure}

Combining the density-temperature points from the Maxwell constructions for the 
cases $x = 0.3$, 0.35, 0.4, 0.45 and 0.5, it is possible to construct a 
three-dimensional plot of the liquid-gas mixture region in the 
density-$x$-temperature space. Figure~\ref{Fig-7} shows the surface of the 
liquid-gas coexistence region; notice that and the critical points now join into 
a {\it critical ridge}. The surface was obtained with the  instruction 
$ListPlot3D$ of the $Mathematica$ package. \\ 

Figure~\ref{Fig-7} indicates that the liquid-gas mixture region ends at a value 
of $x$ smaller than 3, but this is due to a lack of data for lower values of 
$x$. A more insightful estimate of the lower $x$ boundary of the coexistence 
region can be obtained by looking at the trend of the critical temperatures.\\

\begin{figure}  
\begin{center}
\includegraphics[width=3.50in]{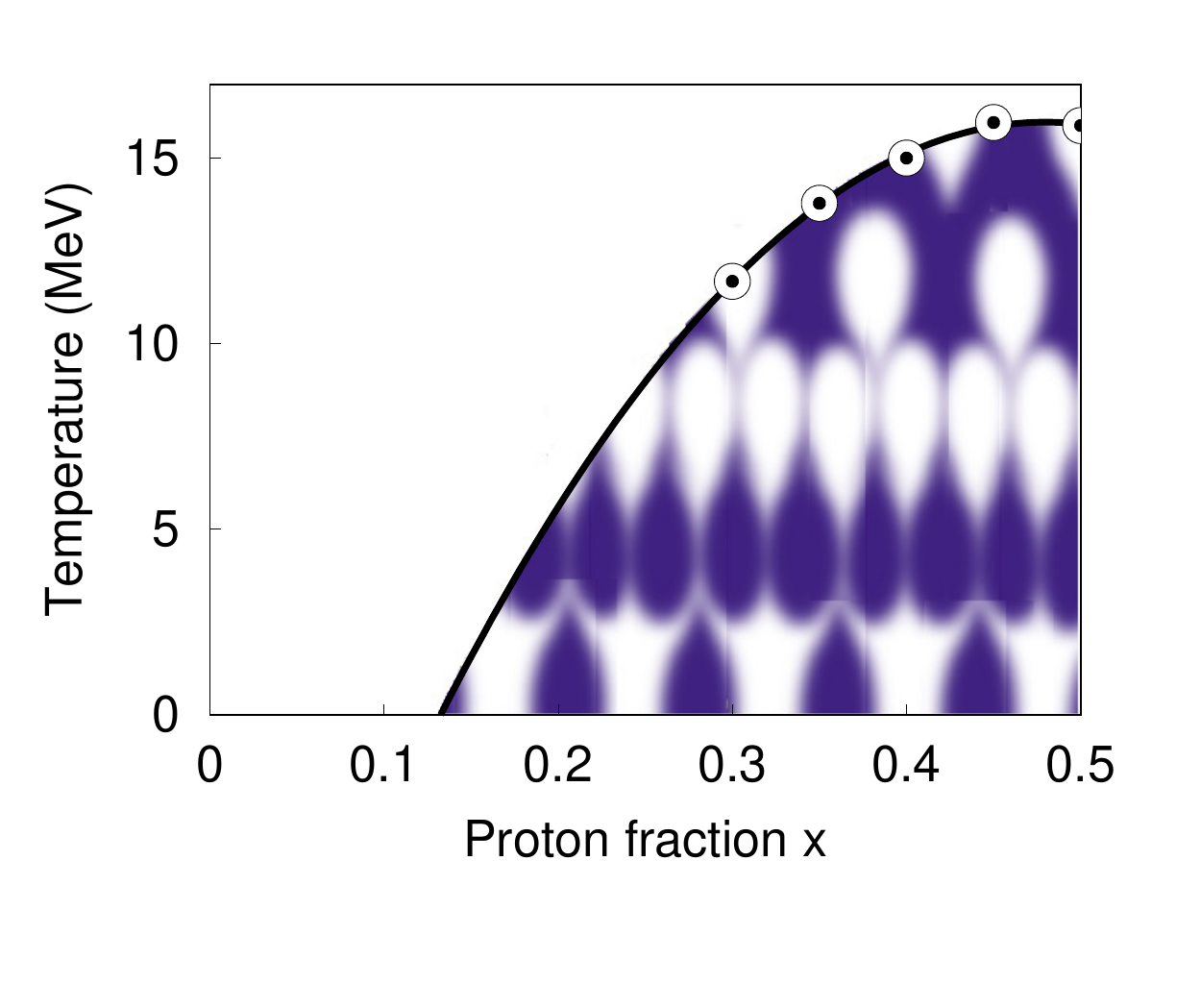}
\end{center}
\caption{\label{Fig-8}Temperature of the critical points as a function of the 
proton fraction $x$. The background image under the curve indicates the region 
where liquid and gas coexist. Extending the trend toward lower values of $x$, 
the curve suggest that the lowest $x$ at which liquid and gas can coexist is 
0.13. }
\end{figure}

Figure~\ref{Fig-8} indicates how the critical temperatures decline for lower 
values of $x$. The points in the figure show the critical temperatures obtained 
from the critical ridge, i.e. from the maxima of the fits used in 
Figure~\ref{Fig-6} for all values of $x$. The continuous line is a parabolic 
fit, which has been extended to lower values of $x$.  An artistic touch was 
added with an Escher-like background indicating the region where bubbles and 
droplets coexist. As the curve reaches the $T$ = 0 value at $x = 0.1324$, we 
conclude that such point is the lower $x$ limit of the liquid-gas mix region, 
and that outside of this $\rho-T-x$ volume all nuclear systems are expected to 
be fully unbound. Although this coexistence region has never been calculated 
before, our findings are in general agreement with the low binding energy of $x 
= 0.3$ matter found in~\cite{Tanihata}, and of unbound pure neutron matter ($x = 
0$) found in~\cite{myers95} at zero temperature.\\

\subsubsection*{Symmetry energy of nuclear matter at intermediate 
temperatures}\label{nseIT}

The symmetry energy can be obtained from the CMD results of $E(T,\rho,x)$ using 
a method first introduced in~\cite{lopez2014} for nuclear matter around liquid 
densities, and in~\cite{lopez2017} for nuclear matter in the liquid-gas mixed 
phase.  For completeness the method for calculating the symmetry energy of 
nuclear matter at intermediate temperatures is presented in 
Appendix~\ref{nse}.\\

Figure~\ref{ESym-Dens} shows $E_{Sym}$ at low densities (in the liquid-gas 
coexistence phase). The figure compares the symmetry energy to the experimental 
points of~\cite{hagel, Kowalski,wada}.  Notice that the CMD-based symmetry 
energies approach non-zero values around the binding energy of alpha clusters in 
the limit of low density, as demanded by Natowitz and coworkers~\cite{hagel}.\\

Similarly, Figure~\ref{ESym-Dens2} presents $E_{Sym}$ expanded up to liquid 
densities to compare with a relativistic Hartree calculation (dashed line 
labeled ``$NL_2$'')  and other field theories calculations at their 
corresponding saturation densities; see~\cite{chen} for complete details. \\

\begin{figure*}[!htbp]
\centering
\subfloat[Symmetry energy al low density\label{ESym-Dens}]{
\includegraphics[width=0.45\columnwidth]
{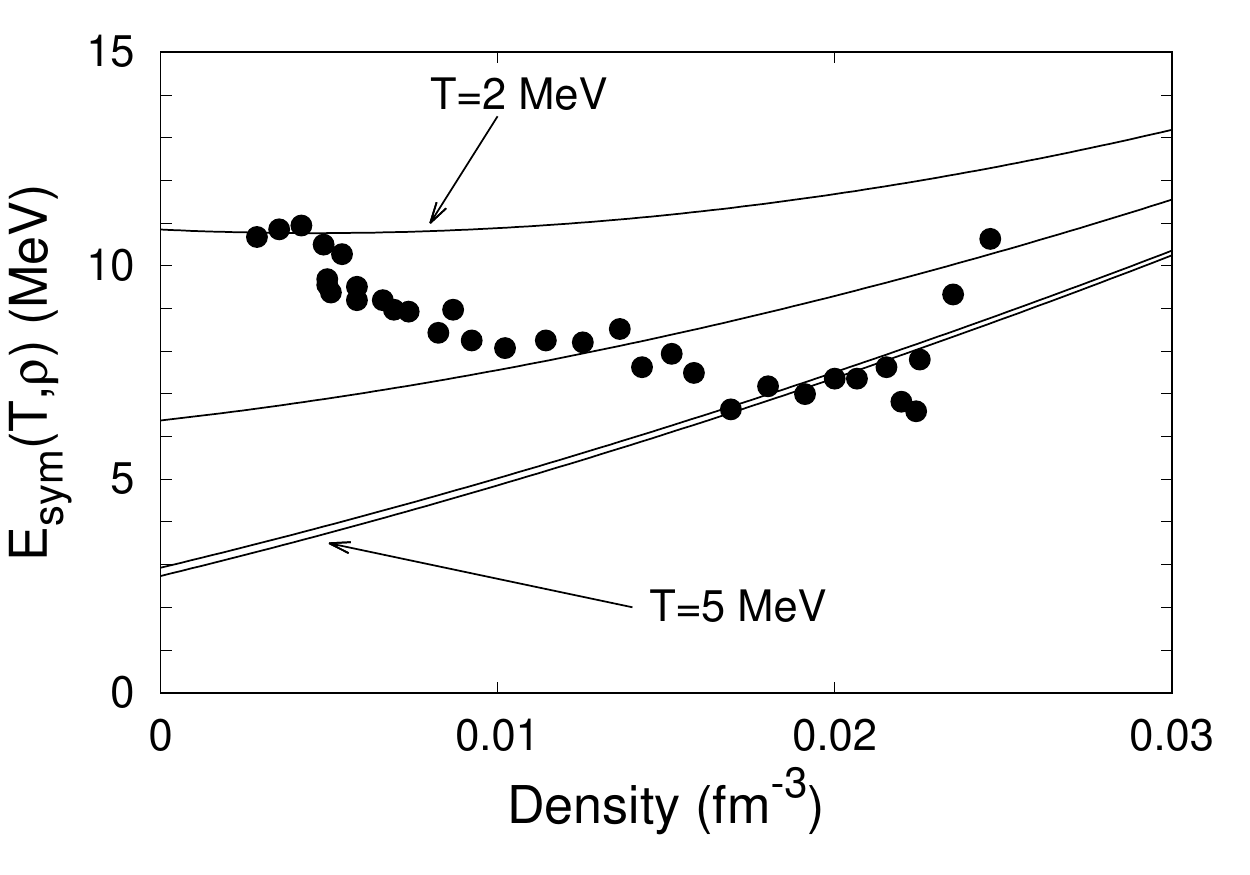}
}
\subfloat[Symmetry energy at sub-saturation densities\label{ESym-Dens2}]{
\includegraphics[width=0.45\columnwidth]
{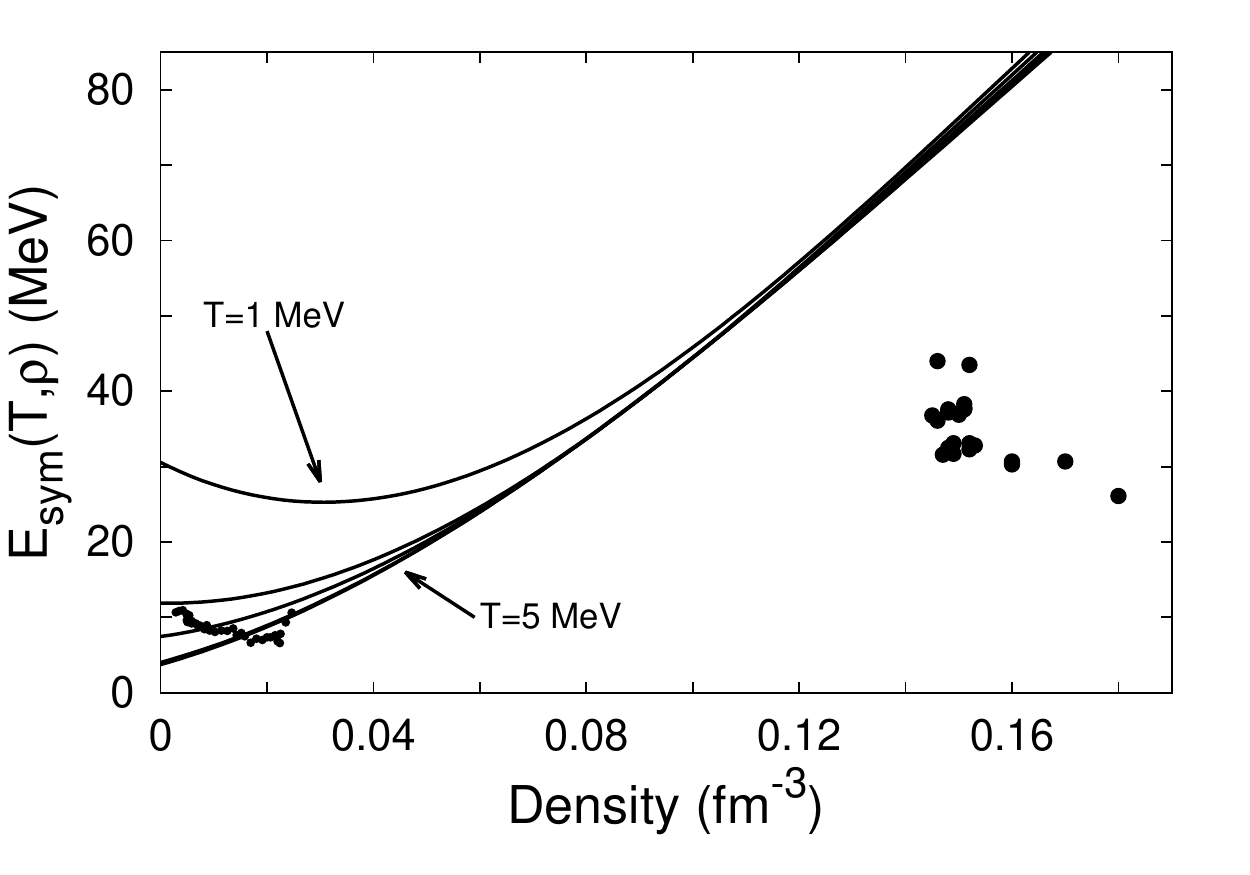}
}
\caption{Symmetry energies obtained from the CMD values of $E(T,\rho,x)$ for 
several temperatures, (a) compared to experimental results at low 
density~\cite{hagel,Kowalski,wada}, and (b) compared to experimental results 
(points on lower left corner) and theoretical predictions (points on right 
side)~\cite{chen}}
\end{figure*}



It should be pointed out that, as discussed in~\cite{lopez2014}), the numerical 
procedure used to obtain $E_{Sym}$ (c.f. Appendix~\ref{nse}) has enough 
flexibility to make the CMD calculation fully agree with the calculations shown 
in Figure~\ref{ESym-Dens2}.  Likewise, some characteristics, such as the 
temperature dependence of $E_{sym}(T,\rho)$ are in agreement with the variations 
of liquid-drop terms and nuclear surface tension in the range of $T\lesssim5$ 
MeV~\cite{randrup} but in disagreement with trends obtained by other 
theories~\cite{horo-s,xu}. Although the CMD symmetry energy is reminiscent of 
relativistic and non-relativistic Hartree calculations~\cite{chen}, there are 
notable differences mostly due to the distinct assumptions of the models, 
see~\cite{lopez2014} for a more complete discussion. \\


\subsection{Nuclear matter at low temperatures}\label{nm-LT}

At low temperatures, $T<1$ MeV, nuclear matter is known to produce crystal-like 
structures around saturation densities, and pasta-like structures at 
sub-saturation densities. In this section we investigate such structures and 
their properties, namely their morphology, phase changes and their phase 
transitions. These studies require specialized tools such as the caloric curve, 
Lindemann coefficient, radial distribution function, Kolmogorov statistic, and 
Minkowski functionals. The calculation of the symmetry energy also needs a 
special procedure based on the one used at intermediate energies (c.f. 
Appendix~\ref{nse}). 

\subsubsection*{Nuclear matter at low temperatures around saturation 
density}\label{nmltsd}

\begin{figure}
\begin{center}
   \includegraphics[width=0.5\columnwidth]{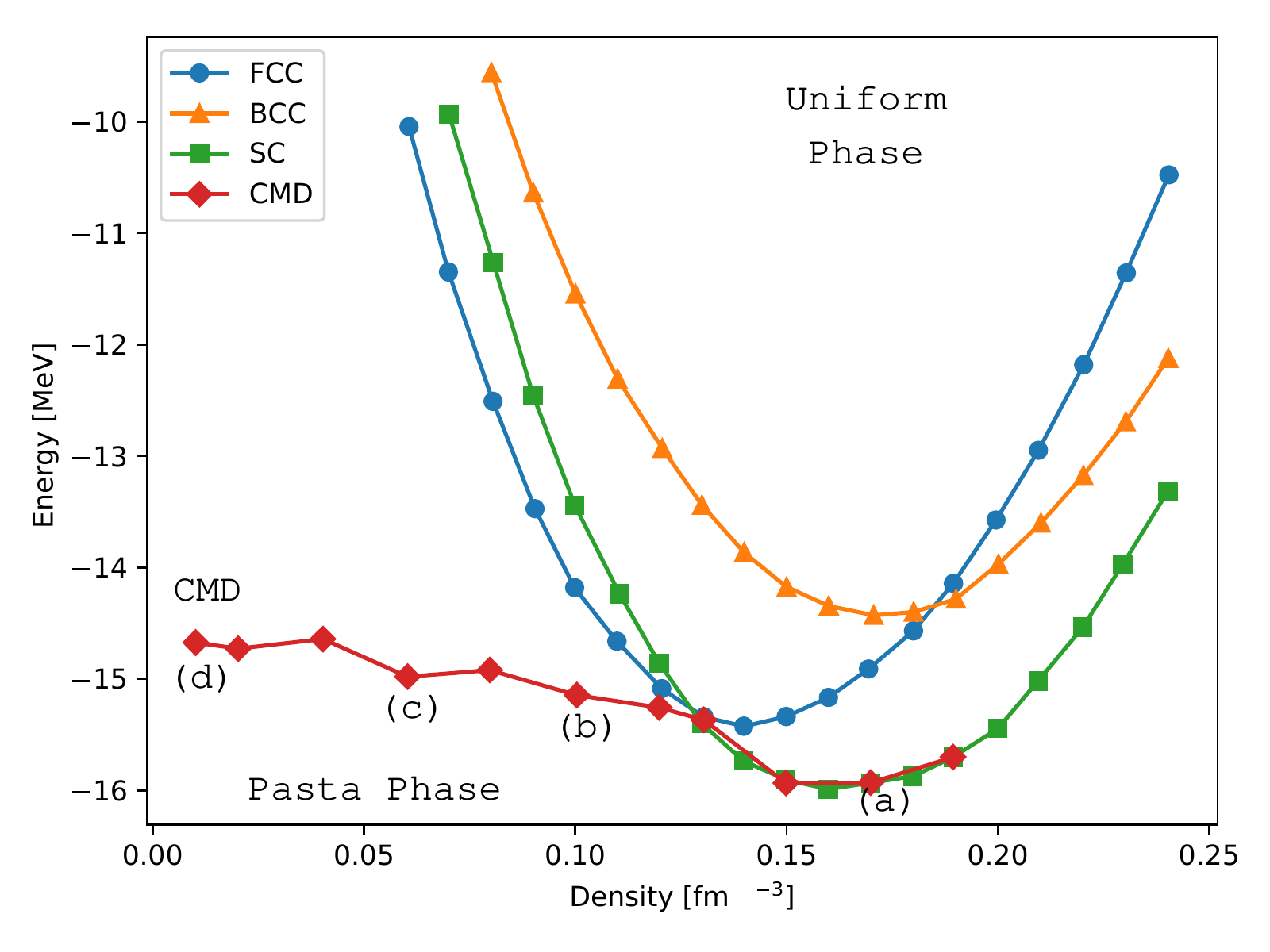}
\caption{Energy per nucleon for isospin symmetric systems in a simple cubic 
(SC), body centered cubic (BCC) and diamond (FCC) crystal lattices. Also plotted 
are the molecular dynamics results at $T = 0.001$ MeV, notice that around 
saturation density the CMD results agree with those of the simple cubic lattice. 
The structures from Figure~\ref{crystals2} that correspond to the CMD points at 
0.01, 0.06, 0.13 and 0.16 fm$^{-3}$ are (a), (b), (c) and (d), 
respectively.}\label{crystals}
\end{center}
\end{figure}

To study the uniform phase at zero temperature, a simple-cubic crystalline 
structure at a given density was constructed and the energy per nucleon 
calculated by direct summation between all nucleons. The dependence of the 
binding energy of the uniform phase on the density is explored by scaling the 
lattice parameter. As shown in Figure~\ref{crystals}, this yields the 
characteristic $\cup$-shaped curve, with a minimum at the saturation nuclear 
density. \\

For comparison, and given that nuclear matter is composed of neutrons and 
protons, systems with crystal geometries similar to those formed by binary 
alloys were constructed. Figure~\ref{crystals} shows a simple cubic lattice in 
which every first neighbor of a proton is a neutron and vice versa (SC), a BCC 
lattice, and a diamond lattice with nucleons arranged so that every first 
neighbor of a protons is a neutron and viceversa (FCC). It is clear that the 
simple cubic is the one with an energy closer to the nuclear binding energy. 
See~\cite{2013} for more details.\\

Since the previous scaling cannot yield non-homogeneous systems, i.e. a phase 
transition, CMD was used with systems at densities around and below saturation. 
Starting from a random positioning of nucleons in a cubic cell under periodic 
boundary conditions, nucleons were endowed with velocities sampled from a 
Maxwell-Boltzmann velocity distribution corresponding to a given initial 
temperature. The systems then were evolved until equilibrium is achieved at a 
high temperature ($T \gtrsim 2$ MeV) and then brought down to the final desired 
temperature of $T = 0.001$ MeV using the Andersen thermostat in small 
temperature steps. After reaching equilibrium, the analysis tools described in 
Appendix~\ref{tools} were used to characterize and visualize the produced 
configurations.\\

\begin{figure}
\begin{center}
   \includegraphics[width=0.5\columnwidth]{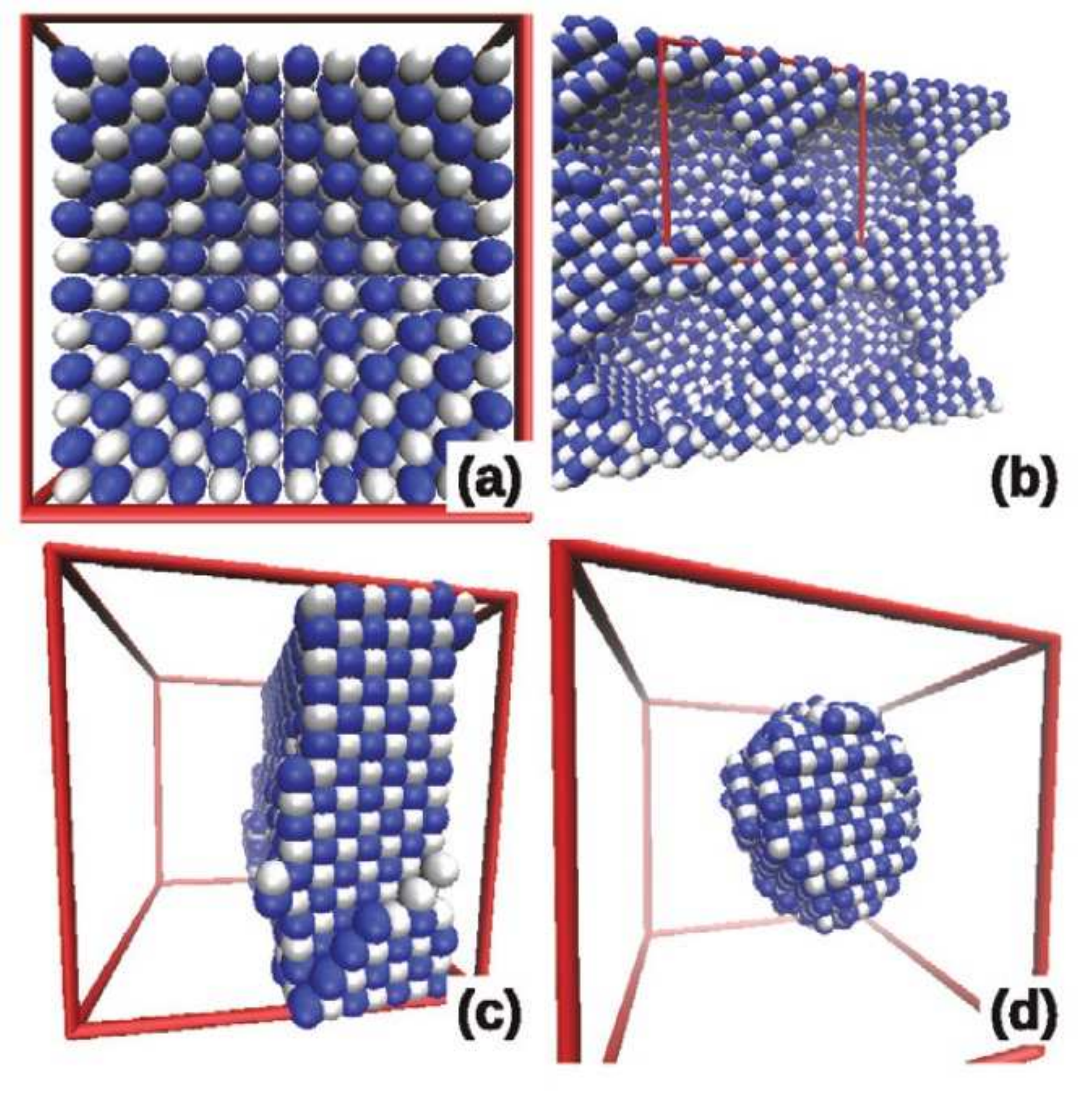}
\caption{Structures produced by CMD at $T = 0.001$ MeV and corresponding to the 
points of Figure~\ref{crystals}. Point (a) corresponds to a regular (B1) 
lattice, while the rest of the points are non-homogeneous 
structures.}\label{crystals2}
\end{center}
\end{figure}

Figure~\ref{crystals} shows the near zero temperature results for symmetric ($x 
= 0.5$) matter. As it can be seen, CMD reproduces the simple cubic (B1) lattice 
calculations, corresponding to the uniform phase, up to a density of $\rho 
\approx 0.13$ fm$^{−3}$, while for lower densities the systems goes into a 
series of non-homogeneous phases. Figure~\ref{crystals2} shows visual 
representations of the structures corresponding to the four densities labeled 
from (a) to (d) in Figure~\ref{crystals}.\\

\begin{figure*}[!htbp]
\centering
\captionsetup[subfigure]{justification=centering}
\subfloat[Energy, pressure and compressibility\label{crystals3}]{
\includegraphics[width=0.45\columnwidth]
{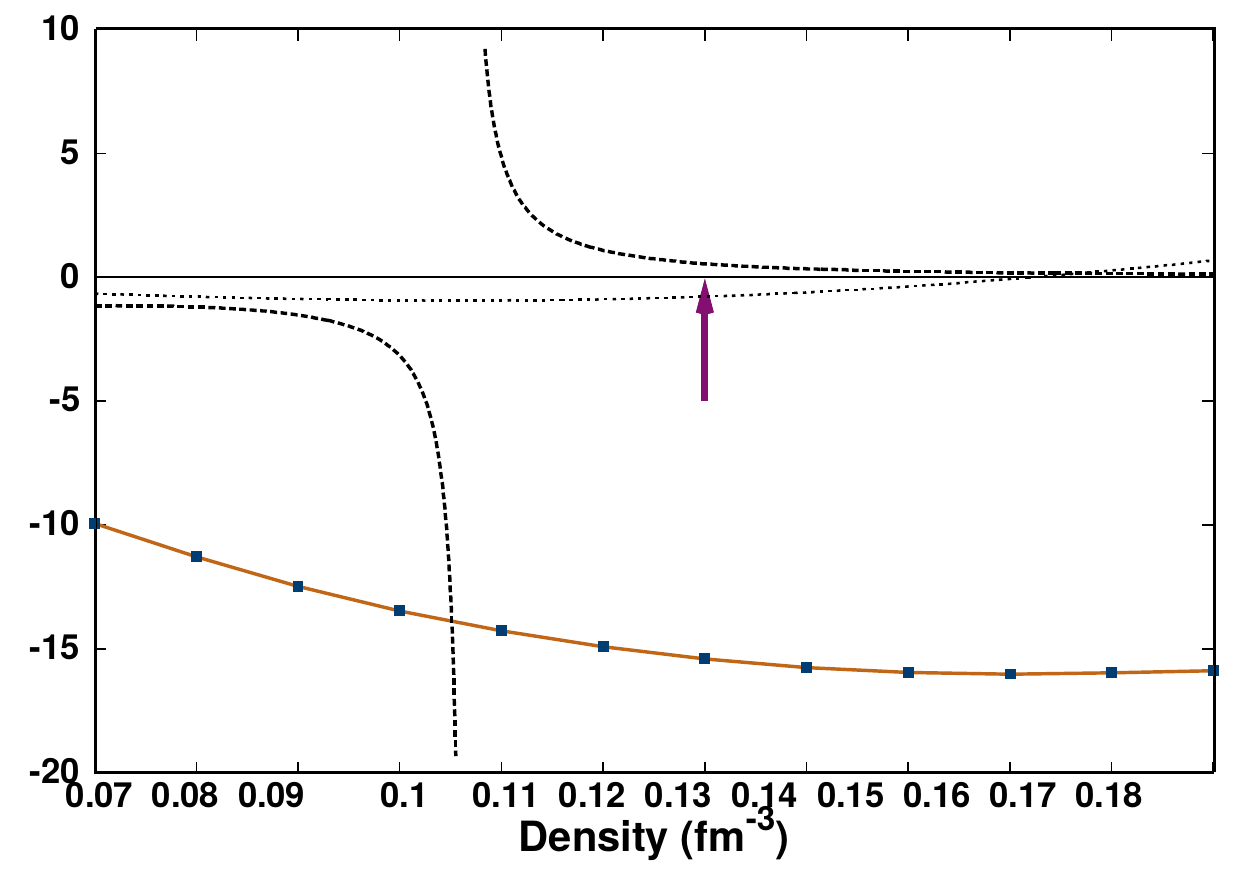}
} 
\subfloat[Binding energy\label{crystals4}]{
\includegraphics[width=0.45\columnwidth]
{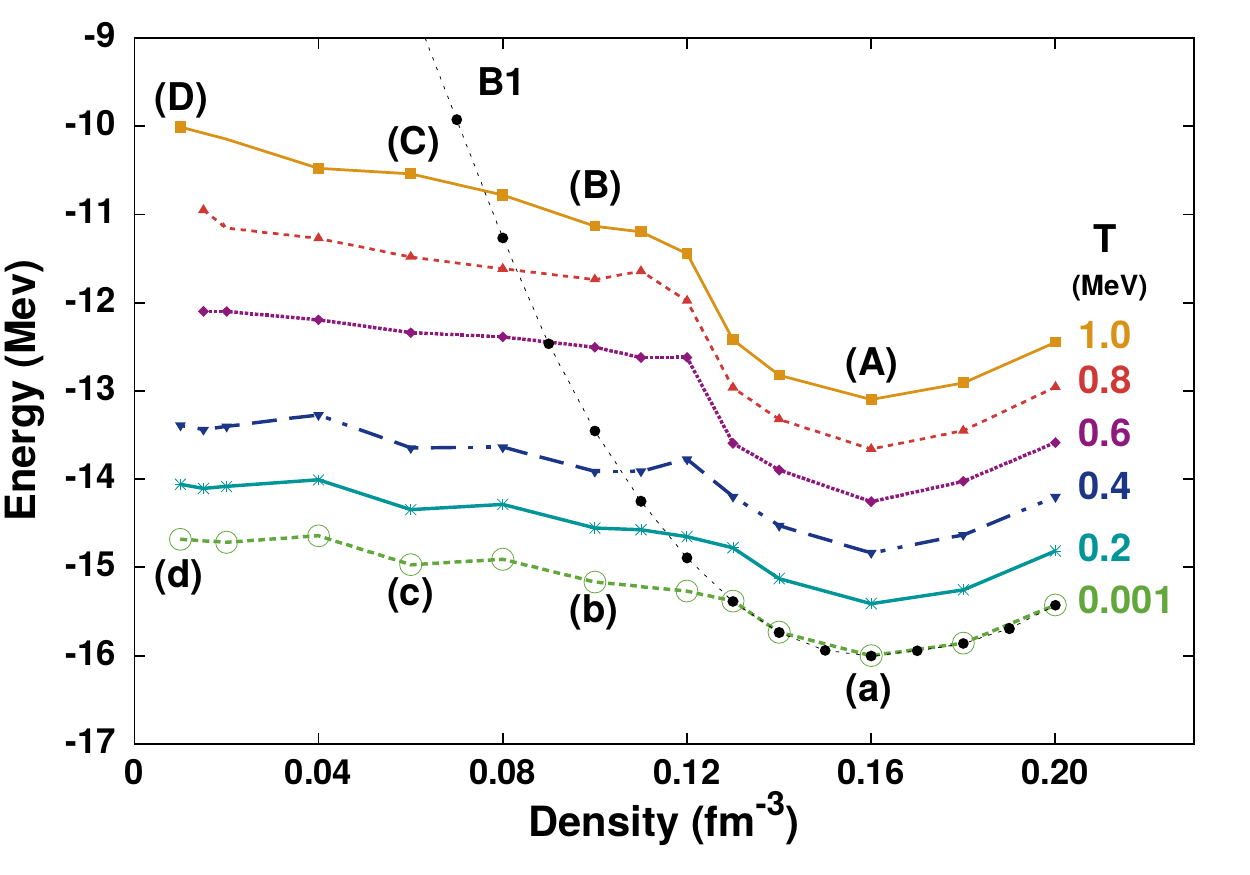}
} 
\caption{(a) Energy (continuous lines with squares), pressure (dashed line) and 
compressibility (dotted line) for the B1 lattice of the Medium potential; the 
arrow points at the density at which the simulated systems departs from 
homogeneity. (b) Binding energy per nucleon for systems obtained with the 
Pandharipande medium potential at the listed temperatures.}
\end{figure*}


The compressibility can be obtained through $K = {9\rho_0^2 
[d^2E/d\rho^2]}_{\rho_0}$, where $E(\rho,T,x)$ is a simple degree-two polynomial 
fit of the B1 curve of Figure~\ref{crystals} around the saturation density.  
Figure~\ref{crystals3} shows the energy, pressure and compressibility for the 
homogeneous B1 lattice. The bulk modulus for the medium Pandharipande potential 
is found to be 283 MeV, comparable to the value of 250 MeV quoted by its 
creators~\cite{pandha}, who used different fitting polynomial.  The pasta-like 
structures are found in a mechanically unstable (negative pressure) density 
region, but well above the divergence in compressibility that would signal the 
thermodynamic instability. \\


\begin{figure}
\begin{center}
   \includegraphics[width=0.5\columnwidth]{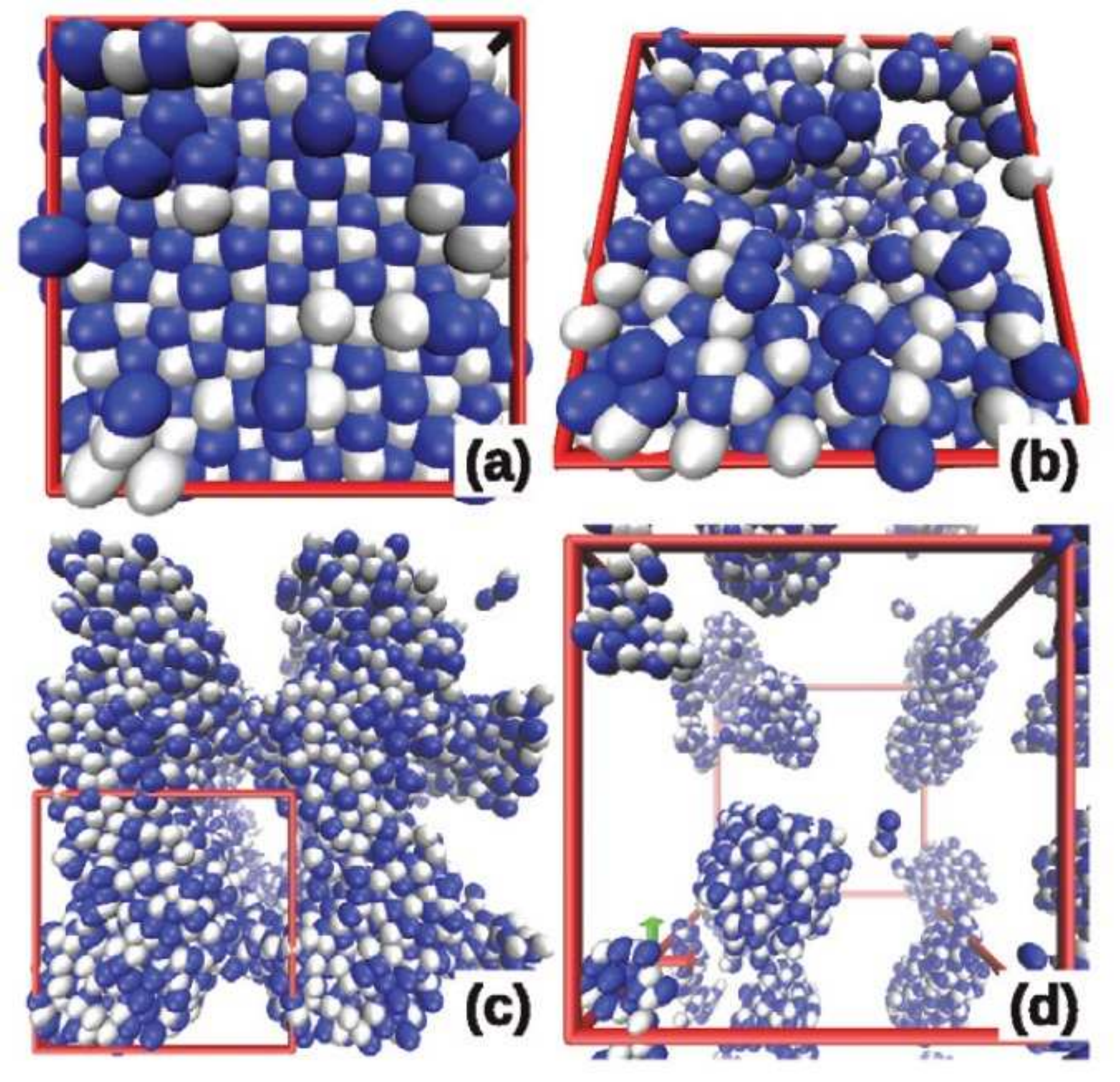}
\caption{Structures corresponding to the labeled points of 
Figure~\ref{crystals4} obtained with the Pandharipande medium potential at $T = 
1.0$ MeV.}\label{crystals5}
\end{center}
\end{figure}

At slightly higher temperatures, between $0.001 MeV \le T \le 1.0$ MeV, the NM 
systems follow the $\cup$ shape characteristic of the uniform $T = 0$ 
crystalline phase at densities $\rho \gtrsim 0.13$ fm$^{−3}$. 
Figure~\ref{crystals4} shows the same type of results as Figure~\ref{crystals}, 
again, at a density $\rho \approx 0.13$ fm$^{−3}$ the systems move away from the 
uniform phase forming non-homogeneous stable arrangements. 
Figure~\ref{crystals5} shows some of those structures at $T = 1.0$ MeV for the 
four densities labeled from (a) to (d) in Figure~\ref{crystals4}.\\

These non-homogeneous structures can be characterized using the mean curvature 
and Euler characteristic. As explained in detail in Appendix~\ref{tools}, the 
Euler characteristic is a topological invariant that describes a structure 
regardless of the way it is bent. Different structures have distinct values of 
these variables and, in general, follow the pattern outlined in 
Table~\ref{tab1}.  As a reference, the perfect crystals formed at $T = 0$ and 
$\rho \approx 0.13$ fm$^{−3}$  (v.g. point (A) in Figures~\ref{crystals} 
and~\ref{crystals4}) are formally uniform and infinite because of the periodic 
boundary conditions imposed, hence they have no surfaces and null Euler 
characteristic. \\

\begin{figure}
\begin{center}
   \includegraphics[width=0.5\columnwidth]{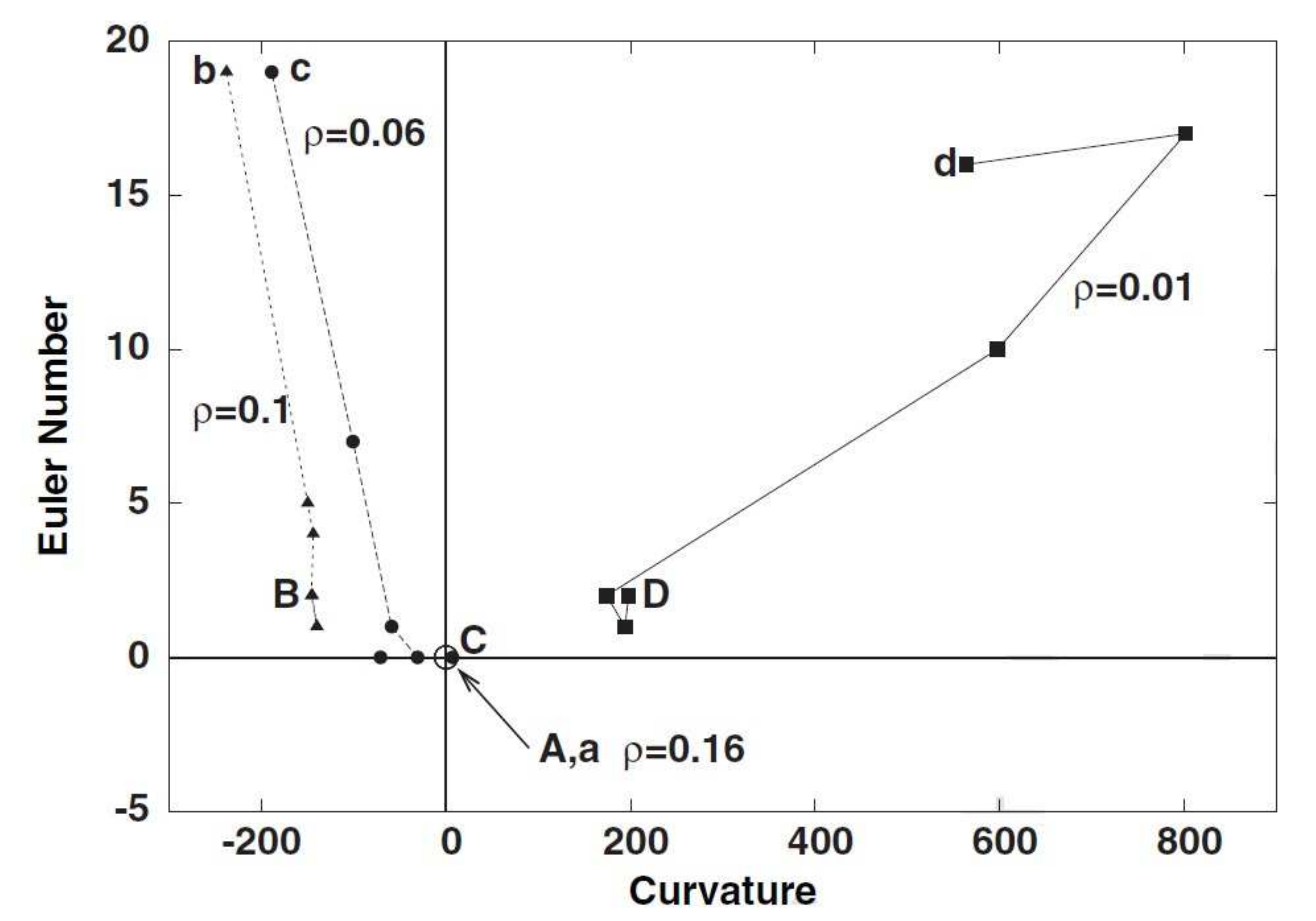}
\caption{Curvature - Euler coordinates of the structures from 
Figure~\ref{crystals4}. The lines connect points with the same densities but 
temperatures varying from $T = 0.001$ MeV to $1.0$ MeV.}\label{crystals6}
\end{center}
\end{figure}

Figure~\ref{crystals6} shows the effect of temperature on the curvature-Euler 
coordinates of the labeled structures of Figure~\ref{crystals4}.  Points on the 
lines have the same densities but their temperatures vary from $T = 0.001, 0.1, 
0.5, 0.6, 0.8$ and $1.0$ MeV. The labels corresponding to those of 
Figure~\ref{crystals4}: cases ``A-a'' at normal density ($\rho_0$) correspond to 
uniform crystalline structures so all have zero curvature and Euler number at 
the temperatures studied; cases ``B-b'' of density $0.1$ fm$^{−3}$ the almost 
spherical bubbles at $T = 0.001$ MeV become distorted at higher temperatures; 
cases is ``C-c'' at $\rho = 0.06$ fm$^{−3}$, go from being a perfect ``lasagna'' 
to a ``jungle-gym'', a complex of lasagna joined by columns (and similar to the 
Schwarz’s Primitive surface of differential geometry); and cases ``D-d'', on the 
other side, go from spherical ``gnocchi'' to deformed droplets. \\

Complementary studies for higher compressibility potentials, the stiff 
Pandharipande potential and the effect of the periodic boundary conditions can 
be found in reference~\cite{2013}.\\

\subsubsection*{Properties of NM pasta-like structures}\label{propNMP}

A big advantage of using CMD to study the pasta-like structures, is that the 
nucleon dynamics can drive the system into states with local free energy minima, 
which abound in complex energy landscapes, and find nontraditional (amorphous, 
sponge-like) structures.  Such states cannot be found by static models due to 
the energy barriers which usually surround local minima~\cite{dorso2014}.\\

To characterize the pasta, its phase changes and to calculate the symmetry 
energy we use the cluster-recognition algorithms, the radial distribution 
function, the Lindemann coefficient, Kolmogorov statistics, Minkowski 
functionals and a numerical method to estimate $E_{sym}$. These techniques are 
reviewed in the Appendices~\ref{nse-LT} and~\ref{tools}.\\

\paragraph{Phases}\label{phases}

The temperature-excitation energy correlation known as the caloric curve, has 
been used in nuclear physics to detect first-order liquid-gas phase 
transitions~\cite{raciti}. The caloric curve simply relates the temperature of 
the system to the energy input needed to reach such temperature. Whenever energy 
is used for something other than heating up the system, like breaking bonds 
while melting a crystal, energy will flow in but the temperature will not 
increase, such discontinuities in the slope of the caloric curve will signal 
phase changes.  In this section the caloric curve is computed for systems with 
three different proton ratios to identify phase changes.  See 
Ref.~\cite{dorso2014} for complete details. \\

\begin{figure*}[!htbp]
\centering
\captionsetup[subfigure]{justification=centering}
\subfloat[Caloric curve\label{cc}]{
\includegraphics[width=0.46\columnwidth]
{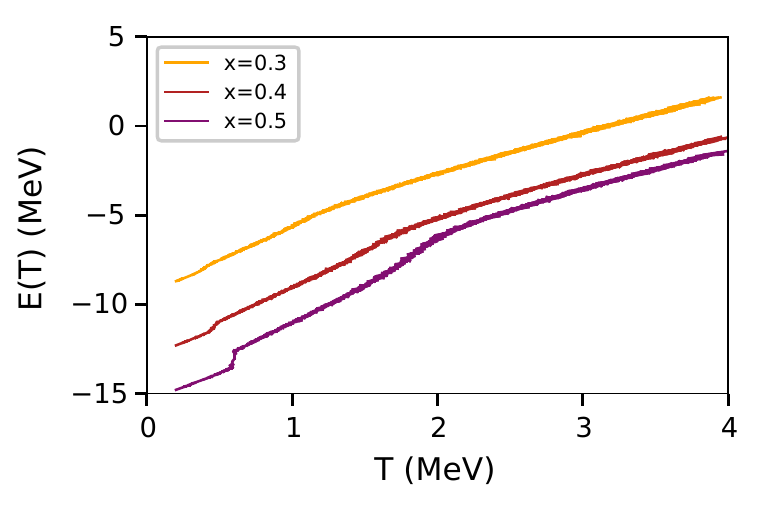}
} 
\subfloat[Lindemann coefficient and caloric curve\label{lin}]{
\includegraphics[width=0.43\columnwidth]
{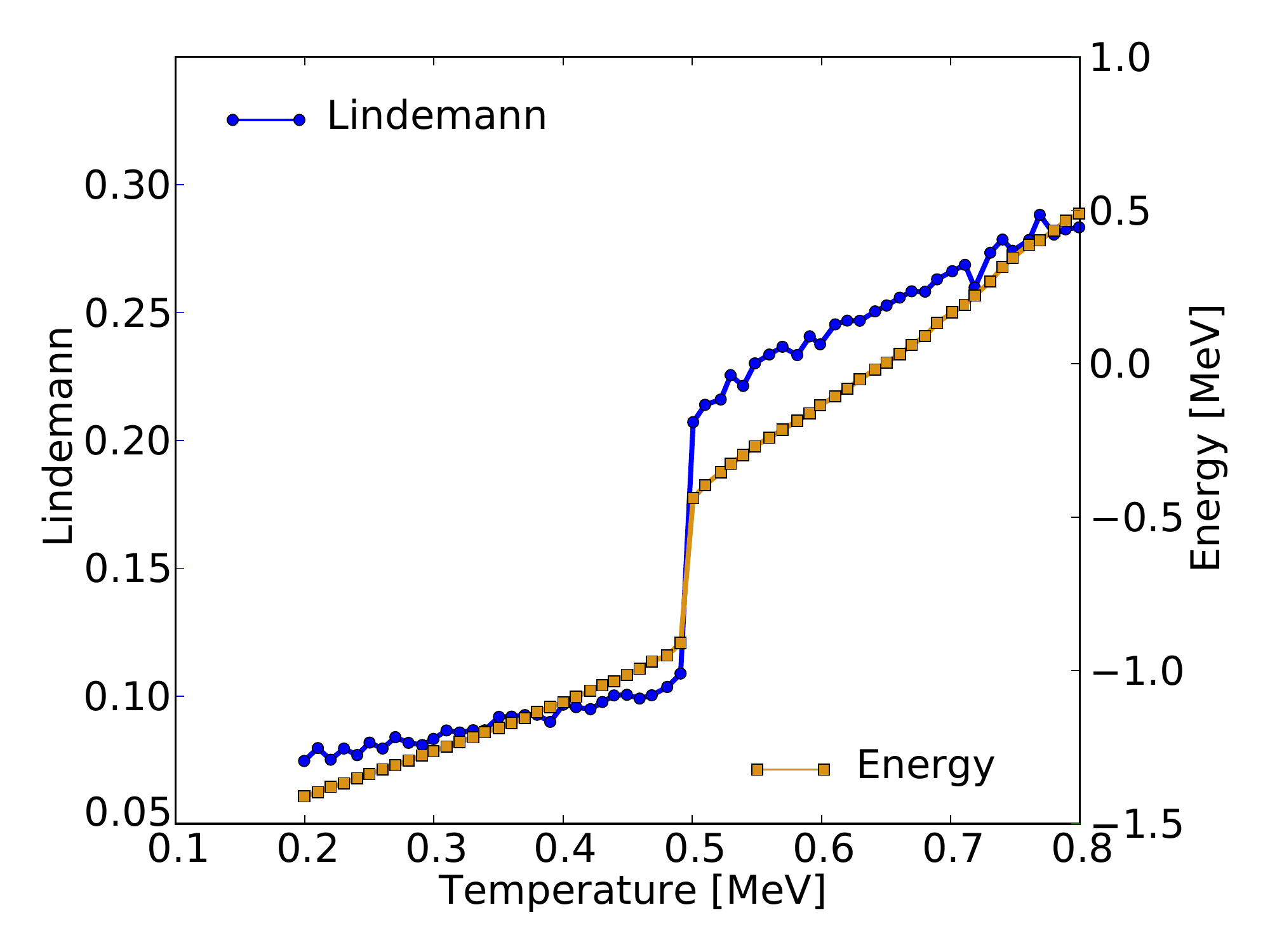}
} 
\caption{(a) The caloric curve for nuclear matter at 
$\rho=0.05\,\mathrm{fm}^{-3}$ and $x = 0.3$, $0.4$, $0.5$, with a total number 
of nucleons of 6000. (b) Lindemann coefficient superimposed on the caloric curve 
for $\rho = 0.05\,$fm$^{-3}$ for a symmetric system. The sudden change in going 
from $T<$ 0.5 MeV to $T>$ 0.5 MeV signals a phase change from nucleons in a 
solid state inside the pasta to nucleons in a liquid phase inside the pasta.}
\end{figure*}

\begin{figure*}[!htbp]
\centering
\captionsetup[subfigure]{justification=centering}
\subfloat[$x=0.5$]{
\includegraphics[width=0.5\columnwidth]
{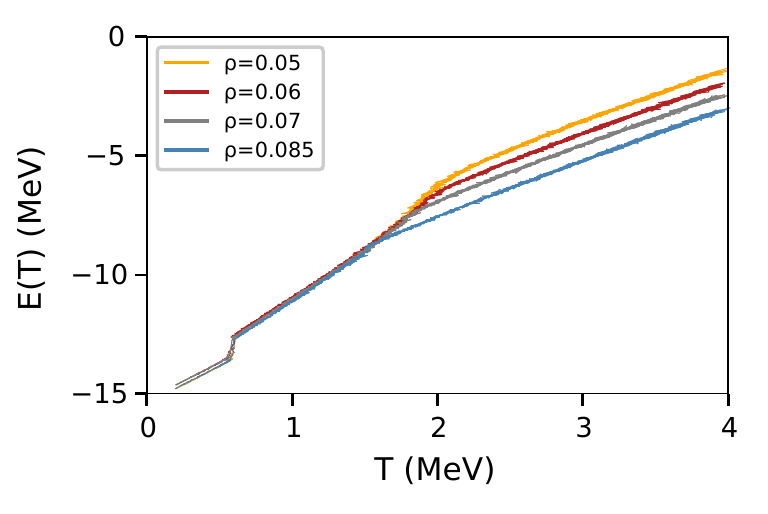}
} 
\subfloat[$x=0.5$]{
\includegraphics[width=0.5\columnwidth]
{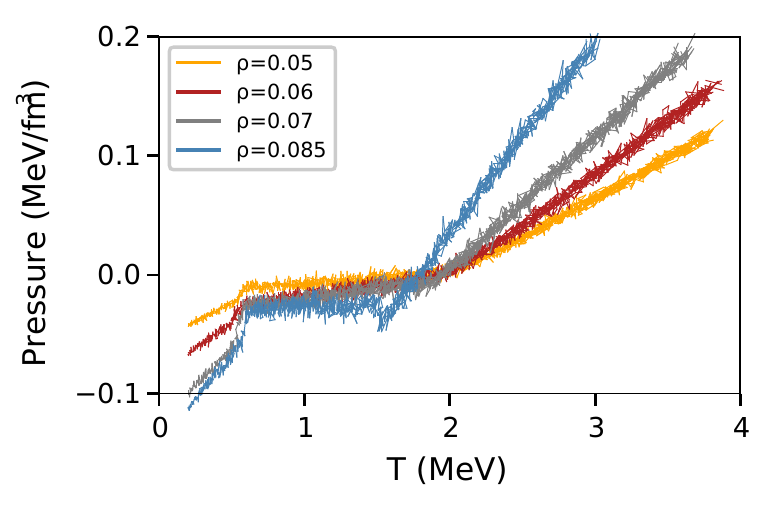}
} 
\\
\subfloat[$x=0.4$]{
\includegraphics[width=0.5\columnwidth]
{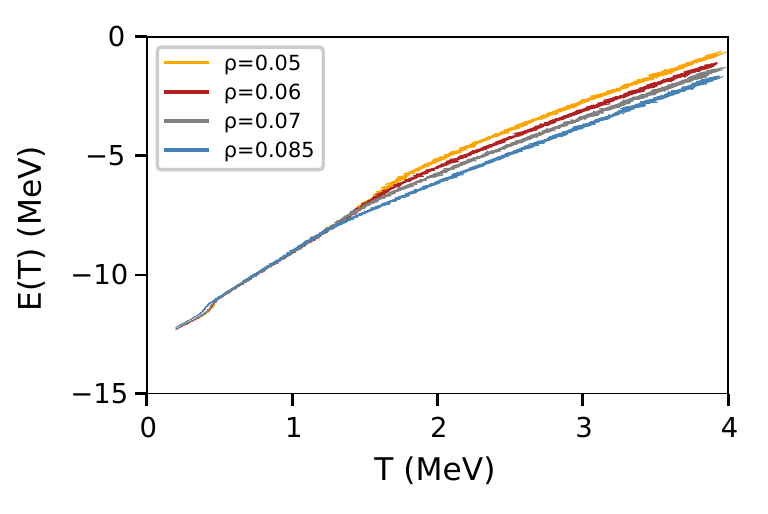}
} 
\subfloat[$x=0.4$]{
\includegraphics[width=0.5\columnwidth]
{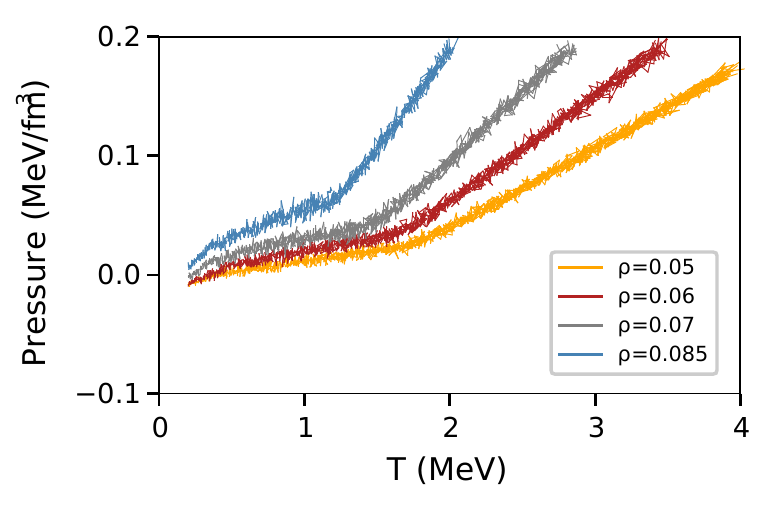}
}
\\
\subfloat[$x=0.3$]{
\includegraphics[width=0.5\columnwidth]
{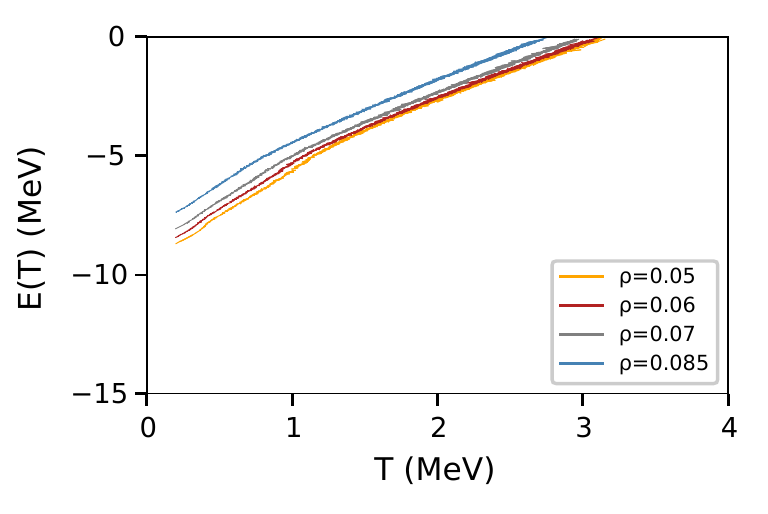}
} 
\subfloat[$x=0.3$]{
\includegraphics[width=0.5\columnwidth]
{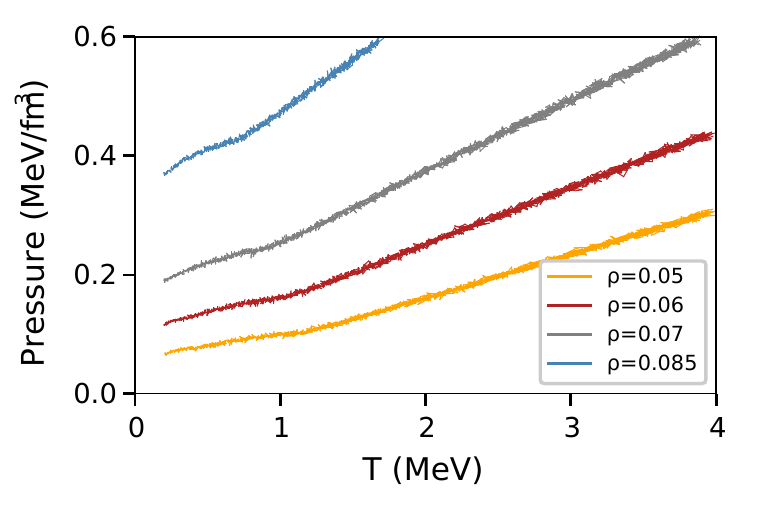}
}
\caption{\label{fig:energy_pressure}   Energy and pressure as a function of 
temperature  for nuclear matter systems with 6000 nucleons at densities 
$\rho$=0.05, 0.06, 0.07 and 0.085$\,$fm${}^{-3}$.  Cases (a) and (b) are for 
symmetric nuclear matter ($x=0.5$) and correspond to the configurations shown in 
Fig.~\ref{fig:pasta}, (c) and (d) are for non-symmetric nuclear matter with 
$x=0.4$, and (e) and (f) for $x=0.3$.}
\end{figure*}

Figure~\ref{cc} shows the caloric curve at $\rho=0.05\,$fm$^{-3}$ for systems of 
6000 nucleons at proton ratios of $x =0.3$, 0.4 and 0.5. Two changes of slope 
are noticeable, a sharp one close to $T=0.5$ MeV, and one less conspicuous just 
before $T=2.0$ MeV; the changes are more pronounced in symmetric matter. Similar 
changes are seen for other cases on the left panels of 
Fig.~\ref{fig:energy_pressure}.\\


The significance of the slope changes of the caloric curve can be investigated 
with the Lindemann coefficient. As explained in the Appendix~\ref{lind}, the 
Lindemann coefficient provides an average estimation of displacement of the 
nucleons, and it can signal a change of mobility and, hence, a phase 
transitions. Figure~\ref{lin} shows the behavior of the Lindemann coefficient as 
a function of the temperature superimposed on the caloric curve of 
Figure~\ref{cc}. It is easy to see that at lower temperatures the nucleon 
mobility is only a fraction of its higher-temperature value; the change at 
$T=0.5$ MeV appears to correspond to a solid-liquid phase transition 
\textit{within the pasta regime}, as concluded in Ref.~\cite{dorso2014}.\\

The pressure-temperature curves can be used to construct phase diagrams. The 
right panels of Fig.~\ref{fig:energy_pressure} show the behavior of the system 
pressure as a function of the temperature. Comparing to the left panels, one can 
see that the $p-T$ curves experience slope changes at similar temperatures as 
the $E-T$ curves. Notice that the pressure of the symmetric nuclear matter 
($x=0.5$) also changes sign, while the pressure of the asymmetric nuclear matter 
($x<0.5$) remains
positive until very low temperatures. It is clear that the former enters into
the metastable regime, while the latter does not.\\

\begin{figure*}[!htbp]
\centering
\captionsetup[subfigure]{justification=centering}
\subfloat[$x=0.5$\label{rad_a}]{
\includegraphics[width=0.5\columnwidth]
{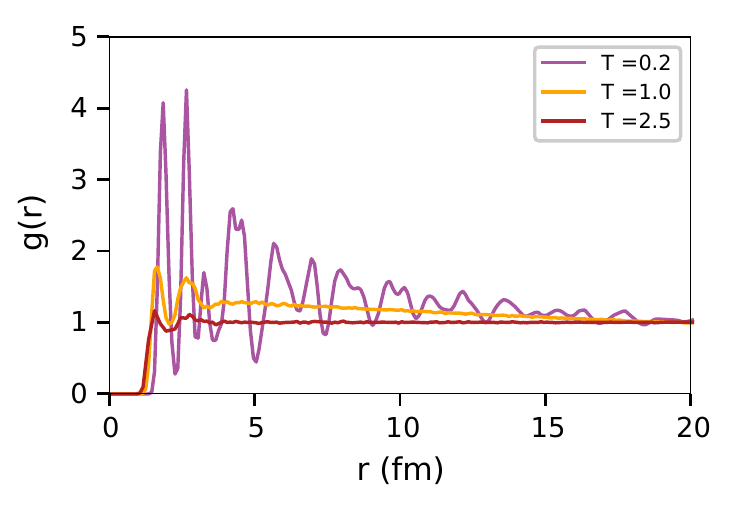}
} 
\subfloat[$x=0.4$\label{rad_b}]{
\includegraphics[width=0.5\columnwidth]
{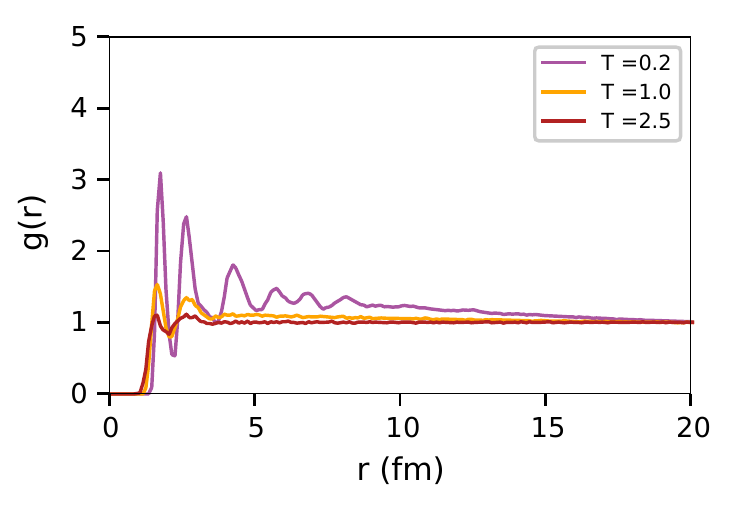}
} 
\caption{\label{rad}Radial distribution function $g(\mathbf{r})$ for the case of 
6000 nucleons of symmetric (a) and  non-symmetric (b) nuclear matter, at $\rho= 
0.085\,$fm$^{-3}$ and $T = 0.2\,$MeV, $1.0\,$MeV and 2.5$\,$MeV. The binning is 
$0.1\,$fm width in a simulation cell of 41.2$\,$fm  of width.}
\end{figure*}

The microscopic structure of the different phases can be explored through the 
radial distribution function $g(\mathbf{r})$, introduced in 
Appendix~\ref{g-for-pasta}. To study the caloric curve results, $g(\mathbf{r})$ 
was calculated in the three temperature regions found, namely, in $T<0.5$ MeV, 
$0.5$ MeV $<T<2.0$ MeV, and $T>2.0$ MeV. Figure~\ref{rad} shows $g(\mathbf{r})$ 
for nuclear matter at $\rho=0.085\,$fm$^{-3}$ and at temperatures in the three 
different regions. Panel (a) show the results for the symmetric ($x=0.5$) case, 
and (b) 
the non-symmetric ($x=0.4$) case.\\

The peaks of $g(\mathbf{r})$ in Figure~\ref{rad} indicate the average position 
of the nearest neighbors. At $T=0.2$ MeV the peaks of panel (a) show large 
correlations at 1.85 fm, 2.65 fm and 3.25 fm; such uniformity of nearest- and 
second-neighbors is characteristic of a solid state. Furthermore, since the 
peaks occur at $a,\sqrt{2}a,\sqrt{3}a$ (where $a$ is the nearest-neighbor 
distance), it indicates that the solid is a simple cubic arrangement.  The 
gradual loss of correlation at $T=1$ MeV and 2 MeV indicate that the phase is 
losing correlations, as expected in melting. Similar behaviors are observed in 
panel (b) for the case of $x = 0.4$. Notice that both the $T=1$ MeV and 2 MeV 
cases show a rapid drop of correlations at distances larger that, say, 4 fm; 
this is expected as such distances lie outside the pasta structure. The results 
from the caloric curves,  $g(\mathbf{r})$, Lindemann coefficient, and the $p-T$ 
curves, together indicate that the \textit{nucleons inside the pasta structures 
exist in different phases.}\\

\begin{figure}  
\begin{center}
   \includegraphics[width=0.5\columnwidth]{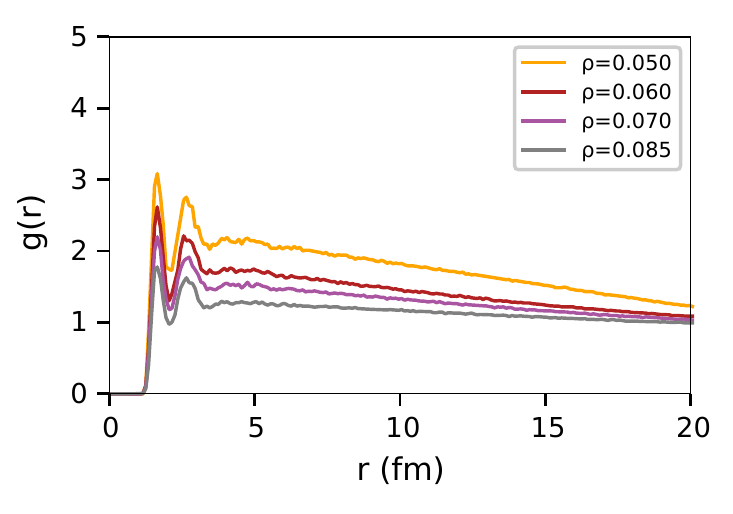}
\caption{Radial distribution function $g(r)$ for the case of 6000 nucleons at 
$T=1.0\,$MeV and
$x=0.5$. The densities are $\rho= 0.05$ (top curve), $0.06$, $0.07$ and 
$0.085\,$fm$^{-3}$ (bottom curve). The binning is $0.1\,$fm width, while the 
simulation cell is approximately 40$\,$fm width.
}\label{rad2}
\end{center}
\end{figure}

\begin{figure*}[!htbp]
\centering
\captionsetup[subfigure]{justification=centering}
\subfloat[\label{fig:rho085_all}]{
\includegraphics[width=0.34\columnwidth]
{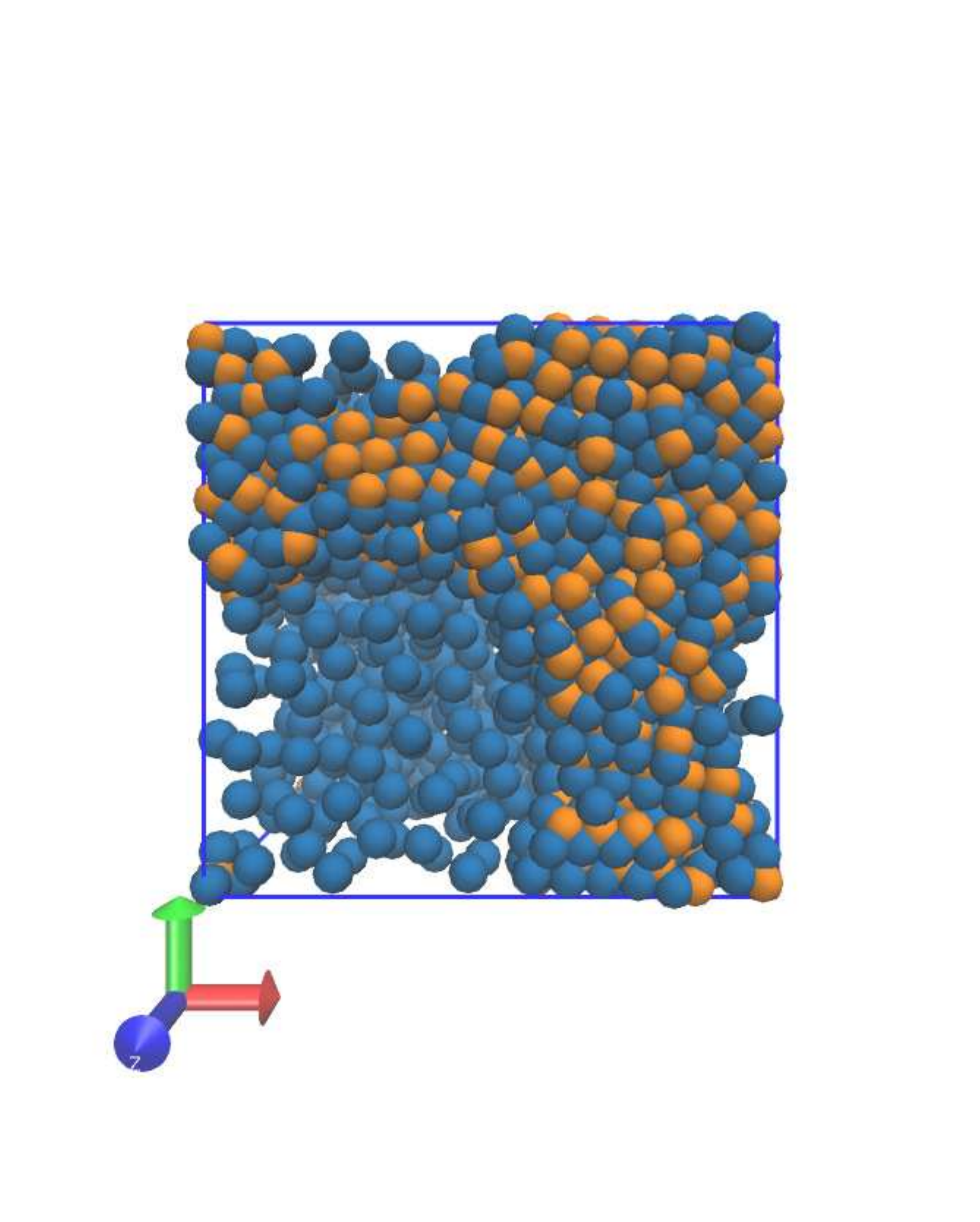}
}
\subfloat[\label{fig:rho085_protons}]{
\includegraphics[width=0.34\columnwidth]
{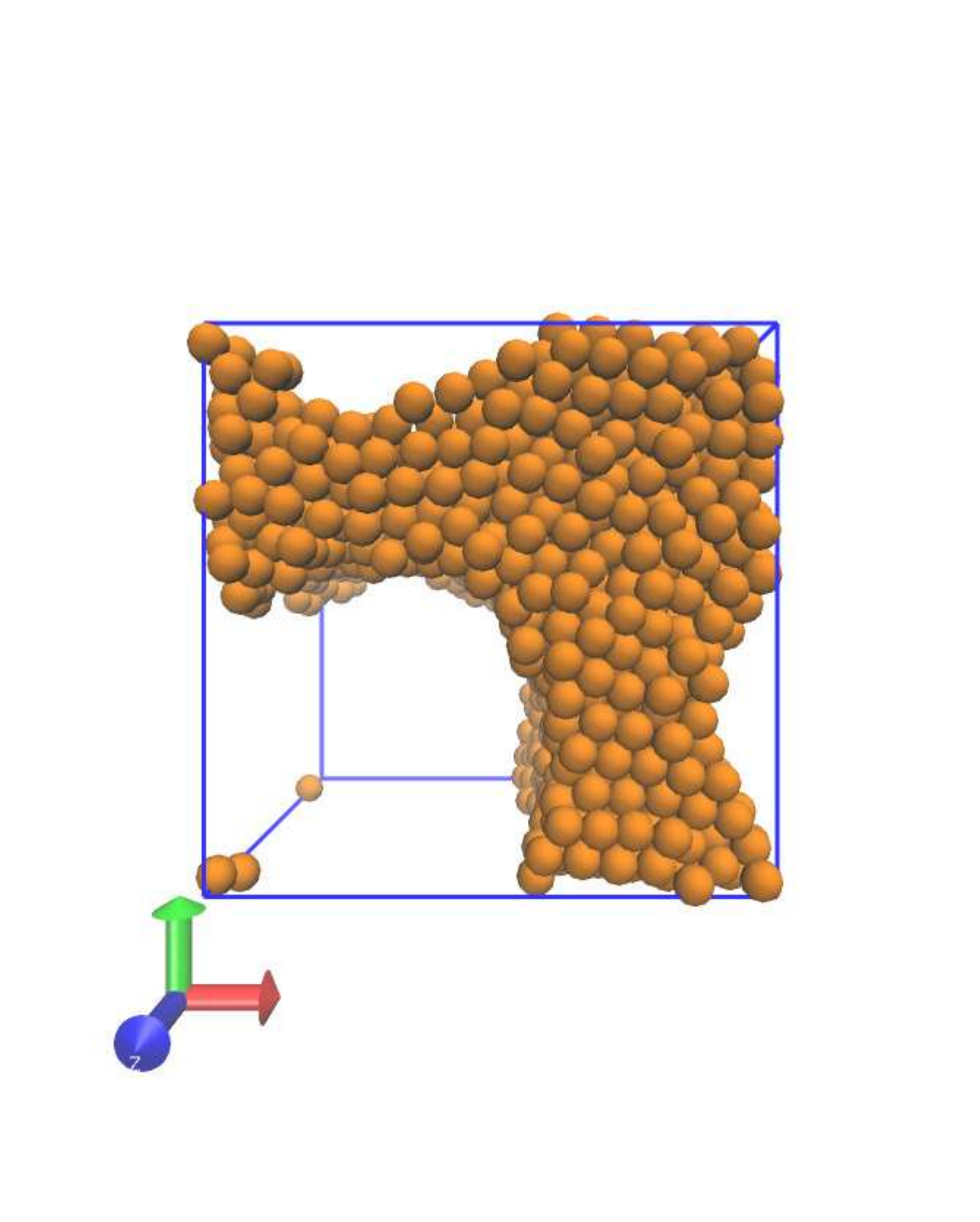}
}
\subfloat[\label{fig:rho085_neutrons}]{
\includegraphics[width=0.34\columnwidth]
{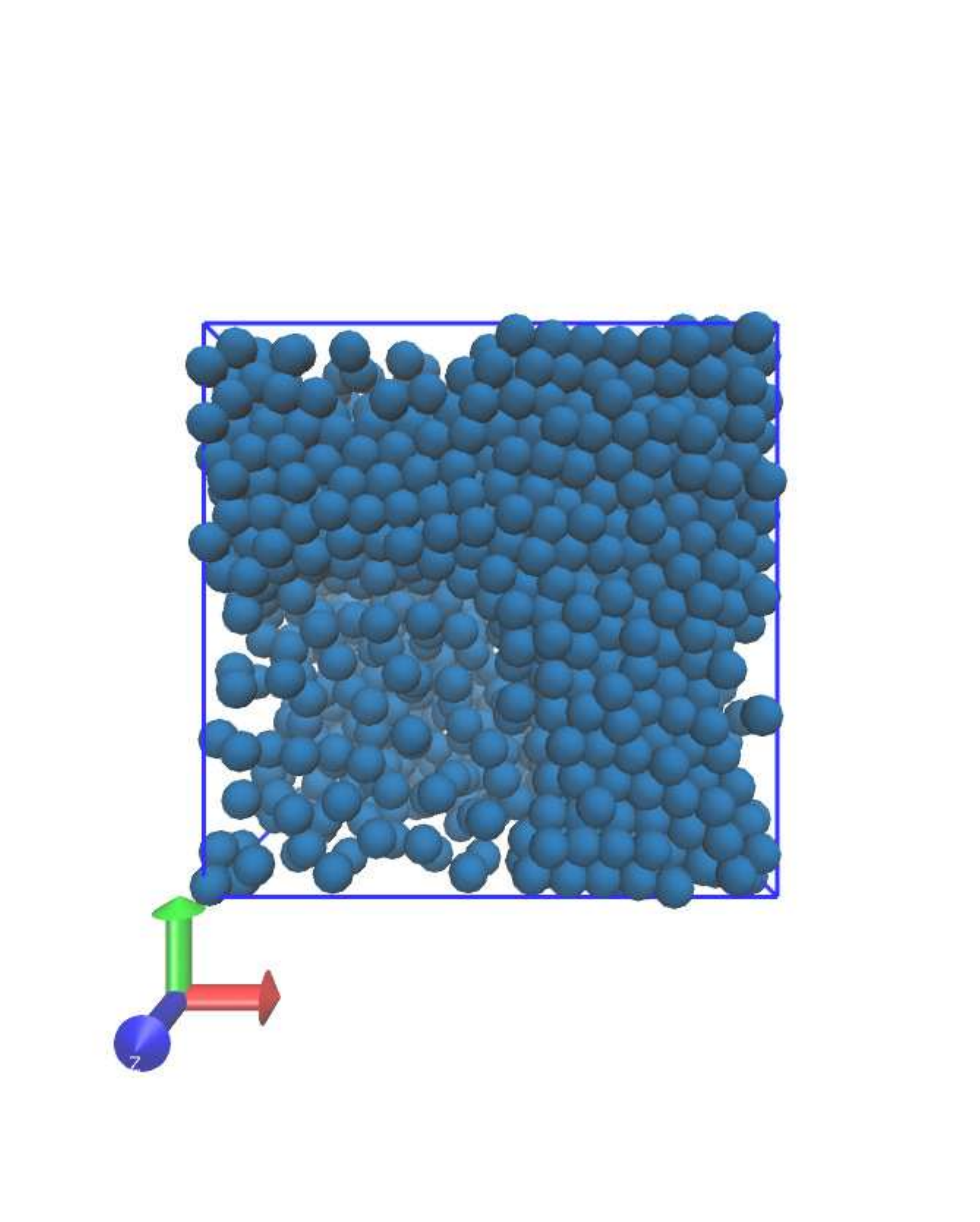}
}
\caption{\label{fig:pasta_decomposed}   Pasta structures for nuclear matter 
systems with 6000 nucleons, $x=0.4$ at $T=0.2\,$MeV and density 
$\rho$=0.085$\,$fm${}^{-3}$. (a) All nucleons (protons in light color and 
neutrons in dark), (b) protons only, and (c) neutrons only.}
\end{figure*}

\begin{figure*}[!htbp]
\centering
\captionsetup[subfigure]{justification=centering}
\subfloat[\label{fig:rho085_all_2}]{
\includegraphics[width=0.34\columnwidth]
{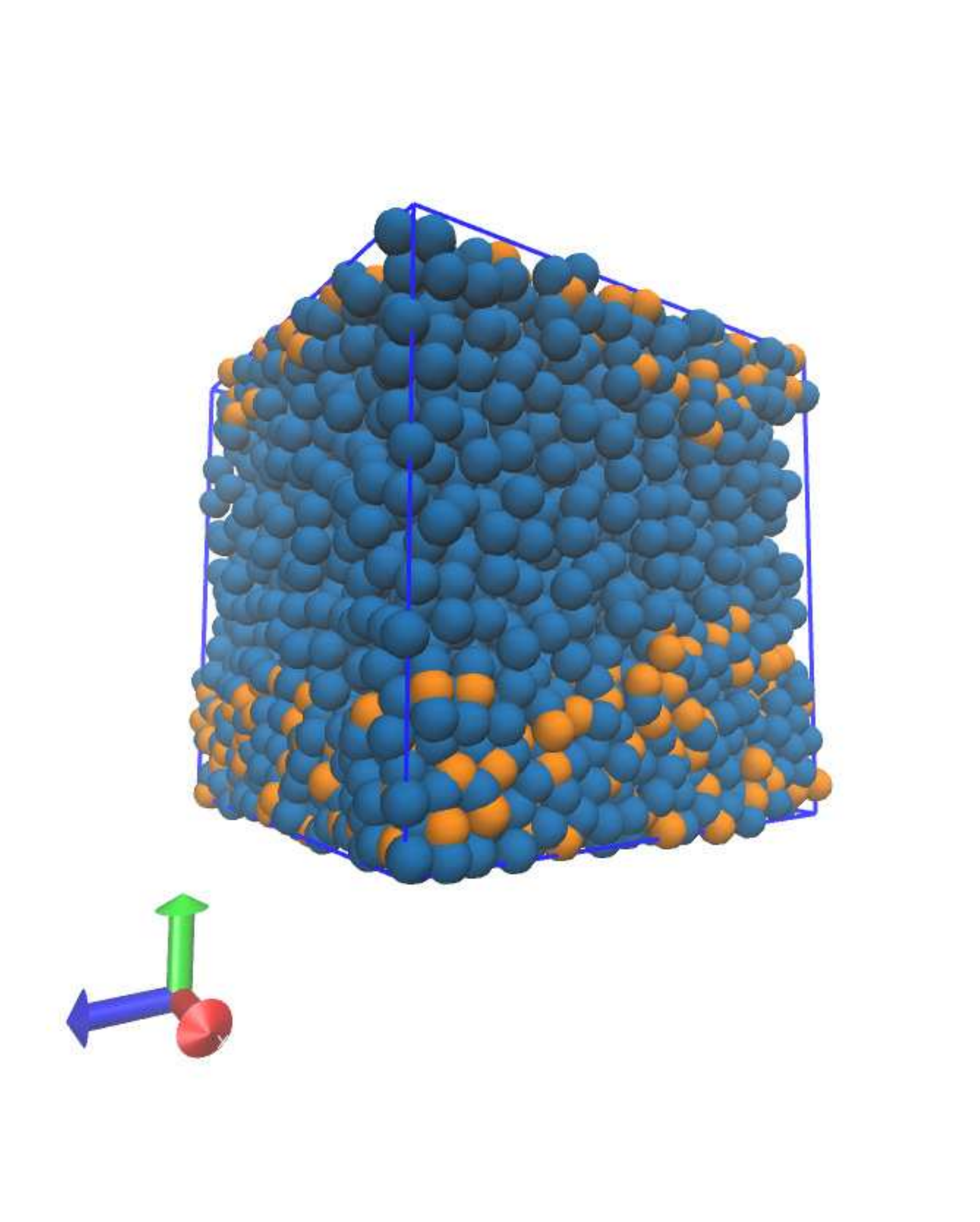}
}
\subfloat[\label{fig:rho085_protons_2}]{
\includegraphics[width=0.34\columnwidth]
{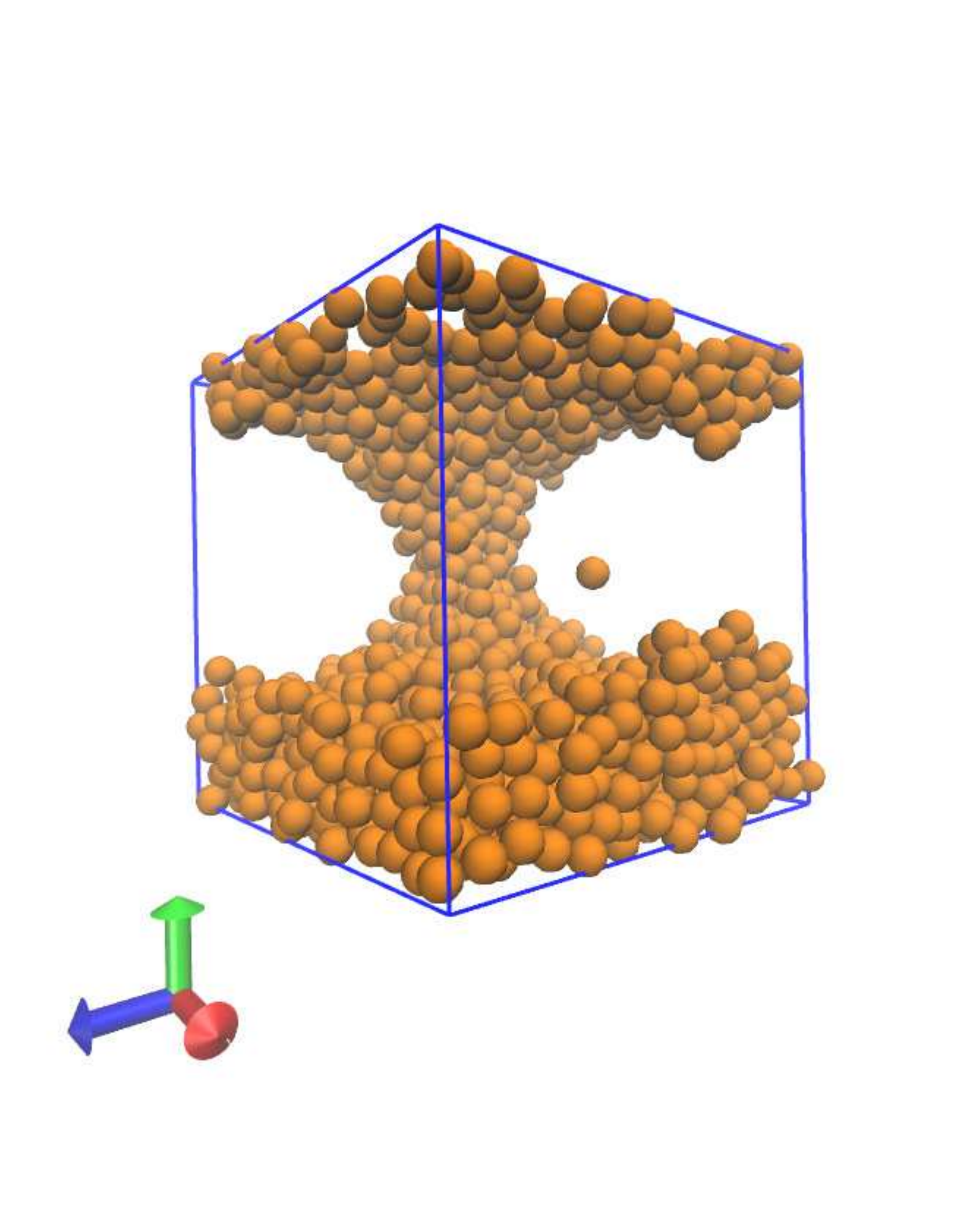}
}
\subfloat[\label{fig:rho085_neutrons_2}]{
\includegraphics[width=0.34\columnwidth]
{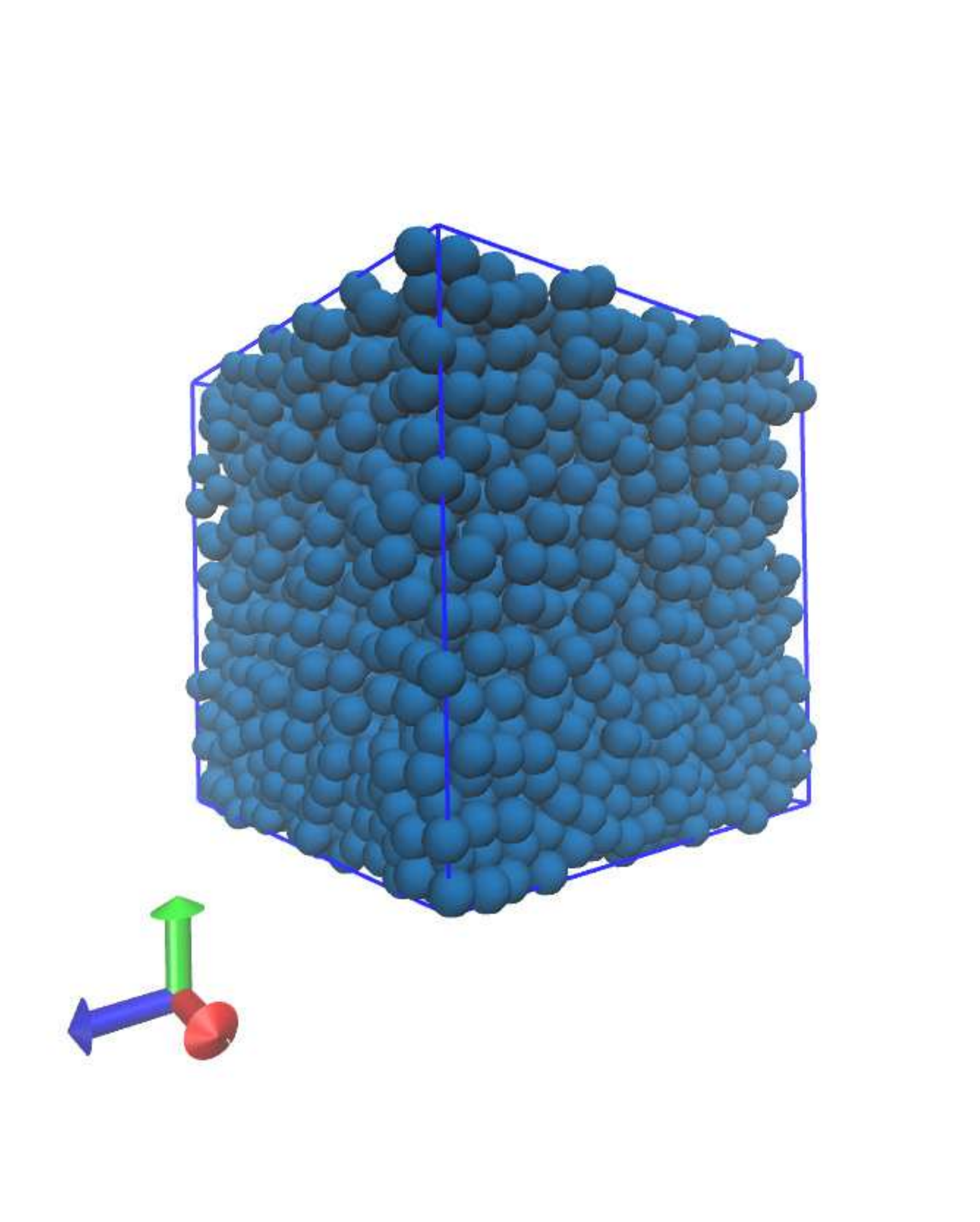}
}
\caption{\label{fig:pasta_decomposed_2}   Pasta structures for nuclear matter 
systems with 6000 nucleons, $x=0.3$ at $T=0.2\,$MeV and density 
$\rho$=0.085$\,$fm${}^{-3}$. (a) All
nucleons (protons in light color and neutrons in dark), (b) protons only, and 
(c) neutrons only.}
\end{figure*}

The argument outlined so far does not hypothesize on the pasta shape. It only 
compares the average density inside the pasta with respect to $N/V$. Thus, it is 
expected to hold on a variety of \textit{pastas}. This was verified for 
different densities, as shown in Fig.~\ref{rad2}. 
Figures~\ref{fig:pasta_decomposed} and \ref{fig:pasta_decomposed_2} show the 
pasta structures that correspond to Figure~\ref{rad}b. Panel (a) shows all 
nucleons (protons in light color and neutrons in dark), (b) protons only, and 
(c) neutrons only. Among other things, this indicates that pasta structures 
exists in non-symmetric nuclear matter.\\

\paragraph{The onset of the pasta}\label{onset} 

One way of determining the onset of the creation of the pasta is by looking at 
the spatial distribution of nucleons during the cooling of a nuclear matter 
system, and by quantifying the geometrical shape of the pasta structure. For 
these tasks we use the Kolmogorov Statistics and the Minkowski Functionals.\\

\begin{figure*}[!htbp]
\centering
\captionsetup[subfigure]{justification=centering}
\subfloat[\label{fig:rho05_x05_t20}]{
\includegraphics[width=3.3in]
{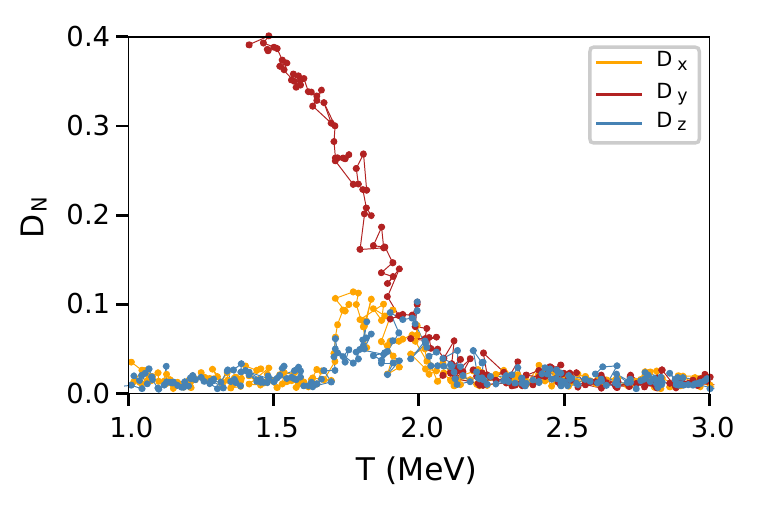}
}
\subfloat[\label{fig:rho05_x05_t20_image}]{
\includegraphics[width=0.33\columnwidth]
{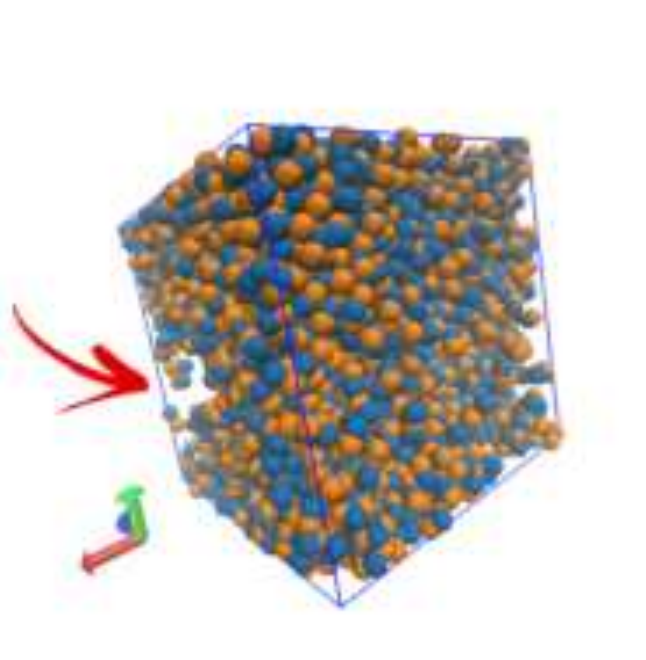}
}
\caption{\label{fig:rho05_x05_t20_all}   (a) The Kolmogorov
1D statistic versus temperature. Data corresponds to the position of 6000 
nucleons
 at $\rho=0.05\,\mathrm{fm}^{-3}$ and $x = 0.5$ (the same configuration as in
Fig.~\ref{fig:rho05} at $T=0.2$ MeV). The simulation cell (with periodic 
boundary conditions) was slowly cooled from $T=4\,$MeV down to $T=0.2\,$MeV. 
$D_x$, $D_y$ and $D_z$ correspond to the $x$, $y$ and $z$ Kolmogorov statistics 
sampled from the simulation cell, respectively. (b) Visualization of the system 
analyzed in (a) at $T=2\,$MeV. The arrow points to the most noticeable bubble 
appearing in the picture.}
\end{figure*}

The spatial characteristics can be quantified by means of the Kolmogorov 
statistics, which measures the difference between a given distribution of nuclei 
and a homogenous one. As the temperature decreases and the pastas are created, 
the spatial distribution of nucleons will depart from a uniform distribution. 
The Kolmogorov statistic is presented in Appendix~\ref{kolm}. In the present 
case we treat the distribution in separate coordinates (more useful for the case 
of planar structures, such as ``lasagnas''),  case for which the Kolmogorov 
statistic is ``distribution-free'' or parameter-free. 
Figure~\ref{fig:rho05_x05_t20_all}  shows the corresponding results for an 
isospin symmetric system at $\rho = 0.05\,$fm$^{-3}$; these results are from a 
single simulation, since the Minkowski statistics can only be applied to raw 
data, not to averaged one; see caption for further details on the simulation 
procedure. \\

As can be seen in Fig.~\ref{fig:rho05_x05_t20_all}, the values of the 1D 
Kolmogorov statistic, i.e. the discrepancies to homogeneous distributions, are 
negligible for temperatures above $2\,$MeV, as expected for energetic particles 
moving around homogeneously. At $T\approx 2\,$MeV, the three statistics 
experience a change in the slope, although $D_x$ and $D_z$ return to negligible 
values as the temperature further decreases. The $D_y$ statistic, instead, 
attains a definite departure from homogeneity for $T<2\,$MeV. A \textit{lasagna} 
or slab-like structure across the $y$-axis occurs at conditions $\rho=0.05$ 
fm$^{-3}$, $T=0.2$ MeV and $x = 0.5$, as presented in Fig.~\ref{fig:rho05}.  \\

\begin{figure*}[!htbp]
\centering
\captionsetup[subfigure]{justification=centering}
\subfloat[\label{fig:rho05_x05_T2_0}$T=2\,$MeV]{
\includegraphics[width=0.33\columnwidth]
{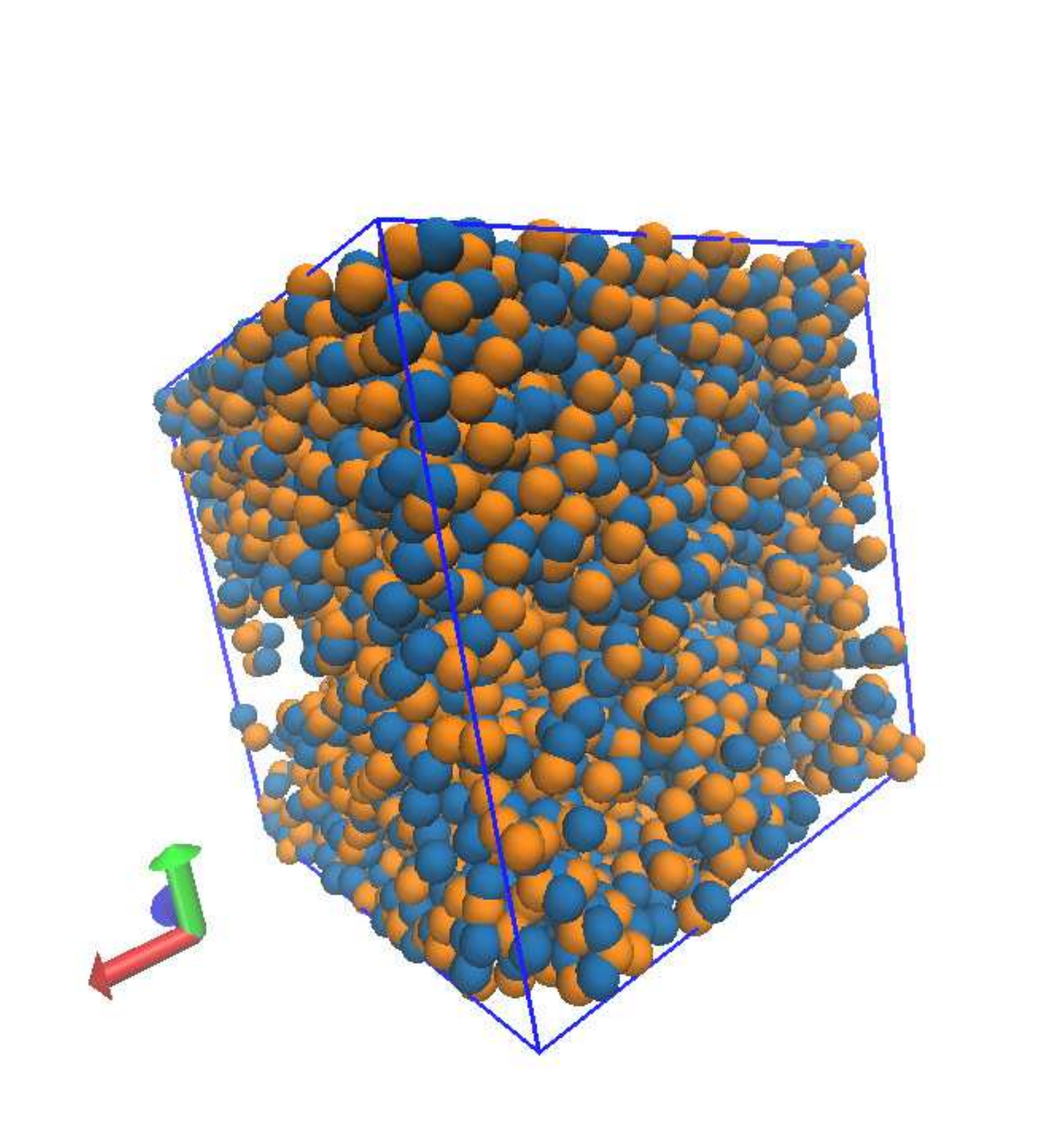}
}
\subfloat[\label{fig:rho05_x05_T1_8}$T=1.8\,$MeV]{
\includegraphics[width=0.33\columnwidth]
{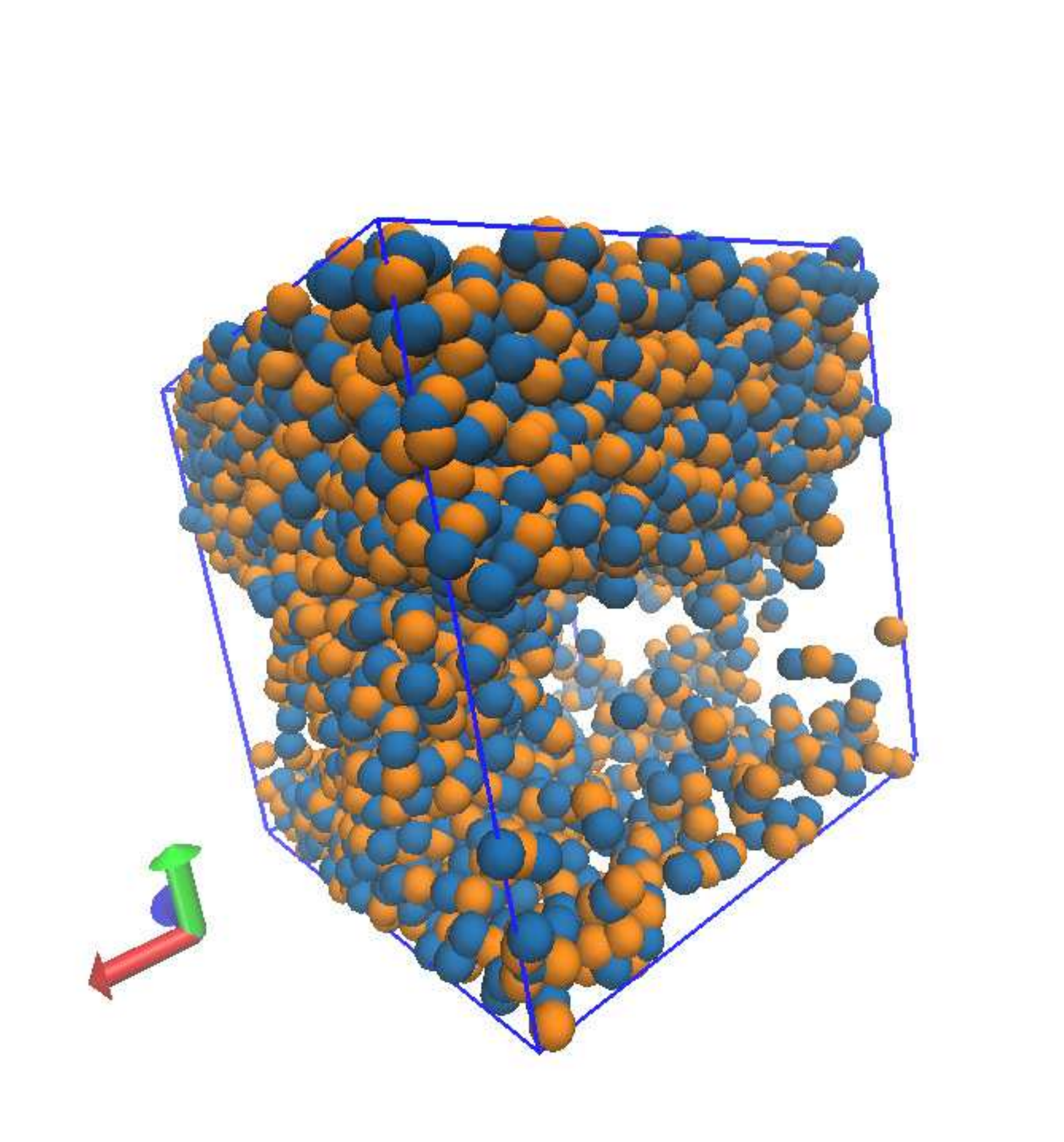}
}
\subfloat[\label{fig:rho05_x05_T1_5}$T=1.5\,$MeV]{
\includegraphics[width=0.33\columnwidth]
{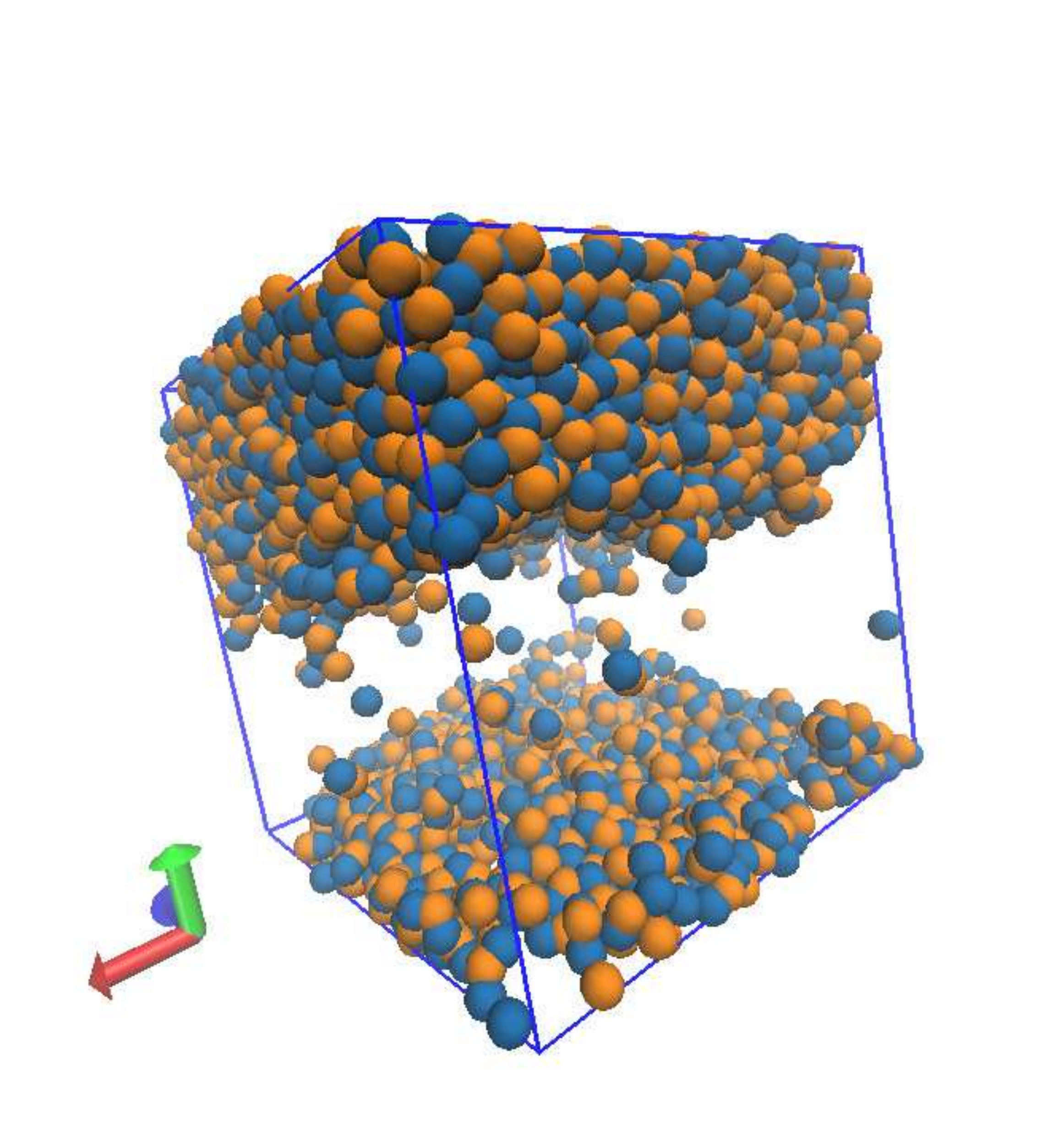}
}
\caption{\label{fig:rho05_x05_seq}   Visualization of 6000
nucleons  at $\rho=0.05\,\mathrm{fm}^{-3}$, $x = 0.5$ (the same configuration
as in Fig.~\ref{fig:rho05}) at three temperatures. (a) A bubble can be seen on 
the left. (b) A
tunnel appears along the $x,y$ plane. (c) The tunnel widens and breaks into two
slabs.  }
\end{figure*}

The 1D Kolmogorov statistic attains the departure from homogeneity at an early
stage of the pasta formation. The arrow in
Fig.~\ref{fig:rho05_x05_t20_image} points to the most noticeable bubble
appearing in the system at $T=2\,$MeV. The bubble-like heterogeneity
also explains the changes in the slope for $D_x$ and $D_z$ at this temperature,
as pictured in Fig.~\ref{fig:rho05_x05_seq}. For decreasing temperatures, the
bubble widens (on the right side of the image due to the periodic boundary
conditions), a tunnel appears (Fig.~\ref{fig:rho05_x05_T1_8}), and at 
$T<1.5\,$MeV 
it finally splits into two pieces while the $(x,y)$ homogeneity gets
restored (Fig.~\ref{fig:rho05_x05_T1_5}) returning $D_x$ and 
$D_z$ back to their negligible values.\\

\begin{figure*}[!htbp]
\centering
\captionsetup[subfigure]{justification=centering}
\subfloat[\label{fig:rho05_pn}$\rho=0.05\,\mathrm{fm}^{-3}$]{
\includegraphics[width=0.5\columnwidth]
{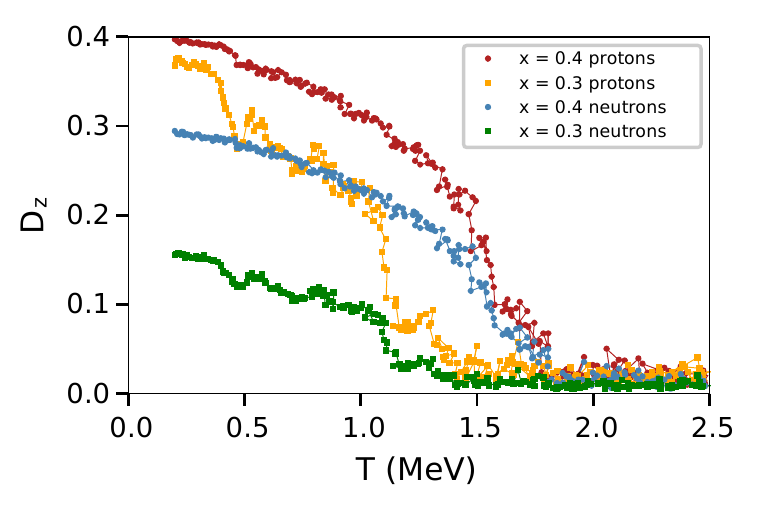}
}
\subfloat[\label{fig:rho085_pn}$\rho=0.085\,\mathrm{fm}^{-3}$]{
\includegraphics[width=0.5\columnwidth]
{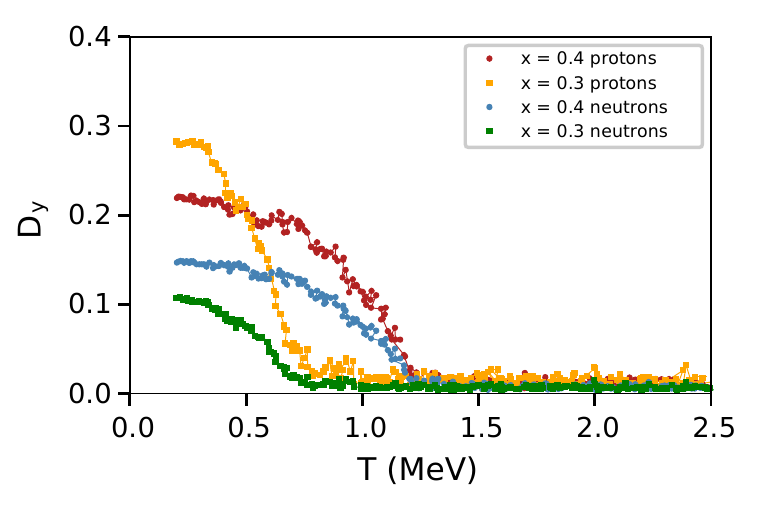}
}
\caption{\label{fig:Kolmogorov_asym}   (a) The Kolmogorov 1D statistic versus 
temperature of a system with 6000 nucleons for $x = 0.3$ and $0.4$. (a) the 
$D_z$ statistic at $\rho=0.05\,\mathrm{fm}^{-3}$, and (b) the $D_y$ statistic at 
$\rho=0.085\,\mathrm{fm}^{-3}$. }
\end{figure*}

\begin{figure*}[!htbp]
\centering
\captionsetup[subfigure]{justification=centering}
\subfloat[\label{fig:rho05_all}]{
\includegraphics[width=0.33\columnwidth]
{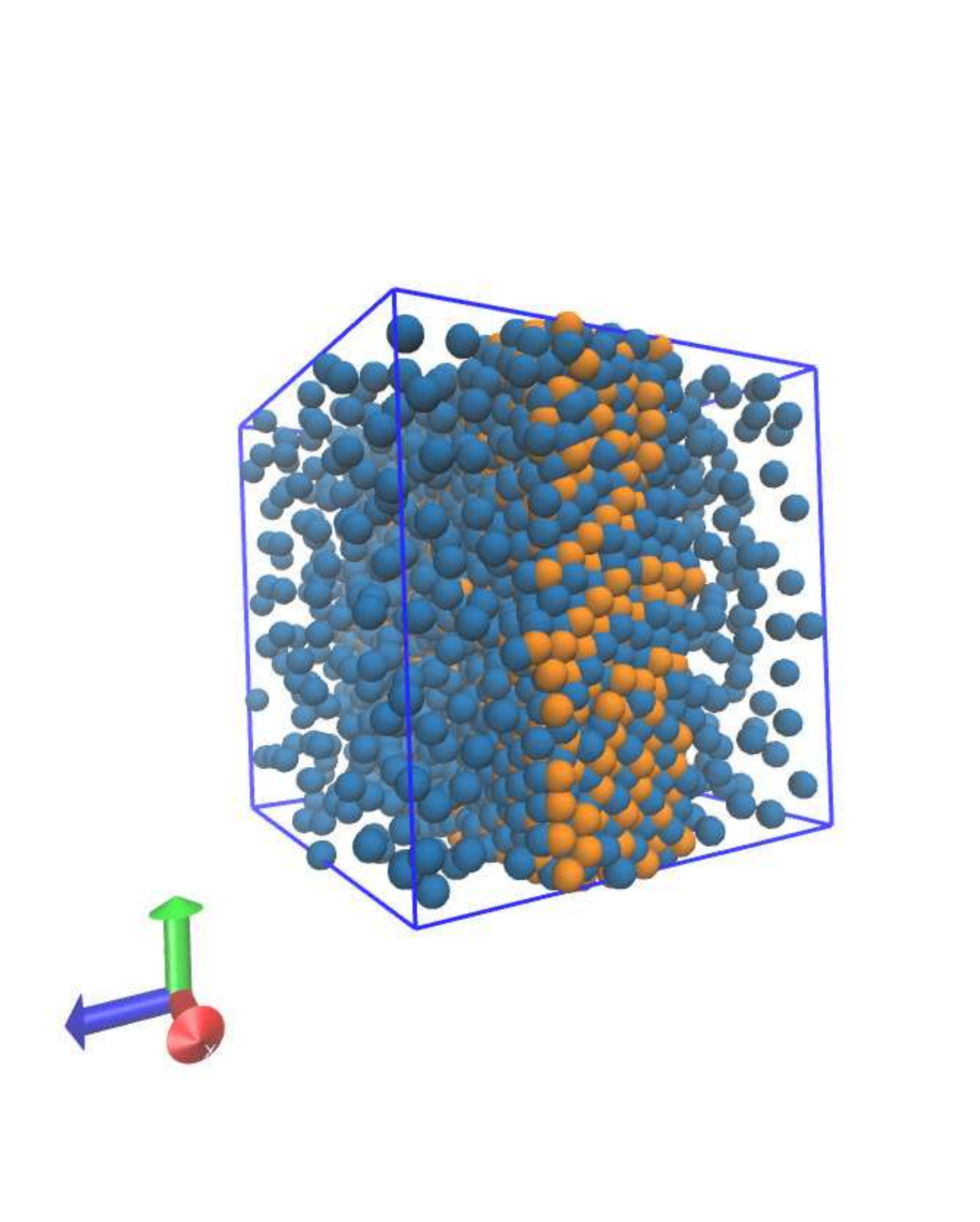}
}
\subfloat[\label{fig:rho05_protons}]{
\includegraphics[width=0.33\columnwidth]
{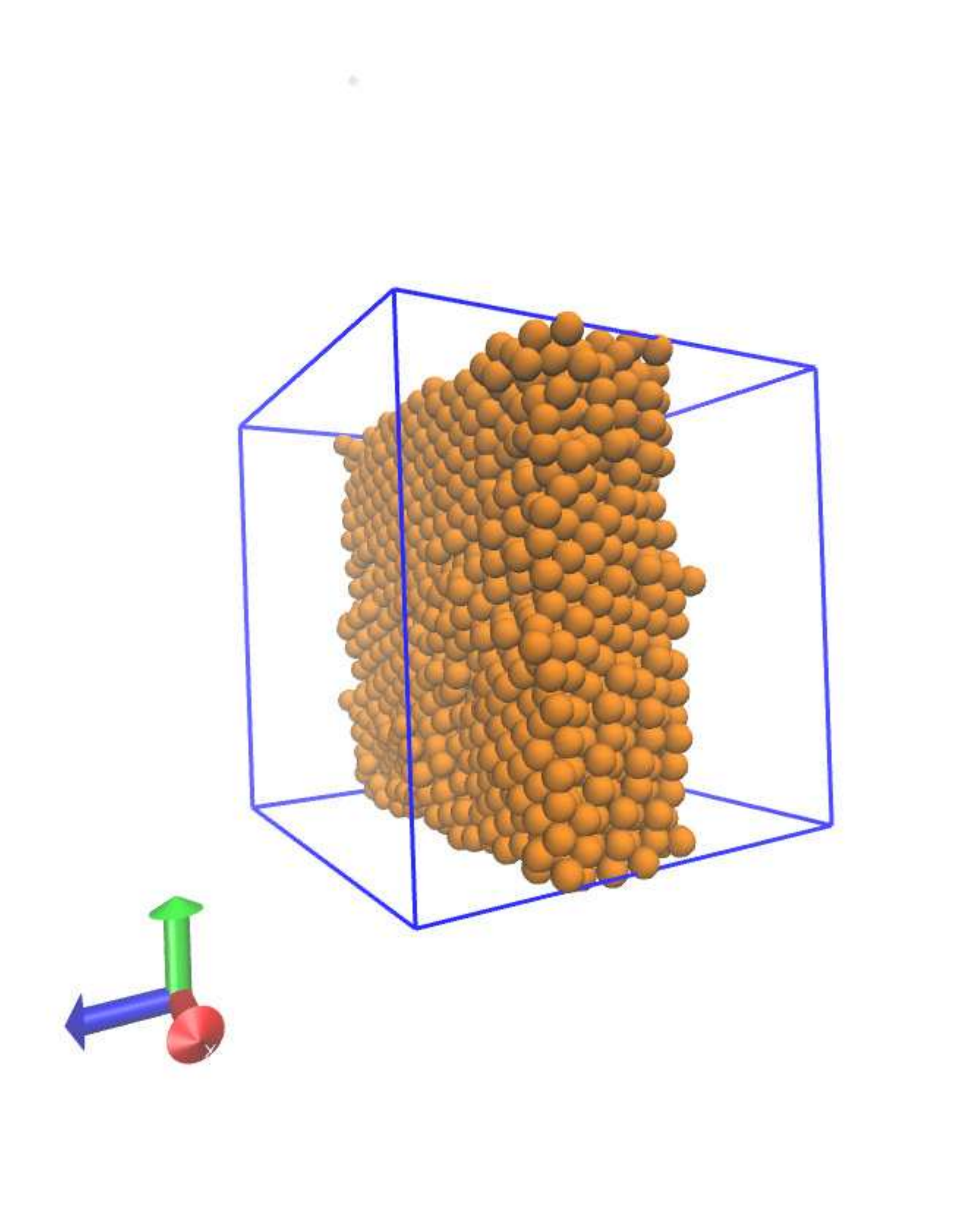}
}
\subfloat[\label{fig:rho05_neutrons}]{
\includegraphics[width=0.33\columnwidth]
{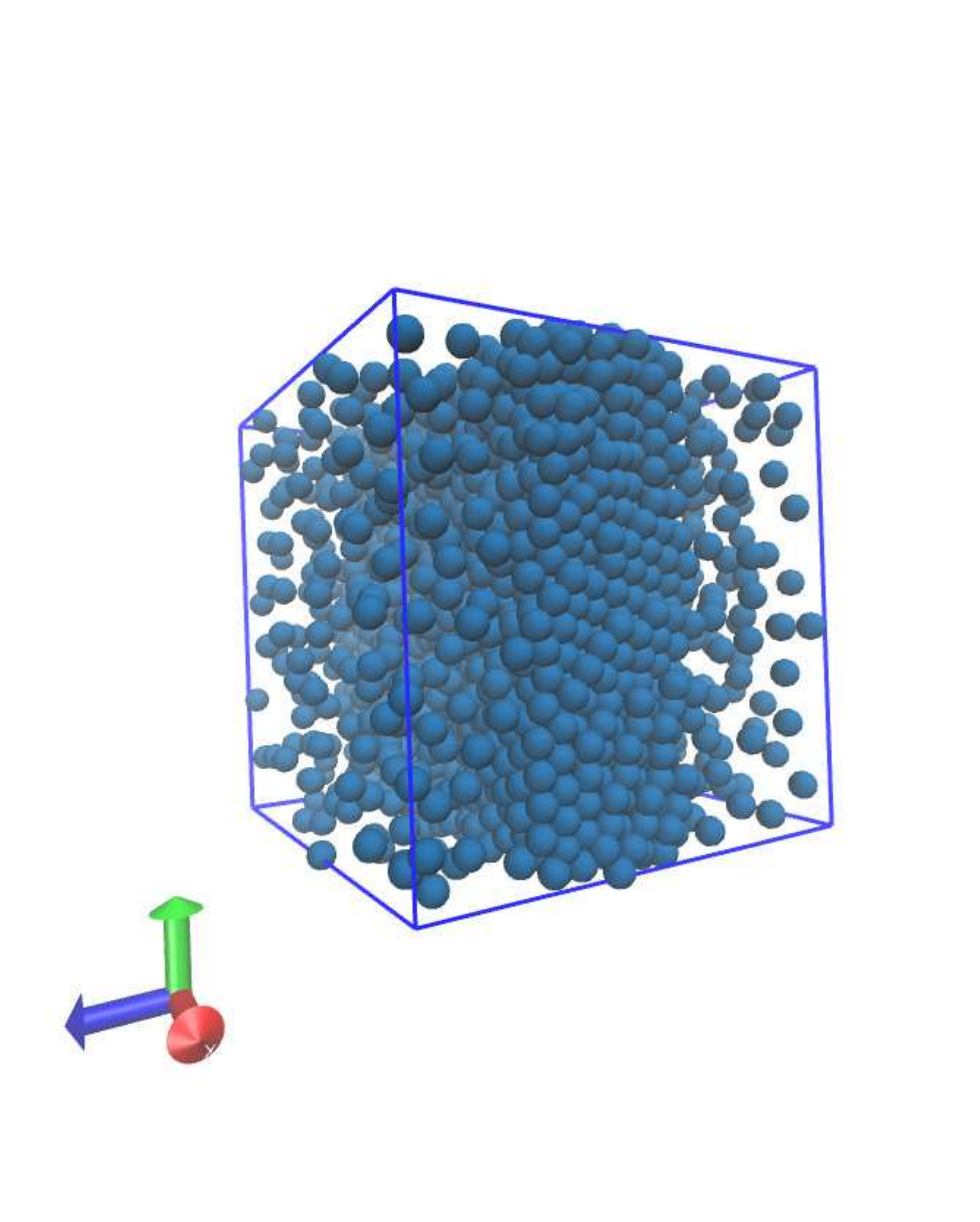}
}
\caption{\label{fig:pasta_decomposed_3} Pasta structures for nuclear matter 
systems with 6000 nucleons with $x=0.4$, $T=0.2\,$MeV and density 
$\rho$=0.05$\,$fm${}^{-3}$. (a) All
nucleons (protons in orange and neutrons in blue). (b) Protons only. (c)
Neutrons only.}
\end{figure*}

\begin{figure*}[!htbp]
\centering
\captionsetup[subfigure]{justification=centering}
\subfloat[\label{fig:rho05_all_2}]{
\includegraphics[width=0.33\columnwidth]
{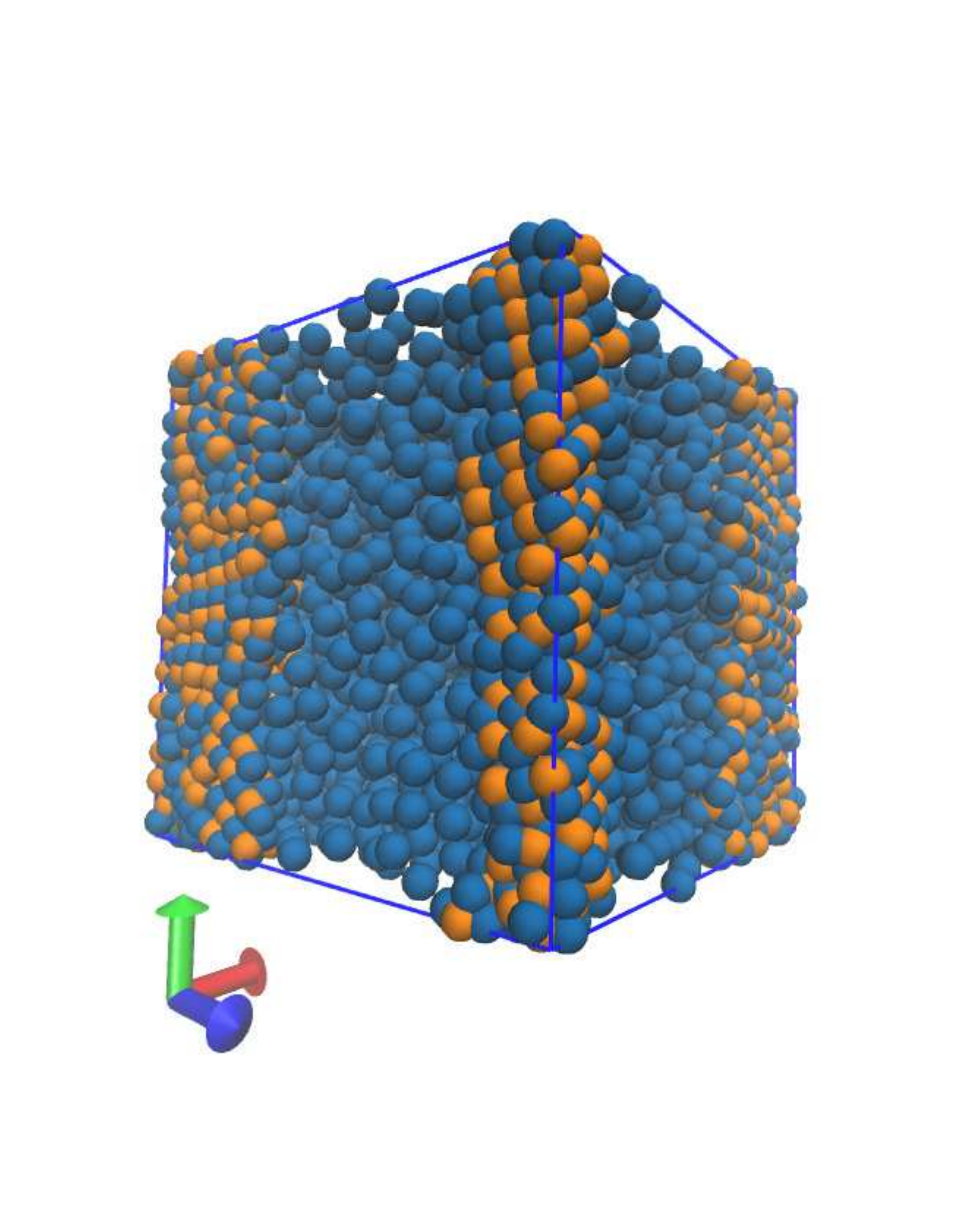}
}
\subfloat[\label{fig:rho05_protons_2}]{
\includegraphics[width=0.33\columnwidth]
{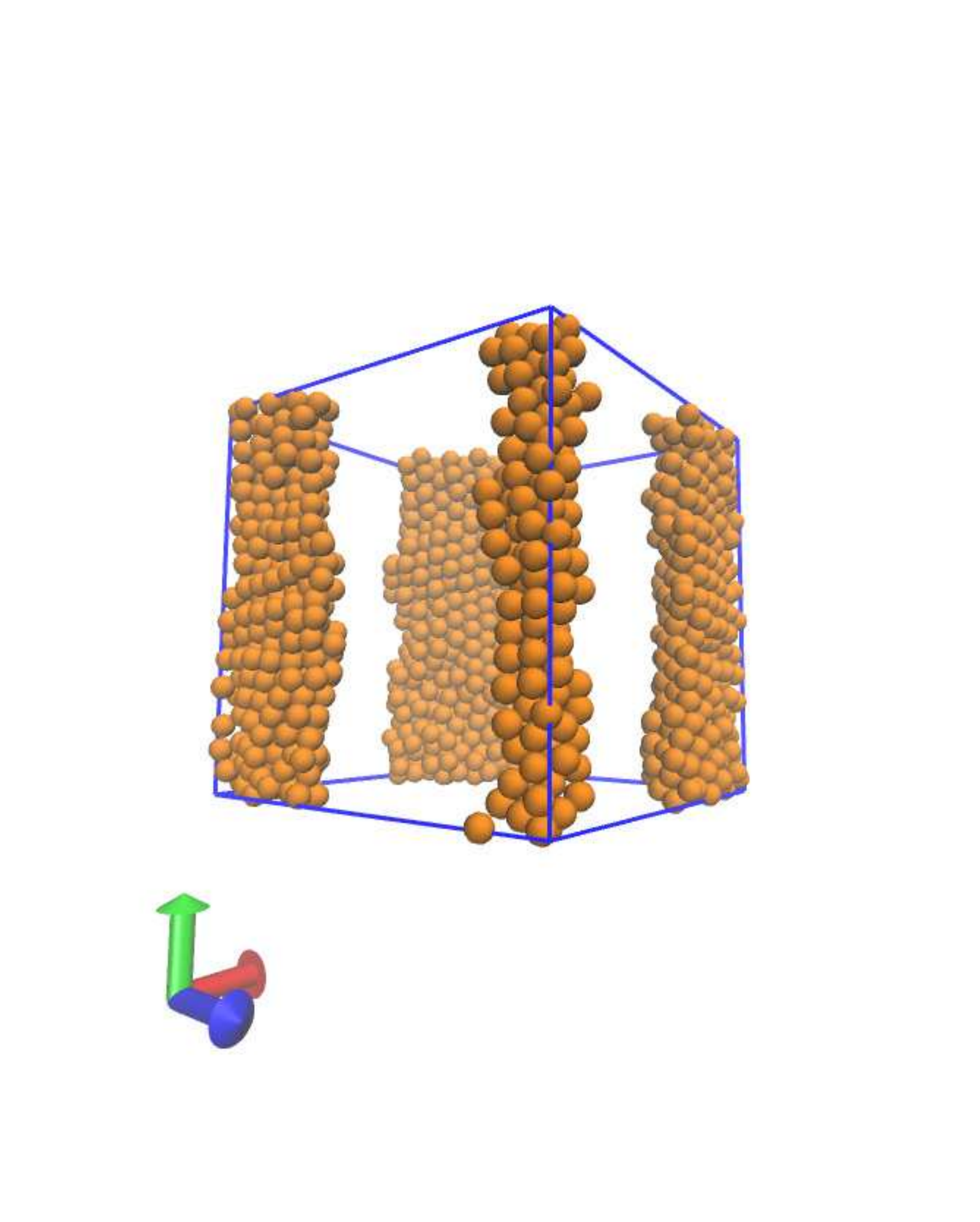}
}
\subfloat[\label{fig:rho05_neutrons_2}]{
\includegraphics[width=0.33\columnwidth]
{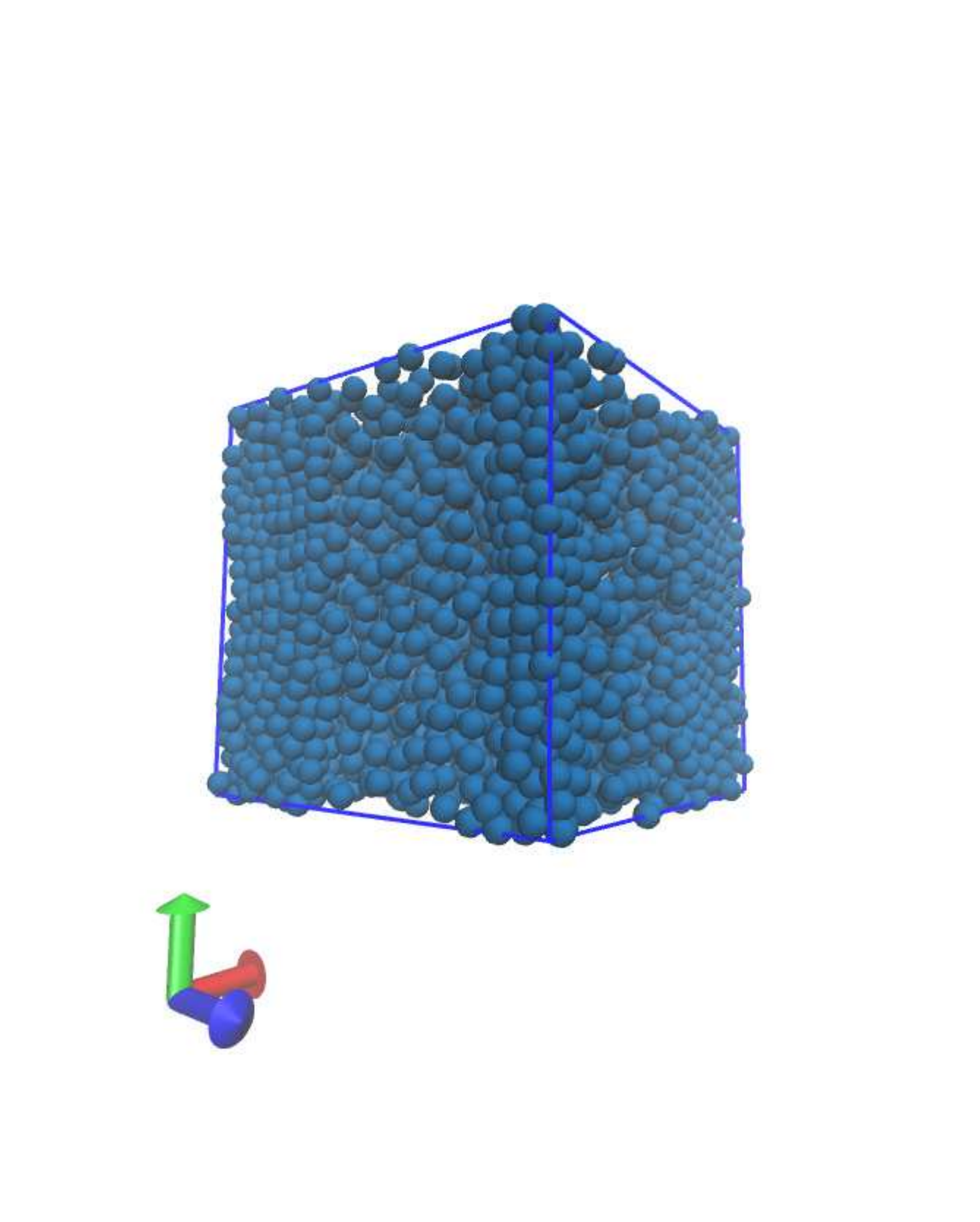}
}
\caption{\label{fig:pasta_decomposed_4} Pasta structures for nuclear matter 
systems with 6000 nucleons at $x=0.3$, $T=0.2\,$MeV and density 
$\rho$=0.05$\,$fm${}^{-3}$. (a) All
nucleons (protons in orange and neutrons in blue), (b) protons only, and (c) 
neutrons only.}
\end{figure*}

We further applied the 1D Kolmogorov statistic on protons and neutrons
separately for non-symmetric nuclear matter systems. Figure
\ref{fig:Kolmogorov_asym} exhibits the most significant 1D statistics for the
densities $\rho=0.05\,\mathrm{fm}^{-3}$ and $\rho=0.085\,\mathrm{fm}^{-3}$,
respectively. The corresponding spatial configurations can be seen in
Figs.~\ref{fig:pasta_decomposed} and \ref{fig:pasta_decomposed_2} for 
$\rho=0.085\,\mathrm{fm}^{-3}$,
and in Figs.~\ref{fig:pasta_decomposed_3} and \ref{fig:pasta_decomposed_4} for 
$\rho=0.05\,\mathrm{fm}^{-3}$.\\

According to Fig.~\ref{fig:Kolmogorov_asym}, the temperature threshold at which
the 1D Kolmogorov statistic becomes significant, decreases with smaller proton
ratios (at fixed density), in agreement with the caloric curves of Section 
\ref{phases}, but more sharply exposed now. Both species, protons and neutrons, 
seem to depart from homogeneity at the same temperature threshold, but the 
departure appears more sharply for protons than for neutrons, which attain 
higher maximum values than the neutrons.  \\

Apparently, neutron-rich systems do not develop bubbles at the same temperature 
as the symmetric systems. The excess of neutrons inhibits the pasta formation 
until lower temperatures where protons manage to form the pasta. The released 
neutrons get distributed along the cell disrupting the pasta structure. \\

\paragraph{Classification of the pasta}\label{minkowski_results}

The Minkowski functionals supply complementary information on the pasta 
structure from the early stage to the solid-like stage; see Table~\ref{tab1} for 
the classification of the pasta in terms of the curvature $B$ and the Euler 
characteristic $\chi$. The accuracy of this information is, however, conditional 
to the correct binning of the simulation cell. Tiny ``voxels'' (that is, a high 
density binning) may produce fake empty voids (artificial bubbles or tunnels) 
and, on the contrary, oversized voxels may yield a wrong structure of the system 
due to the lack of details. Therefore, some effort needs to be spent to 
determine the correct size for the voxels; Appendix \ref{sec:voxel} 
summarizes this procedure.  \\

\begin{figure}  
\begin{center}
   \includegraphics[width=0.5\columnwidth]{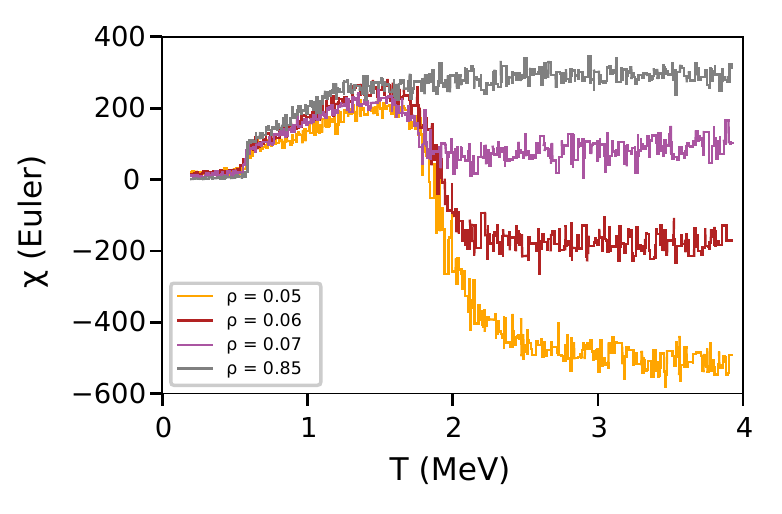}
\caption{The Euler characteristic $\chi$ as a function of
temperature for a system of 6000 nucleons at densities $\rho=0.05$, $0.06$, 
$0.07$, $0.085\,$fm$^{-3}$ and $x=0.5$.}\label{mink_x05}
\end{center}
\end{figure}

The simulation cell was first divided into cubic voxels of edge length 
$d=2.35\,$fm. The Euler characteristic $\chi$ was computed for symmetric nuclear 
matter, according to Eq.~(\ref{eq:chi}) and the results are shown in 
Fig.~\ref{mink_x05}.  As seen in this figure, the Euler characteristic $\chi$ 
for symmetric matter  shows three distinct  regions as a function of 
temperature, one at $T>2\,$MeV, one  at $T<0.5\,$MeV, and a transition one 
between these two. \\

\begin{figure*}[!htbp]
\centering
\captionsetup[subfigure]{justification=centering}
\subfloat[$\rho=0.05$]{
\includegraphics[width=0.45\columnwidth]
{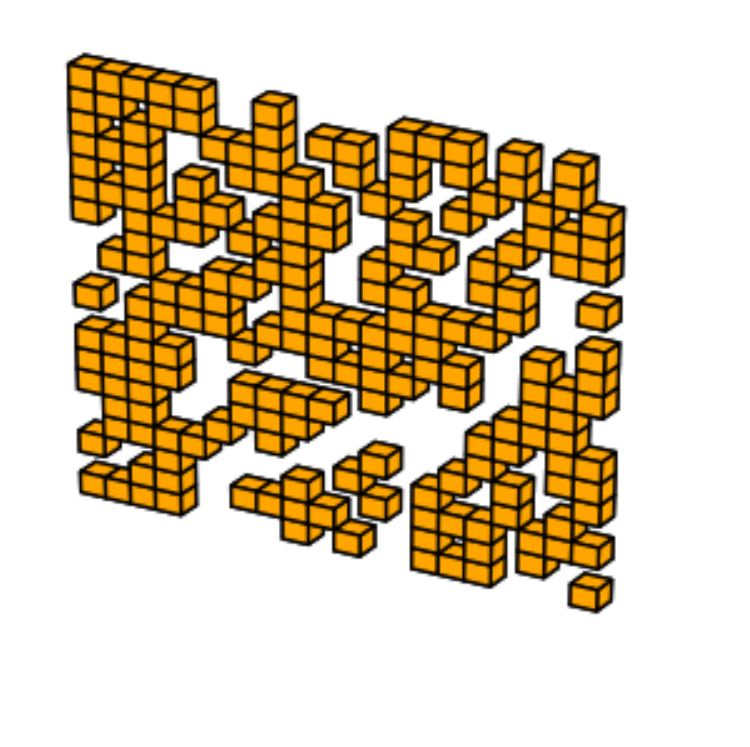}
} 
\subfloat[$\rho=0.06$]{
\includegraphics[width=0.45\columnwidth]
{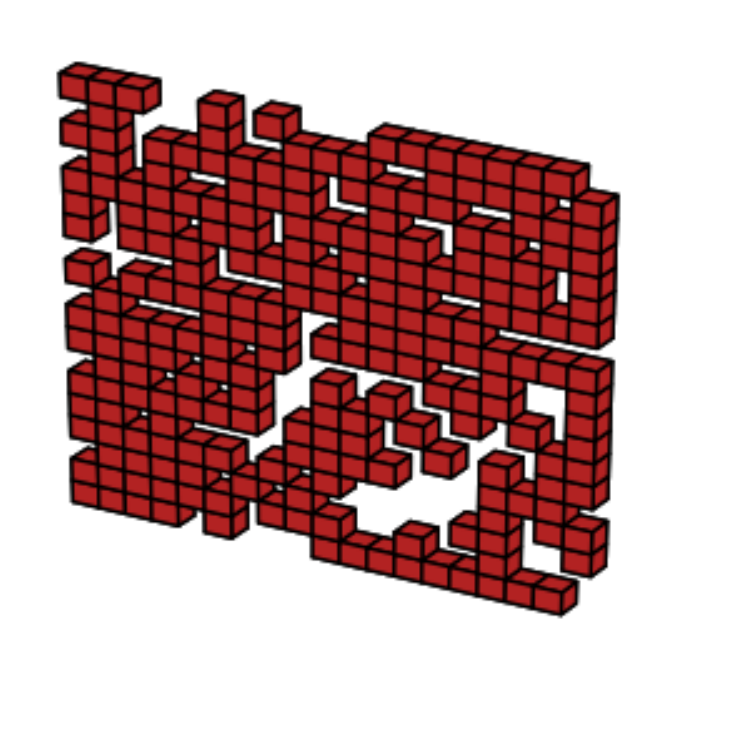}
} \\
\subfloat[$\rho=0.07$]{
\includegraphics[width=0.45\columnwidth]
{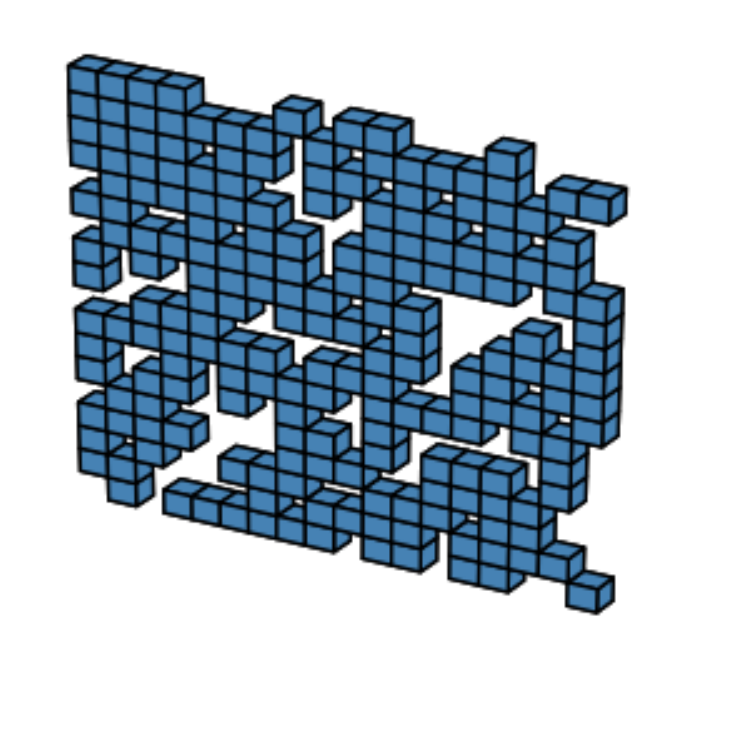}
} 
\subfloat[$\rho=0.085$]{
\includegraphics[width=0.45\columnwidth]
{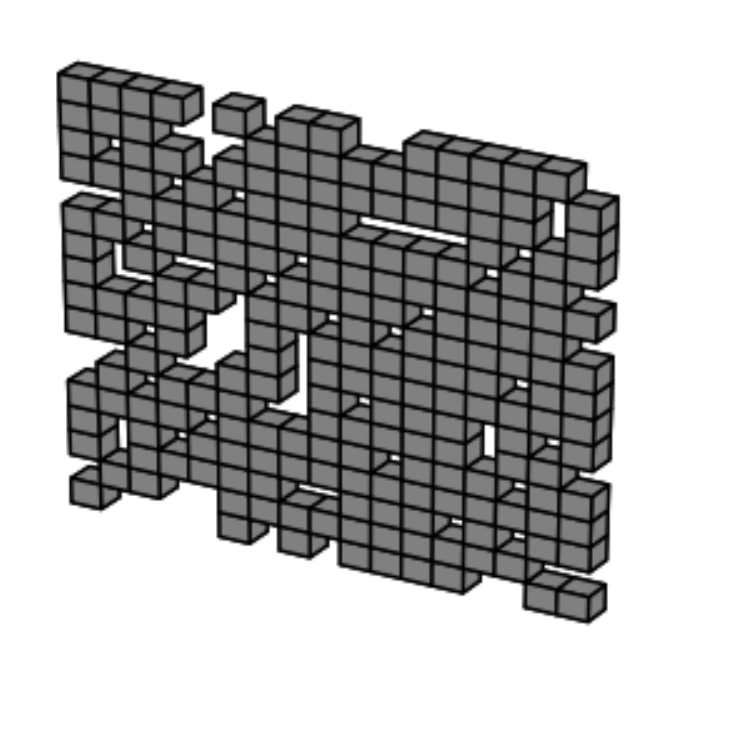}
} 
\caption{\label{fig:mink_slice}   Inside view of discretized
nuclear matter with 6000 nucleons at
$T=2.5\,$MeV and $x=0.5$. The discretization turned the 6000 nucleons into
(approximately) 4300 voxels (edge length $d=2.35\,$fm). Only a slice from the
middle ($y$-plane) of the simulation cell is represented. The labels correspond
to the density values in fm$^{-3}$. The colors are in correspondence with
Fig.~\ref{mink_x05}.}
\end{figure*}

Figure~\ref{mink_x05} shows that at $T>2\,$MeV $\chi$ does not vary much 
although it attains different values depending on the density. At the low 
densities of $\rho=0.05\,$fm$^{-3}$ and $\rho=0.06\,$fm$^{-3}$ $\chi$ exhibits 
negative values which, according to Eq.~(\ref{eq:chi}), indicate that the 
nucleons are sparse enough to form tunnels and empty regions across the cell. At 
higher densities, however, $\chi$ becomes positive indicating that tunnels begin 
to fill forming cavities and isolated regions. This is confirmed by 
Fig.~\ref{fig:mink_slice} which shows an inside view of the discretized nuclear 
matter at $T=2.5\,$MeV, $x=0.5$ and for the four densities under consideration.  
\\

The complementarity of $\chi$ over other measures can be seen by comparing 
it to, for instance, the results of the Kolmogorov statistics 
(Section~\ref{onset}). As seen in Section~\ref{subsec:esym}, as the nucleons get 
distributed uniformly at high temperatures neither the 1D nor the 3D Kolmogorov 
statistic capture the qualitative difference between a tunnel-like and a 
cavity-like scenarios. Both landscapes may not exhibit noticeable 
heterogeneities, and thus, they 
appear to be essentially the same from the point of view of the Kolmogorov 
statistic. \\

The energy, however, distinguishes between aforementioned different scenarios.
From the comparison between the caloric curves introduced in
Section~\ref{phases} and the current Euler characteristic, one can see that both
magnitudes are density-dependent in the high temperature regime. While
$\chi$ increases for increasing densities, the (mean) energy per nucleon
diminishes (see Figs.~\ref{fig:energy_pressure} and~\ref{mink_x05}). Thus,
the less energetic configuration (say, $\rho=0.085\,$fm$^{-3}$) appears to be
a cavity-like (or small bubble-like) scenario from the point of view of $\chi$ 
(with voxels size of $d=2.35\,$fm). \\

The Euler characteristic $\chi$ exhibits a dramatic change at $T\approx 
2\,$MeV. 
This is associated to the departure from homogeneity at the early stage of 
the pasta formation, as already mentioned in Section~\ref{minkowski_results}. 
It also agrees with the results from quantum models shown in 
Ref.~\cite{sonoda}). Notice that the $\chi$ values for the examined densities 
join into a single pattern for $T<2\,$MeV, in agreement with the behavior of 
the 
energy seen in Fig.~\ref{fig:energy_pressure}.  \\

\begin{figure*}[!htbp]
\centering
\captionsetup[subfigure]{justification=centering}
\subfloat[$\rho=0.05$]{
\includegraphics[width=0.45\columnwidth]
{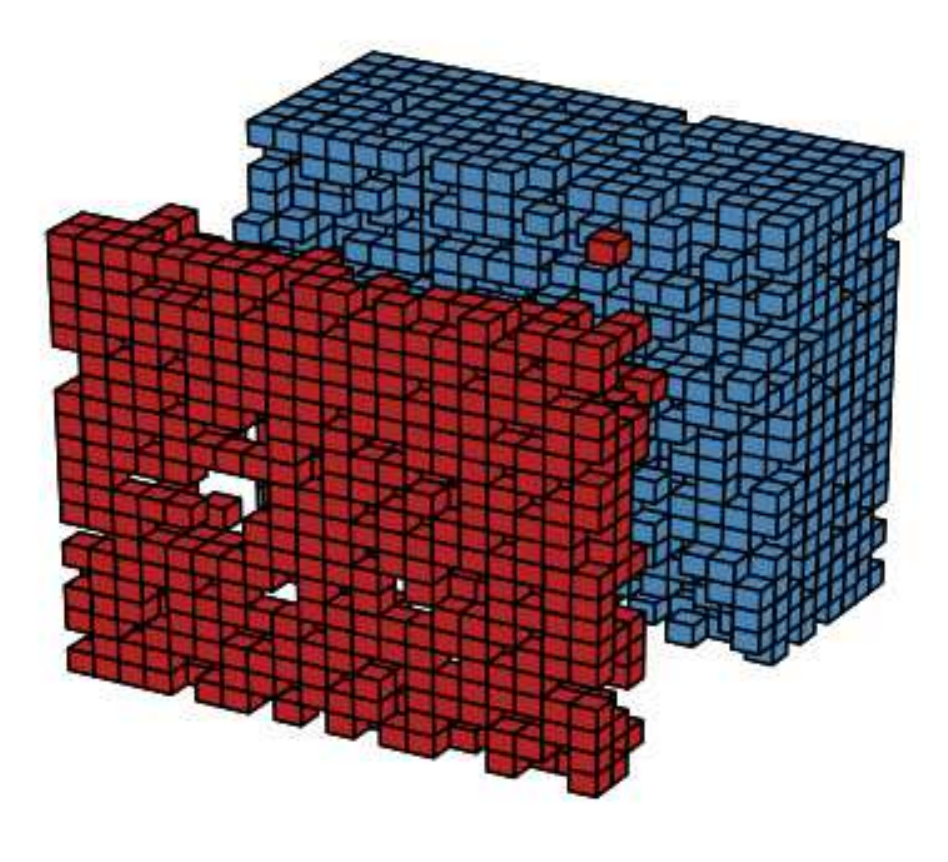}
} 
\subfloat[$\rho=0.085$]{
\hspace{5mm}\includegraphics[width=0.45\columnwidth]
{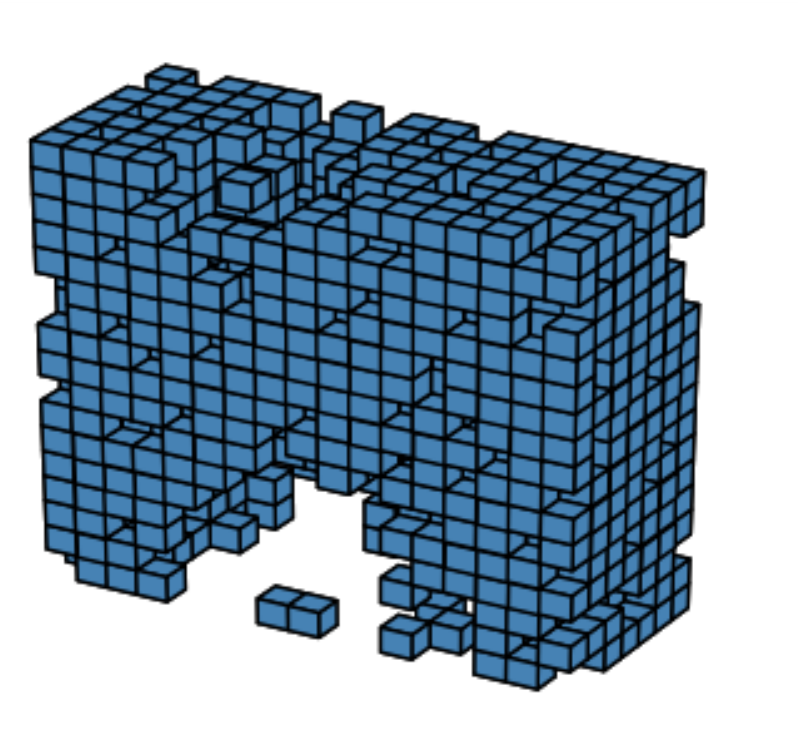}
} 
\caption{\label{fig:mink_slice_2}   Inside view of
discretized nuclear matter with 6000 nucleons
at $T=1.5\,$MeV and $x=0.5$. The discretization turned the 6000 nucleons into
(approximately) 4300 voxels (edge length $d=2.35\,$fm). (a) Density
$\rho=0.05\,$fm$^{-3}$. Two colors are used for a better view 
of the gap in the middle. Many small pieces also lie across the gap (not 
visible). (b) Density $\rho=0.085\,$fm$^{-3}$. Only a mid-slice of seven voxels 
thick
is shown for practical reasons. }
\end{figure*}

It should be emphasized that although all the examined densities share the same
$\chi$ pattern for $T<2\,$MeV, their current morphology may be quite
different. Figure~\ref{fig:mink_slice_2} illustrates two such situations,
see caption for details. It seems, though, that whatever the morphology,
these are constrained to be equally energetic (see 
Fig.~\ref{fig:energy_pressure}). \\

\begin{figure*}[!htbp]
\centering
\captionsetup[subfigure]{justification=centering}
\subfloat[$\rho=0.05$\label{fig:mink_func_a}]{
\includegraphics[width=0.5\columnwidth]
{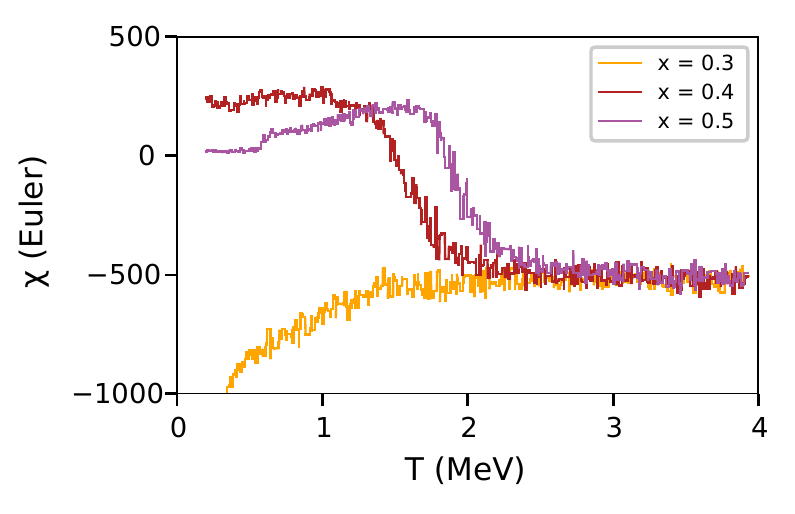}
} 
\subfloat[$\rho=0.085$\label{fig:mink_func_b}]{
\includegraphics[width=0.5\columnwidth]
{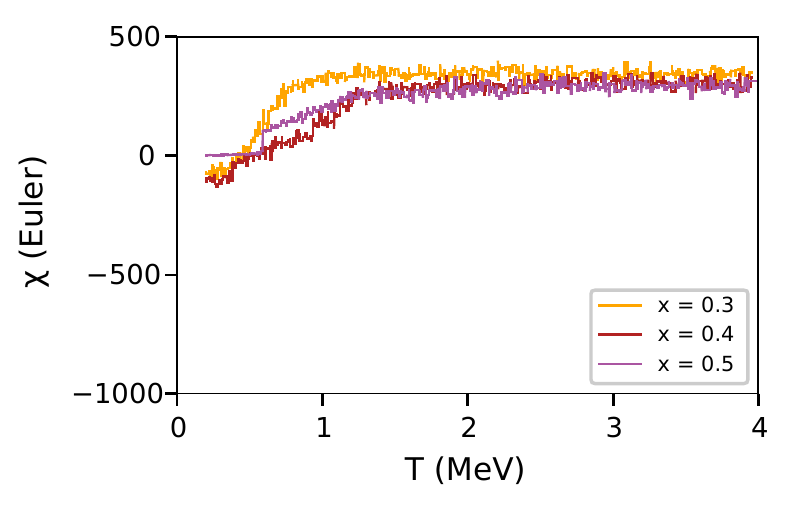}
} 
\caption{\label{fig:mink_func} The Euler characteristic as
a function of temperature for $x=0.5$, $0.4$ and $0.3$. The voxel's edge length
is $d=2.35\,$fm. The total number of nucleons is $6000$. (a)
$\rho=0.05\,$fm$^{-3}$ and (b) $\rho=0.085\,$fm$^{-3}$. }
\end{figure*}

Extending the $\chi$ study for non-symmetric nuclear matter appears to confirm 
the complexity observed in Sections~\ref{phases} and \ref{onset}. The global 
pressure does not present negative values for $x=0.4$ and $x=0.3$ at 
temperatures above the solid-like state. Neither noticeable bubbles nor other 
heterogeneities could be detected at the early stage of the pasta formation 
(say, $T\approx 2\,$MeV). We are now able to confirm these results through the
$\chi$ functional. Fig.~\ref{fig:mink_func} shows the Euler characteristic for 
two different densities and $x$= 0.3, 0.4 and 0.5. \\

\begin{figure*}[!htbp]
\centering
\captionsetup[subfigure]{justification=centering}
\subfloat[$x=0.4$\label{fig:mink_slice_3_04}]{
\includegraphics[width=0.45\columnwidth]
{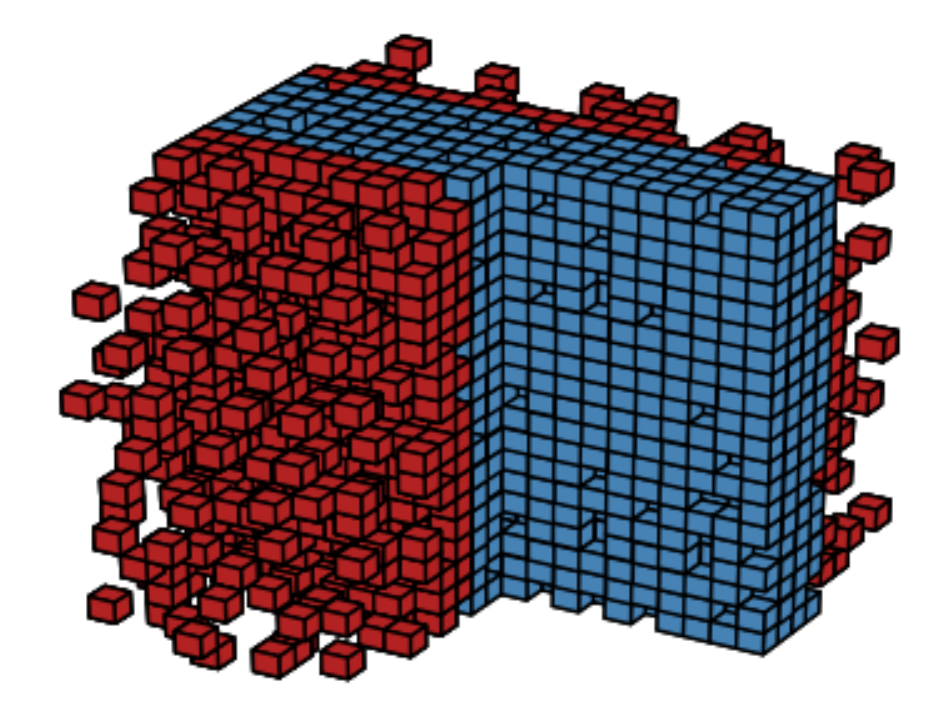}
} 
\hspace{5mm}\subfloat[$x=0.3$\label{fig:mink_slice_3_03}]{
\includegraphics[width=0.45\columnwidth]
{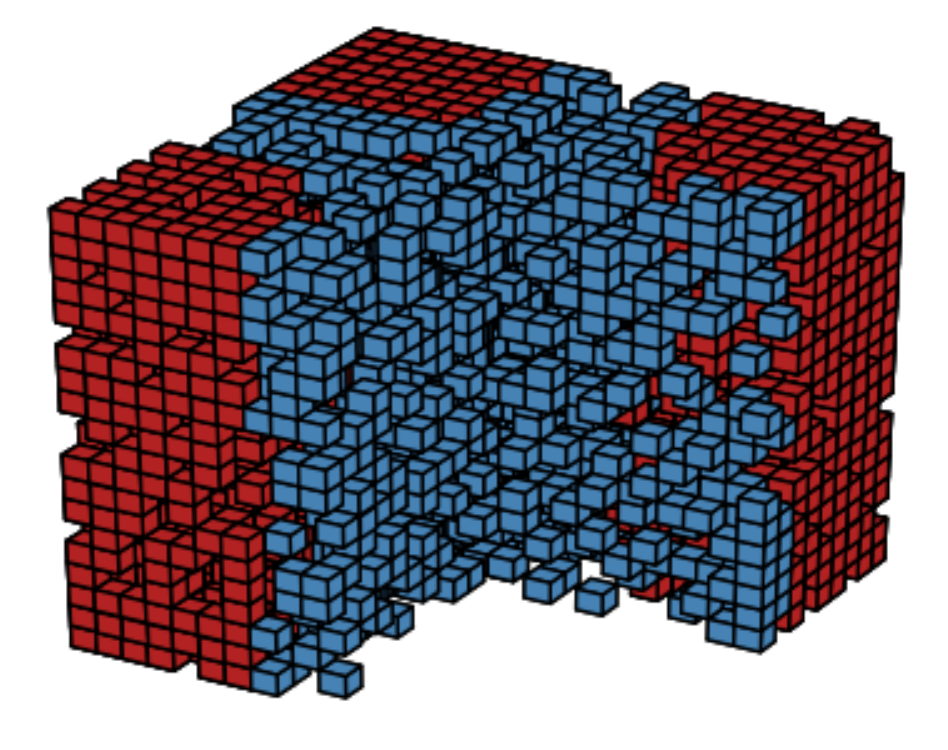}
} 
\caption{\label{fig:mink_slice_3}   Inside view of
discretized nuclear matter with 6000 nucleons
at $T=1.0\,$MeV, and $\rho=0.05\,$fm$^{-3}$. A quarter slice has been cut out
for a better view of the inner most region of the cell. The discretization
turned the 6000 nucleons into (approximately) 3800 voxels (edge length
$d=2.35\,$fm). (a)  Non-symmetric nuclear matter for $x=0.4$. The red color
corresponds to the region mostly occupied by neutrons (compare with
Fig.~\ref{fig:pasta_decomposed_3}). (b) Non-symmetric nuclear matter for
$x=0.3$. The blue color corresponds to the region mostly occupied by neutrons
(compare with Fig.~\ref{fig:pasta_decomposed_4}). }
\end{figure*}

Fig.~\ref{fig:mink_func_a} shows three distinct behaviors of $\chi$ at 
$\rho=0.05\,$fm$^{-3}$. The case of symmetric nuclear matter ($x=0.5$)  was 
already analyzed above. The curve for $x=0.4$ appears left-shifted with respect 
to the symmetric case, in agreement with our previous
observation that proton ratios of $x<0.5$ frustrate for a while the pasta 
formation (see Section~\ref{onset}). In spite of that, the pattern for $x=0.4$ 
achieves a higher positive value at lower temperatures than the symmetric case 
indicating that the tunnel-like scenario ($\chi<0$) switched to a bubble-like or 
an isolated-structure scenario ($\chi>0$). It can be verified from 
Fig.~\ref{fig:mink_slice_3_04} that this is actually occurring at $T\approx 
1\,$MeV; many isolated structures may be visualized in red, while no tunnels 
seem to be present in the blue region (see caption for details). \\

The curve of $x=$ 0.3 in Fig.~\ref{fig:mink_func_a} does not change sign nor 
increases in value at lower temperatures, as opposed to the other two curves. 
The fact that $\chi < 0$ for all of the examined temperatures indicates that 
tunnel-like landscapes are the relevant ones. 
Fig.~\ref{fig:mink_slice_3_03} illustrates this scenario: the heavy tunnel-like 
region (highlighted in blue color) is mostly occupied by neutrons, as shown in 
Fig.~\ref{fig:pasta_decomposed_4}, 
indicating that repulsive forces between neutrons dominate in a large fraction 
of the cell, and thus producing a positive global pressure all along the 
examined temperature range, as noticed in Fig.~\ref{fig:energy_pressure}. 
Fig.~\ref{fig:mink_func_a} also show that $\chi$ for $x=0.3$ decreases in 
magnitude with at lower temperatures; this corresponds to the departure from the 
spacial homogeneity detected in Fig.~\ref{fig:rho05_pn}, and found by the 
Kolmogorov statistic and Euler 
functional to occur at $T\approx 1.5\,$MeV. \\

At the higher density of $\rho=0.085\,$fm$^{-3}$ the behavior of $\chi$ is 
substantially different. Fig.~\ref{fig:mink_func_b} shows the temperature 
dependence of $\chi$ for $x=$ 0.3, 0.4 and 0.5; both isospin asymmetric cases 
are qualitatively similar to the symmetric one. An inspection of the voxels' 
configuration (not shown) confirms that tunnels become relevant at very low 
temperatures; this can be checked from the configurations presented in 
Figs.~\ref{fig:pasta_decomposed} and \ref{fig:pasta_decomposed_2}.\\

Recall from \ref{sec:voxel} that the chosen binning procedure minimizes the 
existence of spurious voids, that is, voxels that appear to be empty but they 
are really not. This criterion involves 
\textit{all} the nucleons in the simulation cell. Thus, we are not able to 
extend the Minkowski analysis to protons or neutrons separately, as we did with 
the Kolmogorov statistic. In this context, this would produce an incorrect 
counting of voids (or tunnels). \\

In summary, at $T\approx 2\,$MeV the Euler characteristic $\chi$ of 
isospin-symmetric low-density systems ($\rho< 0.06\,$fm$^{-3}$) shows drastic 
changes from negative to positive values indicating a transition from a 
void-dominated regime to one with bubbles and isolated regions. Higher density 
systems ($\rho>0.06\,$fm$^{-3}$), in spite of always having $\chi > 0$, also 
increase the value of their $\chi$ at this temperature, reaching a common 
maximum for all densities at $T\approx 1.5\,$MeV. This maximum corresponds to 
the formation of bubbles or isolated regions, and indicates the formation of the 
pasta near the solid-liquid transition; recall that the Kolmogorov statistic was 
able to detect the pasta formation since the bubbles or isolated regions stage. 
\\

For isospin asymmetric systems the low-density ($\rho\approx 0.05\,$fm$^{-3}$) 
growth of $\chi$ is also observed but only for $x=$ 0.4 and 0.5; systems at $x=$ 
 0.3 have $\chi<0$ at all temperatures. At higher densities ($\rho\approx 
0.08\,$fm$^{-3}$) the Euler characteristic is always positive for all 
temperatures. \\

Table~\ref{tab1} classifies the pasta according to the sign of $\chi$ and the 
curvature $B$, and it was a goal of the present study to extend this 
classification for isospin asymmetric systems, but our results indicate that 
this labeling 
becomes meaningless for the non-symmetric case. For a given temperature, the 
$\chi$ 
functional attains positive or negative values depending on the isospin content 
and the density of the system. In general, the excess of neutrons obscures the 
pasta structures for the protons and, thus, the early stage of the pasta 
formation 
(that is, the formation of bubbles or isolated regions) is not detectable. In 
spite of this, we observe that the system departs from homogeneity 
at $T\sim 1.5\,$MeV (see, for example, Fig.~\ref{fig:rho05_pn}).  \\

\paragraph{Symmetry energy in the pasta}\label{subsec:esym}

We now study the symmetry energy of nuclear matter in the pasta region. As 
mentioned before, at a given temperature the energy $E(\rho,T)$ shows three 
distinct behaviors as a function of the density: the pasta region for densities 
below $0.085\,$fm$^{-3}$, the crystal-like region for densities above 
$0.14\,$fm$^{-3}$, and an intermediate region in between the first two. In what 
follows we will focus on the symmetry energy in the pasta region. \\

\begin{figure}
\begin{center}
   \includegraphics[width=0.5\columnwidth]{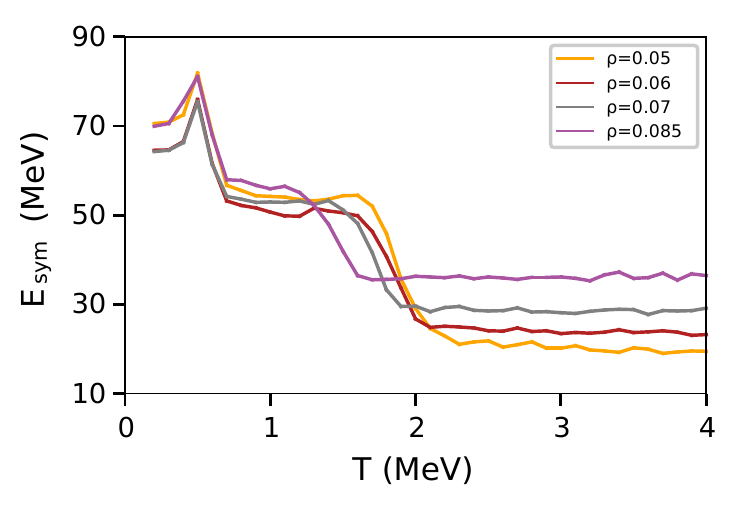}
\caption{Symmetry energy as a function of the temperature for four density 
values, as indicated in the insert. The system corresponds to nuclear matter 
with 6000 nucleons. 
}\label{fig:esym}
\end{center}
\end{figure}

Fig.~\ref{fig:esym} shows the computed symmetry energy as a function of the 
temperature for four densities. The $E_{sym}$ was computed through the fitting 
procedure outlined in Appendix~\ref{nse-LT}; an analysis of the goodness of the 
fitting can be found in Ref.~\cite{dor2018}.  \\

Figure~\ref{fig:esym} shows several distinct regions for $E_{sym}$. In cooling, 
a liquid system with $T>2\,$MeV starts with a low value of $E_{sym}$. Upon 
entering the $T<2\,$MeV region and until $T\approx 1.5$ MeV, the symmetry energy 
increases in magnitude while the liquid pasta is formed.  Its value stabilizes 
in an intermediate value in the warm-to-low temperature range of $0.5$ 
MeV$<T<1.5$ MeV where the liquid pasta exists. At $T<0.5$ MeV, when the 
liquid-to-solid phase transition happens within the pasta, the symmetry energy 
reaches its highest 
value. We now look at these  stages in turn. \\

In the temperature range, $T>2\,$MeV, Figure~\ref{fig:esym} shows that $E_{Sym}$ 
has higher values for higher densities. Since a similar relationship is 
maintained by the Euler characteristic $\chi$ (c.f. Figure~\ref{mink_x05}), it 
is possible that there may be a connection between the symmetry energy and the 
morphology of the system. Remembering from Section~\ref{minkowski_results} that 
at those temperatures higher densities are associated to cavity-like or isolated 
regions, and lowest densities with tunnel-like structures, it is possible that 
$E_{sym}$ increases as the 
tunnels become obstructed and cavities or isolated regions prevail. \\ 

\begin{figure*}[!htbp]
\centering
\captionsetup[subfigure]{justification=centering}
\subfloat[$\rho=0.05$]{
\includegraphics[width=0.49\columnwidth]
{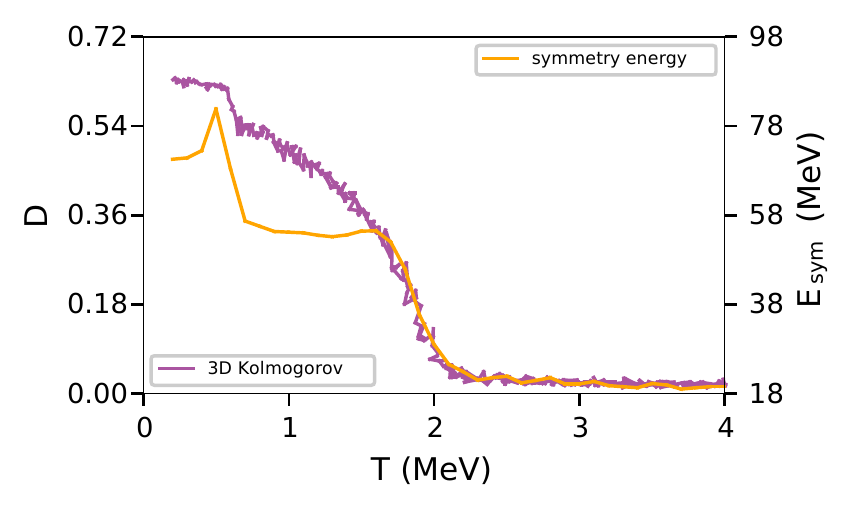}
} 
\subfloat[$\rho=0.06$]{
\includegraphics[width=0.49\columnwidth]
{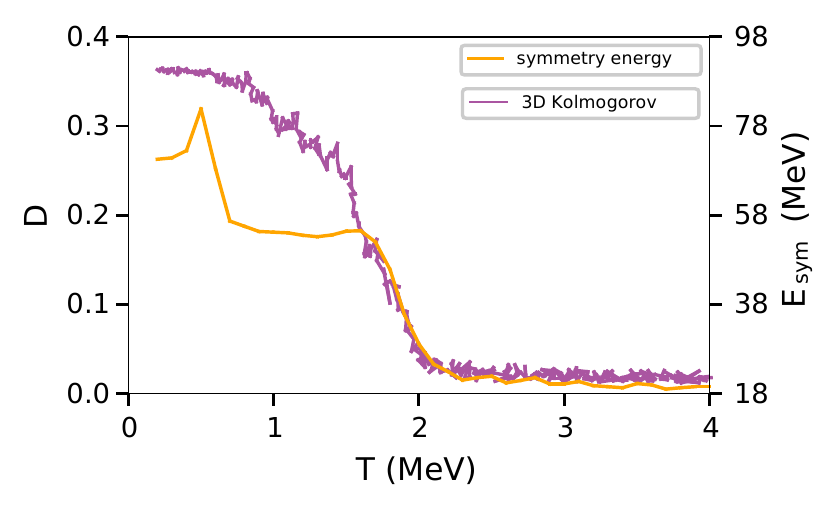}
} 
\\
\subfloat[$\rho=0.07$]{
\includegraphics[width=0.49\columnwidth]
{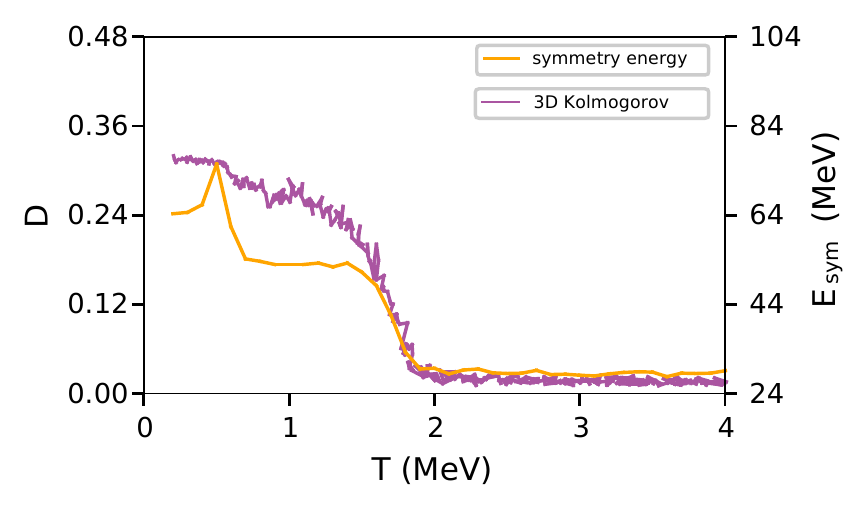}
} 
\subfloat[$\rho=0.085$]{
\includegraphics[width=0.49\columnwidth]
{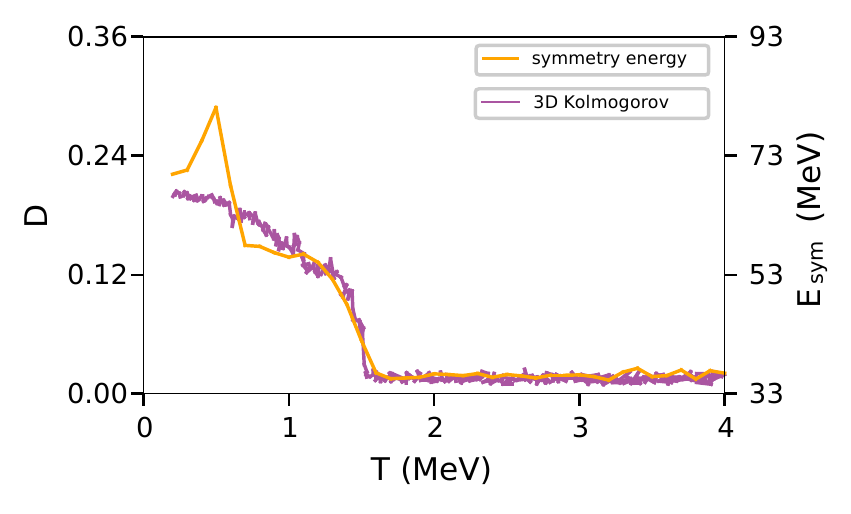}
} 
\caption{\label{fig:esym_kolmogorov}   Symmetry energy (right scale) and 3D 
Kolmogorov statistic (left scale) as a function of the temperature, for nuclear 
matter with 6000 nucleons at $x=0.5$ and the densities indicated below each plot 
(units fm$^{-3}$). As in Fig.~\ref{fig:rho05_x05_t20_all}, the Kolmogorov 
statistic was computed from a slowly cooled simulation cell, with temperature 
varying from $T=4\,$MeV down to $T=0.2\,$MeV.}
\end{figure*}

Between $1.5$ MeV $<T<2$ MeV the symmetry energy varies in a way resembling the 
variation of the 3D Kolmogorov statistic, $D$, during the pasta formation stage. 
Fig.~\ref{fig:esym_kolmogorov} compares the variation of $E_{sym}$ with that of 
$D$ as a function of temperature for the four densities of interest, and finds a 
good match between both quantities; one can then conclude that the variation of 
$E_{sym}$ can be also be associated to the changes in the morphology of the 
nuclear matter structure and, furthermore, that as the pasta is formed during 
cooling, the symmetry energy increases in magnitude. \\

The region between $0.5$ MeV $<T<1.5$ MeV corresponds to the pasta structures 
filled with liquid nuclear matter, and $E_{sym}$ changes at around $T\approx 
0.5$ MeV, at the same temperature at which the caloric curve and the Lindemann 
coefficient undergoes similar changes (cf. Fig.~\ref{lin}), indicating the phase 
transition between a liquid pasta ($T>$ 0.5 MeV) and a 
solid pasta ($T<$ 0.5 MeV). Based on this, one can conclude that the symmetry 
energy attains its largest value in the solid crystal-like phase. \\

Finally, it is worth mentioning that the values of $E_{sym}$ attained in this 
section are not directly comparable to those of Section~\ref{nseIT} for the 
liquid-gas coexistence region and compared to experimental results. This is the 
case because, first, Figure~\ref{ESym-Dens} was obtained at higher temperatures 
($2$ MeV $< T < 5$ MeV) and lower densities ($\rho<$ 0.03$\,$fm$^{-3}$), and, 
second, it corresponds to an homogeneous medium while the present one uses data 
from a pasta-structured system. In spite of these differences, it is reassuring 
that the values of $E_{sym}$ obtained in the present study for the highest 
temperatures used, $T>2$ MeV, are within the range of values calculated in 
Section~\ref{nseIT} for the highest densities considered in such study, namely 
0.05 fm${}^{-3}<\rho<$ 0.06 fm${}^{-3}$. \\


\subsection{Summary of nuclear matter properties }\label{nm-s}

\subsubsection*{Summary of nuclear matter properties at intermediate 
temperatures}\label{nm-s-IT}

In Section~\ref{bulk} the bulk properties of nuclear matter were obtained along 
with their isospin dependence. It was determined that isospin symmetric and 
asymmetric matter can be self bound for low values of T. As $T$ increased from 1 
to 5 MeV their saturation densities varied from $\rho_0=0.16$ fm$^{-3}$ to 
$0.12$ fm$^{-3}$ for isospin symmetric matter, and from $\rho_0\approx 0.12$ 
fm$^{-3}$ to $0.09$ fm$^{-3}$ for matter with $x=0.4$. Likewise, isospin 
asymmetry reduces the compressibility at saturation density by about $30\%$ to 
$50\%$ as $x$ drops from 0.5 to 0.3 and as $T$ increases from 1 to 5 MeV.\\

In Section~\ref{phasesIT} it was shown that nuclear matter can exist in liquid 
and gaseous phases and there are transitions between them. The liquid phase was 
identified at low temperatures for all isospin asymmetries studied by the $\cup$ 
shapes of the energy $E(T,\rho)$.  At sub-saturation densities the 
energy-density curves of $x=0.4$ and 0.5 signaled a transition to a liquid-gas 
mixture phase, while $x=0.3$ systems maintains phases and transitions among them 
at $T = 1$ MeV but become unbound at all densities at T = 3 MeV and higher. The 
addition of the $x$ dependence to the phase diagram alloed its representation in 
Section~\ref{phasediag} in the 3D space of $T, \rho, x$; the liquid-gas 
coexistence region extends, approximately, to densities of up to $\rho_0/2$, 
temperatures of up to 16 MeV, and isospin content as low as of 13\%. \\

The symmetry energy of nuclear matter at intermediate temperatures at low 
densities was calculated in Section~\ref{nseIT}, and compared satisfactorily to 
experimental data. $E_{Sym}$ is susceptible to the clusterization that occurs in 
phase transitions.\\

Besides corroborating previous studies, this work extends some of their results 
to other values of isospin content and non-zero temperatures. Findings that we 
believe are new are the temperature variation of the saturation density and 
compressibility for isospin asymmetric matter, certain details of the existing 
phases at $x=0.3$ and 0.4, the extension of the phase diagram into the isospin 
axis, as well as a new procedure to estimate $E_{sym}(T,\rho)$ from kinetic 
simulations. Finally, it is worth mentioning that the CMD model indeed helps to 
understand the role of isospin on several nuclear properties. \\

\subsubsection*{Summary of nuclear matter properties at low 
temperatures}\label{nm-s-LT}

In Section~\ref{nm-LT} the formation of the pasta-like structures in nuclear 
matter, its phase transitions and its symmetry energy. were investigated.  In 
section~\ref{nmltsd} the crystalline structure of NM was studied at zero and 
near-zero temperatures. Using scaling and CMD it was found that the potentials 
used yield a simple cubic structure. CMD indicated the departure from the 
crystalline state to a non-homogeneous pasta at sub-saturation densities; such 
structures are maintained up to $T \approx 1$ MeV; this was evidenced by the use 
of curvature and Euler characteristic. \\

In Section~\ref{propNMP} it was determined that the pasta exist in isospin 
symmetric and non-symmetric systems. Furthermore, solid-to-liquid phase 
transitions within the pasta were detected. Using the radial distribution 
function, the Lindemann coefficient, Kolmogorov statistics, Minkowski 
functionals the morphology of the pastas was succesfully linked to the phases 
and phase transformations. Nucleons can exist in liquid and crystal phases 
inside the pasta structures. It was also found that non-isospin-symmetric pasta 
depart from the classification of symmetric nuclear matter presented in 
Table~\ref{tab1}. \\

The symmetry energy in pastas was studied in Section~\ref{subsec:esym}.  
connection to the morphology of the pasta and to the phase transitions. 
$E_{Sym}$ showed different values as a function of the temperature and density, 
i.e. in the different phases of nucleons inside the pasta.  $E_{sym}$ is 
associated to the morphology of the nuclear matter structure, and it attains its 
largest value in solid pastas. \\

In conclusion, classical molecular dynamics simulations show the formation of 
pastas in isospin symmetric and non-symmetric systems. The computational tools 
developed and applied, although not perfect, demonstrated their usefulness to 
detect the in-pasta phase transitions first seen in Ref.~\cite{dorso2014}, and 
to extend the calculation of the symmetry energy of Ref.~\cite{lopez2017} 
to lower temperatures, and connect its value to the structure and thermodynamics 
of the neutron-rich pasta.  \\

The bulk properties of nuclear matter, its phases and phase transitions, along 
with the behavior of $E_{sym}$ is indicative of similar phenomena present in 
neutron star matter (NSM). Neutron star matter, however, is a system different 
than pure nuclear matter; the connection of NM and NSM will be investigated in 
Section~\ref{Electron-gas}, and the behavior of NSM in Section~\ref{nsm}.\\


\newpage

\section{Electron gas: connecting nuclear matter with neutron star 
matter}\label{Electron-gas}

As explained in the Introduction (c.f. Section~\ref{intro}), nuclear matter is 
connected to neutron star matter by means of an electron gas that fills all 
space between nucleons. In addition to neutrons and protons, neutron star crusts 
also contain electrons which fill the space between nucleons. That is, neutron 
star matter can be thought of as nuclear matter embedded in an electron gas. \\

Many authors (see e.g.~\cite{horo_lambda,horo-s,Horo2004,horo-2006}) believe the 
pasta is nothing but frustrated structures formed by a competition between 
nuclear and Coulomb forces; obviously, as seen in the section devoted to nuclear 
matter (Section~\ref{Nuclear-Matter}), this is not necessarily the case; nuclear 
repulsive $nn$ and $pp$ and attractive $np$ forces form pastas as well. Thus, 
before embarking on a full study of neutron star matter, we find convenient to 
dissect the effect an electron gas has on the nuclear matter pastas. \\

As it will be explained below, to introduce an electron gas in a CMD study of NM 
pasta, the addition of an screened Coulomb potential is necessary. This section 
focuses on the effects the strength and range of such interaction have on the 
morphology of the pastas. \\

At this point it is worth mentioning that, as it will be seen in 
Section~\ref{screen}, the pasta-like structures calculated without the proper 
screening length introduce spurious effects due to the size of the cell and form 
only one "pasta like" structure per cell; such ``pastas'' are not a true nuclear 
pasta property, but their use in the analysis does not invalidate the study of 
the effect of the strength of the screened Coulomb potential. \\

\subsection{The strength of $V_C$}

To mimic neutron star matter it is necessary to include the Coulomb effect of an 
electron gas in the nuclear matter studied in Section~\ref{Nuclear-Matter}. CMD 
is specially well suited for this study as it provides a microscopic view of the 
structures formed. The electron cloud introduces an screening effect on the 
Coulomb potential of the protons, which in turn modifies the pasta structures 
seen in Section~\ref{phasesIT}.  The techniques to introduce the Coulomb 
interaction of an electron gas into CMD are presented in 
Appendix~\ref{cmd_star-2}.\\

The effect of an electron gas on cold nuclear structures has been studied 
before.  In 2003 a static liquid-drop model was used, and it was found that its 
main effect was to extend the range of densities where bubbles and clusters 
appear~\cite{30}. Later, in a 2005 study a density functional method was used 
(at zero temperature) to find that the density region in which the pasta exists 
becomes broader when the electron gas is taken into 
account~\cite{Maruyama-2005}. More recently, Monte Carlo simulations, which 
approximated the Coulomb interaction via an Ewald 
summation~\cite{horo_lambda,Horo2004,P15,P2012}, determined the lowest proton 
fractions that are compatible with $\beta$ equilibrium in neutron star crust 
environments (at $T=1.0 \ MeV$).\\

It is worth mentioning that a system with a coulomb term of just one sign (as 
would be the case of nuclear matter if protons are considered as carrying 
charge) is unstable and thermodynamically non additive. But the inclusion of the 
electrons which render the coulomb term of finite range via screening  
transforms the system into stable and thermodynamically additive, and as such 
the energy will be additive as well as the entropy.

In this Section we study the effect the strength electron gas has on the 
formation of the nuclear pasta at non-zero temperatures through the calculation 
of the pasta structures with and without effect of the electron gas as well as 
with {\it softened interactions} which --of course-- do not exist in nature. 
Full details of this study can be found in ~\cite{lopram2015}.\\

\begin{figure}[h]  
\begin{center}
\includegraphics[width=5.5in]{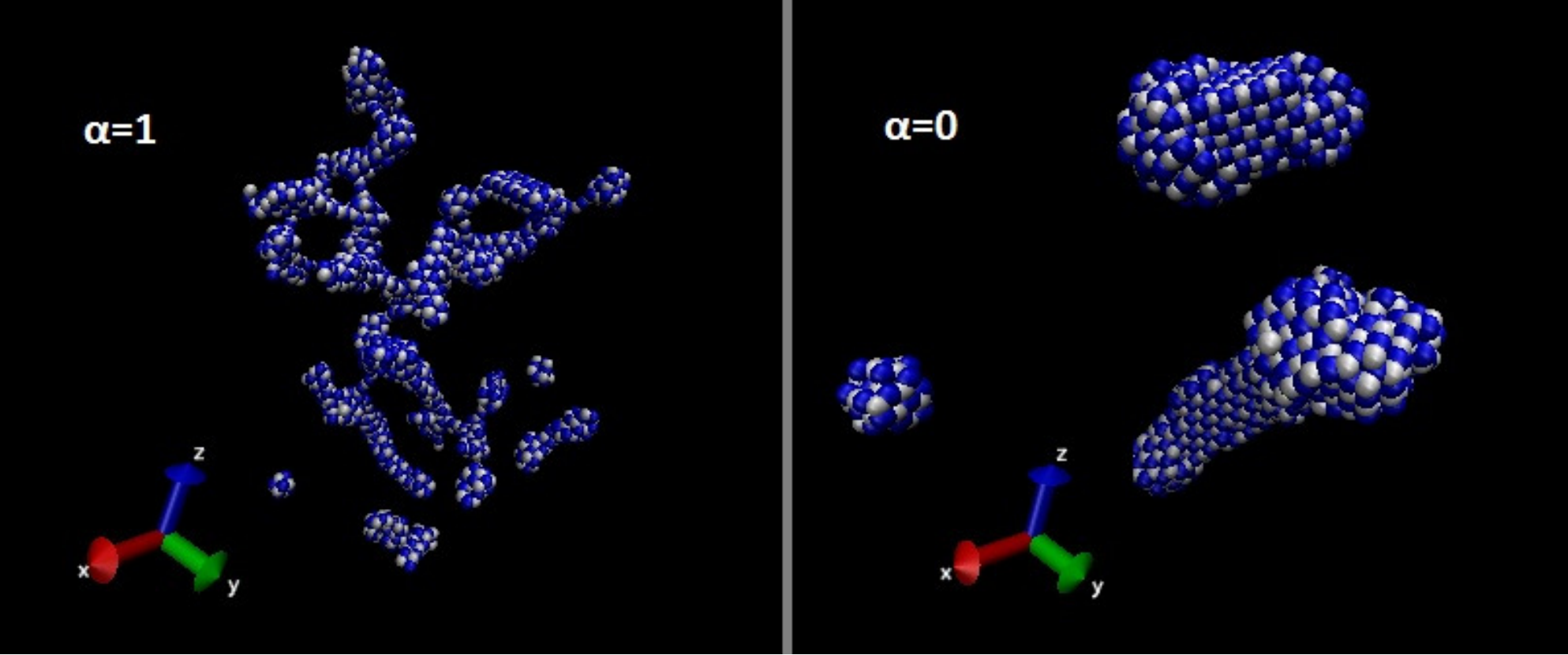}
\end{center}
\caption{Pasta structures obtained with ($\alpha$=$1$) and without 
($\alpha$=$0$) the proton-electron gas interaction for symmetric matter 
($x=0.5$) at density $\rho=0.015 \ fm^{-3}$ and temperature $T=0.1 \ MeV$.} 
\label{x5t1d015}
\end{figure}

\subsubsection*{Simulation procedure}\label{procedure}

To study nuclear matter embedded in a degenerate electron gas, the electron gas 
density is first taken as to produce an overall neutral ($\beta$-equilibrated) 
system. The electron gas effectively screens out the positive charges of the 
protons, resulting in a more local effect. This is taken into account through 
the use of a Thomas-Fermi screened Coulomb potential, $V_{C}=(q^2/r) 
e^{-r/\lambda}$; this potential is introduced in Appendix~\ref{cmd_star-2}.\\

\begin{figure}[b]  
\begin{center}
\includegraphics[width=4.55in]{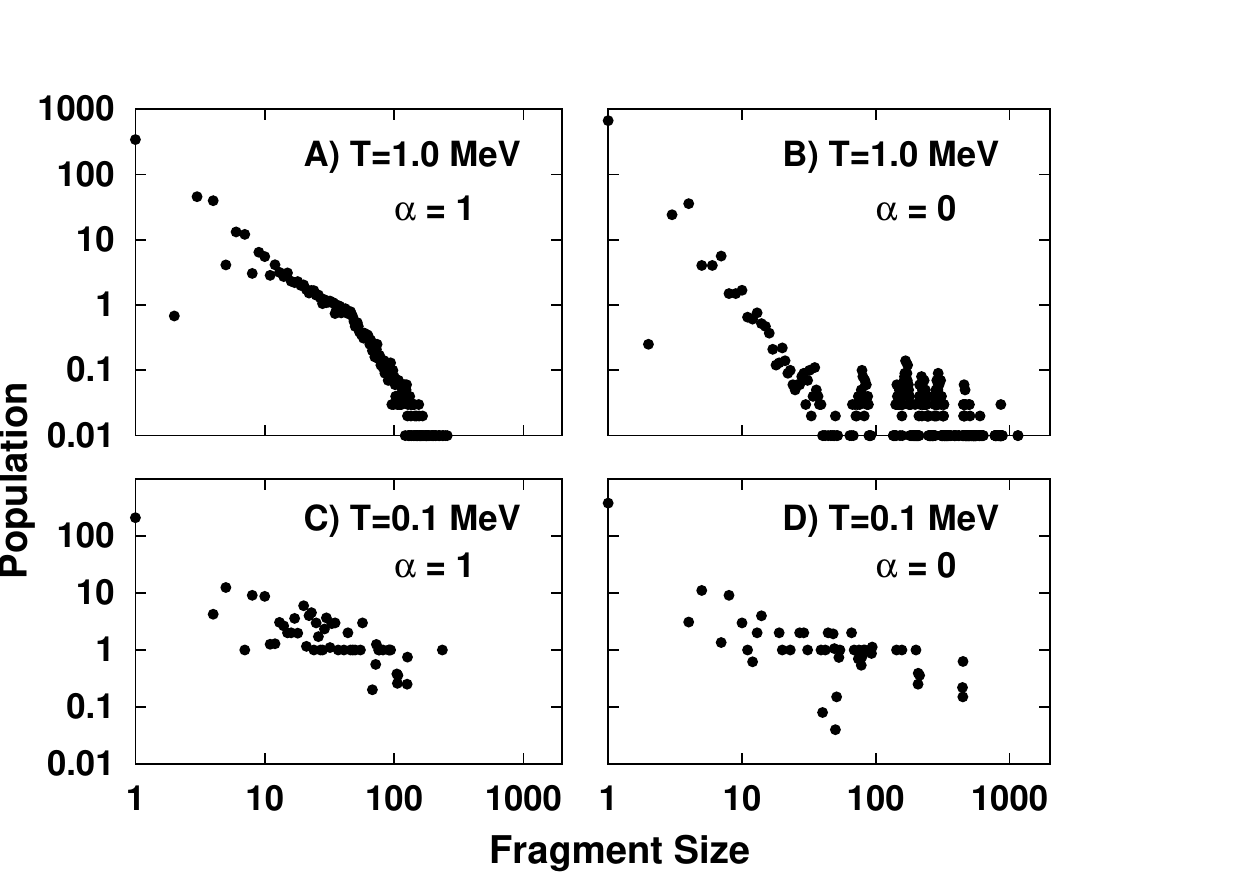}
\end{center}
\caption{Effect of the screened potential on the fragment size multiplicity.  
Plots show the distribution of cluster sizes observed in 200 configurations of 
$x=0.3$ nuclear matter at density $\rho=0.015 \ fm^{-3}$ and temperatures $T=1.0 
\ MeV$ (top) and $T=0.1 \ MeV$ (bottom). The figure on the left panel correspond 
to configurations with Coulomb interaction, and those
on the right to the case without such potential.}
\label{massDistx3D015}
\end{figure}

We focus on CMD simulations of isospin symmetric systems of 1000 protons and 
1000 neutrons, and isospin asymmetric systems comprised of $1000$ protons and 
$2000$ neutrons, i.e. of $x=0.3$. Nucleons are placed in cubical boxes with 
densities between $0.01 \ fm^{-3} \le \rho \le  \rho_0$, temperatures between 
$0.1 \ MeV \le T \le 1.0 \ MeV$ and periodic boundary conditions; these 
conditions correspond to semi-frozen states where pastas are known to exist in 
NM. In particular, we produce pastas as in the previous sections, and compare 
their structures obtained with and without the electron gas as well as with 
varying strengths of it. In summary, pasta structures were obtained at 
subsaturation densities ($0.015 \ fm^{-3} \le \rho \le 0.072 \ fm^{-3}$) and low 
temperatures ($0.1 \ MeV \le T \le 1.0 \ MeV$) for symmetric ($x=0.5$) and 
asymmetric ($x=0.3$) cases. Altogether $200$ simulations were carried out per 
each combination of $\{\rho,T,x\}$.  

To see the effect of the electron gas, each case was ``cooked'' repeatedly with 
a screened Coulomb potential with a varying amplitude, i.e. with $\alpha V_C$, 
where $0 \le \alpha \le 1$.  For a fair comparison, all corresponding cases
with different values of $\alpha$ were produced with identical initial 
conditions of $\{x,\rho,T\}$ and started off from the same initial random 
configuration.  As an illustration, Figure~\ref{x5t1d015} shows two 
corresponding structures obtained with and without the electron gas.\\

\begin{figure}[t]  
\begin{center}
\includegraphics[width=3.2in]{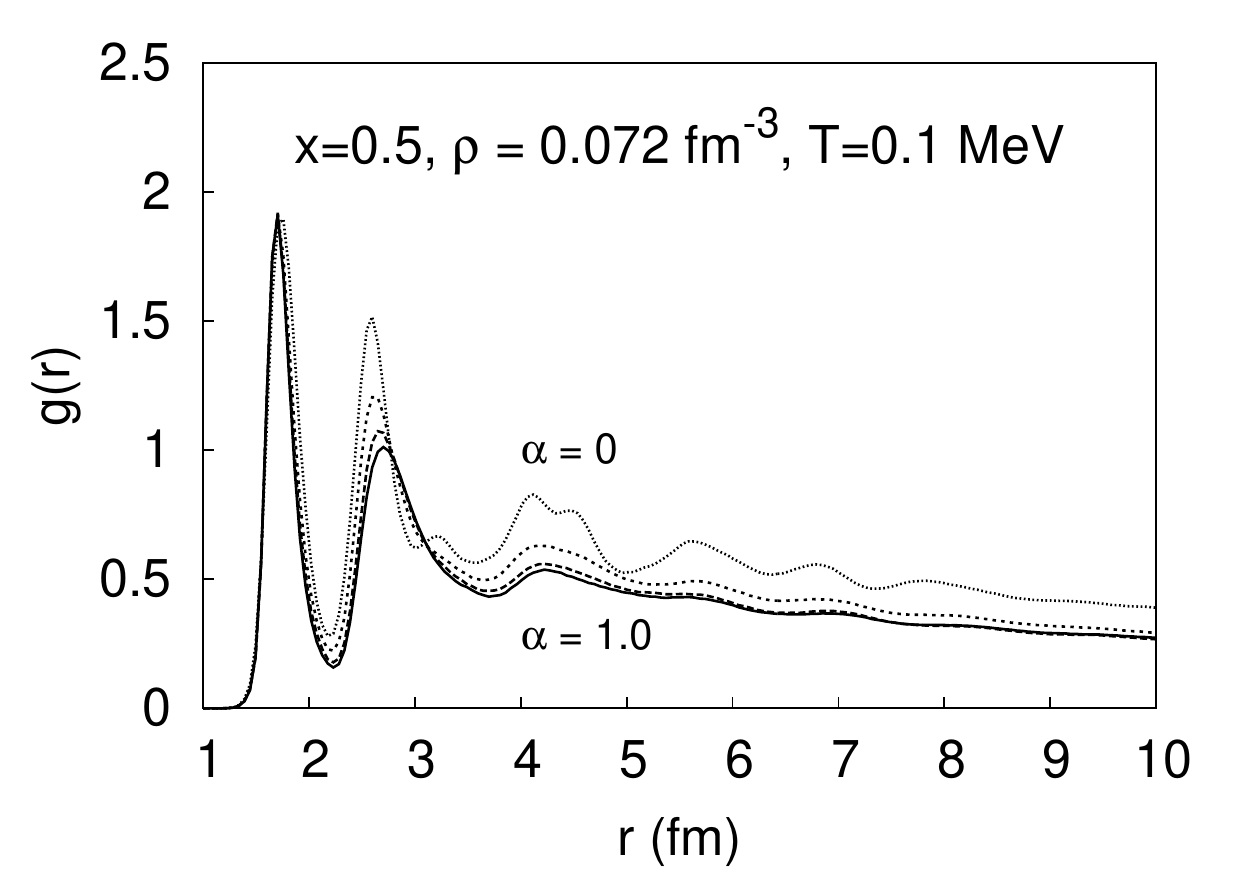}
\end{center}
\caption{Examples of the radial correlation function for varying strengths of 
the Coulomb potential: $\alpha=1$ (full Coulomb), $0.8$, $0.2$, and $0$ (without 
Coulomb).}\label{Radial}
\end{figure}

\begin{figure}[t]  
\begin{center}$
\begin{array}{cc}
\includegraphics[width=3.4in]{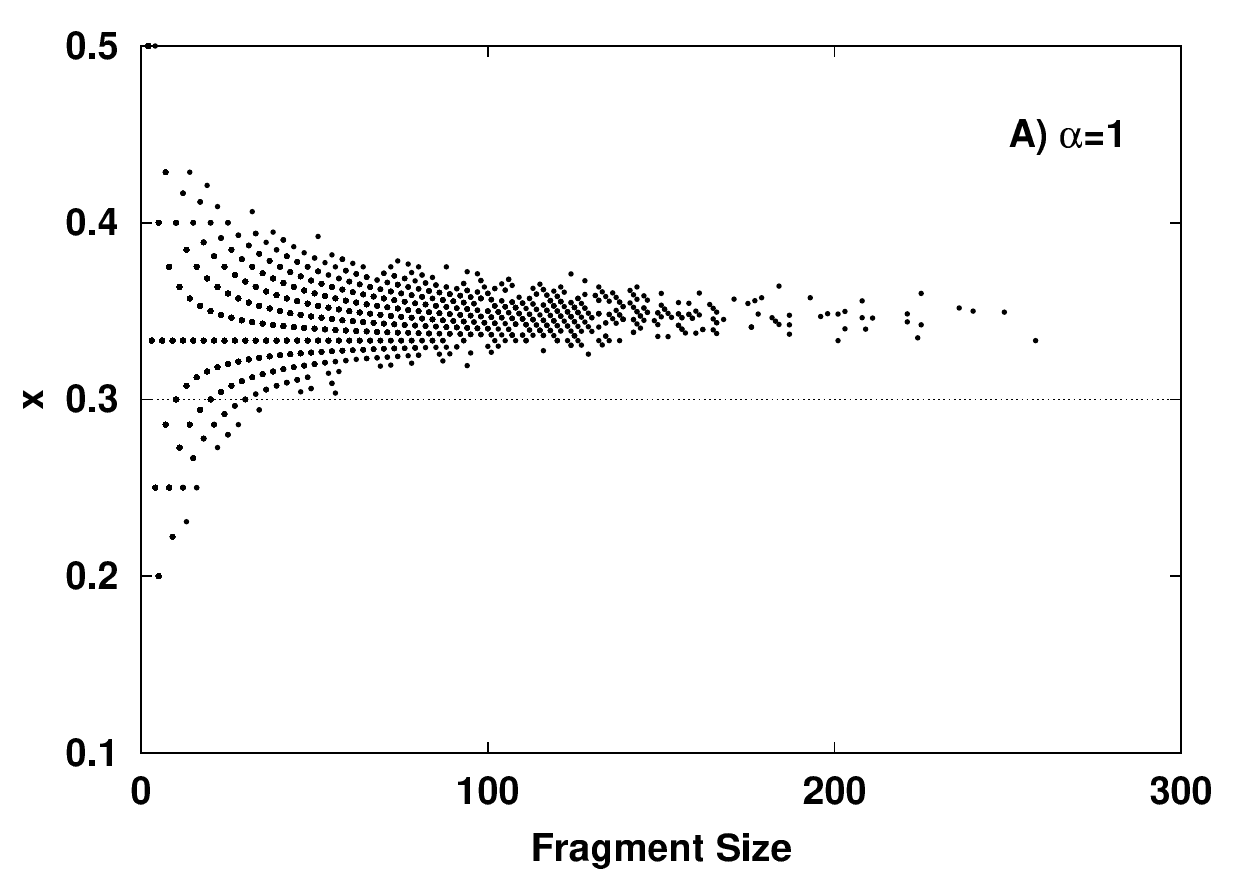} 
\includegraphics[width=3.4in]{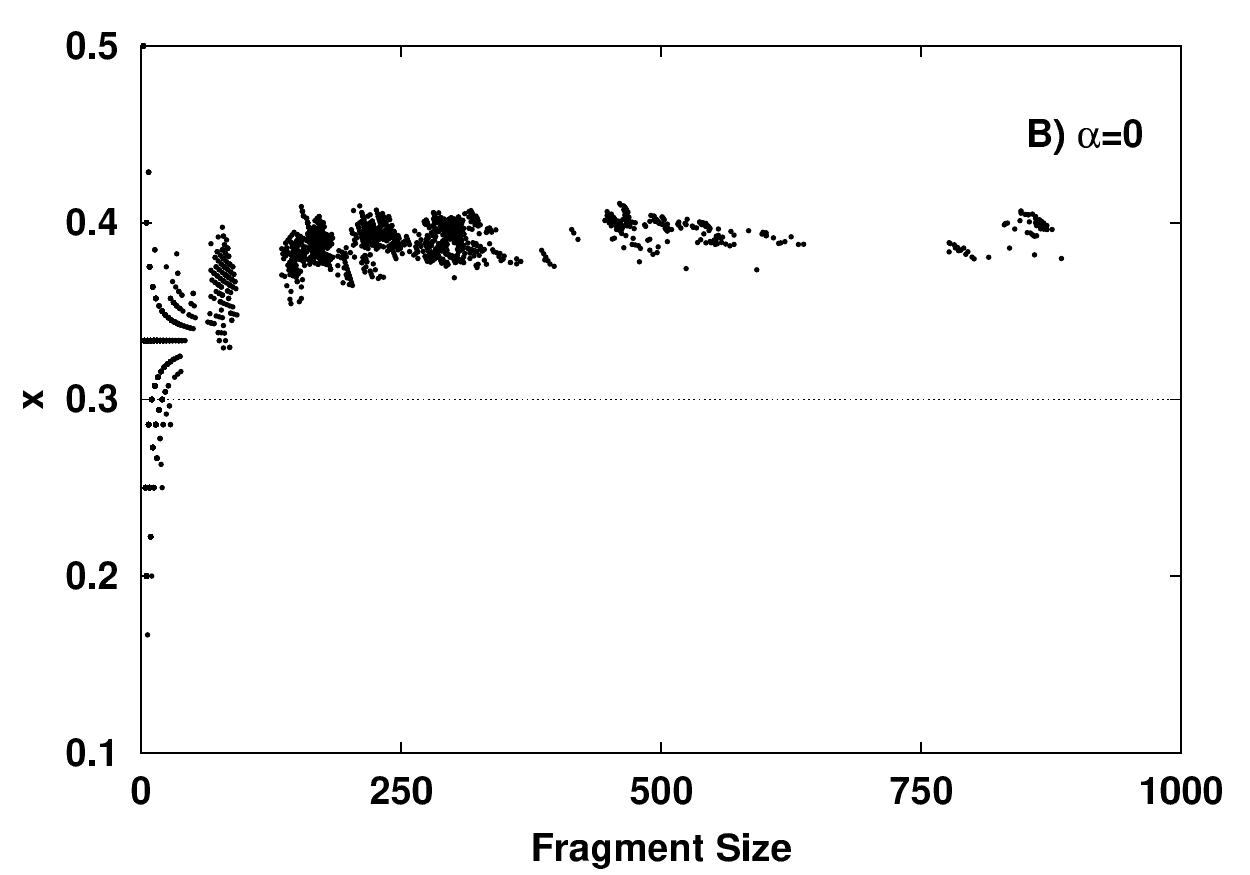}
\end{array}$
\end{center}
\caption{$x$ content of the clusters formed in $200$ configurations of 
asymmetric matter ($x=0.3$) at density $\rho=0.015 \ fm^{-3}$ and temperature 
$T=1.0 \ MeV$. The top panel shows the case with Coulomb and the bottom one the 
case without such interaction. Notice that the abscissas have different scales.} 
\label{XfragX3D015T1}
\end{figure}

We use several tools for inspecting the pastas produced. At the densities and 
temperatures selected, the nucleons do not have large mobilities and self-bound 
cumuli can be detected with a simple ``Minimum Spanning Tree'' ($MST$) 
algorithm, which finds particles closer to each other than a clusterization 
radius set to $3.0 \ fm$, and yields information about the fragment 
multiplicity.  Again, we limit the cluster analysis to the central cell ignoring 
the neighboring cells of the periodic boundary condition, and thus eliminate 
possible clusters that may continue into other cells and can be, in principle, 
infinite in size. \\

At a microscopic level, the dynamics of the nucleons can be quantified through 
their average displacement as a function of the ``time'' steps of the 
simulation.  Likewise, the microscopic stability of the clusters can be gauged 
through the ``persistency''~\cite{lopezlibro} which measures the tendency of 
members of a given cluster to remain in the same cluster. Another interesting 
descriptor is the isospin content $x$ of each cluster produced.  Other global 
characterization tool is the pair correlation function, and the Minkowsky 
functionals.\\

\subsubsection*{The effect of the strength of $\alpha V_C$}\label{Coulomb}

\begin{figure}[h]  
\begin{center}
\includegraphics[width=3.in]{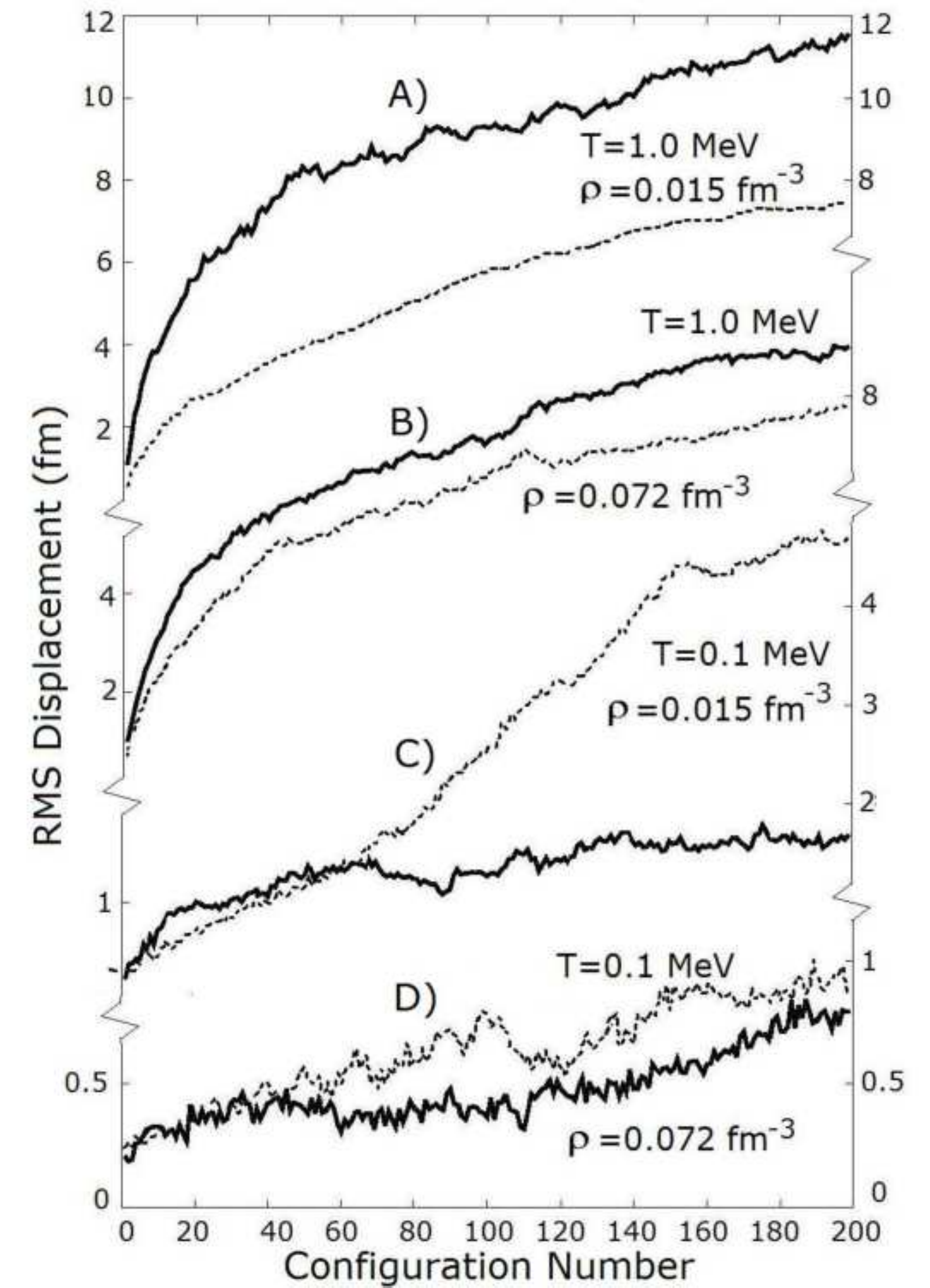}
\end{center}
\caption{$RMS$ displacement of nucleons during the evolution of the stationary 
ergodic process with Coulomb (continuous line) and without Coulomb (dashed line) 
as labeled; see text for details. Notice that at low temperatures the mobility 
with Coulomb is smaller than without Coulomb, while at higher temperatures the 
effect is the opposite.} \label{desplx5T1}
\end{figure}

\begin{figure}[h]  
\begin{center}
\includegraphics[width=3.in]{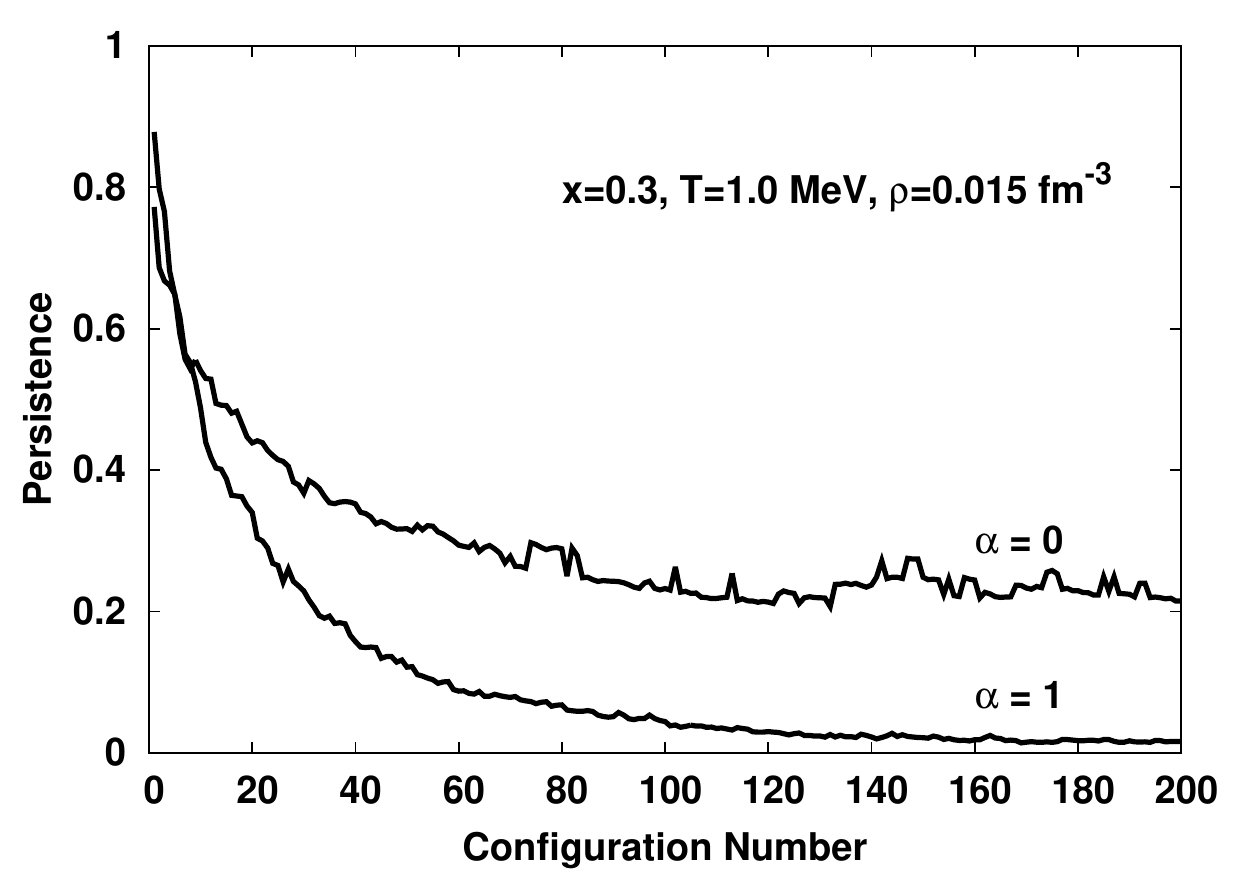}
\end{center}
\caption{Persistence with and without Coulomb as it evolves
through 200 configurations of the stationary ergodic process.}
\label{persis}
\end{figure}

The presence of an electron gas affects the fragment size multiplicity.  
Figure~\ref{massDistx3D015} shows the cluster multiplicity observed in 200 
configurations of
asymmetric matter with and without the Coulomb interaction, at $x=0.3$, 
$\rho=0.015 \ fm^{-3}$ and $T=0.1$ and $1.0$ MeV. The increase of the number of 
fragments of sizes $A \gtrsim 500$ as $\alpha$ goes from $\alpha=1$ to 
$\alpha=0$ underline the role the electron gas has on determining the mass 
distribution.\\

\begin{table}[ht]
\centering  
\caption{Classification Curvature - Euler}
\begin{tabular}{c|c|c|cc|cc} 
\hline \hline                        
 & $\rho$ & $T$ & \multicolumn{2}{c|}{$\alpha=1$} & 
\multicolumn{2}{c}{$\alpha=0$} \\
& ($fm^{-3}$) & ($MeV$)  & Curvature & Euler & Curvature & Euler \\
[0.5ex]
A & 0.072 & 1.0 & -14.5 & 8.9 & -0.2  & 0.47 \\
B & 0.072 & 0.1 & -3.18 & -1.58 &  0.245   &  -0.19 \\
C & 0.015 & 0.1 & 14.76 & 0.89 &  4.28 &   0.35 \\
D & 0.015 & 1.0 & 18.43 &  -0.17  & 2.28 &  0.86 \\
E & 0.015 & 1.0 & 75.75 & -61.2 & 91.76 & -24.3 \\
F & 0.015 & 0.1 & 90.03 & 45.4   & 95.3 &  15.9 \\
G & 0.072 & 1.0 & -24.74 & 44.7 & -37.6 & 60.2 \\
H & 0.072 & 0.1 & -37.38 & 79.8 & -49.28 & 93.3 \\
[1ex]      
\hline 
\end{tabular}
\label{table2}
\end{table}

The change of the inner structure can be quantified through the use of the 
radial distribution function. Figure~\ref{Radial} shows the $g(\mathbf{r})$ of 
symmetric structures at $x=0.5$, $\rho=0.072 \ fm^{-3}$ and $T=0.1 \ MeV$ and 
for varying strengths of the Coulomb potential, namely $\alpha=1$ (full 
Coulomb), $0.8$, $0.2$, and $0$ (without Coulomb).  The electron gas appears not 
to have an effect on the nearest neighbor distances.\\

\begin{figure}[t]  
\begin{center}$
\begin{array}{cc}
\includegraphics[width=3.4in]{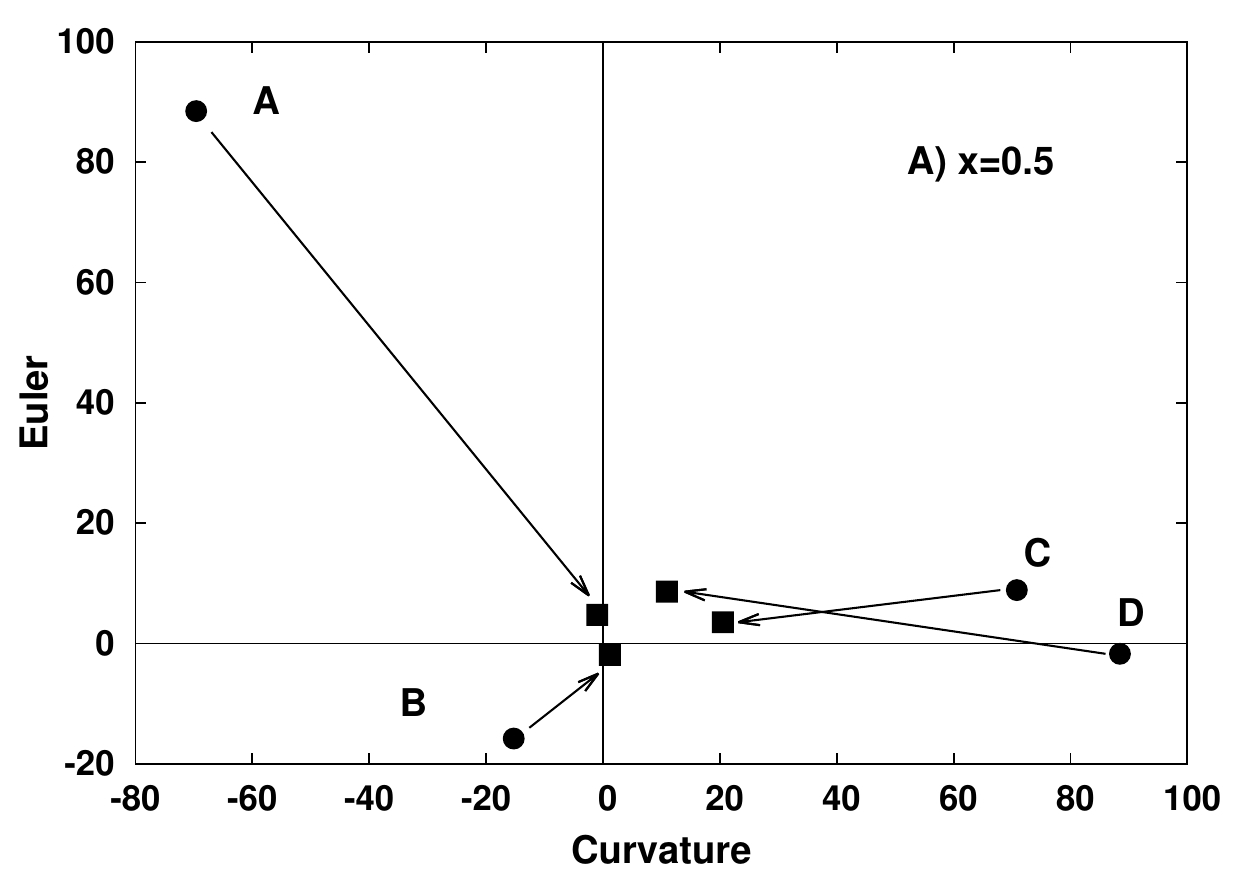} 
\includegraphics[width=3.4in]{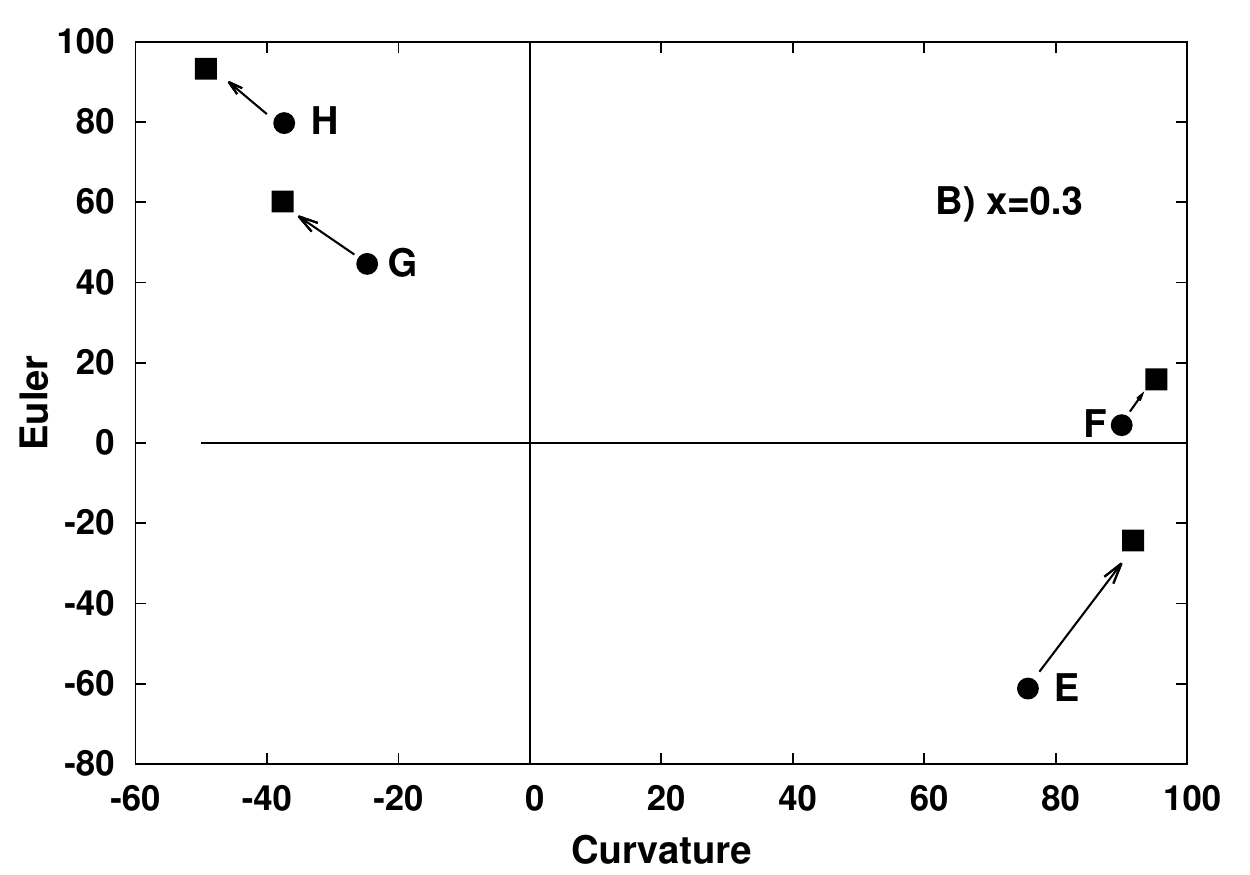}
\end{array}$
\end{center}
\caption{Average values of the Curvature and Euler numbers of the
structures listed in Table~\ref{table2}; circles correspond to
structures with Coulomb and squares to structures without Coulomb,
arrows indicate the average displacement of the structures as
$\alpha$ goes from $1$ to $0$. } \label{curv-euler}
\end{figure}

The isotopic content on the cluster is also affected by the presence of the 
electron gas. Figure~\ref{XfragX3D015T1} shows the $x$ content of the clusters 
formed in $200$ configurations of
asymmetric matter with $x=0.3$, $\rho=0.015 \ fm^{-3}$ and $T=1.0 \ MeV$. The 
electron gas clearly enhances the isotopic content from $x=0.3$ up to $x\approx 
0.4$, as well as the maximum fragment size which grows up to $A\approx 800$.\\

As suspected in~\cite{Maruyama-2005}, the interactions between protons and 
electrons are responsible for the rearrangement of protons in varying degrees.  
Figure~\ref{desplx5T1} shows the $RMS$ displacement of nucleons with and without 
Coulomb during the evolution of the ergodic process.  Each curve is made of 200 
points, each representing the average displacement with respect to the original 
configuration of the $2000$ nucleons in systems with $x=0.5$, $\rho=0.015$ and 
$0.72 \ fm^{-3}$ at $T=0.1$ and $1.0 \ MeV$). At low temperatures Coulomb 
enhances nucleon mobility more than at high temperatures.\\

Figure~\ref{persis} shows the ``persistence'', (i.e. the tendency of nucleons to 
remain in the same cluster), with and without Coulomb, as it evolves through 200 
configurations of the stationary ergodic process, measured with respect to the 
original configuration of $3000$ nucleons with  $x=0.3$, $\rho=0.015$ and $T=1.0 
\ MeV$.  As with the mobility, Coulomb enhances the transfer of nucleons thus 
decreasing the persistence as shown by the two curves.\\

The Euler characteristic and mean curvature introduced in Appendix~\ref{mink} 
can demonstrate the effect of the electron gas on the shape of the pastas. The 
mean curvature and the Euler characteristic were calculated from the digitized 
polyhedra constructed for each of the 200 nuclear structures obtained with and 
without Coulomb at each value of $\{x,\rho,T\}$. The values of the curvature 
varied approximately from -6,000 to 12,000 and those of the Euler characteristic 
from -700 to 1,000. Table~\ref{table2} shows the curvature-Euler number as 
percentages of the maximum values obtained, and Figure~\ref{curv-euler} shows 
their location on the curvature-Euler plane as well as their changes in position 
as the Coulomb strength is diminished; the standard deviations are of the size 
of the points used in the plots.\\


\subsection{The screening length}\label{screen}

A crucial parameter of $V_C$ is the screening length, $\lambda$. Here we study 
the effect this parameter has on the properties of the pasta within the 
framework of CMD. In particular, we study pasta properties with values of 
$\lambda$ ranging from $\lambda=0$ to 20 fm with densities $\rho=0.005$, 0.03, 
0.05, 0.08 fm$^{-3}$, and $\lambda=30\,\text{fm}$ and $\lambda=50\,\text{fm}$ 
with $\rho=0.005\,\text{fm}^{-3}$, where ``gnocchi'' are formed. The cut-off 
length was chosen at $r_c=\lambda$.  \\
 
\subsubsection*{Critical screening length}\label{lambda_c}

As observed in previous works~\cite{horo13,2013}, in absence of any Coulomb 
interaction (equivalent to $\lambda=0$), pasta-like structures exist, although 
only one per cell, which indicates that the structure is limited by the periodic 
boundary conditions imposed on the box. When there is Coulomb interaction, 
however, the competition between opposing interactions gives rise to a 
characteristic length which shapes the pasta structures. By increasing the value 
of $\lambda$, starting from $0\,\text{fm}$, we determine the critical 
$\lambda_c$ at which the pasta structures go from the artificial \emph{one 
structure per cell}, to the more realistic case of more than one structure per 
cell. \\

A way to determine $\lambda_c$ is by inspecting the nucleons pressure, i.e. the 
mechanical stability of the structures formed. The pressure is computed by the 
virial equation~(\ref{PandE}) with $\left< \rho T \right>={N k_B T}/V$, applying 
it only to the nucleons and not to the electron gas. \\

\begin{figure}[h!]  \centering
\includegraphics[width=0.5\columnwidth]{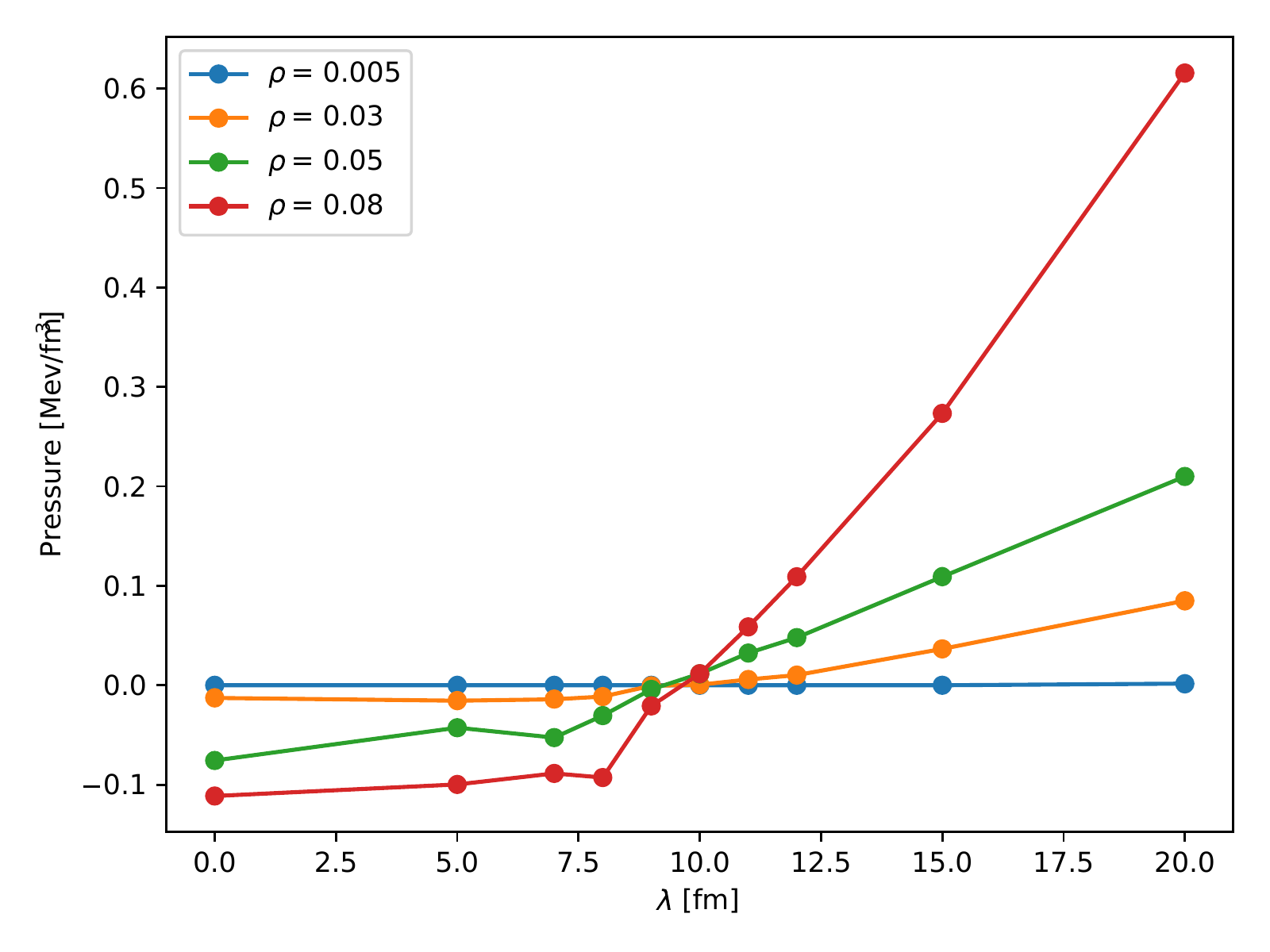}
\caption{Pressure as a function of $\lambda$ for different
  densities. We see that for $\lambda<10\,\text{fm}$, the pressure is
  negative, implying that periodic boundary conditions are affecting
  the morphology of the solution.}
\label{fig:pre}
\end{figure}

Negative pressure indicates unstable structures kept in place by the periodic 
replicas. In this case the overall effective interaction is mostly attractive 
and periodic boundary conditions play a major role in shaping the structure. The 
pastas formed under negative pressure are artificial and can only exist under 
periodic boundary conditions (see~\cite{binder,2013}). Figure~\ref{fig:pre} 
shows that for all $\lambda<10$ fm the pressure is negative.\\

Positive nucleon pressure, on the other hand, indicates that the structures form 
due to the physical interactions and not to the boundary conditions. Indeed in 
the positive-pressure regime the density fluctuations are of smaller sizes than 
the size of the cell, and the morphology of the structures changes drastically. 
Figure~\ref{fig:pre} shows that, for those conditions, the pressure becomes 
positive for $\lambda>10$ fm.\\

\begin{figure*}[!htbp]
\centering
\subfloat[Surface.]{
\includegraphics[width=0.3\columnwidth]
{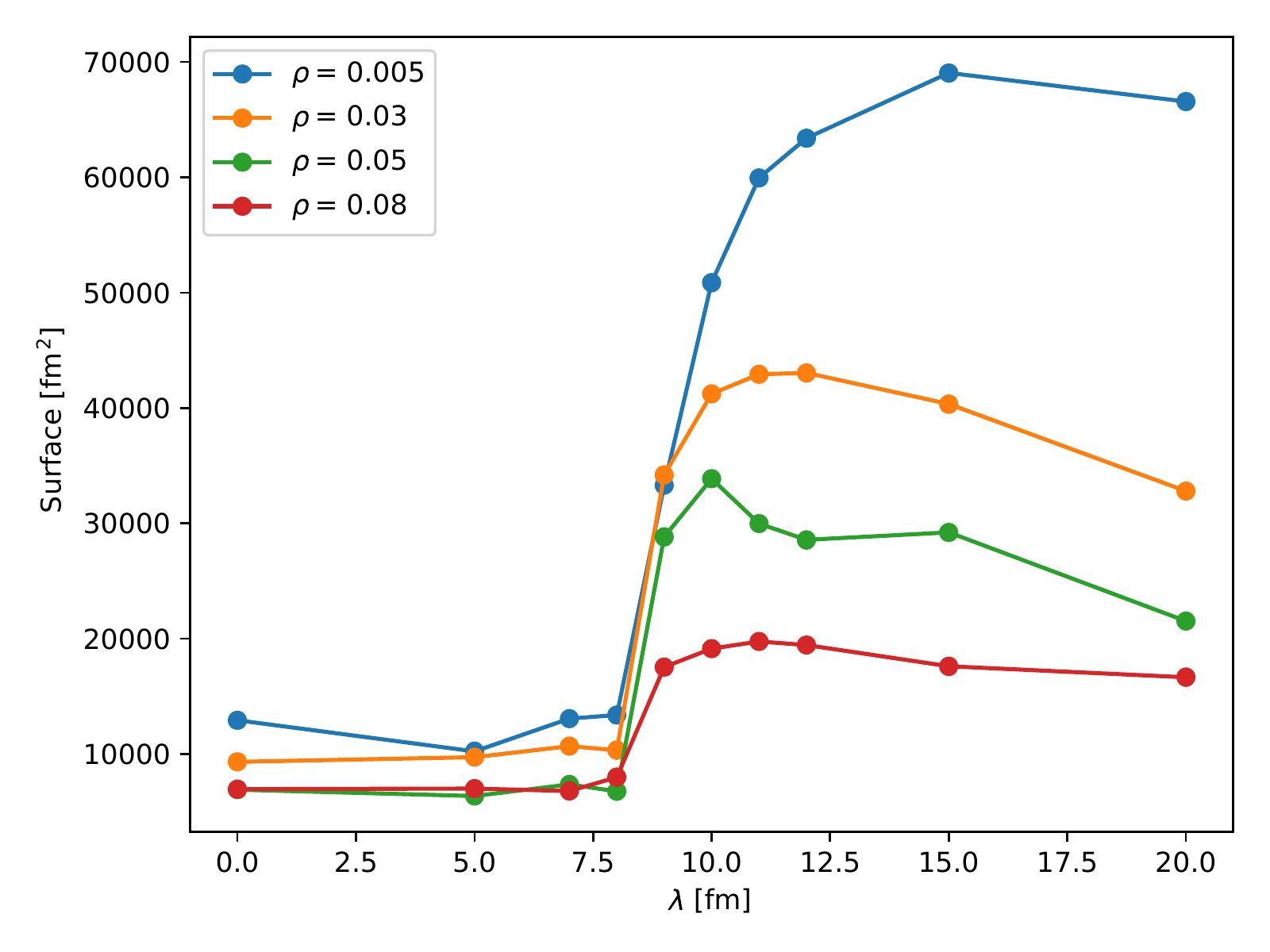}
}
\subfloat[Mean Breadth.]{
\includegraphics[width=0.3\columnwidth]
{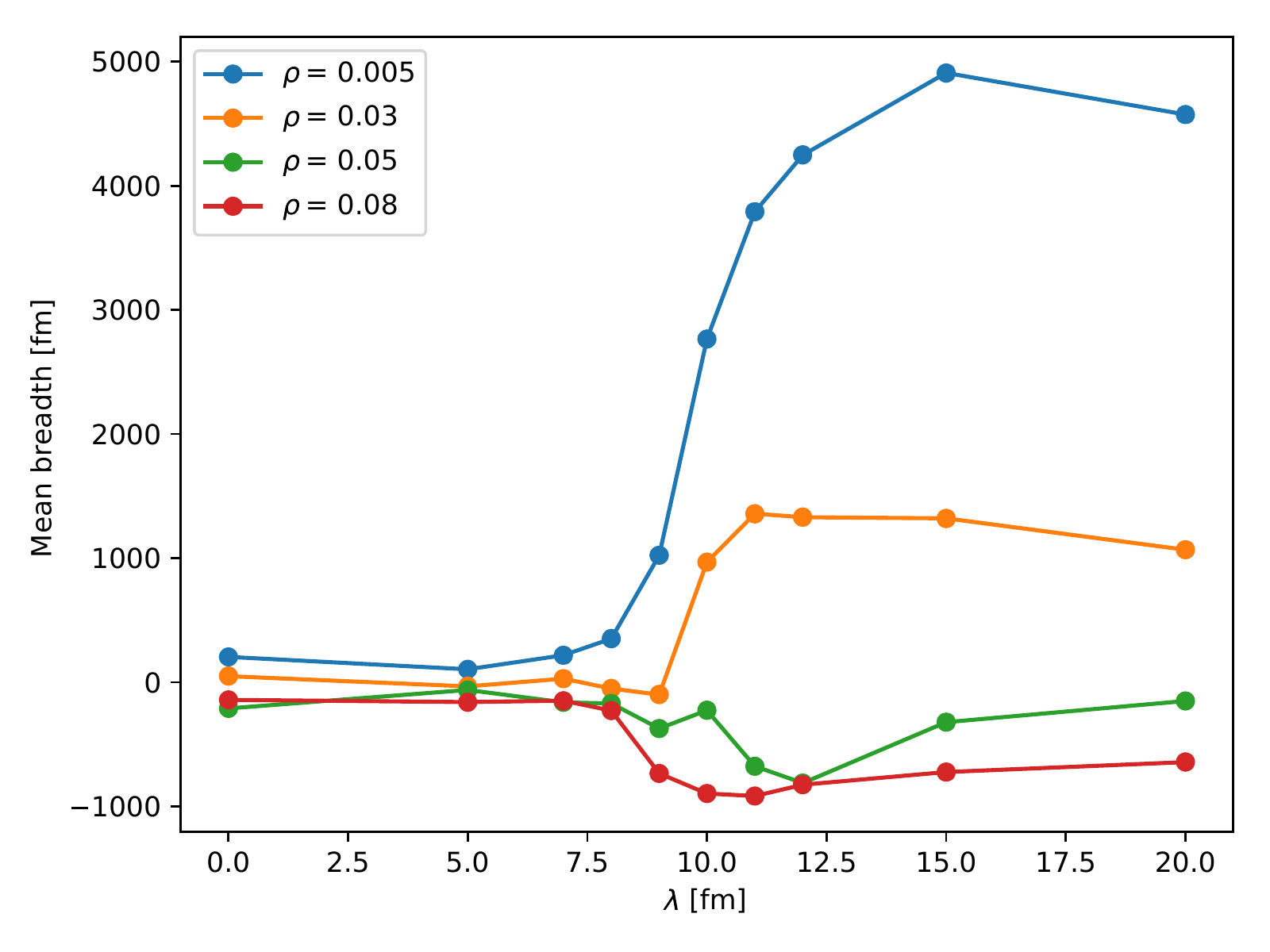}
}
\subfloat[Euler number.]{
\includegraphics[width=0.3\columnwidth]
{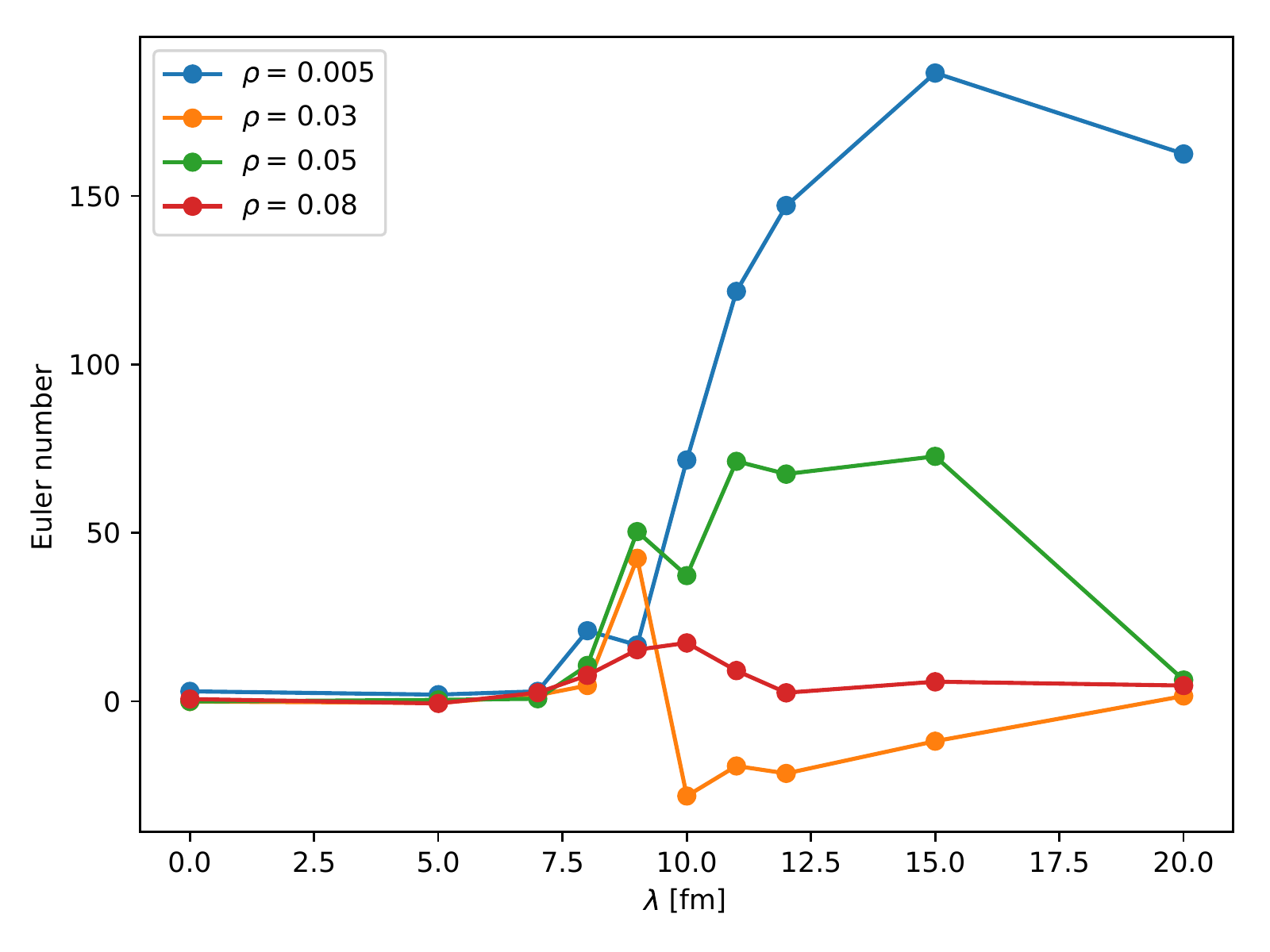}
}
\caption{Minkowski functionals dependence with $\lambda$. We can see that there 
is a transition regime between $\lambda=7\,\text{fm}$ and 
$\lambda=15\,\text{fm}$, where the Minkowski functionals are
  changing.}
\label{fig:minkowski}
\end{figure*}

$\lambda_c$ can also be determined by the effect of $\lambda$ on the morphology 
of the pastas, as quantified by means of the Minkowski functionals. The plot in 
Figure~\ref{fig:minkowski} shows the surface, mean breadth and Euler number for 
ground-state structures ($T=0$) as a function of $\lambda$.  Between $\lambda=7$ 
fm and $\lambda=10$ fm all three of the Minkowski functionals change drastically 
before reaching well defined values. \\


The changes of the Minkowski functionals indicate changes in the pasta 
structures. As stated in table~\ref{tab1}, the lasagna and spaghetti have Euler 
characteristics $\chi\approx 0$, the gnocchi $\chi_{gn} \approx 2$, and a system 
with $N_{gn}$ gnocchi $\chi \approx 2\cdot\,N_{gn}$, and the surface changes 
correspondingly as well. The mean breadth, on the other hand, should be positive 
for spaghetti and gnocchi, zero for lasagna, and negative for tunnels.  The 
structures obtained with $\lambda < 10$ fm do not exhibit the full range of 
values the Minkowski functionals should have.\\

\subsubsection*{Evolution of the pastas as a function of $\lambda$} 
\label{pasta-bup}

\begin{figure*}[!htbp]
\centering
\subfloat[$\rho=0.03\,\text{fm}^{-3}$, $\lambda=0\,\text{fm}$.]{
\includegraphics[width=0.3\columnwidth]
{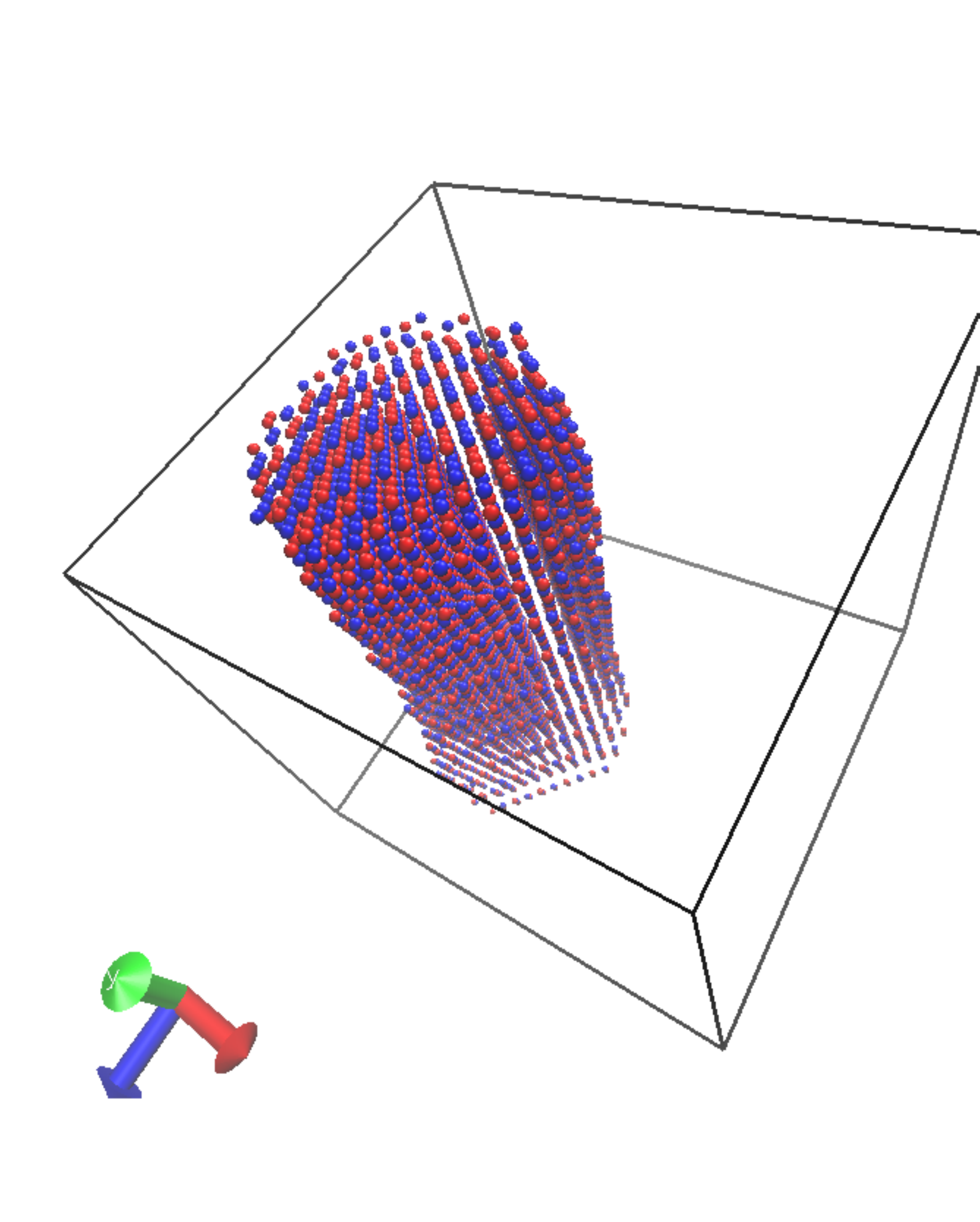}
}
\subfloat[$\rho=0.03\,\text{fm}^{-3}$, $\lambda=10\,\text{fm}$.]{
\includegraphics[width=0.3\columnwidth]
{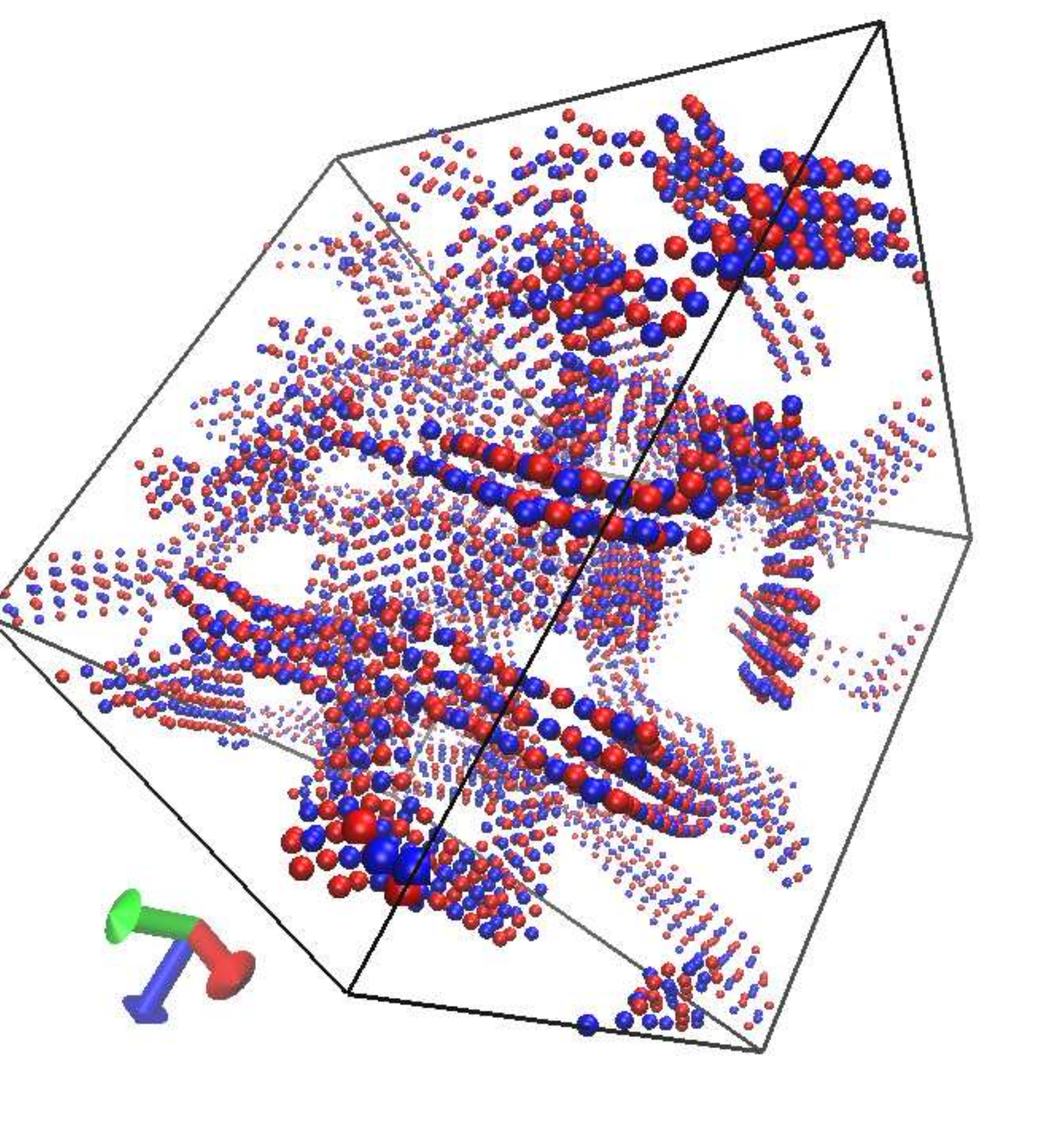}
}
\subfloat[$\rho=0.03\,\text{fm}^{-3}$, $\lambda=20\,\text{fm}$.]{
\includegraphics[width=0.3\columnwidth]
{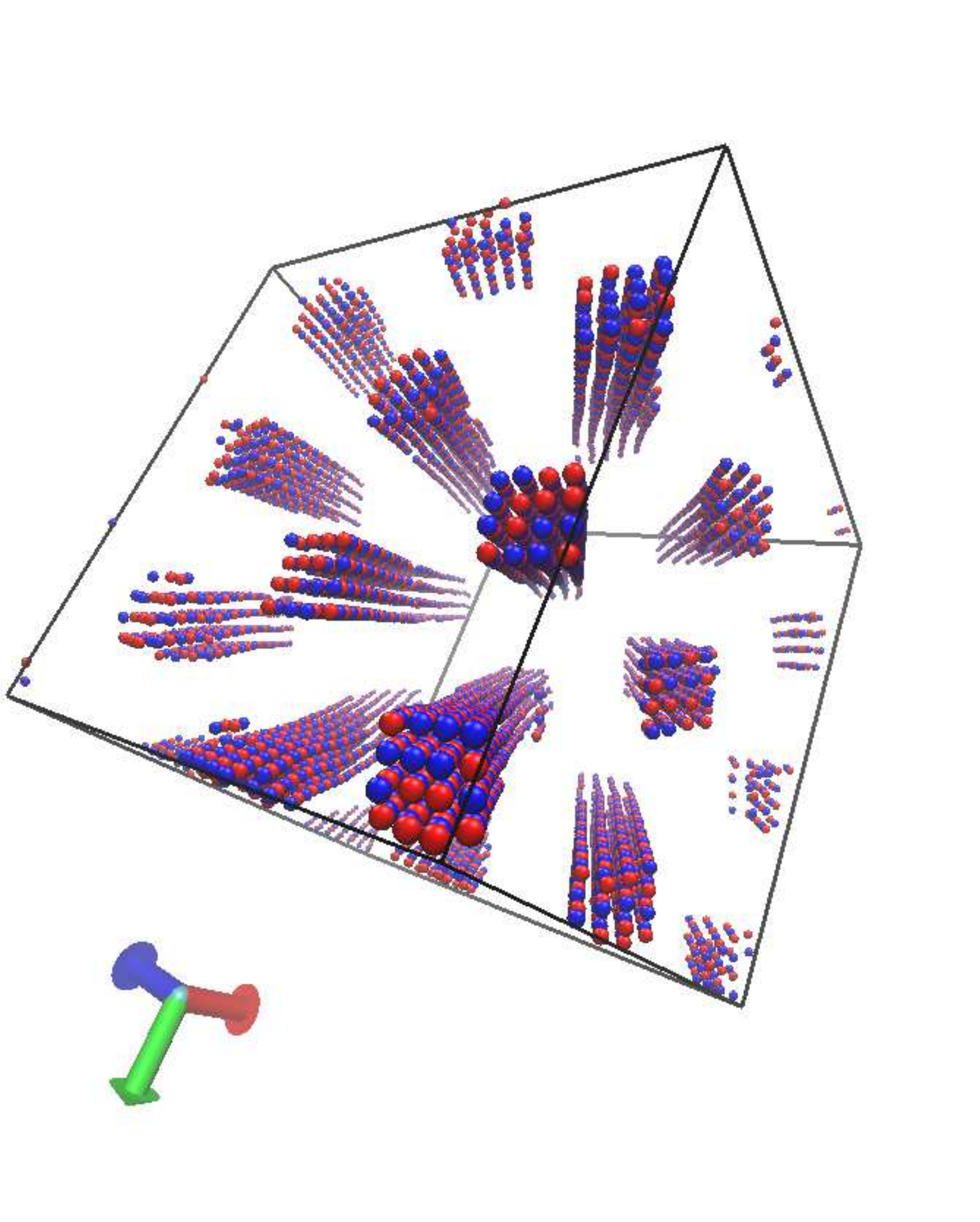}
}
\caption{Difference between pasta with and without Coulomb
  interaction. We can see that the Coulomb interaction splits up the
  pasta, converting one structure per cell to multiple structures per
  cell.}
\label{fig:w-wo-coulomb}
\end{figure*}

\begin{figure*}[!htbp]
\centering
\subfloat[$\rho=0.05\,\text{fm}^{-3}$, $\lambda=0\,\text{fm}$.]{
\includegraphics[width=0.3\columnwidth]
{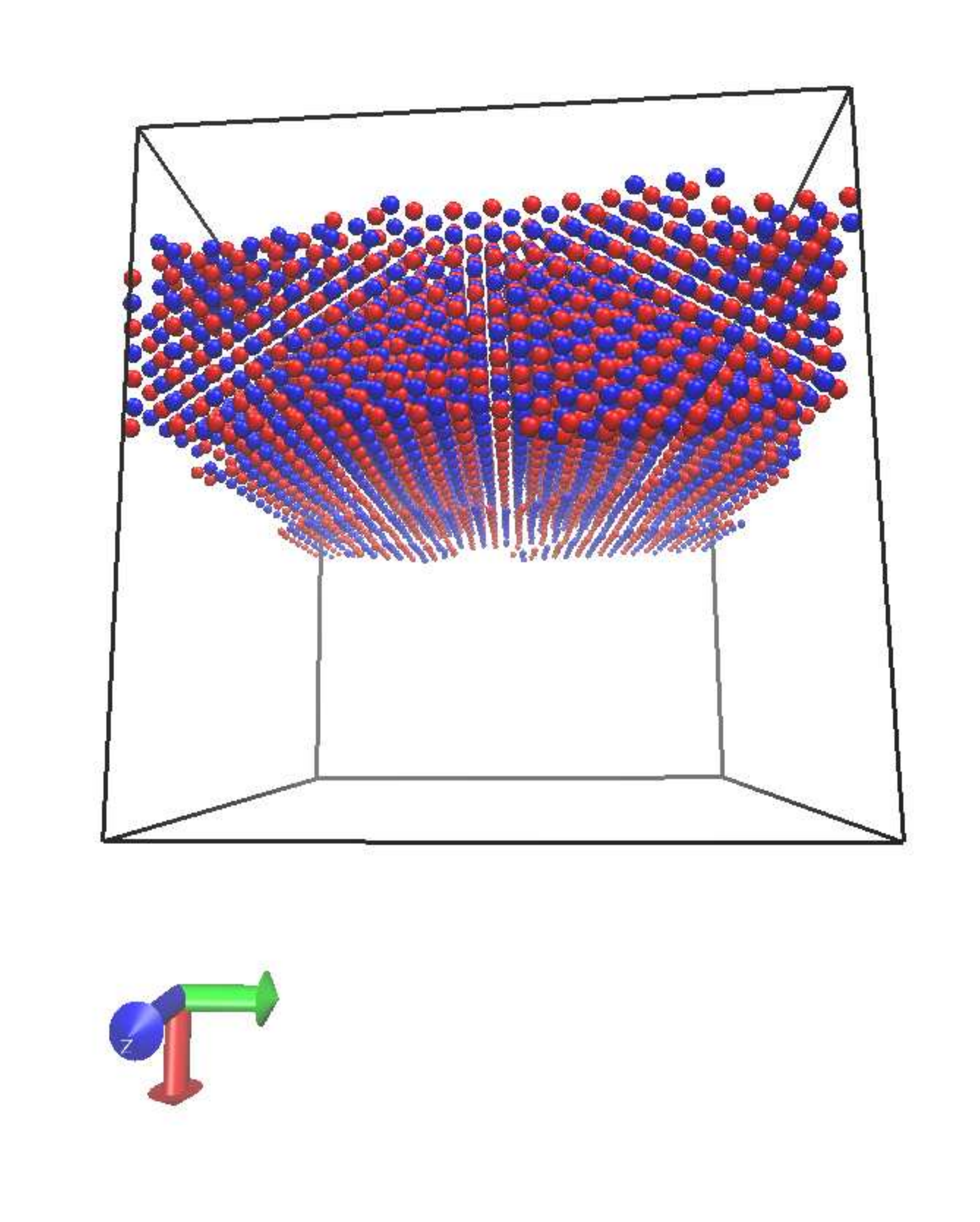}
}
\subfloat[$\rho=0.05\,\text{fm}^{-3}$, $\lambda=10\,\text{fm}$.]{
\includegraphics[width=0.3\columnwidth]
{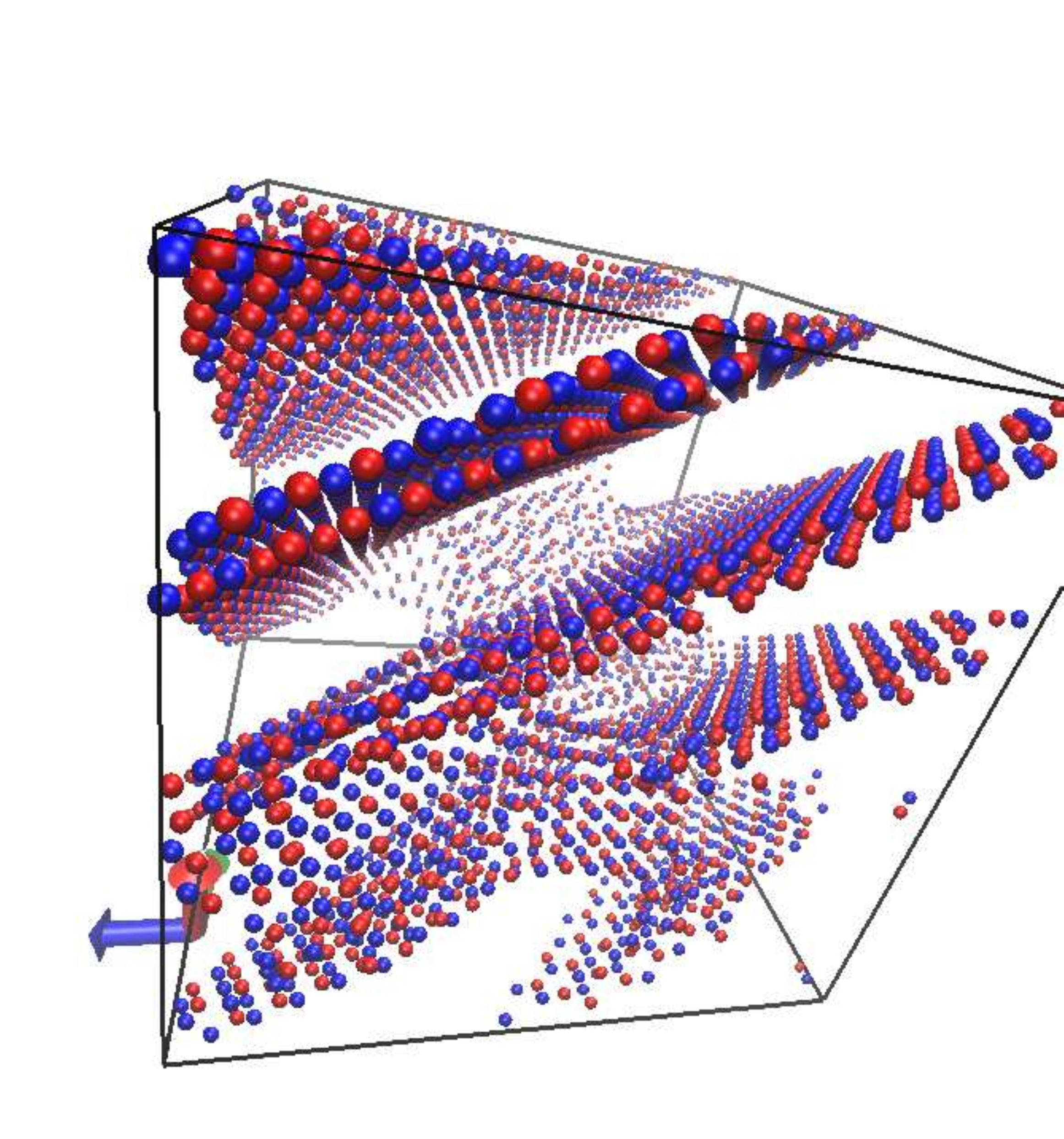}
}
\subfloat[$\rho=0.05\,\text{fm}^{-3}$, $\lambda=20\,\text{fm}$.]{
\includegraphics[width=0.3\columnwidth]
{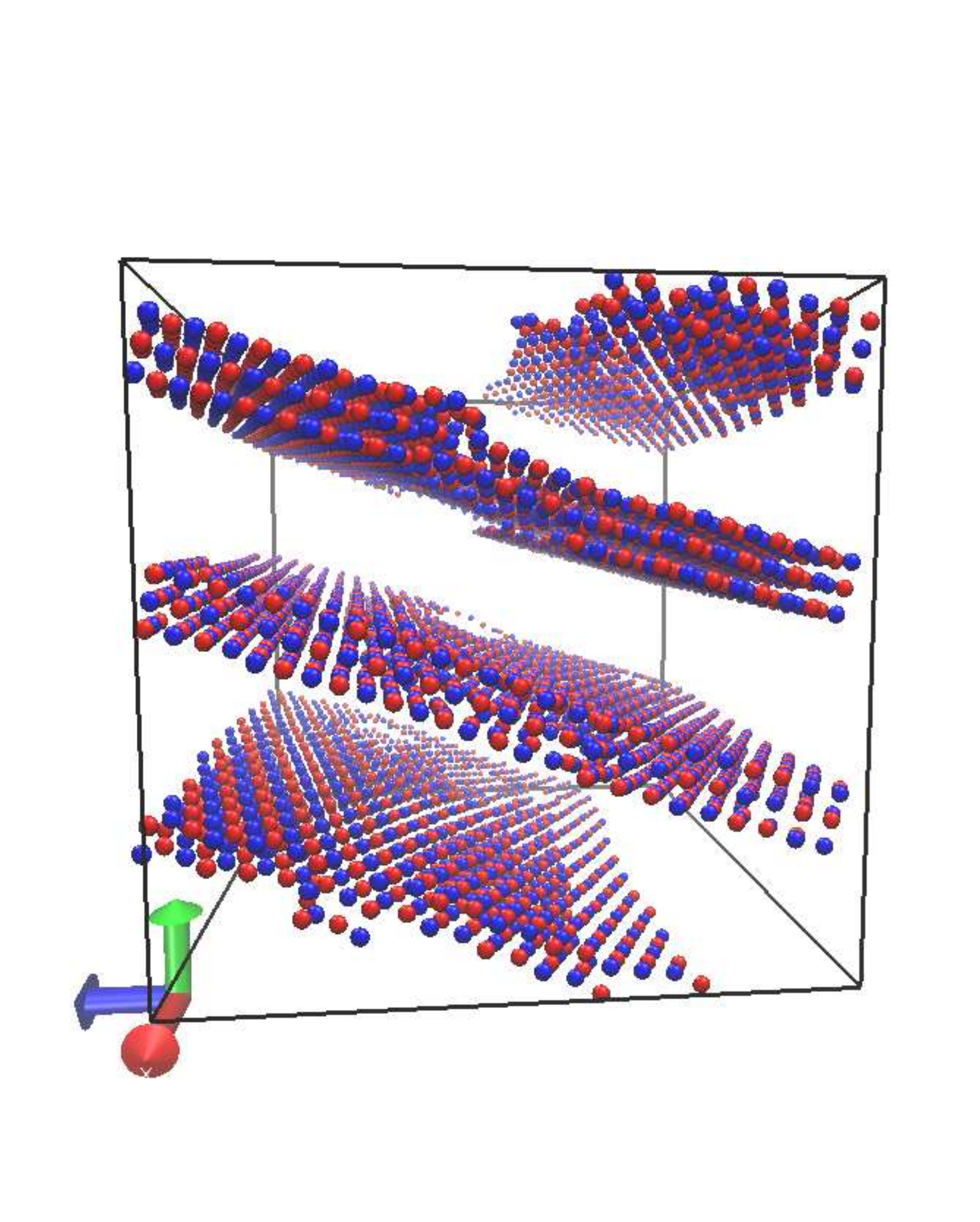}
}
\caption{Same as Figure~\ref{fig:w-wo-coulomb}}
\label{fig:w-wo-coulomb2}
\end{figure*}

\begin{figure*}[!htbp]
\centering
\subfloat[$\rho=0.08\,\text{fm}^{-3}$, $\lambda=0\,\text{fm}$.]{
\includegraphics[width=0.3\columnwidth]
{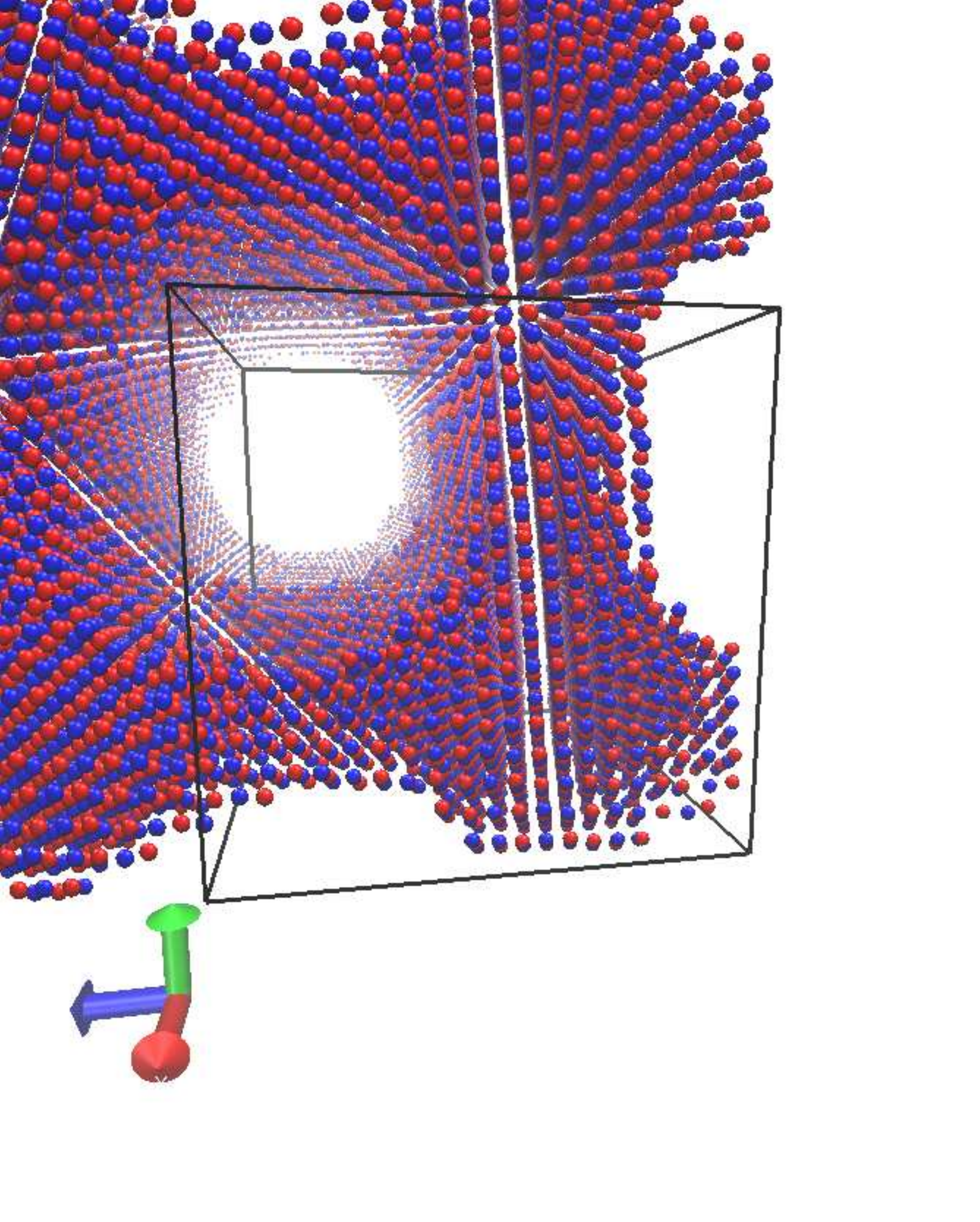}
}
\subfloat[$\rho=0.08\,\text{fm}^{-3}$, $\lambda=10\,\text{fm}$.]{
\includegraphics[width=0.3\columnwidth]
{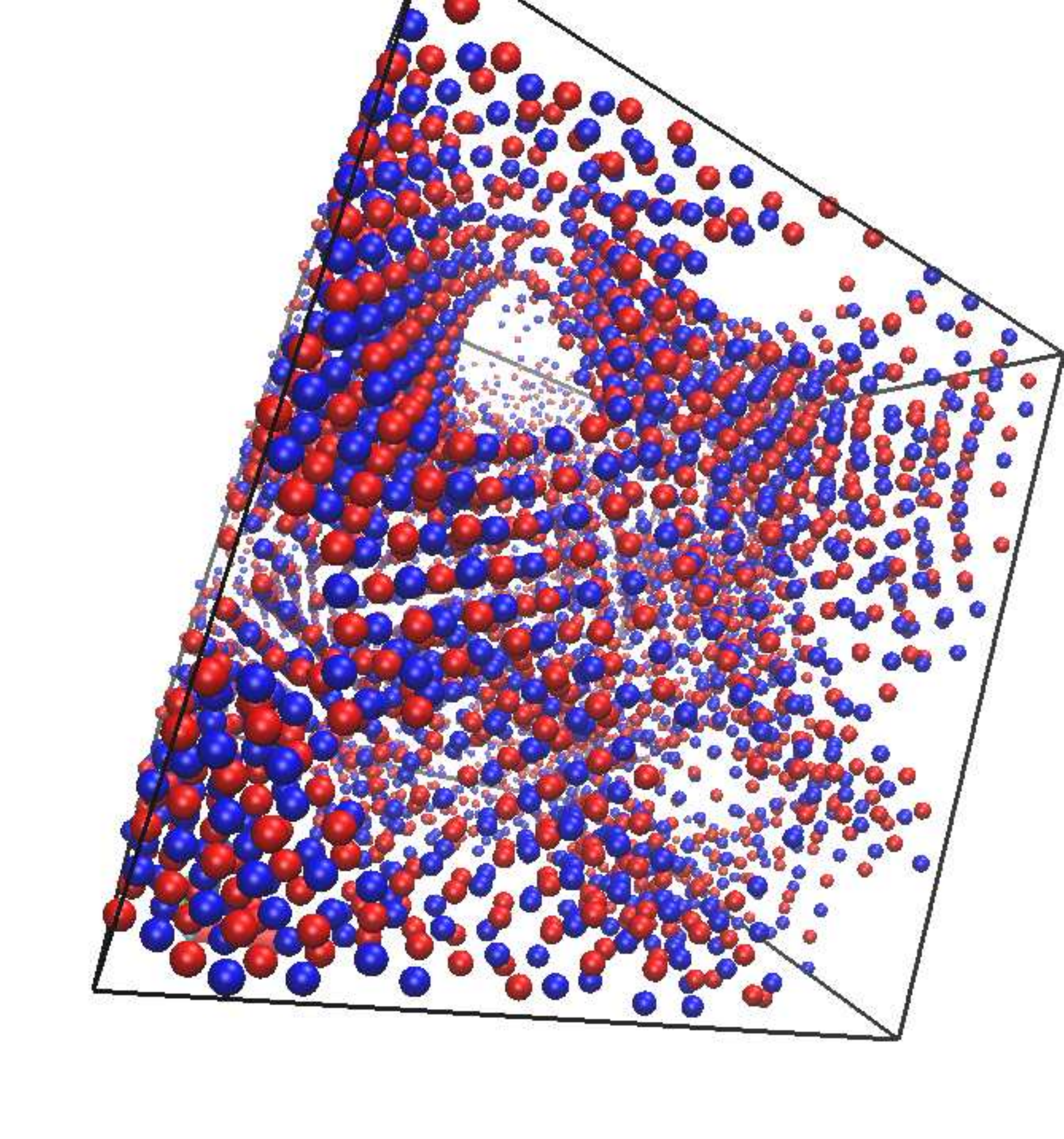}
}
\subfloat[$\rho=0.08\,\text{fm}^{-3}$, $\lambda=20\,\text{fm}$.]{
\includegraphics[width=0.3\columnwidth]
{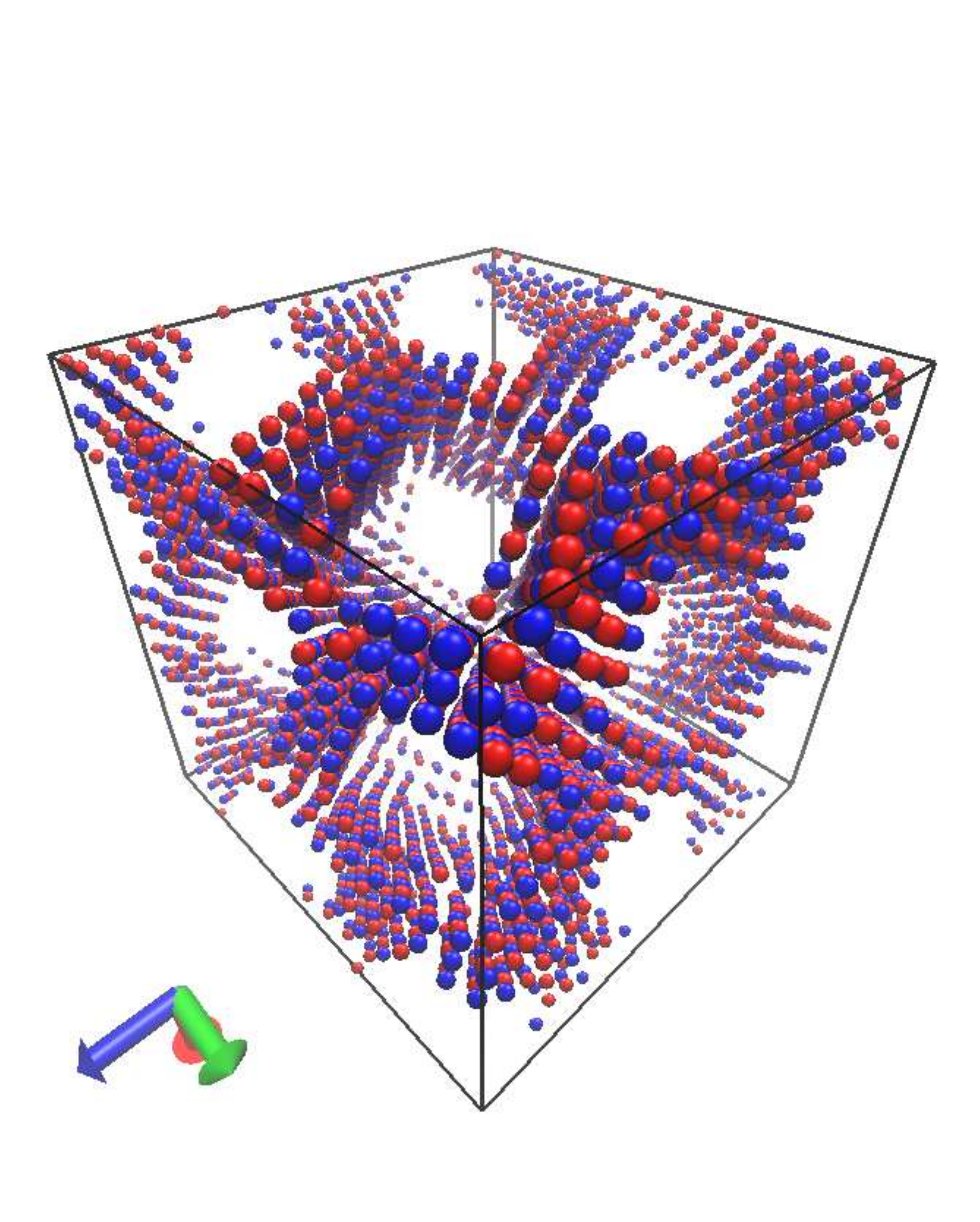}
}
\caption{Same as Figure~\ref{fig:w-wo-coulomb}}
\label{fig:w-wo-coulomb3}
\end{figure*}

Figures~\ref{fig:w-wo-coulomb}-\ref{fig:w-wo-coulomb3} show ground-state 
structures for $\lambda=0$, 10 and 20 fm, and three different densities. While 
the structures at $\lambda=0$ and 20 fm are recognizable pasta structures, those 
obtained at $\lambda=10$ fm are more exotic, probably due to the very rough 
energy landscape at this value of $\lambda$. \\

\begin{figure*}[!htbp]
\centering
\subfloat[$\lambda=0\,\text{fm}$.]{
\includegraphics[width=0.3\columnwidth]
{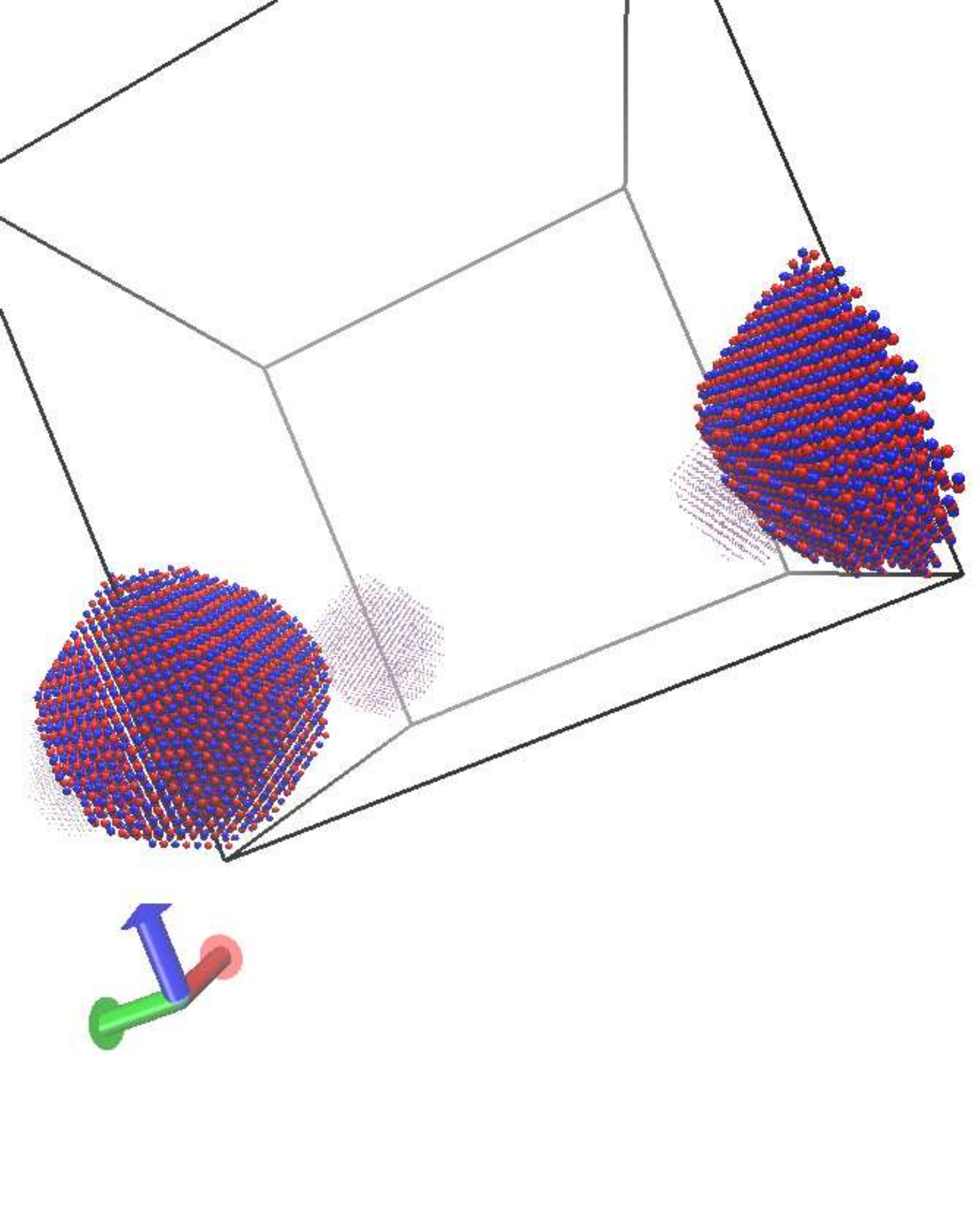}
}
\subfloat[$\lambda=10\,\text{fm}$.]{
\includegraphics[width=0.3\columnwidth]
{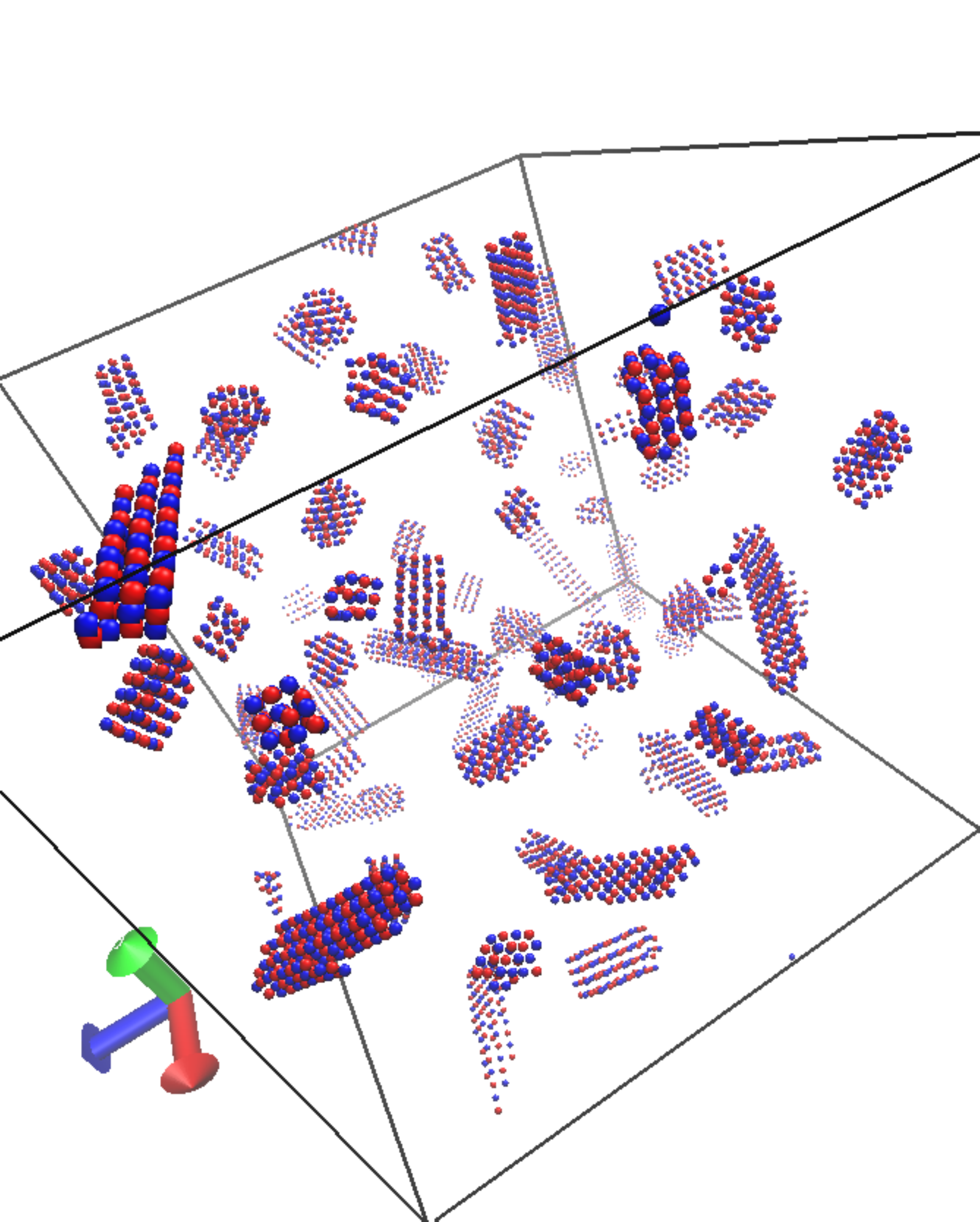}
}
\subfloat[$\lambda=20\,\text{fm}$.]{
\includegraphics[width=0.3\columnwidth]
{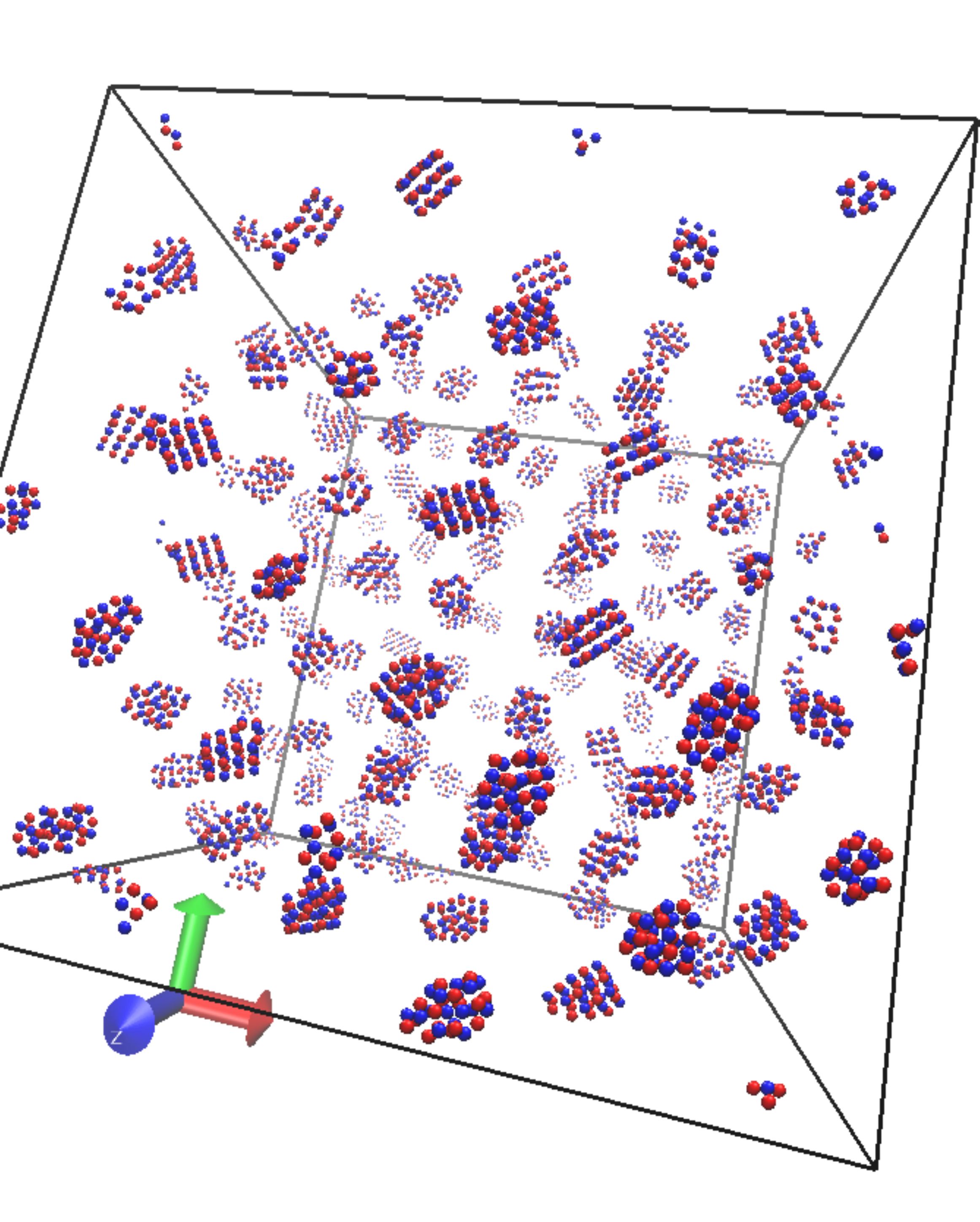}
}
\caption{Different structures got while varying the $\lambda$
    parameter, for $\rho=0.005\text{fm}^{-3}$. In the transition
    regime, we find, at $\lambda=10\,\text{fm}$, that the structure
    breaks down to many \emph{short-spaghetti}-like parts.}
\label{fig:gnocchi}
\end{figure*}

\begin{figure}[h] 
\begin{center}
\includegraphics[width=0.5\columnwidth]{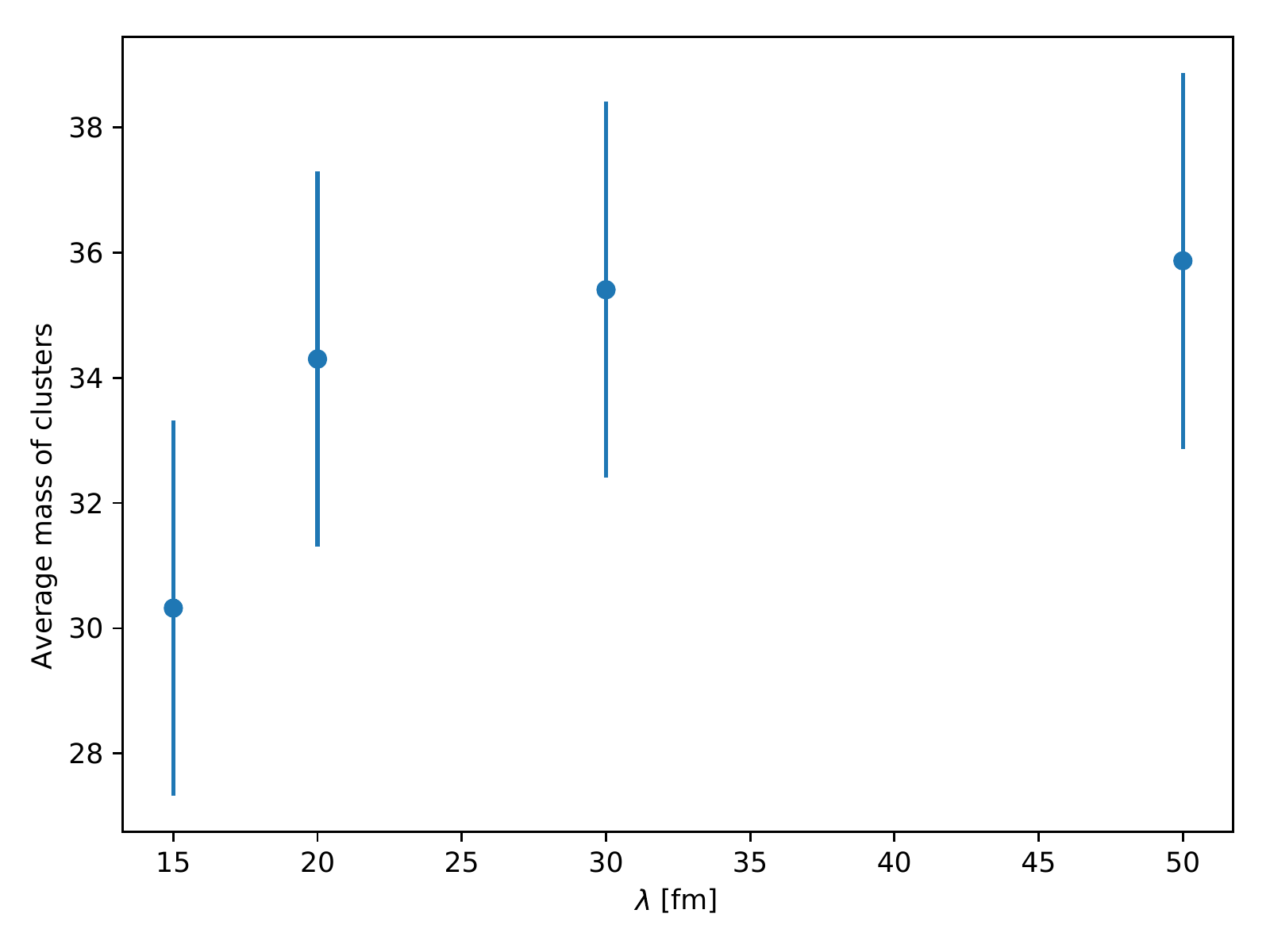}
\end{center}
\caption{Average size of nuclei depending on the screening length. We can see 
that, when considered the standard deviation, the mass remains the 
same.}\label{fig:gnocchi_mass}
\end{figure}

More information about the $\lambda_c$ can be obtained from the size of the 
gnocchi obtained for various values of $\lambda$. Figure~\ref{fig:gnocchi} shows 
that at $\rho=0.005$ fm$^{-3}$, there is a single gnocco for $\lambda<10$ fm, 
but several for $\lambda$ between 15 fm to 20 fm. Likewise, 
Figure~\ref{fig:gnocchi_mass} indicates that the average size of the gnocchi 
gets reduced while its number increases reaching a stable value for $\lambda 
\geq 20$ fm. This larger number of clusters explains the changes of the 
Minkowski functionals. \\

\begin{figure*}[!htbp]
\centering
\subfloat[Multiple lasagna.]{
\includegraphics[width=0.5\columnwidth]
{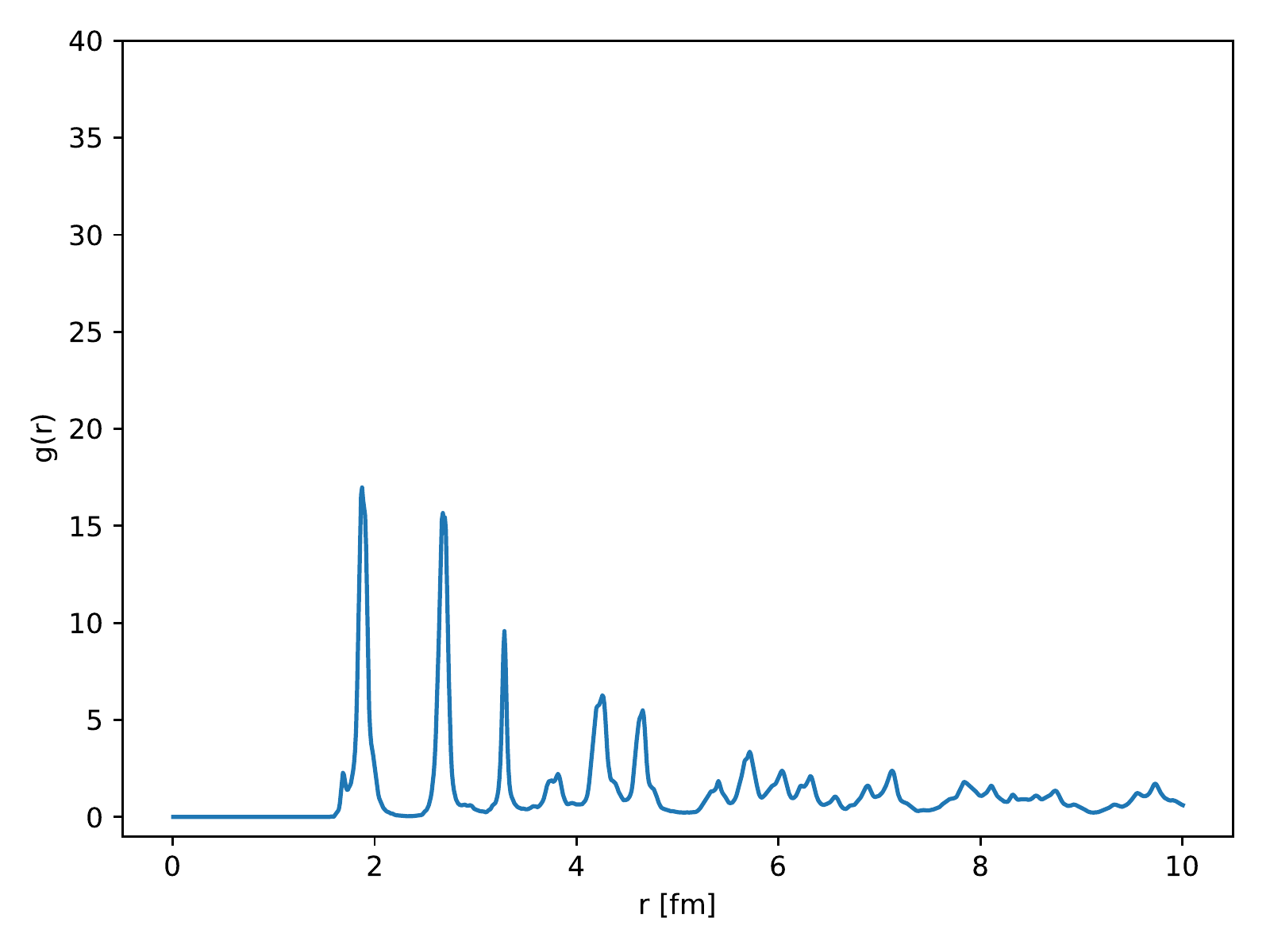}
}
\subfloat[Single lasagna.]{
\includegraphics[width=0.5\columnwidth]
{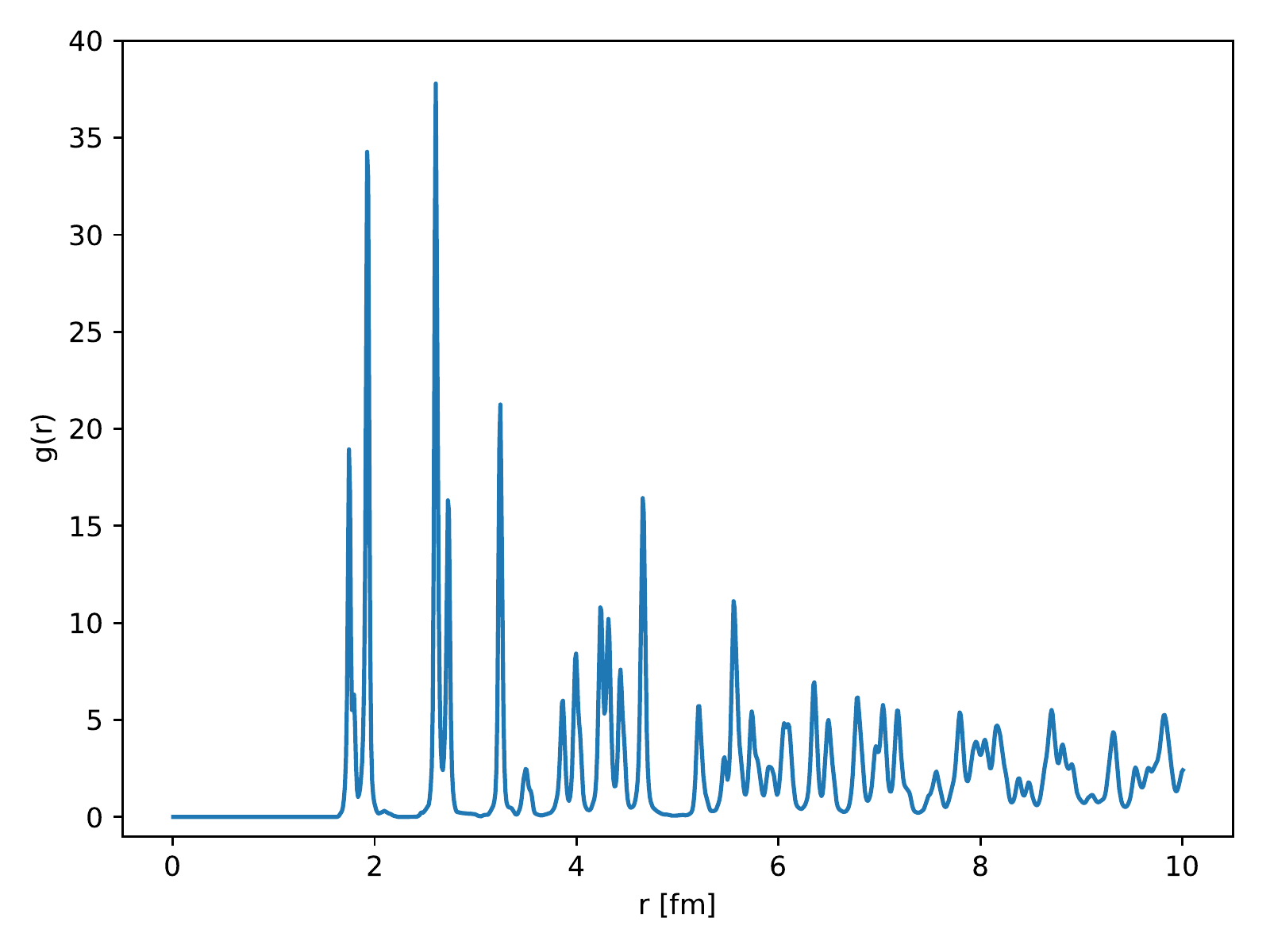}
}
\caption{Examples of the radial correlation function for 
$\rho=0.05\,\text{fm}^{-3}$ and two screening lengths: (a) 
$\lambda=20\,\text{fm}$, and (b) $\lambda=0\,\text{fm}$. Please notice the 
difference in the y-scales of the graphs.}
\label{fig:gofr}
\end{figure*}

Figure~\ref{fig:gofr} examines the $\lambda=0$ and 20 fm cases at $\rho=0.05$ 
fm$^{-3}$ with the radial distribution function, $g(\mathbf{r})$. It is clear 
that with Coulomb ($\lambda = 20$ fm) the pastas have a more ordered neighboring 
structures than without Coulomb ($\lambda = 0$).\\

\subsection{Summarizing the electron gas}\label{sum-e-gas}

The effect of the electron gas on symmetric and neutron-rich matter was studied 
at low densities and temperatures by varying the Coulomb interaction strength 
and screening lengths.  Its effect on the fragment size multiplicity, the 
inter-particle distance, the isospin content of the clusters, the nucleon 
mobility, and on the modification of the topological shape was studied.\\

The most general result is the existence of the  nuclear pasta structures even 
without the presence of the electron gas.  As seen in Section~\ref{nm-LT}, the 
existence of the pastas, which are usually associated with the presence of the 
competing interactions of the long range Coulomb potential and the short range 
nuclear force, can exist even in the absence of the electron gas (i.e. when 
$\alpha \rightarrow 0$).  These pseudo-pastas are the result of the competition 
between the attractive $V_{np}$, and the repulsive $V_{nn}$ interactions; this 
same effect has been seen in other potentials~\cite{P2012} as was reported 
in~\cite{2013}. It must be remembered that, if not calculated with the proper 
screening lengths, the pseudo-pastas contain effects due to the cell size.\\

The main effect of the strength of the Coulomb interaction of an electron gas is 
to allow the system to display true pastas structures, and distribute matter 
more and form less-compact objects.  At a microscopic scale, Figure~\ref{Radial} 
shows that varying the strength of the Coulomb interaction does not change the 
inter-particle distance, but certainly decreases a bit the $x$ content of the 
fragments as can be seen in Figure~\ref{XfragX3D015T1}.  In agreement with this, 
the increment of nucleon mobility produced by Coulomb (cf. 
Figure~\ref{desplx5T1}) gets reflected in a reduction of the persistence (cf. 
Figure~\ref{persis}), and of the curvature and Euler characteristic of the 
structures, ``compactifying'' symmetric matter into spaghetti and lasagna type 
structures as shown in Figure~\ref{curv-euler}, although the effect is less 
pronounced in for $x=0.3$ (bottom panel).\\

Likewise, our study of the effect of the screening length of the Coulomb 
interaction determined that there is a critical screening length $\lambda_c$ at 
which the structures change drastically. For $\lambda<\lambda_c$, the Coulomb 
interaction is barely acting, a negative pressure produces only one structure 
per cell indicating that the structures formed are due to finite size effects. 
For $\lambda>\lambda_c$, on the other hand, the pressure becomes positive and 
the systems present density fluctuations smaller than the cell, and the 
morphology of the structures stabilize and cease to depend on $\lambda$. For the 
Pandharipande potential $\lambda_c$ lies between $10\,\text{fm}$ and 
$15\,\text{fm}$ depending on the density. \\


\newpage

\section{Neutron star matter}\label{nsm}

In this Section we extend the study of nuclear matter of Section~\ref{nm-LT} to 
the realm of neutron star matter. We use CMD to study the existence of 
pasta-like structures and their possible phase changes. We focus the 
investigation at saturation and sub-saturation densities, low temperatures, and 
proton fractions in the range of 10$\%$ to 50$\%$ and. We also study the 
behavior of $E_{Sym}$ in the different structures found.\\

We continue using CMD as presented in Appendix~\ref{CMD-NSM}. Although the 
parameters $\mu_r$, $\mu_a$, $\mu_0$ and $V_r$, $V_a$, $V_0$ were first set by 
Pandharipande for cold nuclear matter~\cite{pandha}, a recent 
improvement~\cite{dor2018}, here named \textit{New Medium}, reproduces the cold 
nuclear matter binding energies more accurately and, thus, is used in this study 
of NSM. The corresponding values are summarized in Table~\ref{table_parameter}. 
Figs.~\ref{fig:nn} and\ref{fig:np} contrasts these potentials with those of 
Pandharipande Medium potentials.\\

As explained in Section~\ref{Electron-gas}, the neutron star matter contains 
protons and neutrons embedded in an electron gas, which introduces a screening 
effect on the Coulomb potential of the protons. The screened Coulomb potential 
is included in these calculations using the Thomas-Fermi  approximation as 
explained in Appendix~\ref{cmd_star-2}.  It might be appropriate at this point 
to emphasize that the  exponential cut-off renders the Coulomb effective 
interaction, short ranged. In this way the energy and the entropy is additive 
(i.e. energy scales with the number of particles).\\

\subsection{Symmetric neutron star matter}\label{sym_energy}

We first study the case of symmetric neutron star matter, i.e with $x = z/A= 
0.5$, focusing on the caloric curve and the binding energy. In most of the cases 
presented in this section the total number of nucleons in the primary cell was 
$N=4000$, interacting through the New Medium model (c.f. 
Appendix~\ref{cmd_star-1}).

\subsubsection*{The caloric curve}\label{subsec:sym_energy_1}

For starters we calculate the internal energy of symmetric neutron star matter. 
This involves averaging the kinetic and potential energy of each nucleon in the 
system, c.f. Appendix~\ref{CMD-LT}. For full details see~\cite{dor2019}\\

As explained in Section~\ref{nm-LT}, the caloric curve can be used to detect 
phase changes. The caloric curve is the relationship between the internal energy 
and the temperature, and changes in the slope of the $E-T$ curve can be used as 
indicators of phase transitions. It is convenient to remember that, as seen in 
Section~\ref{propNMP}, nuclear matter maintains a liquid-like structure for 
temperatures larger than about 2.0 MeV, and around $T=0.5$ MeV it transforms 
into a pseudo-pasta with nucleons frozen much like in crystalline structures. 
The results  for neutron star matter are somewhat different. \\

\begin{figure*}[!htbp]
\centering
\subfloat[Internal energy \label{fig:eos_0}]{
\includegraphics[width=0.45\columnwidth]
{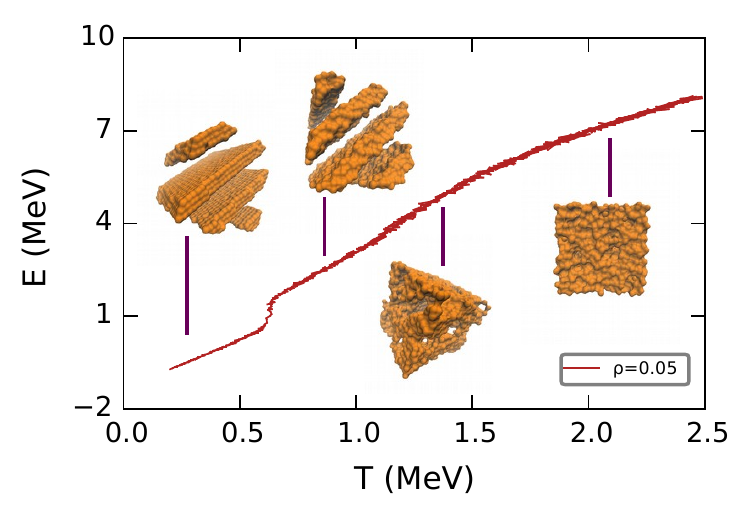}
}
\subfloat[Internal energy \label{fig:eos_1}]{
\includegraphics[width=0.46\columnwidth]
{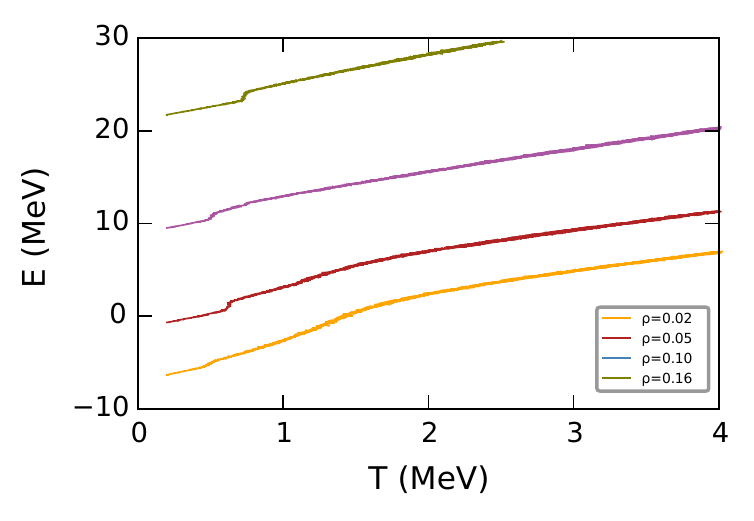}
}
\caption{(a) Internal energy per nucleon for symmetric neutron star matter 
($x=0.5$) as a function of the bath temperature, for $\rho=0.05\,$fm$^{-3}$, and 
(b) for for $\rho=0.02$, 0.05, 0.10, 0.16 fm$^{-3}$.}
\end{figure*}



Figure~\ref{fig:eos_0} presents the caloric curve for the case of $\rho=0.05$ 
fm$^{-3}$, and Figure~\ref{fig:eos_1} shows the caloric curves for densities 
$\rho=$0.02, 0.05, 0.10 and 0.16 fm$^{-3}$ in the extended temperature range of 
0.2 to 4 MeV.  As it can be seen more clearly in Figure~\ref{fig:eos_0}, the 
slope of the internal energy exhibits a change at around $T\simeq 1.5\,$MeV, and 
a sharper change at $T\simeq 0.5\,$MeV. Although these jumps are not as 
pronounced as those found for NM, they happen at around the same temperatures 
and can be taken as the same type of transitions.\\

We claim that the smooth change at $T\simeq 1.5\,$MeV signals the onset of the 
topological phase transition, (i.e. a bubble appears). A difference with NM is 
that this change in slope appears more 
pronounced for smaller densities ($\rho \leq 0.05$ fm$^{-3}$), and tends to 
disappear for larger densities. As this result is different than the NM case, we 
believe it is due to the presence of the electron gas. \\

On the other hand, the sharper jump observed at $T\simeq 0.5\,$MeV both in 
Figs.~\ref{fig:eos_0} and ~\ref{fig:eos_1} is maintained throughout the density 
range studied and, in fact, becomes more pronounced for larger densities. We 
believe this discontinuity in the derivative of the $E-T$ curve signals the 
change from amorphous pasta to crystalline pasta already identified in 
Section~\ref{Electron-gas} for NM  
with the Pandharipande Medium model.\\

\begin{figure}
\begin{center}
   \includegraphics[width=0.5\columnwidth]{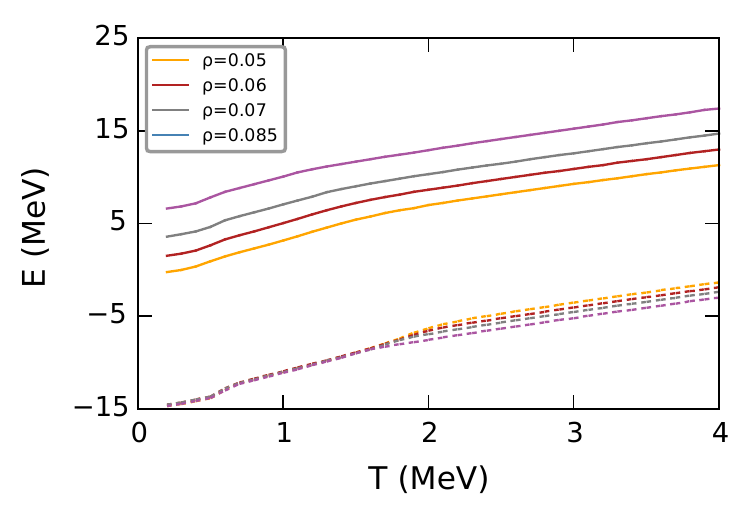}
\caption{ Internal energy per nucleon for symmetric neutron star matter 
($x=0.5$) as a function of temperature. The mean density for each profile is 
indicated in the inset in fm$^{-3}$. The dashed lines correspond to the 
Pandharipande Medium model (nuclear matter), while the continuous lines 
correspond to the New Medium potential with the screened Coulomb 
potential.}\label{fig:eos_3}
\end{center}
\end{figure}

The noticeable differences between the caloric curve of nuclear matter and that 
of neutron star matter are shown in Figure~\ref{fig:eos_3}. The figure compares 
the NSM results (continuous lines) against the NM curves (dashed lines) for the 
cases of $\rho=0.05$, 0.06, 0.07 and 0.085 fm$^{-3}$ and temperatures from 0.2 
to 4 MeV. The NSM lines remain separated and parallel throughout the range of 
explored temperatures, whereas the NM curves merge with one another at low 
temperatures. Noteworthy, the NM highest density curve ($\rho=0.085\,$fm$^{-3}$) 
corresponds to the lowest energy, while the opposite is true for NSM where the 
higher density curve has the higher energy of all the curves shown.\\ 

\begin{figure}[h!]  \centering
\includegraphics[width=0.5\columnwidth]{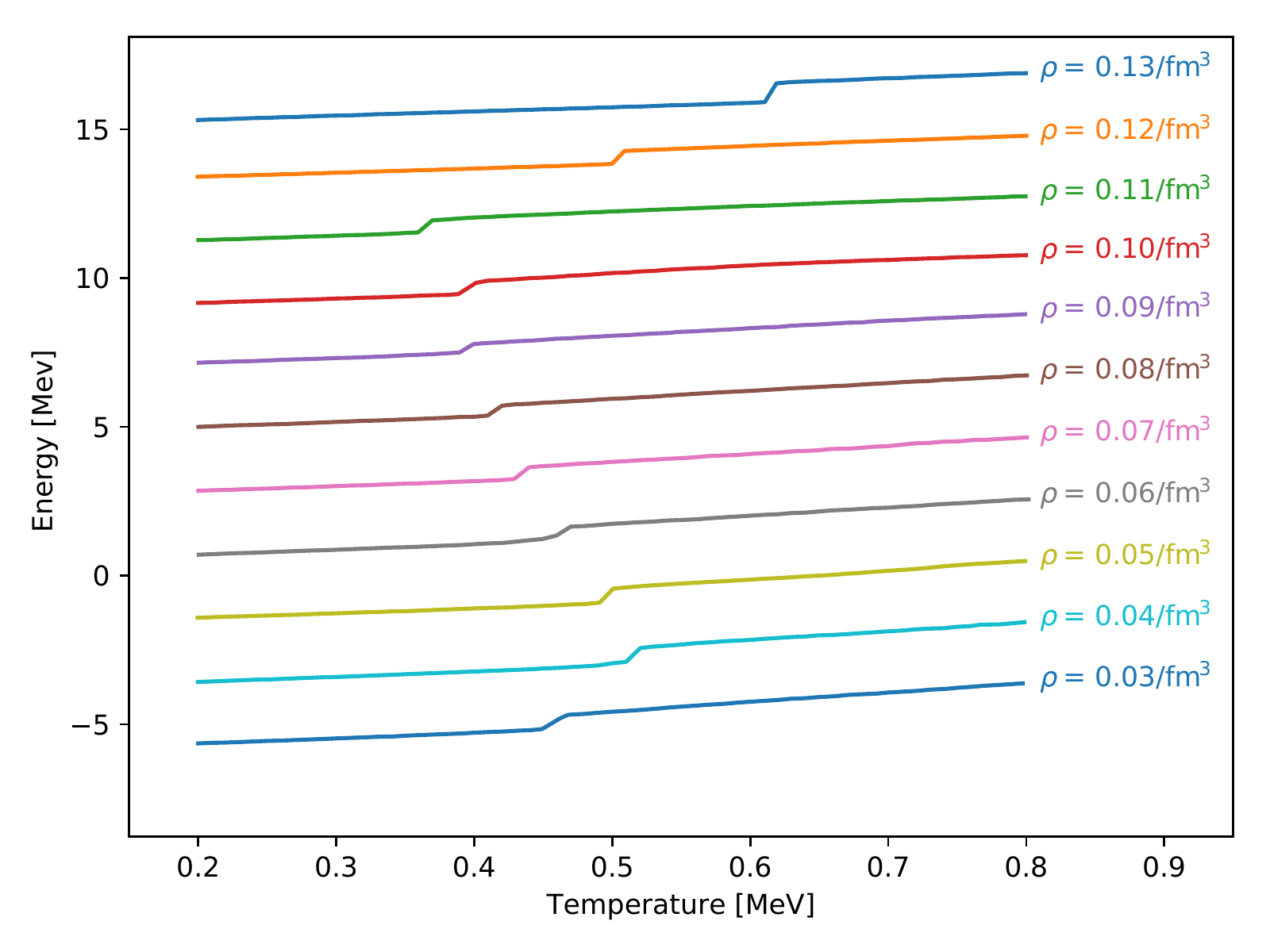}
\caption{Energy as a function of temperature for different densities. We see 
that there is a discontinuity in the range of   $T=0.35$ MeV to $0.65$ MeV, 
depending on the density, a signal of a first-order phase transition. In the 
figure, densities range from $\rho=0.03$ fm$^{-3}$ and $\rho=0.13$ fm$^{-3}$, in 
steps of $\Delta\rho=0.01\,\text{fm}^{-3}$ upwards.}
\label{fig:energy}
\end{figure}

Figure~\ref{fig:energy} shows more examples of the caloric curve for several 
densities. Each of these densities exhibit a discontinuity in the energy at 
certain temperatures, a signal of a first order phase transition.  This 
transition can be confirmed and further characterized as a solid-liquid phase 
transition by looking at the Lindemann coefficient. The Lindemann coefficient 
for $\rho=0.05\,\text{fm}^{-3}$ as a function of temperature, can be seen in 
figure~\ref{lin}, along the energy. This figure shows that the discontinuities 
in Lindemann coefficient and in energy are at the same temperature. These two 
factors are, effectively, the signature of a solid-liquid phase transition.\\

\subsubsection*{Energy - density}\label{subsec:sym_energy_2}

As seen in the case of NM (c.f. Section~\ref{Nuclear-Matter}), the $E-\rho$ 
curve can signal the saturation density (i.e. the minimum of the curve), bound 
and unbound states (positive or negative energies), the range of densities of 
the liquid phase (span of the ``$\cup$'' shape), and changes of phase, among 
others. For the case of NSM, however, the situation is different.\\

\begin{figure}[!htbp]
\begin{center}
   \includegraphics[width=0.5\columnwidth]{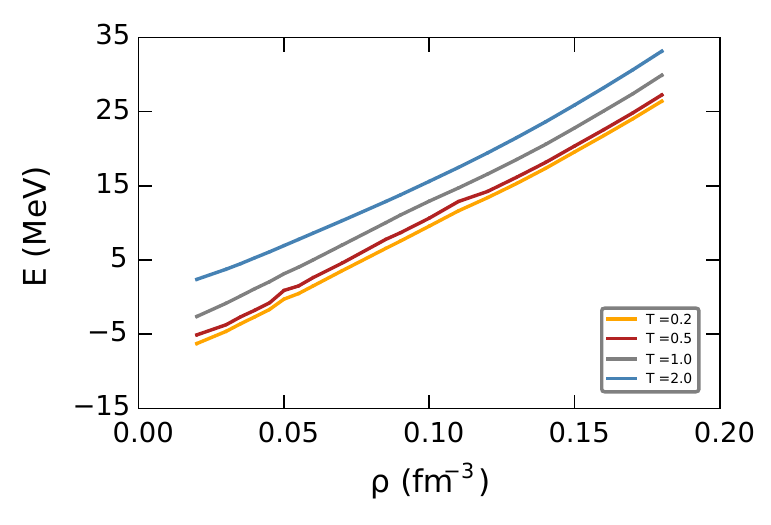}
\caption{\label{fig:eos_2}  Internal energy per nucleon for 
symmetric neutron star matter ($x=0.5$) as a function of the mean density at 
fixed temperatures. The corresponding bath temperature for each is indicated in 
the inset.}
\end{center}
\end{figure}

Figure~\ref{fig:eos_2} shows the energy as a function of the average density at 
$T = 0.1$, 0.2, 0.5, 1.0 and 2.0 MeV.  At a difference from the usual $\cup$ 
shape of symmetric nuclear matter, the NSM curves do not exhibit absolute 
minima, indicating that NSM at these temperatures and densities does not have an 
equilibrium point nor a saturation density. Furthermore, at most densities the 
NSM system appears to be unbound, except for $T\leq 1.0\,$MeV and $\rho\leq 
0.05\,$fm$^{-3}$. The difference between NM and NSM is the effect of the cloud 
of electrons which, with its exponential cut-off of the order of 20 fm, 
introduces a short range Coulomb screening which produces smaller aggregates of 
nucleons, as seen in Section~\ref{screen}.\\

\subsubsection*{The Minkowski functionals} \label{subsec:sym_topology}

As before, the morphology of the pasta structures will be studied with the 
Minkowski functionals, namely the volume, surface area, Euler characteristic 
$\chi$, and integral mean curvature B. The calculation of the Minkowski 
functionals requires the binning of nucleons into ``voxels'', as described in 
the Appendix~\ref{sec:voxel}. \\

\begin{figure}
\begin{center}
   \includegraphics[width=0.5\columnwidth]{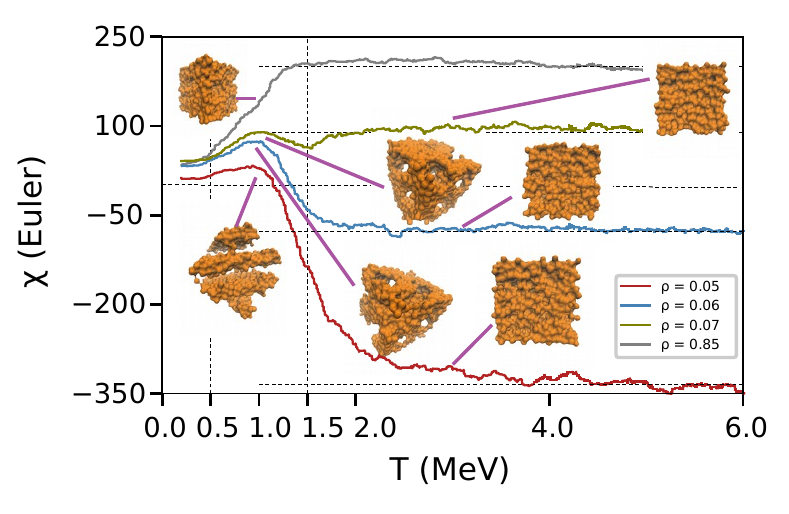}
\caption{ The Euler characteristic  $\chi$ obtained for isospin symmetric 
($x=0.5$)  NSM systems as a function of temperature. The data has been smoothed 
with a moving average procedure.}\label{fig:minkowski_rho}
\end{center}
\end{figure}

Figure~\ref{fig:minkowski_rho} shows the Euler characteristic as a function of 
the temperature. $\chi$ has a clear change of behavior at $T\simeq 1\,$MeV. 
Although this happens at a lower temperature than its NM counterpart (see 
Figure~\ref{mink_x05}), it also appears to be associated to the early stage of 
the pasta formation.  Furthermore, the $\chi$ values for the examined densities 
almost join into a single pattern for $T\leq 0.5\,$MeV, much like those of NM at 
the same temperature but less pronounced; this could also be related to the 
change of slope found in the caloric curve at $T\simeq 0.5$ MeV.\\

A look at the sign of $\chi$ can yield information about the morphology of the 
structure. Figure~\ref{fig:minkowski_rho} indicates that the lower density 
systems attain negative values of $\chi$ at, say, $T>1$ MeV. Since values of 
$\chi < 0$ corresponds to cases where the number of tunnels overcome the number 
of voids and isolated regions, see Eq.~(\ref{eq:chi}), it seems that low density 
configurations tend to be more cavity-like (more tunnels), but as density 
increases tunnels fill up yielding more compact structures.\\

\begin{figure}  
\centering 
\includegraphics[width=0.5\columnwidth]{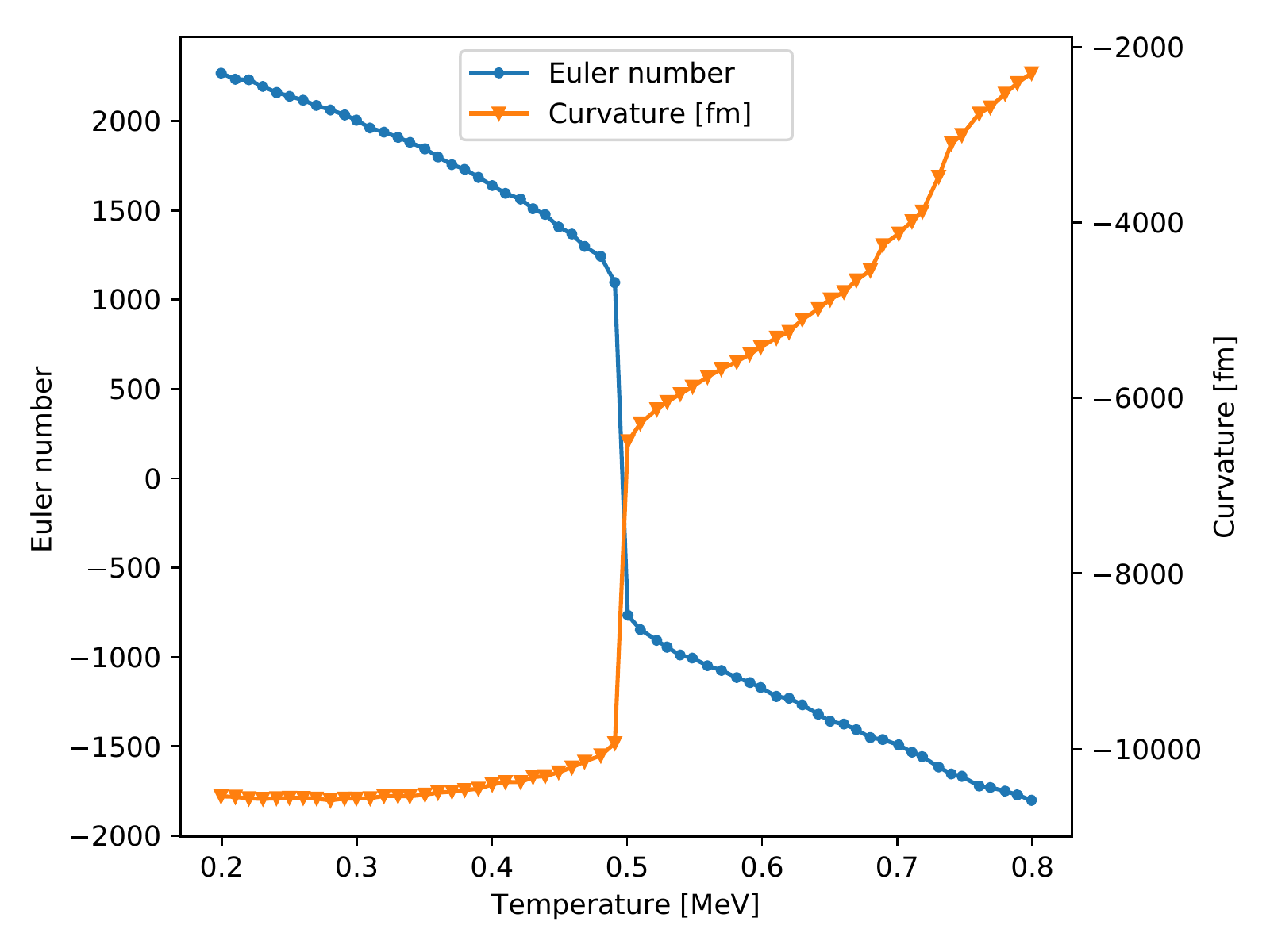}
  \caption{Euler number and mean breadth for $\rho=0.05\,\text{fm}^{-3}$.
    We observe a sharp transition for both Minkowski functionals.}
  \label{fig:euler-curv}
\end{figure}

When we look at the Minkowski functionals, particularly the Euler characteristic 
and the mean breadth, we can see that there is again a critical temperature at 
which both the Euler characteristic and the mean breadth show a sharp 
transition. We show, as an example, these magnitudes as a function of 
temperature for density $\rho=0.05$ fm$^{-3}$ in Figure~\ref{fig:euler-curv}. As 
this transition is signaled by morphological observables, we conclude that this 
transition is morphological.\\

\begin{figure}
  \centering
  \includegraphics[width=0.5\columnwidth]{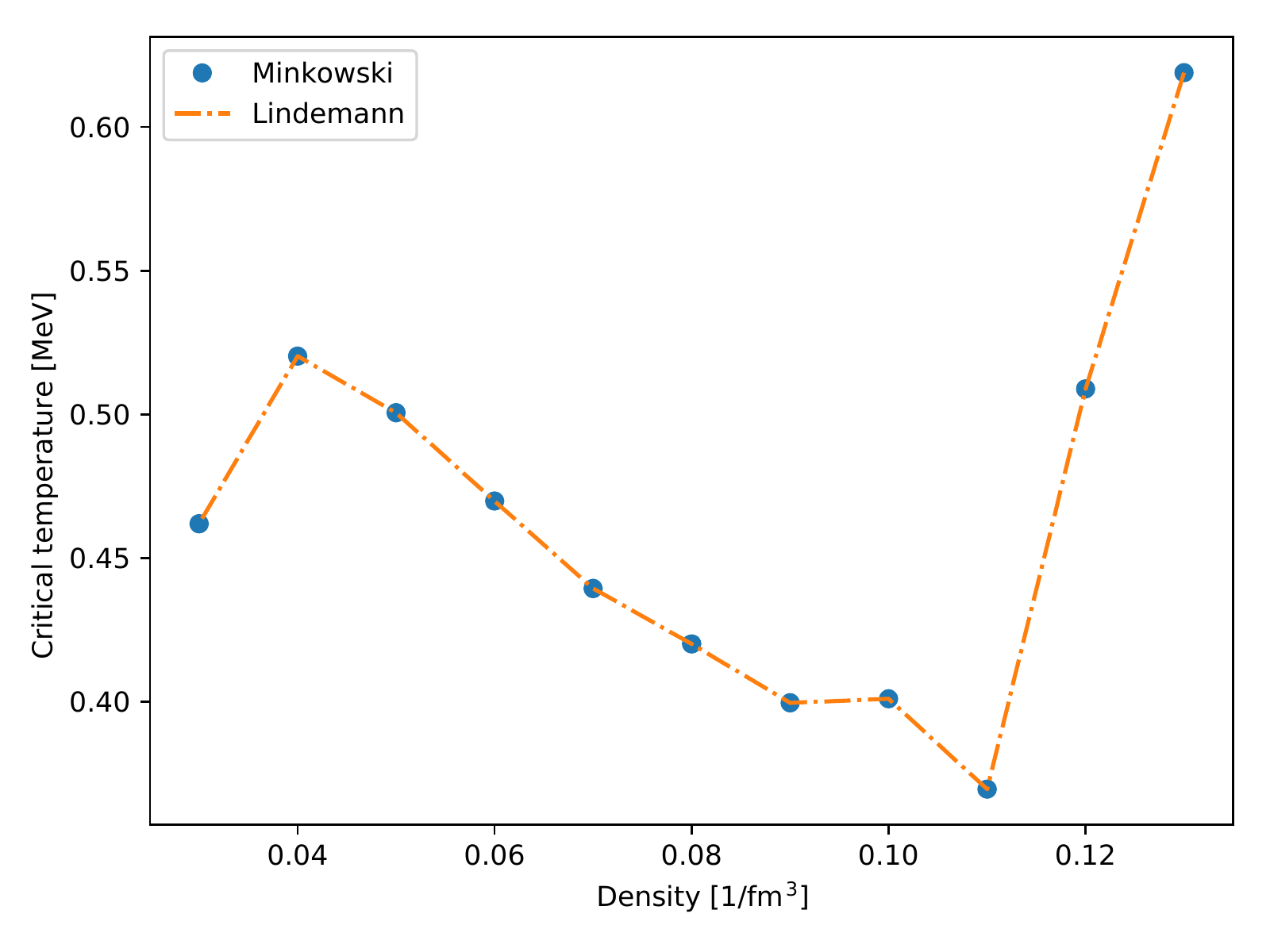}
  \caption{Critical temperature as a function of density. We see the
    overlap between the Minkowski and the Lindemann critical
    temperature.}
  \label{fig:critical_temperature}
\end{figure}

These signals of a solid-liquid phase transition (energy and Lindemann's 
coefficient discontinuity) and morphological transition (Minkowski functionals 
discontinuity) point at the same transition temperature, as can be seen in the 
phase diagram of Figure~\ref{fig:critical_temperature}.  This means that as the 
systems are cooled down at fixed volume, they undergo a thermodynamical and a 
morphological phase transition, and they do so at the same temperature.\\

In summary, and comparing to NM, the introduction of the coulomb screened 
potential appears to smooth out the phase transitions. The Euler functional 
$\chi$ experiences a somewhat smooth change along $1-2\,$MeV, in correspondence 
with the energy changes mentioned in Section~\ref{subsec:sym_energy_1}. Thus, 
the pasta forming process may be located at this temperature range, in a similar 
fashion as in nuclear matter systems. At temperatures below $T\simeq 0.5$ MeV 
the (inner) nucleons in the pasta structure freeze into a solid state. 

\subsubsection*{The radial distribution function}\label{subsec:rdf}

\begin{figure}
\begin{center}
   \includegraphics[width=0.5\columnwidth]{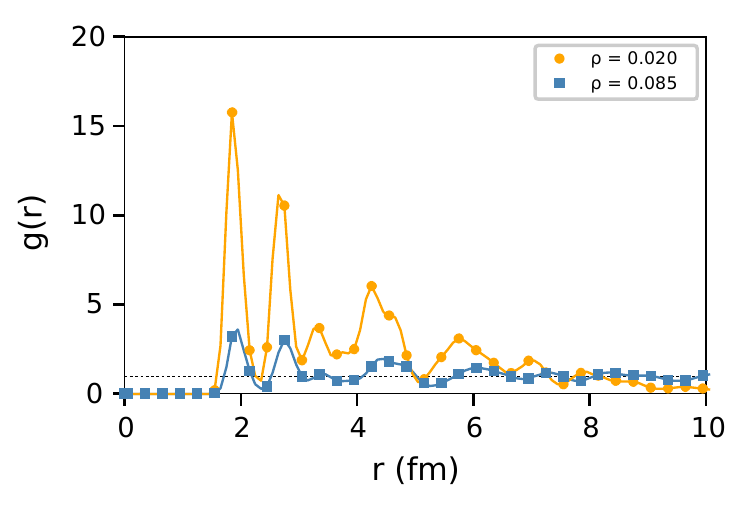}
\caption{Radial distribution function $g(\mathbf{r})$ for nucleons 
corresponding to a symmetric neutron star matter system of $N=4000$ nucleons 
and 
$T=0.2$ MeV. The system mean density is indicated in the inset in fm$^{-3}$. 
The horizontal line at $g(\mathbf{r})=1$ corresponds to the asymptotic limit 
expected for 
infinite systems.}\label{fig:gr_rho02_rho085_all}
\end{center}
\end{figure}

Here we use the radial distribution function to explore the phase 
transformations detected by the caloric curve. 
Figure~\ref{fig:gr_rho02_rho085_all} shows $g(\mathbf{r})$ for systems with 
$\rho=0.02$ and  0.085 fm$^{-3}$; the lower density case shows more pronounced 
nearest-neighbor peaks than the high density one, and it corresponds to a more 
crystalline phase. Looking at Figure~\ref{fig:eos_3}, we see that the highest 
internal energy is associated to a more regular distributions of nucleons within 
the pasta regime.\\

\begin{figure*}[!htbp]
\centering
\subfloat[Radial distribution function for $\rho=0.03$ fm$^{-3}$ ]{
\includegraphics[width=0.24\columnwidth]
{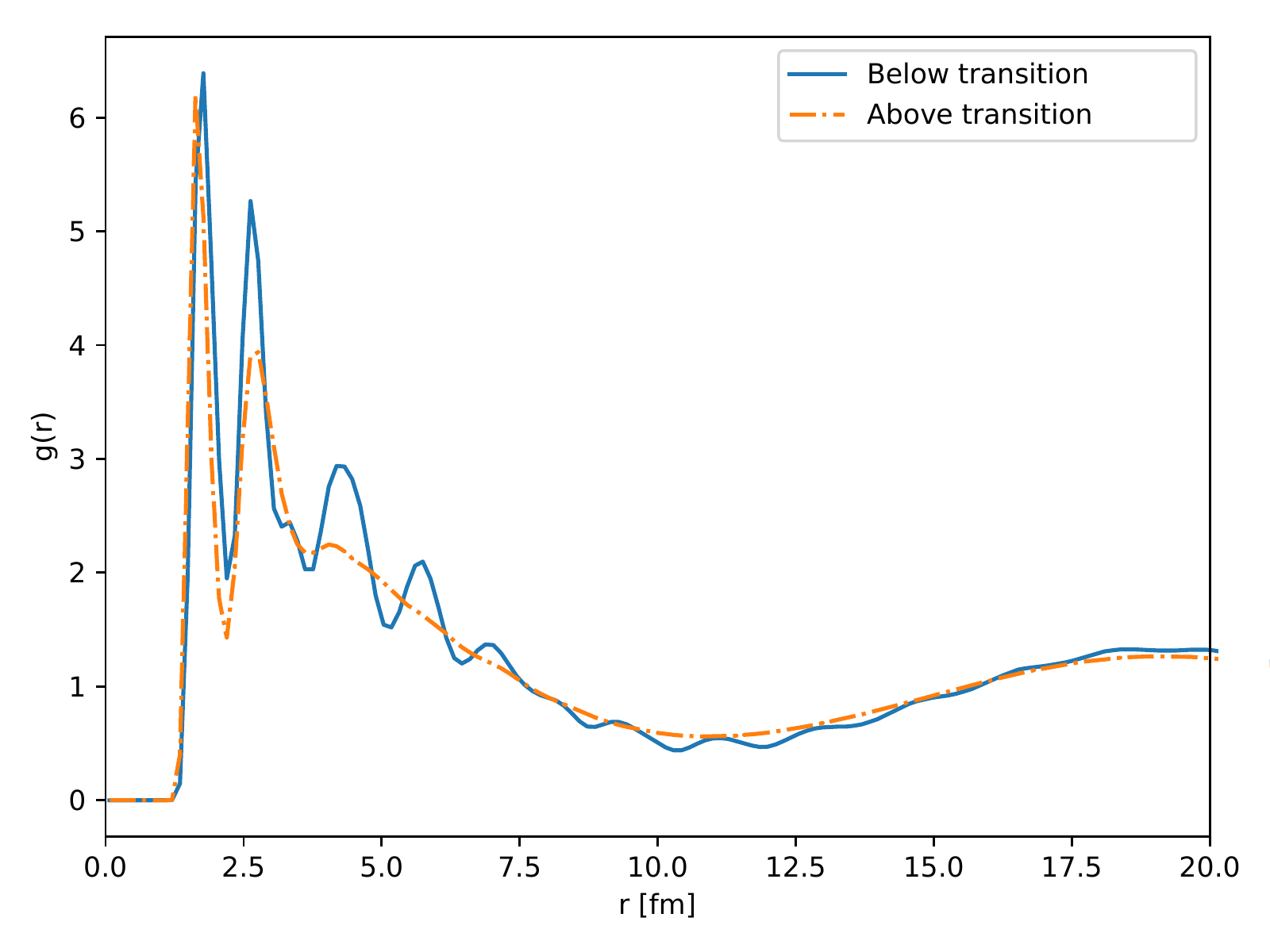}
}
\subfloat[Snapshot of the system in the liquid phase for 
$\rho=0.03\,\text{fm}^{-3}$ ]{
\includegraphics[width=0.24\columnwidth]
{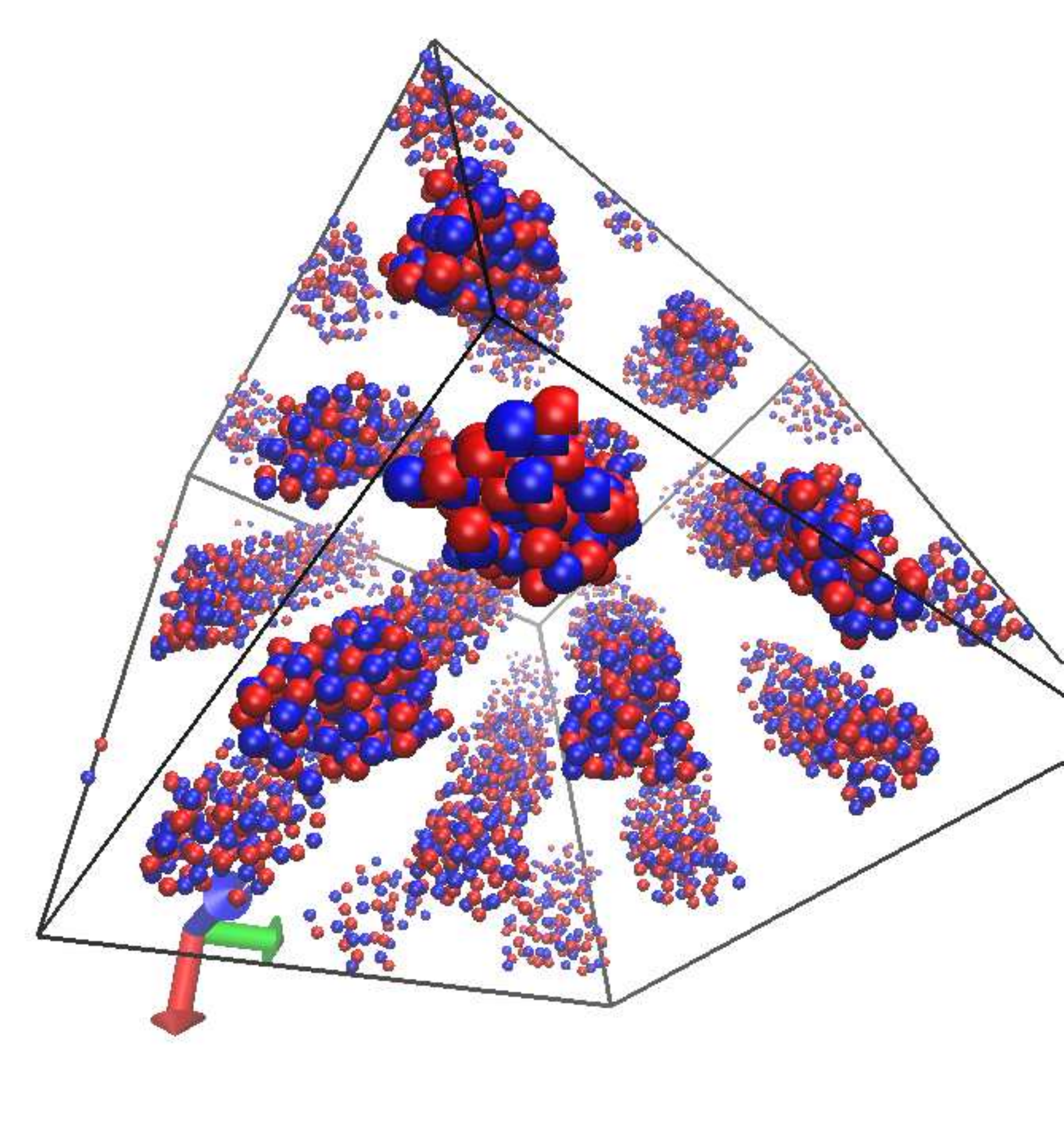}
}
\subfloat[Radial distribution function for $\rho=0.05$ fm$^{-3}$ ]{
\includegraphics[width=0.24\columnwidth]
{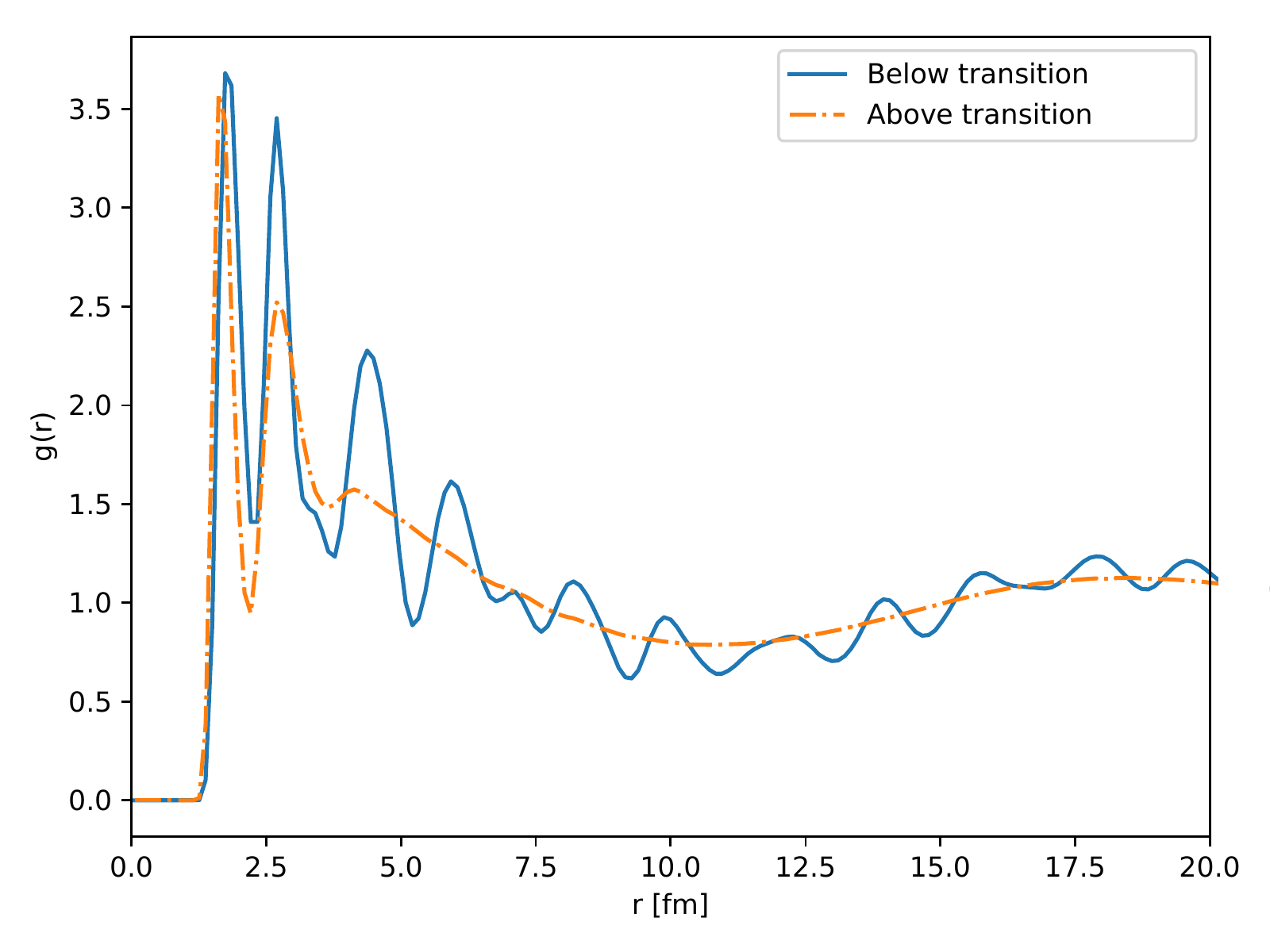}
}
\subfloat[Snapshot of the system in the liquid phase for 
$\rho=0.05\,\text{fm}^{-3}$ ]{
\includegraphics[width=0.24\columnwidth]
{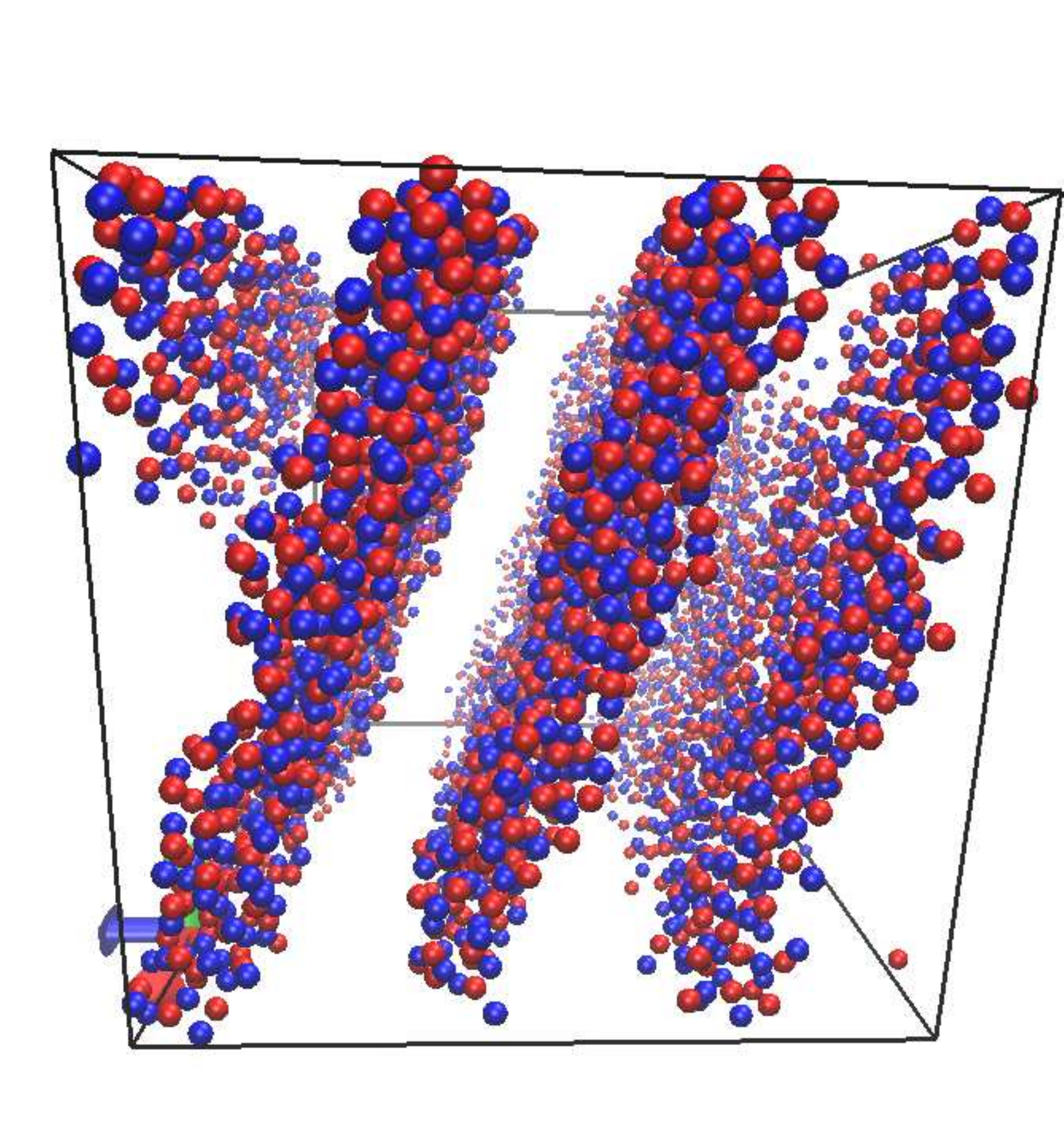}
}
 \caption{Radial distribution function for different densities, both below and 
above the transition temperature, and snapshots of the system in the liquid 
phase. Although the first peaks of the distribution are in the same position for 
both temperatures, the following peaks, which exhibit a long-range order typical 
of solids, are only present below the transition temperature.} \label{fig:rdf}  
\end{figure*}

In figure~\ref{fig:rdf} we show the radial distribution function for three 
different densities: $\rho=0.03$ fm$^{-3}$ (spaghetti), $\rho=0.05$ fm$^{-3}$ 
(lasagna) and $\rho=0.08$ fm$^{-3}$ (tunnels), just above and below the 
transition temperature, as well as a snapshot of the system at the high 
temperature phase. Since the first peaks (corresponding to the nearest 
neighbors) are at the same position regardless of the temperature, we conclude 
that the short range order is present both above and below the transition. 
However, the peaks for third and higher order neighbors, distinctive of solid 
phases, disappear as the temperature is increased through the transition. \\

\begin{figure*}[!htbp]
\centering
\subfloat[Below the transition.]{
\includegraphics[width=0.45\columnwidth]
{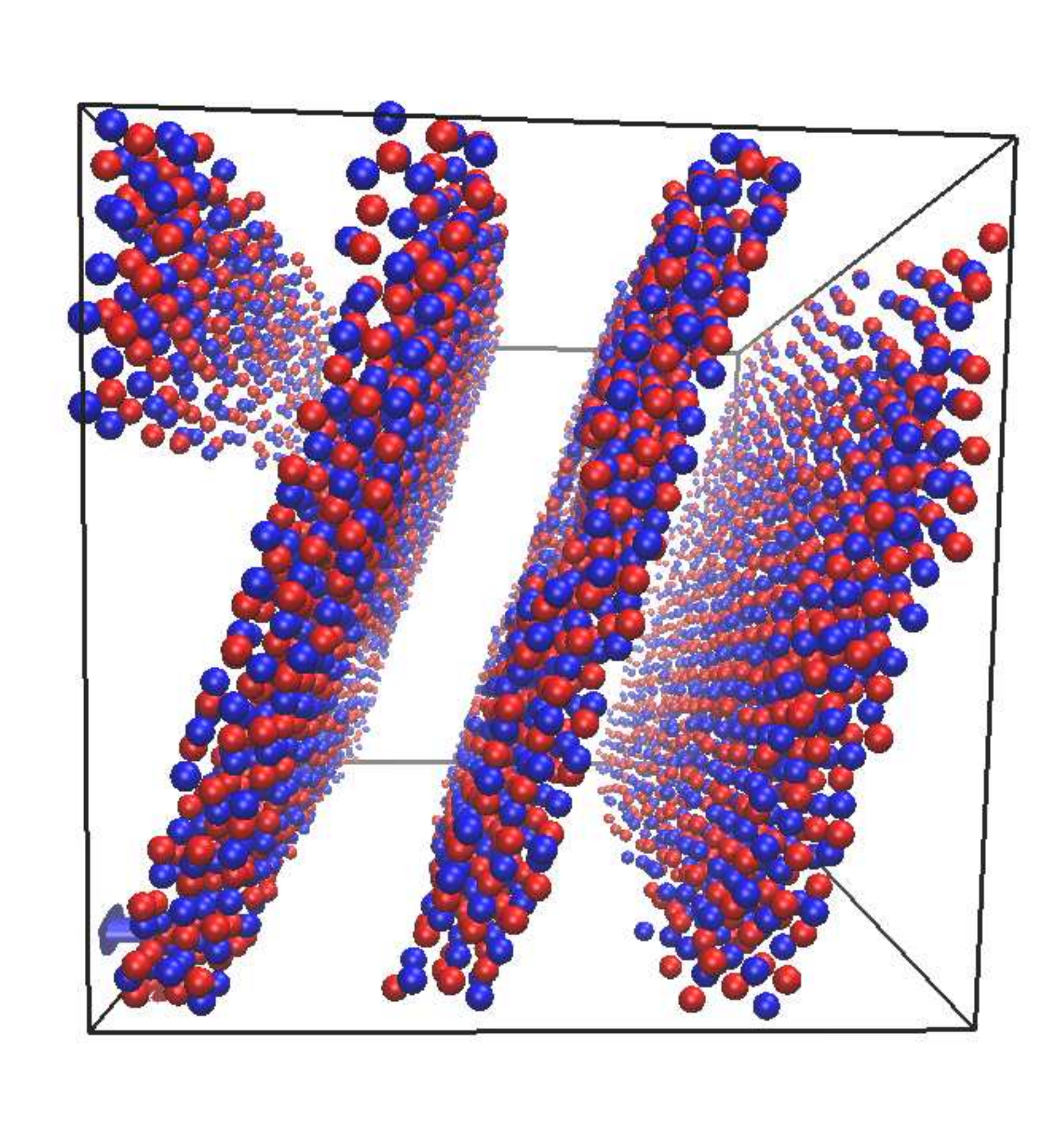}
}
\subfloat[Above the transition.]{
\includegraphics[width=0.45\columnwidth]
{morph_0-05_0-50.pdf}
}
\caption{Spatial distribution for $\rho=0.05\,\text{fm}^{-3}$, both above and 
below the transition temperature. The structures are similar, but much more 
disordered above the transition.}
\label{fig:morph}
\end{figure*}

\begin{figure*}[!htbp]
\centering
\subfloat[Usual \emph{lasagna}\label{fig:frame_rho01_x05}]{
\includegraphics[width=0.21\columnwidth]
{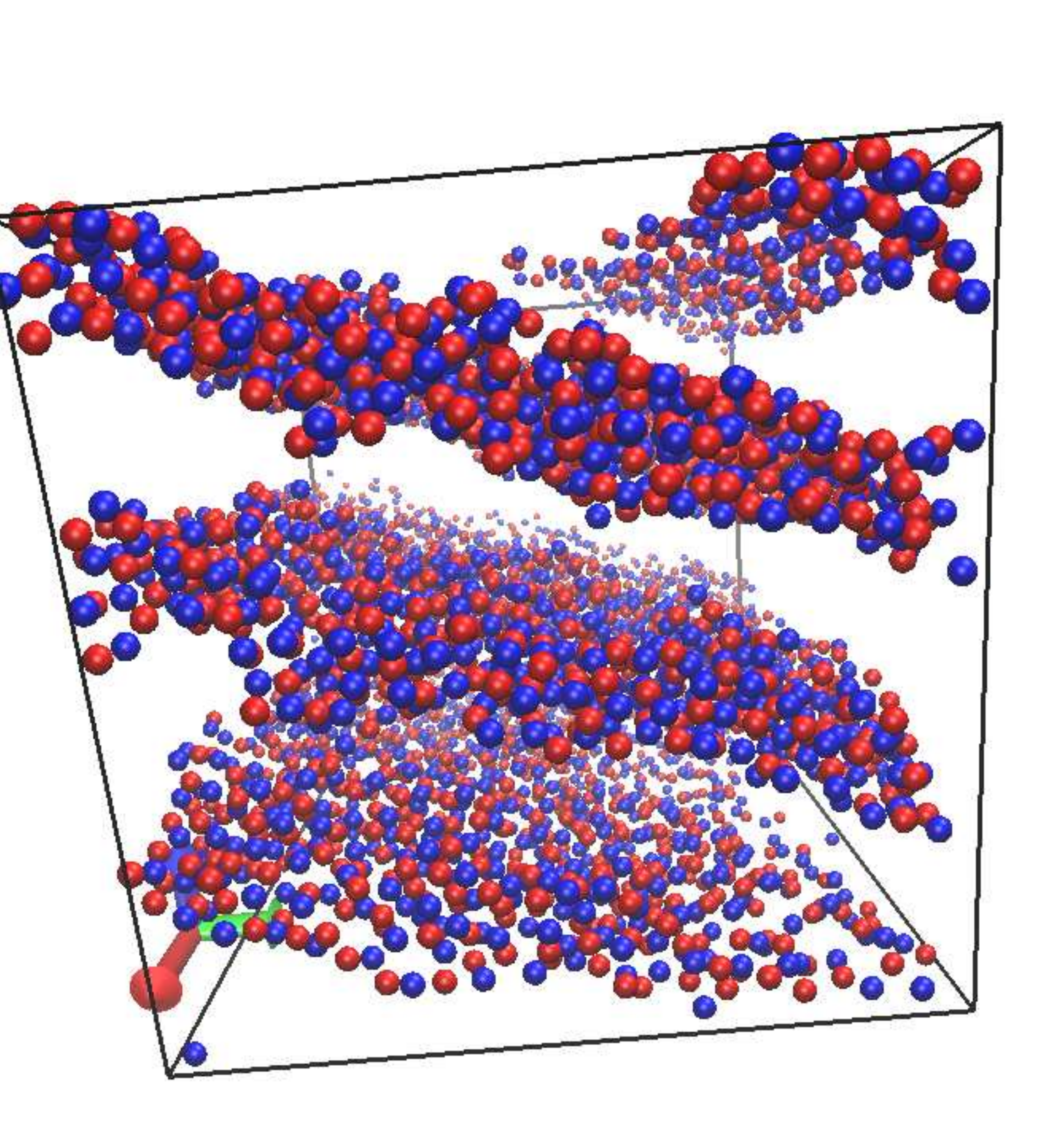}
}
\subfloat[Intertwined \emph{lasagna}\label{fig:frame_rho04_x05}]{
\includegraphics[width=0.21\columnwidth]
{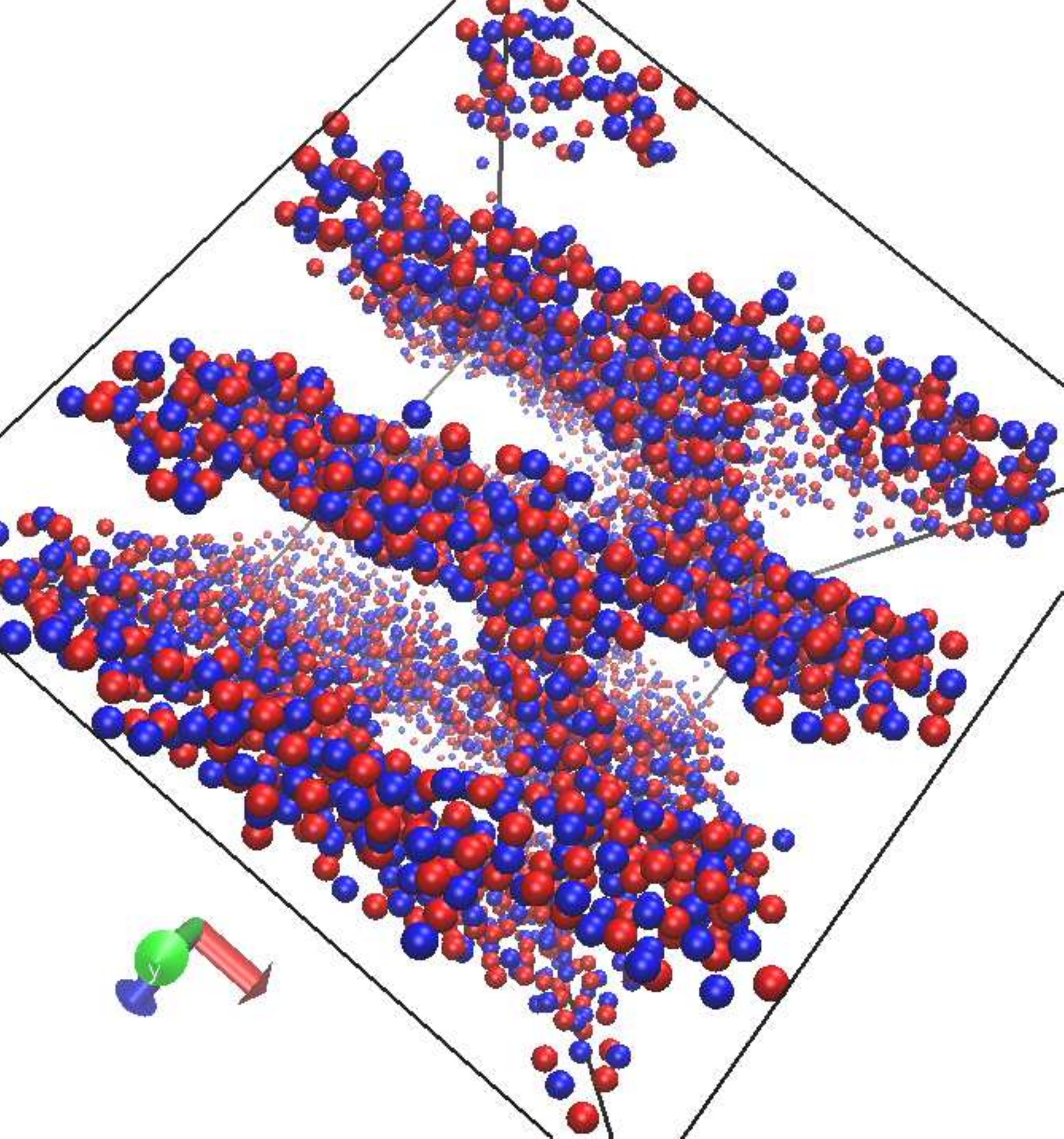}
}
\subfloat[Intertwined \emph{lasagna}\label{fig:frame_rho085_x05_a}]{
\includegraphics[width=0.21\columnwidth]
{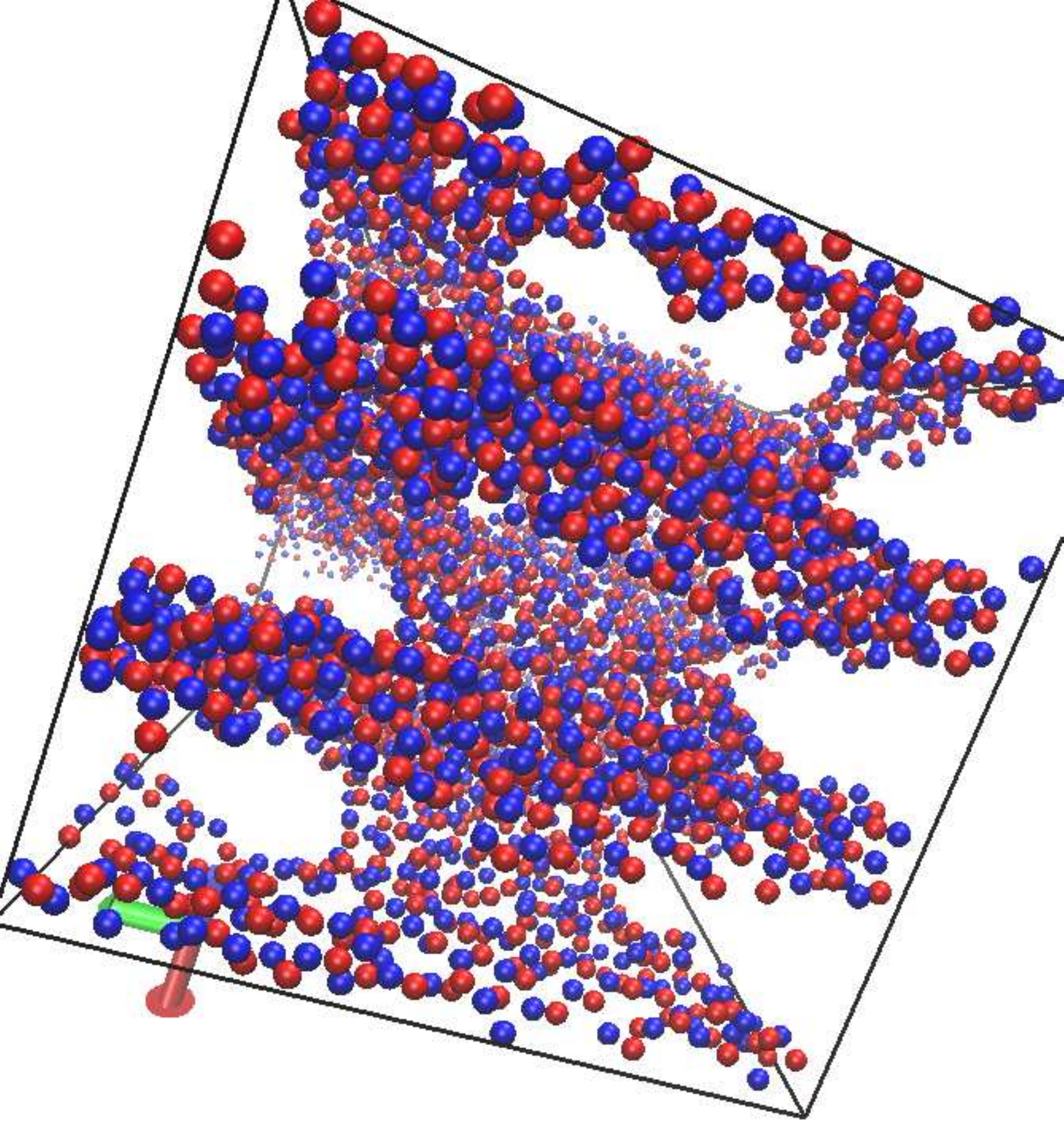}
}
\subfloat[Unusual pasta shape\label{fig:frame_rho085_x05_a}]{
\includegraphics[width=0.21\columnwidth]
{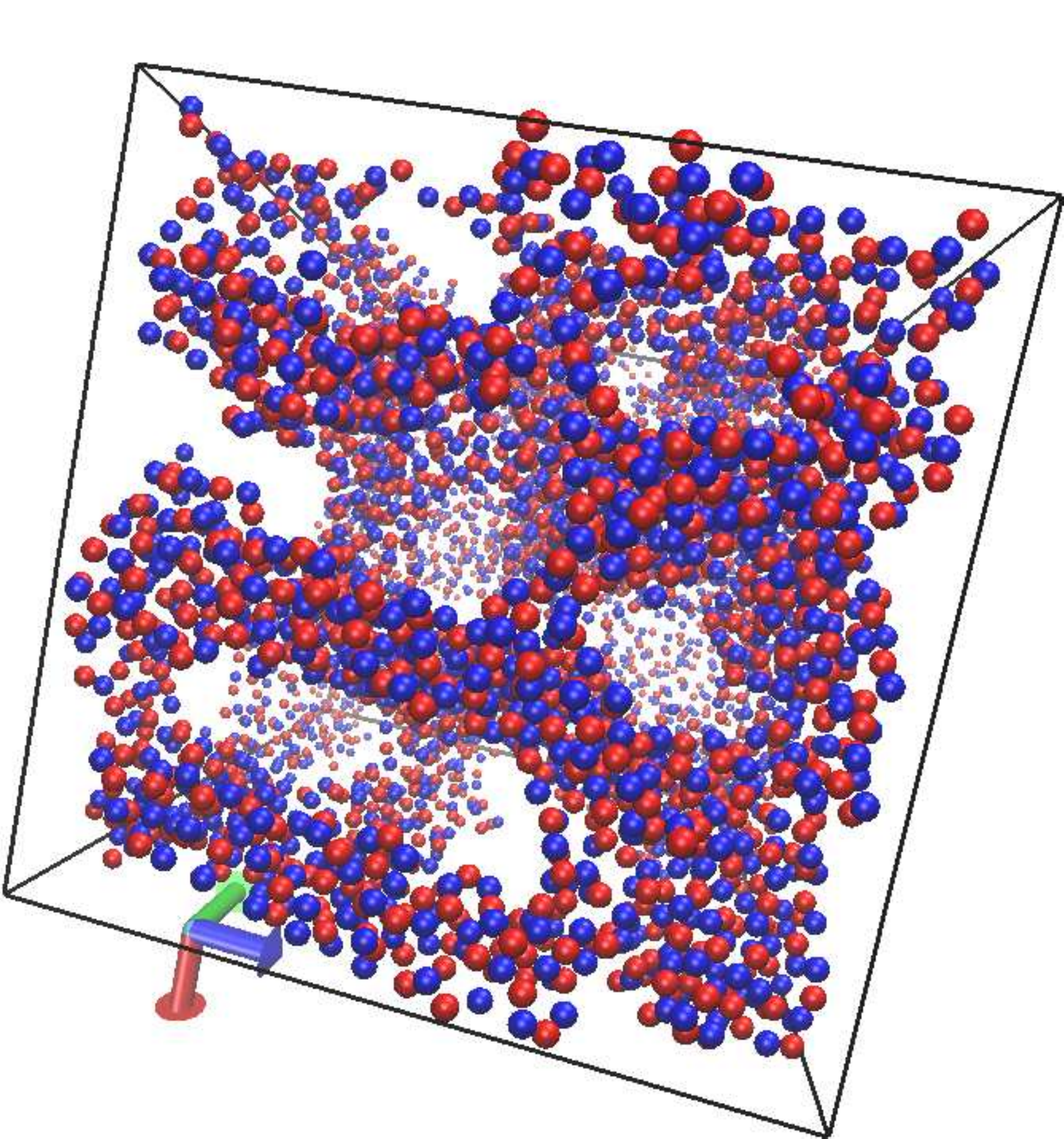}
}
\caption{Spatial distribution for $\rho=0.05$ fm$^{-3}$ for different initial 
conditions at $T = 0.6$ MeV.}\label{fig:cool_morph}
\end{figure*}

On top of the disappearance of the long range order characteristic of solids, 
another feature becomes evident from figure~\ref{fig:rdf}. As temperature 
increases through the solid-liquid transition, a very-long range modulation in 
the pair correlation function survives. This very long range ordering is 
characteristic of the pasta phases. In Figure~\ref{fig:morph}, a visual 
representation of the spatial configuration for $\rho=0.05$ fm$^{-3}$ is shown, 
for temperatures both below and above the transition. In it, we show that not 
only the solid phase has the usual pasta shape, but the liquid phase preserves 
it. Below the transition, we have frozen pasta. Just above it, nucleons may flow 
but confined to a certain pasta or pasta-like structure.  In 
Figure~\ref{fig:cool_morph} the usual lasagna is observed, but also intertwined 
lasagnas and other structures that are not of the usual pasta type. \\

\begin{figure*}[!htbp]
\centering
\subfloat[$\rho=0.02$\label{fig:frame_rho01_x05}]{
\includegraphics[width=0.36\columnwidth]
{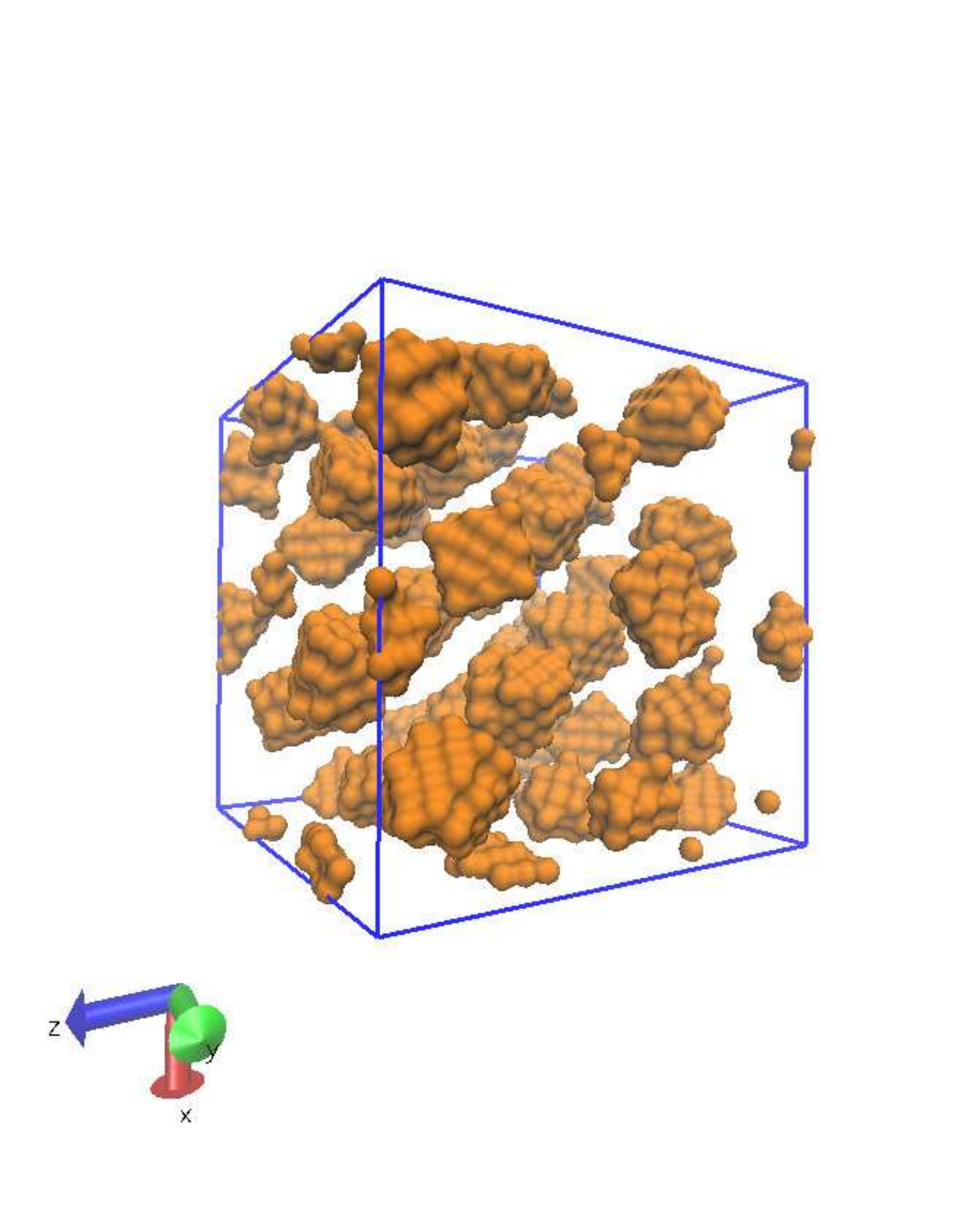}
}
\subfloat[$\rho=0.04$\label{fig:frame_rho04_x05}]{
\includegraphics[width=0.36\columnwidth]
{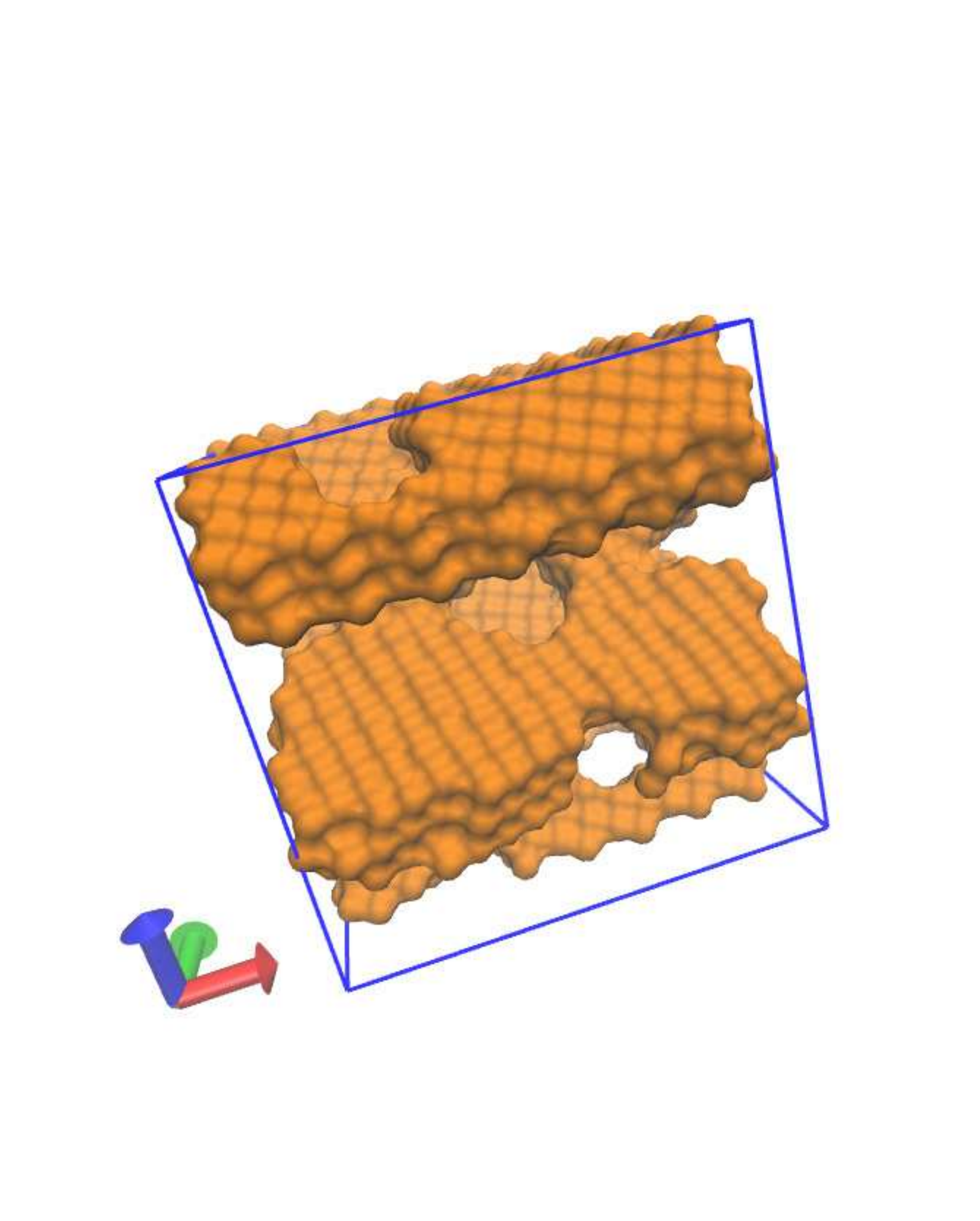}
}
\subfloat[$\rho=0.085$\label{fig:frame_rho085_x05_a}]{
\includegraphics[width=0.36\columnwidth]
{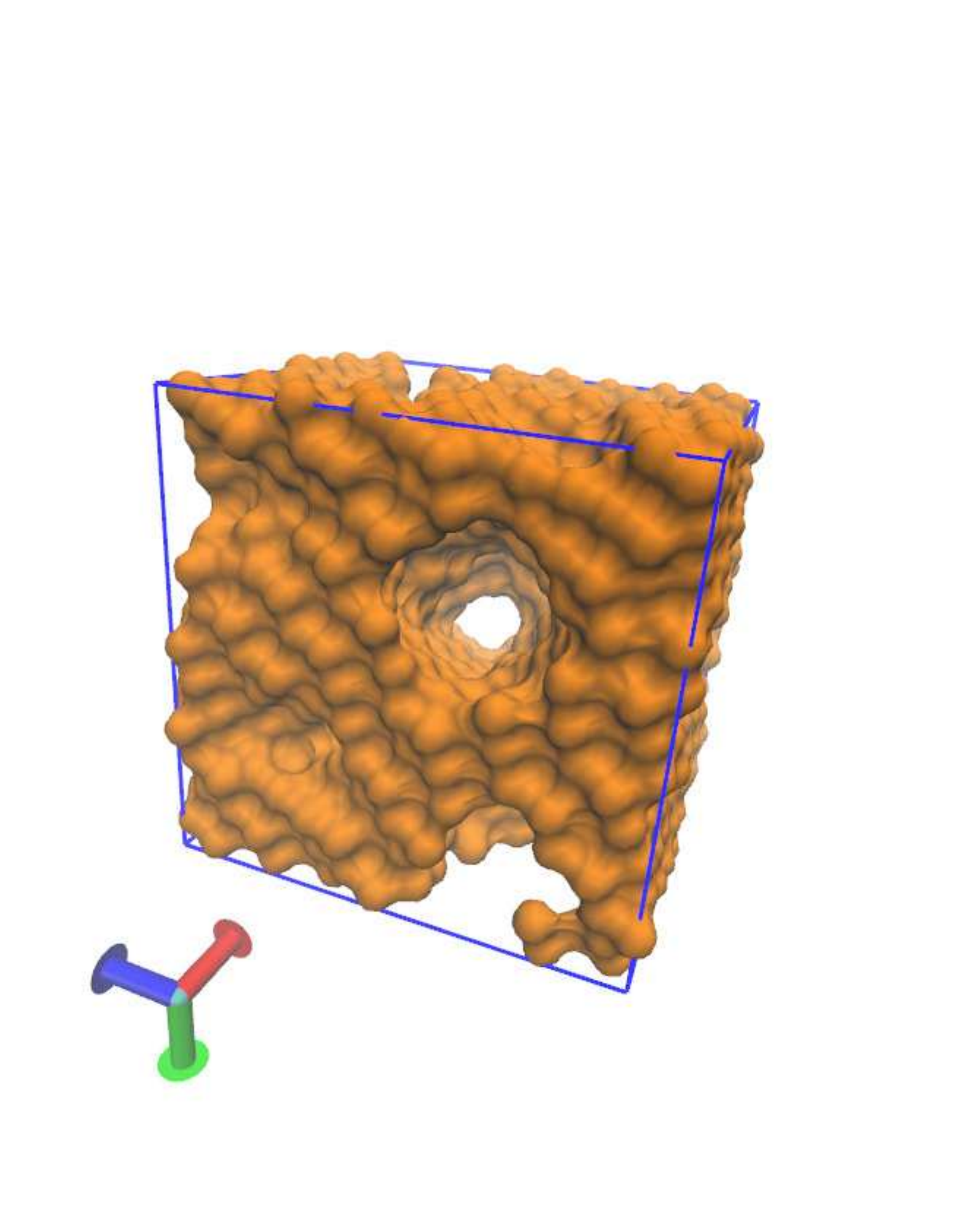}
}
\caption{\label{fig:frame_rho01_rho04_rho085_sym}  Surface 
representation for the protons at the indicated densities. The snapshot was 
taken at $T=0.2$ and $x=0.5$ for a system of $4000$ nucleons.  }
\end{figure*}

Figure~\ref{fig:frame_rho01_rho04_rho085_sym} shows the proton arrangements for 
three density situations. Figure~\ref{fig:frame_rho085_x05_a} indicates that the 
$\rho=0.085$ fm$^{-3}$ situation is highly homogeneous inside the occupied 
regions. These occupied regions split into smaller pieces, according to  
Figs.~\ref{fig:frame_rho04_x05} and \ref{fig:frame_rho01_x05}, forming lasagnas 
and gnocchis, respectively. Thus, as the density diminishes, the structures 
break into smaller pieces, and the internal energy decreases (see 
Figure~\ref{fig:eos_2}). \\

Likewise, at $\rho=0.02$ fm$^{-3}$, Figure~\ref{fig:gr_rho02_rho085_all} shows 
that $g(\mathbf{r})$ has strong neighbor correlations at $r\leq6\,$fm $^{-3}$ 
and almost no correlation ($g(\mathbf{r}) \approx 0$) for $r\geq 8$ fm$^{-3}$. 
That is, strong correlations exist between neighbors belonging to the same 
gnocco, and no correlation at large separation distances between gnocchis (see 
Figure~\ref{fig:frame_rho01_x05}). For the larger density, $\rho=0.085$ 
fm$^{-3}$, $g(\mathbf{r})$ tends to $1$ as expected for homogeneous systems 
(c.f. \ref{fig:frame_rho085_x05_a}).\\

\subsubsection*{Summary of Section~\ref{sym_energy}}\label{subsec:sum}

We may summarize the Section as follows. The internal energy for symmetric 
neutron star matter increases monotonically for either increasing temperatures 
and densities; it does not appear to have a saturation point.  These effects do 
not occur for nuclear matter, and they are due to the Coulomb interation of the 
electron gas as can be seen clearly by comparing Figure~\ref{fig:eos_1} to the 
nuclear matter case. Likewise, the introduction of the Coulomb potential affects 
the system morphology; as the density diminishes, the pasta structures split 
into smaller structures; these are true pasta structures, as opposed to the 
one-structure-per-cell pseudo-pastas observed in NM.

\subsection{Non-symmetric neutron star matter}\label{subsec:asym}

We now turn to a study of NSM with proton fractions more like those in in 
neutron crusts. All simulations in this Section consist of systems with $N=4000$ 
nucleons under periodic boundary conditions and interacting through the New 
Medium potential and, whenever used, the binning distance is $d=2.35\,$fm (see 
Appendices~\ref{cmd} and~\ref{tools} for details). 

\subsubsection*{The internal energy}\label{subsec:asym_energy}

\begin{figure*}[!htbp]
\centering
\subfloat[$\rho=0.04$\label{fig:eos_4}]{
\includegraphics[width=0.5\columnwidth]
{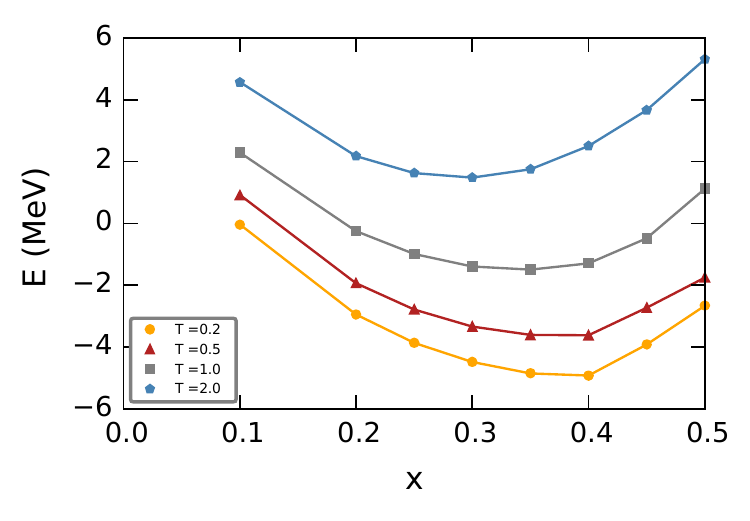}
}\\
\subfloat[$\rho=0.085$\label{fig:eos_5}]{
\includegraphics[width=0.5\columnwidth]
{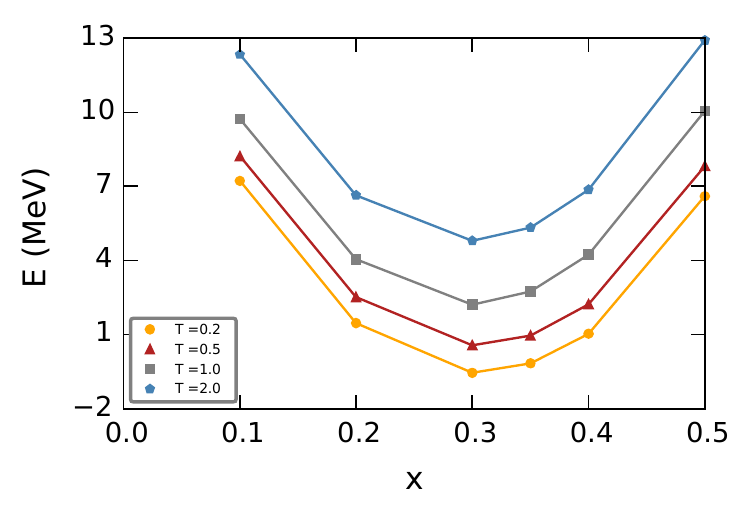}
}
\caption{\label{fig:eos_x} 
Internal energy per nucleon for neutron star matter systems with density (a) 
$\rho=0.04$  fm$^{-3}$ and (b) $\rho=0.085$ fm$^{-3}$. }
\end{figure*}

We start the study of non-isospin-symmetric NSM by exploring the dependence of 
the internal energy on the isospin content. Figures~\ref{fig:eos_4} and 
\ref{fig:eos_5} show the isothermal energies as a function of $x$ for systems 
with densities $\rho=0.04$ fm$^{-3}$ and $\rho=0.085$ fm$^{-3}$. Each of the 
curves exhibits a $\cup$ shape indicating the existence of a minimum of the 
energy at a certain value of $x$. At $\rho=0.04$ fm$^{-3}$ the observed minima 
appear to shift from $x=0.4$ at $T=0.2$ MeV to $x=0.3$ at $T=2.0$ MeV, while at 
$\rho=0.085$ fm$^{-3}$ the minima remain at $x=0.3$ at all temperatures. This 
finding indicates that systems with the freedom of exchanging their content of 
neutrons and protons would favor a specific isospin ratio $x$ depending on the 
local temperature and density of the system.\\

\begin{figure*}[!htbp]
\centering
\subfloat[$T=0.2$\label{fig:eos_6}]{
\includegraphics[width=0.5\columnwidth]
{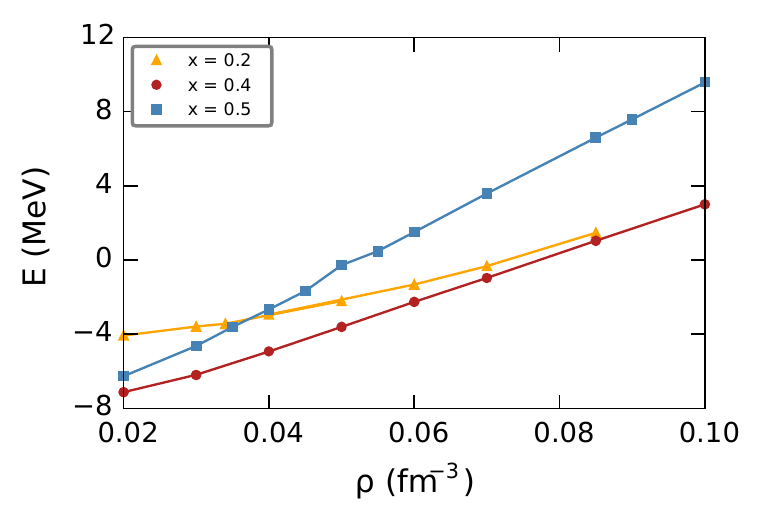}
}\\
\subfloat[$T=1.0$\label{fig:eos_7}]{
\includegraphics[width=0.5\columnwidth]
{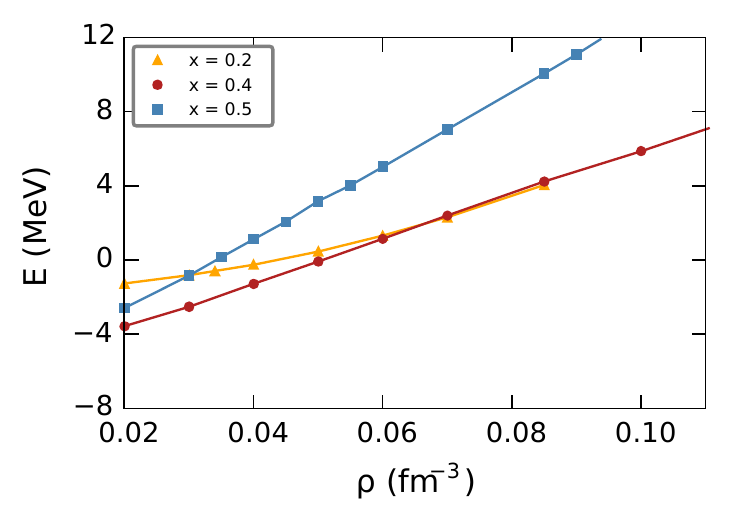}
}
\caption{\label{fig:eos_rho}  Internal energy per nucleon for 
neutron star matter systems of $N=4000$ nucleons at three proton fractions and 
interacting through the New Medium model. (a) Temperature $T=0.2\,$MeV. (b) 
Temperature $T=1.0\,$MeV. }
\end{figure*}

Complementary information can be obtained from the density dependence of the 
energy.  Figure~\ref{fig:eos_rho} shows the corresponding curves for $x=0.2$, 
0.4 and 0.5 at $T=0.2$ MeV and 1.0 MeV. It is interesting to notice that while 
the $x=0.4$ and 0.5 curves appear to have similar monotonically decreasing 
behaviors at all densities, the $x=0.2$ case deviates at low densities; as 
explained in~\cite{dor2018}, this is due to the $\cup$ shape of the $x$ 
dependence of the energy seen in Figures~\ref{fig:eos_4} and \ref{fig:eos_5}.\\

\subsubsection*{The associated morphology} \label{subsec:asym_topology}

The Euler characteristic $\chi$ and the radial distribution function 
$g(\mathbf{r})$ are now used to study the evolution of the pasta structures. \\

\begin{figure*}[!htbp]
\centering
\subfloat[$\rho=0.04$\label{fig:minkowski_asym_rho04}]{
\includegraphics[width=0.45\columnwidth]
{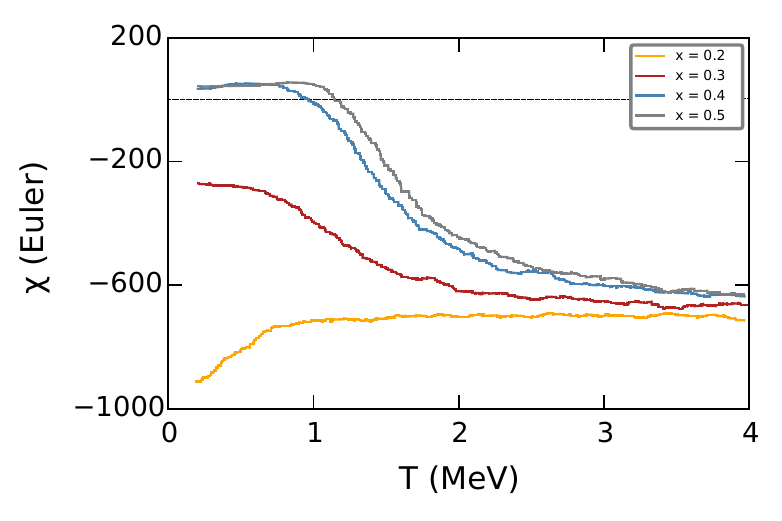}
}
\subfloat[$\rho=0.085$\label{fig:minkowski_asym_rho085}]{
\includegraphics[width=0.45\columnwidth]
{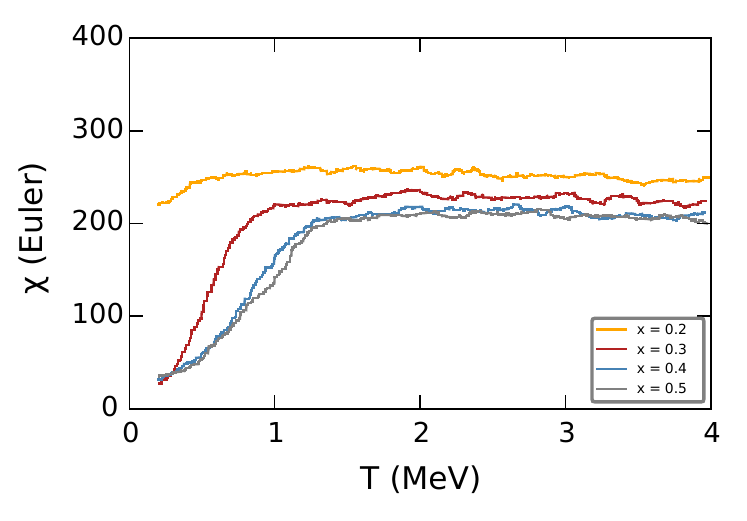}
}
\caption{The Euler characteristic $\chi$ as a function of temperature, for two 
densities and for four proton fractions as indicated in the inset. The 
horizontal line corresponds to the null level, and the data has been smoothed 
with a moving average procedure.}
\label{fig:minkowski_asym} 
\end{figure*}

As explained in Appendix~\ref{tools} and in~\ref{nmltsd} for NM, the Euler 
characteristic $\chi$ can indicate changes of structure. 
Figure~\ref{fig:minkowski_asym} plots $\chi$ as a function of temperature for 
four proton fractions and two densities. Changes of behavior can be seen at 
around $T\approx1$ MeV. Figure~\ref{fig:minkowski_asym_rho04} shows that at 
$\rho=0.04$ fm$^{-3}$ and $T\ge2$ MeV, the system appears dominated by voids and 
tunnels independent of the isospin content, while at $T\le 1$ MeV more compact 
objects appear at all $x$, except for $x=0.2$ which disperses even more.  
Figure~\ref{fig:minkowski_asym_rho085} shows that at the higher density 
($\rho=0.085$ fm$^{-3}$) the system attains a more compact structure at all 
isospin contents and for $T\ge 1$ MeV, becoming less pronounced at lower 
temperatures ($T\le 1$ MeV).\\

\begin{figure*}[!htbp]
\centering
\subfloat[$x=0.2$\label{fig:frame_rho085_x02}]{
\includegraphics[width=0.34\columnwidth]
{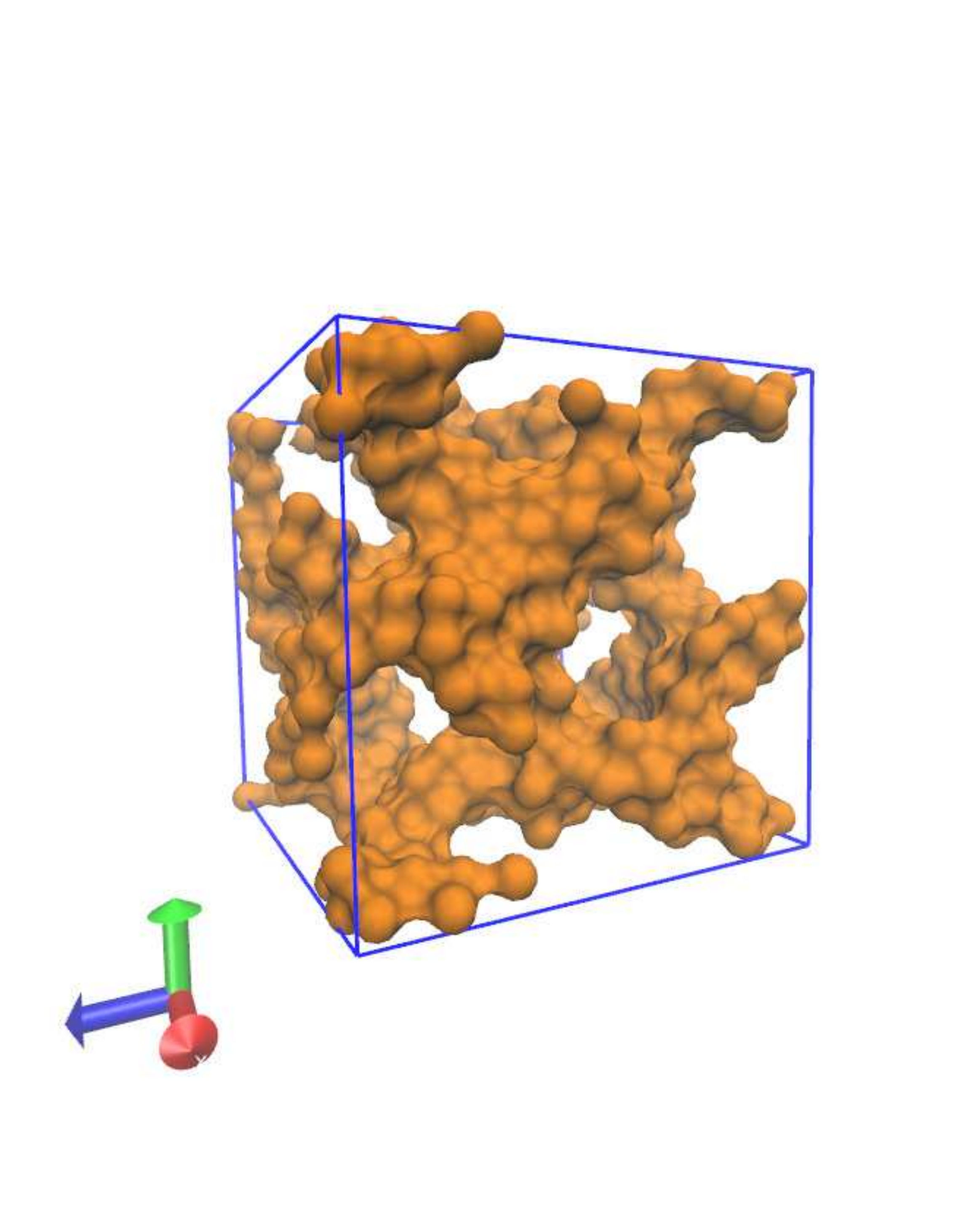}
}
\subfloat[$x=0.4$\label{fig:frame_rho085_x04}]{
\includegraphics[width=0.34\columnwidth]
{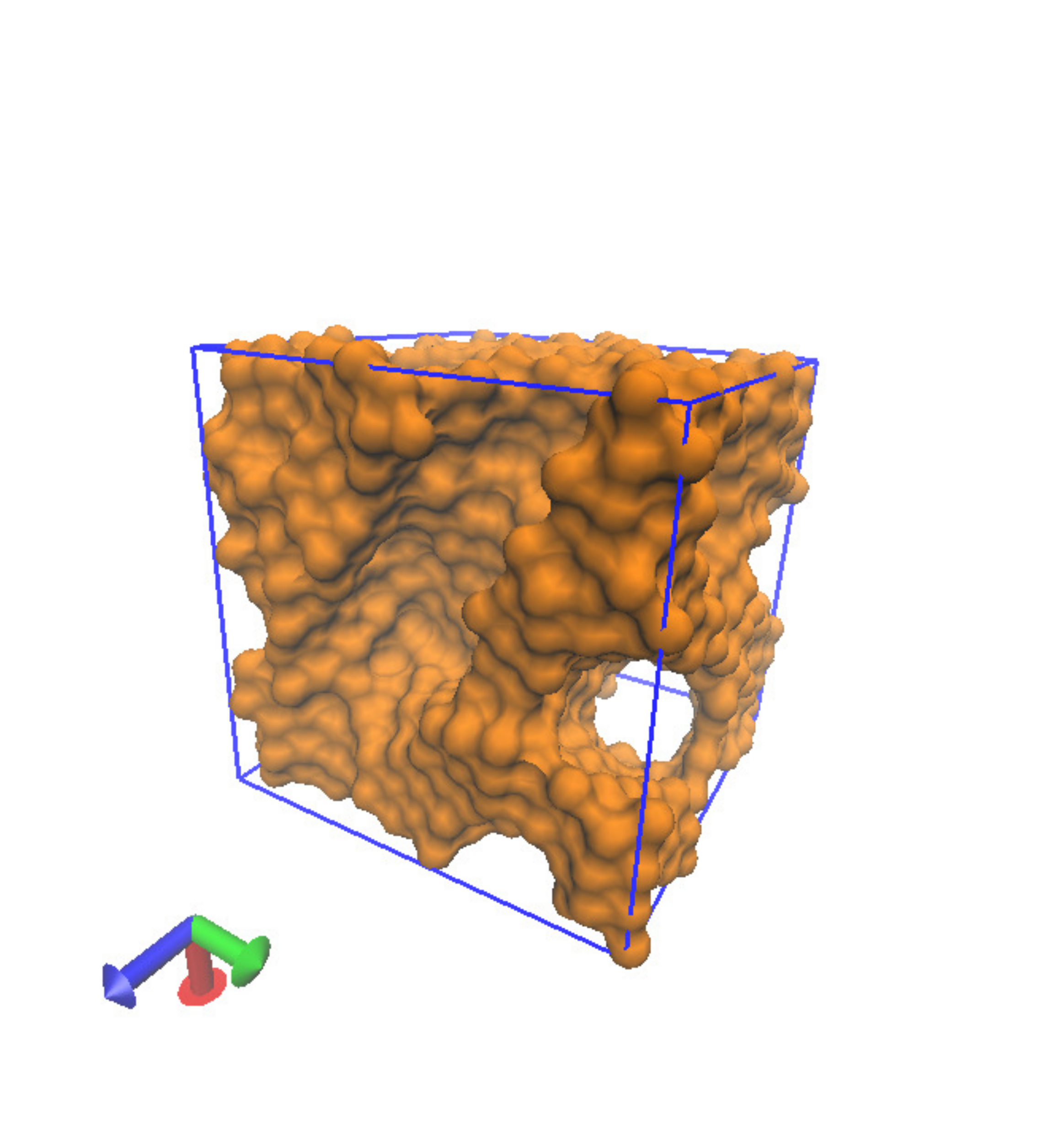}
}
\subfloat[$x=0.5$\label{fig:frame_rho085_x05}]{
\includegraphics[width=0.34\columnwidth]
{./lammps_rho085_x05_protons_surf_star.pdf}
}
\caption{Surface representation for the distribution of protons at the indicated 
proton fractions, $T=0.2$ MeV and $\rho=0.085$ fm$^{-3}$.}
\label{fig:frame_rho085_asym}   
\end{figure*}

Figure~\ref{fig:frame_rho085_asym} shows the position of the protons for the 
cases of $x=0.2$, 0.4 and 0.5 at $T=0.2$ MeV and $\rho=0.085$ fm$^{-3}$. These 
structures correspond to the $T=0.2$ MeV points of the $\chi$ curves in 
Figure~\ref{fig:minkowski_asym_rho085}. Notice that no major changes in the 
percentage of void to filled volumes are observed as it is difficult to 
appreciate such changes in compact structures.\\

\begin{figure*}[!htbp]
\centering
\subfloat[$x=0.1$\label{fig:frame_rho04_x01}]{
\includegraphics[width=0.34\columnwidth]
{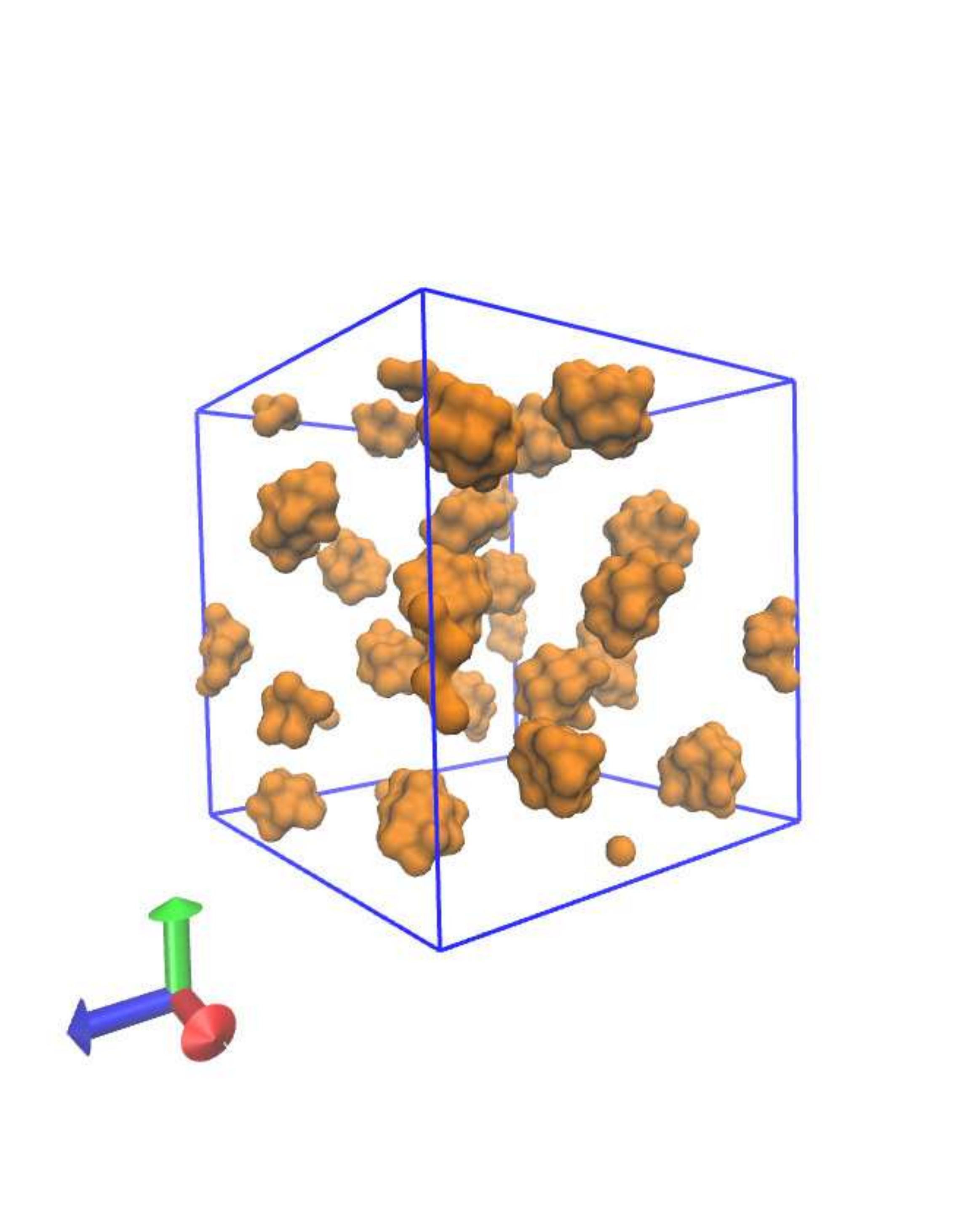}
}
\subfloat[$x=0.2$\label{fig:frame_rho04_x02}]{
\includegraphics[width=0.33\columnwidth]
{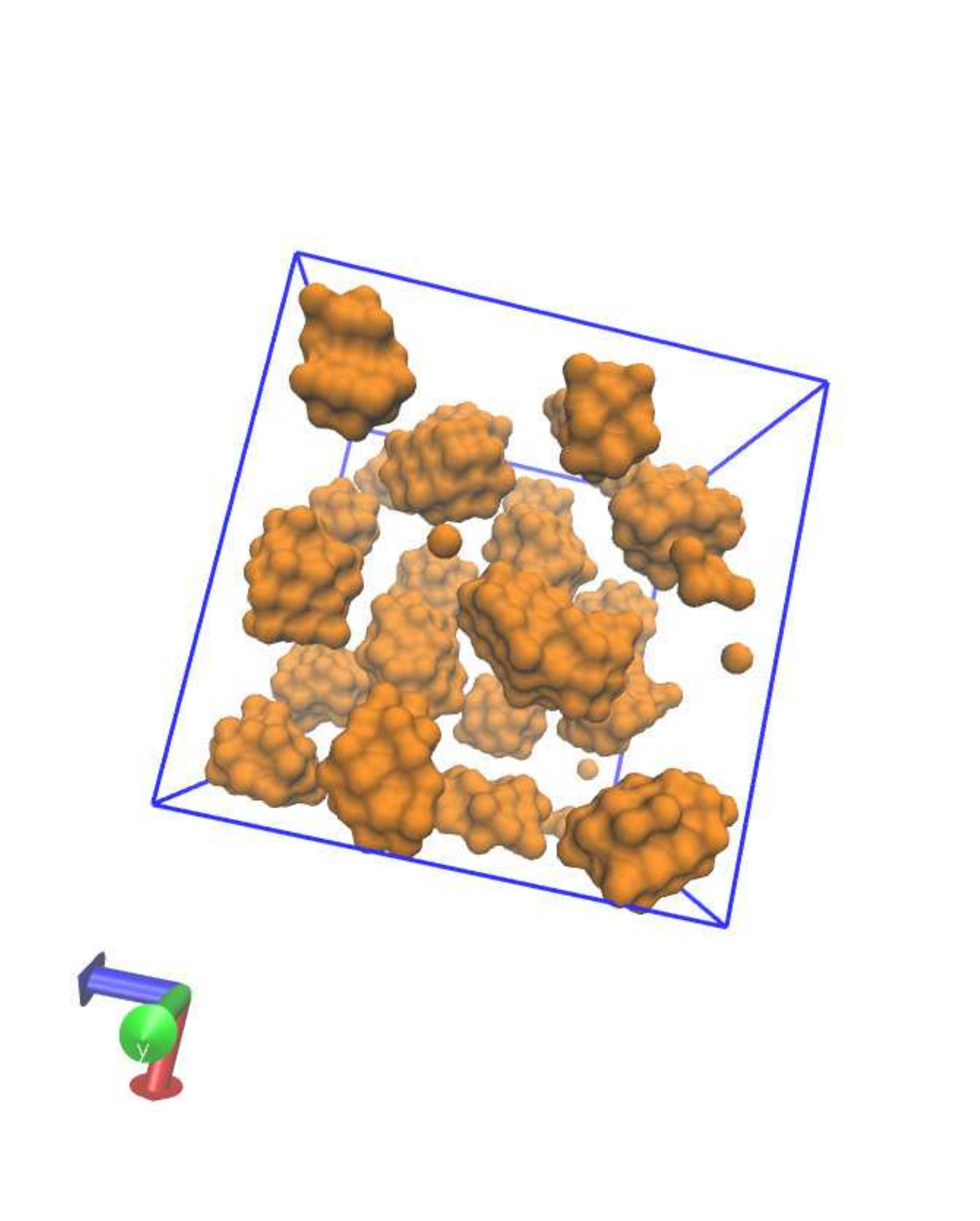}
}
\subfloat[$x=0.3$\label{fig:frame_rho04_x03}]{
\includegraphics[width=0.33\columnwidth]
{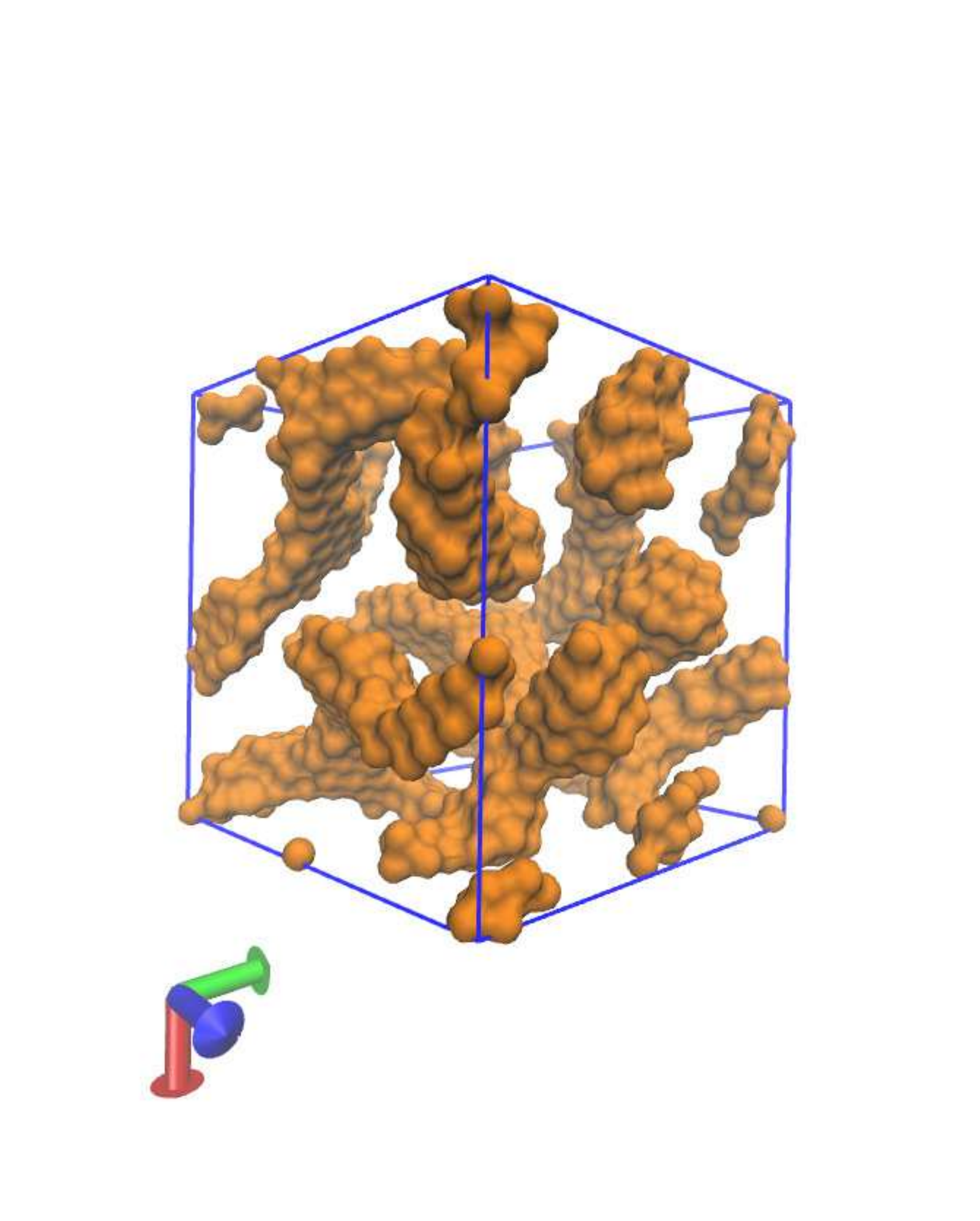}
} \\
\subfloat[$x=0.4$\label{fig:frame_rho04_x04}]{
\includegraphics[width=0.33\columnwidth]
{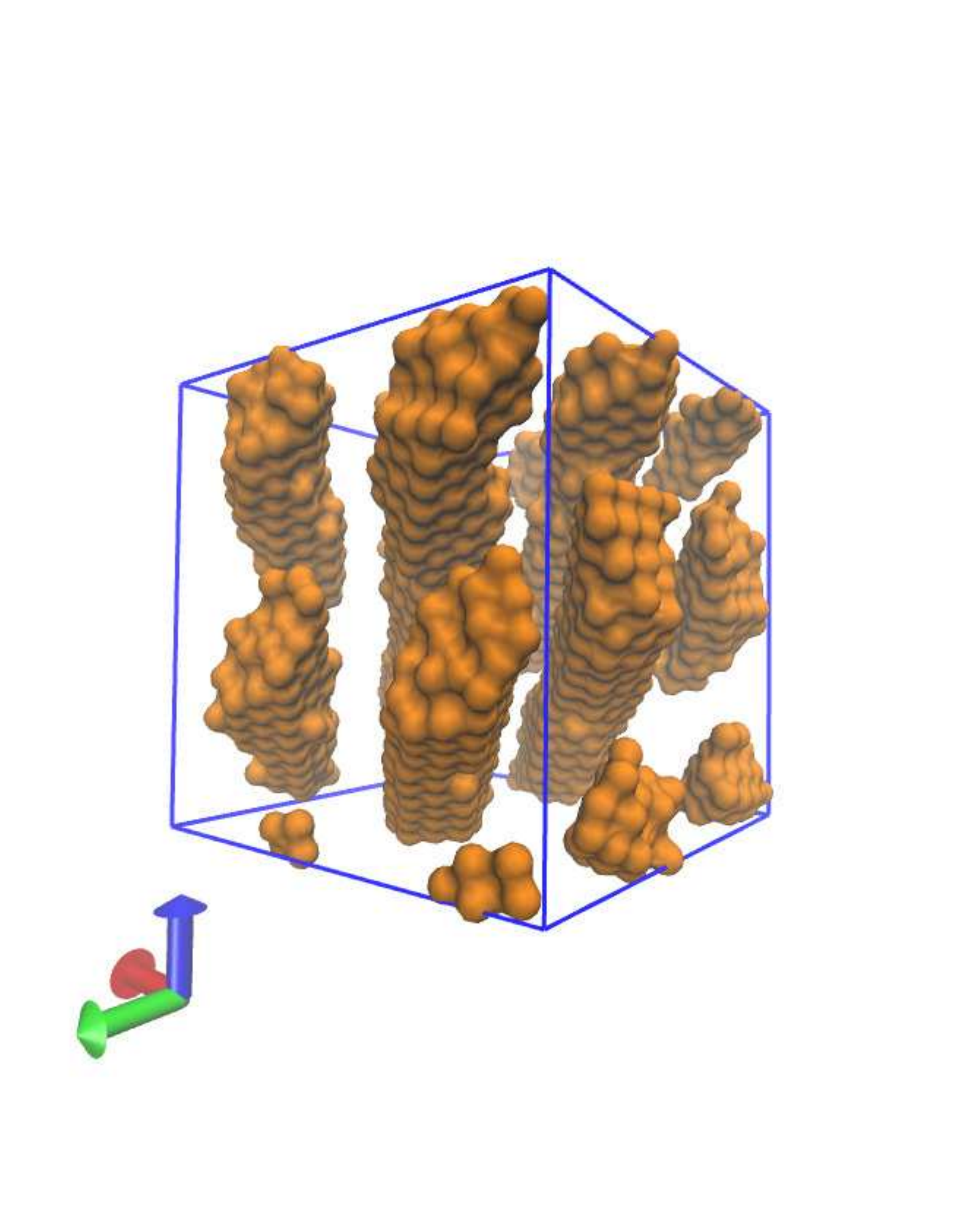}
}
\subfloat[$x=0.45$\label{fig:frame_rho04_x045}]{
\includegraphics[width=0.33\columnwidth]
{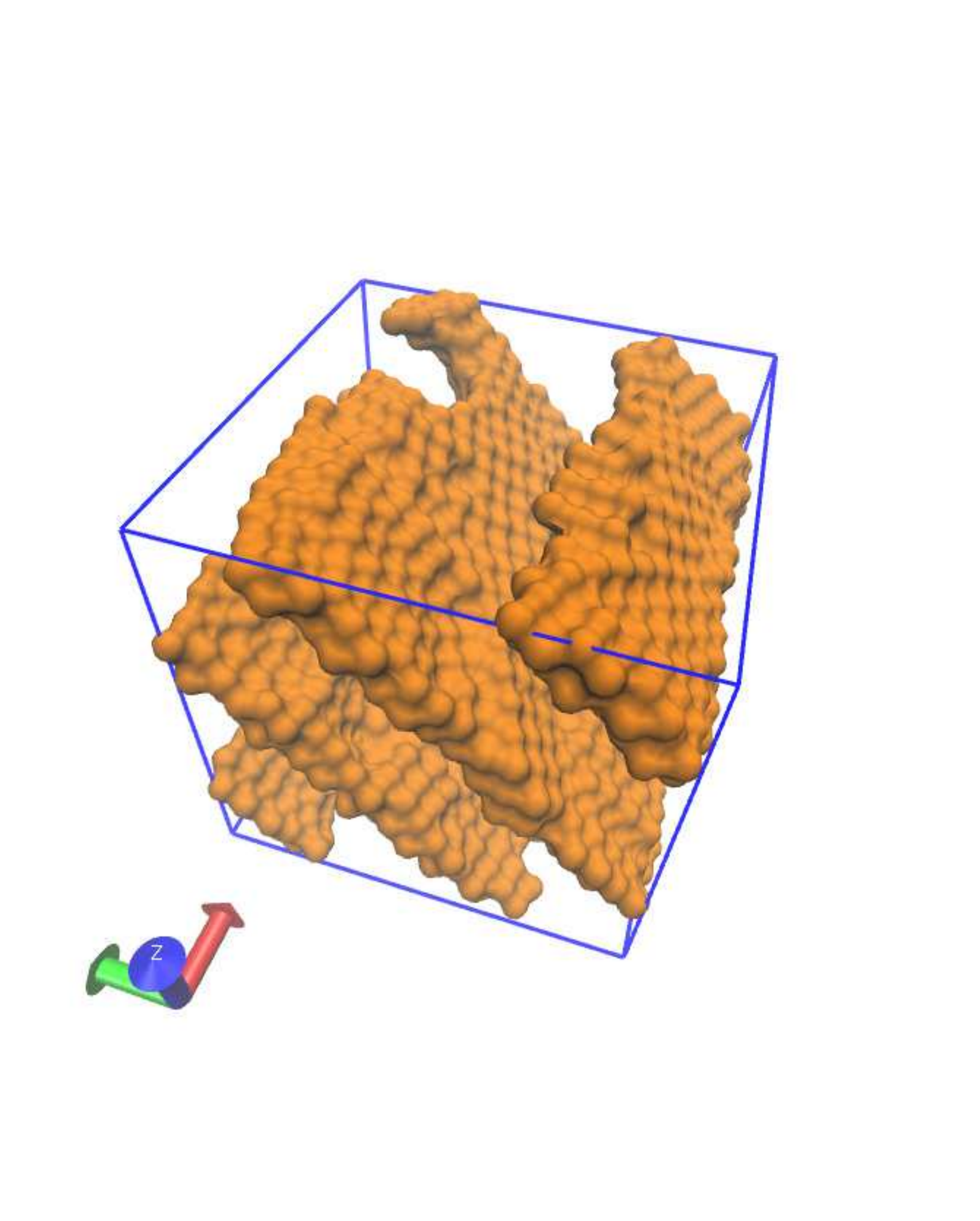}
}
\subfloat[$x=0.5$\label{fig:frame_rho04_x05_b}]{
\includegraphics[width=0.33\columnwidth]
{./lammps_rho04_x05_protons_surf_star.pdf}
}
\caption{\label{fig:frame_rho04} Surface representation for the 
protons at the indicated fractions, $T=0.2$ MeV and $\rho=0.04$ fm$^{-3}$.}
\end{figure*}

The situation is different at lower densities. Figure~\ref{fig:frame_rho04} 
shows the protons for the cases of $x=0.1$, 0.2, 0.3, 0.4, 0.45 and 0.5 at 
$T=0.2$ MeV and $\rho=0.04$ fm$^{-3}$. Comparing 
Figures~\ref{fig:frame_rho04_x02}, \ref{fig:frame_rho04_x04} and 
\ref{fig:frame_rho04_x05_b} to the the $T=0.2$ MeV values of $\chi$ in 
Figure~\ref{fig:minkowski_asym}, one can see the how the voids and tunnels 
decrease as $x$ goes from 0.2 to 0.5.\\

\begin{figure*}[!htbp]
\centering
\subfloat[$\rho=0.02$\label{fig:gr_rho01_all}]{
\includegraphics[width=0.45\columnwidth]
{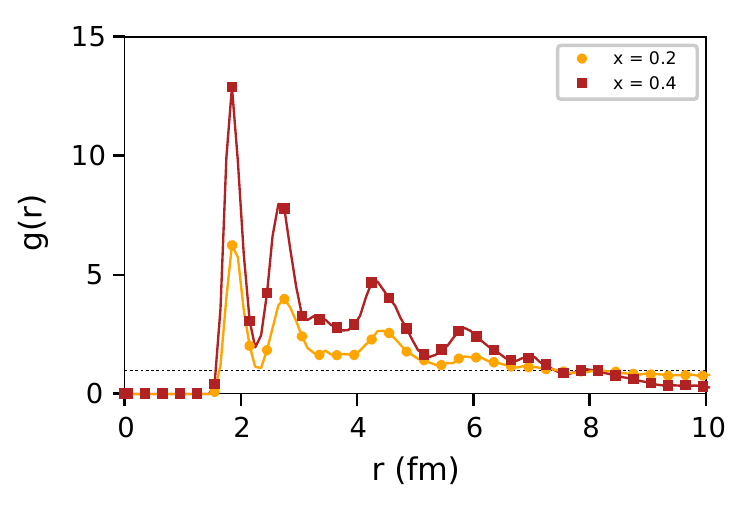}
}
\subfloat[$\rho=0.085$\label{fig:gr_rho085_all}]{
\includegraphics[width=0.45\columnwidth]
{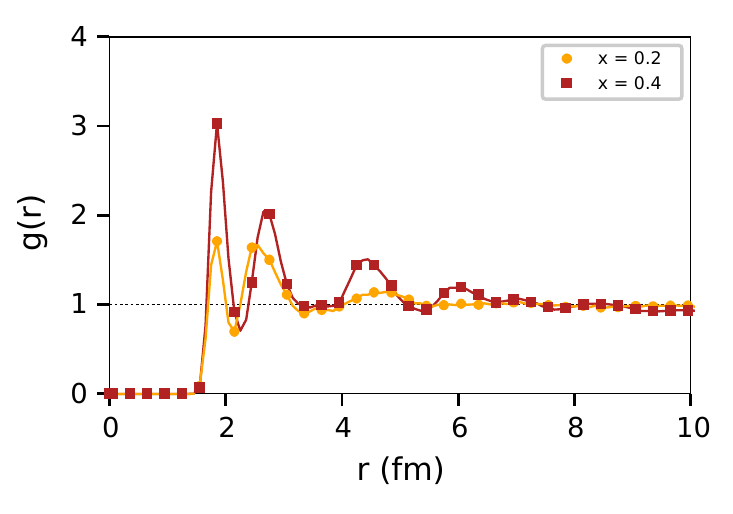}
}
\caption{\label{fig:gr_rho_asym}  Radial distribution function 
$g(\mathbf{r})$ for NSM (a) $\rho=0.02$ and (b) $\rho=0.085$, $T=0.2$ MeV, and 
various proton fractions. The black horizontal line is a view guide for 
$g(\mathbf{r})=1$.}
\end{figure*}

The radial distribution function is shown in Figure~\ref{fig:gr_rho_asym} for 
two representative proton fractions and densities. No qualitative differences 
can be distinguished between the $x=0.2$ and the $x=0.4$ cases. Neither can be 
distinguished between these and the symmetric case shown in 
Figure~\ref{fig:gr_rho02_rho085_all} for $\rho=0.085$. Thus, the overall 
nearest-neighbor distances are not affected by the proton fractions.\\

The lack of change of $g(\mathbf{r})$ in Figure~\ref{fig:gr_rho085_all} can be 
understood by looking at the structures formed. 
Figure~\ref{fig:frame_rho085_asym} shows the location of protons for three 
formations at $\rho=0.085$ fm$^{-3}$; the hollow spaces correspond to
regions occupied by neutrons and voids. The number of hollow regions increases 
as $x$ decreases, but maintaining the average inter-particle distances without 
much change. \\

The radial distribution functions for the lower density $\rho=0.02\,$fm$^{-3}$ 
are shown in Figure~\ref{fig:gr_rho01_all}. Both profiles correspond to 
gnocchi-like structures, similar to those in Figure~\ref{fig:frame_rho01_x05} 
for symmetric matter (not shown). Figure~\ref{fig:gr_rho01_all} and 
Figure~\ref{fig:gr_rho02_rho085_all} indicate that as $x$ diminishes, the 
profiles become somewhat smoother.  \\

\begin{figure*}[!htbp]
\centering
\subfloat[$x=0.2$\label{fig:gr_rho01_x02_separated}]{
\includegraphics[width=0.45\columnwidth]
{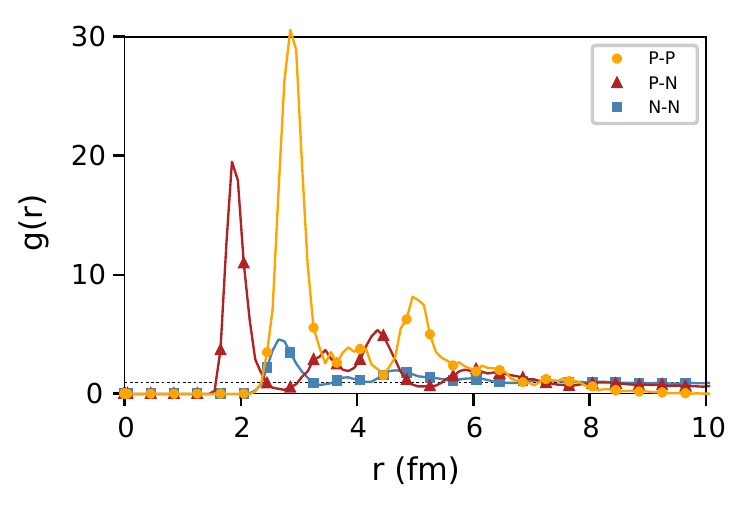}
}
\subfloat[$x=0.4$\label{fig:gr_rho01_x04_separated}]{
\includegraphics[width=0.45\columnwidth]
{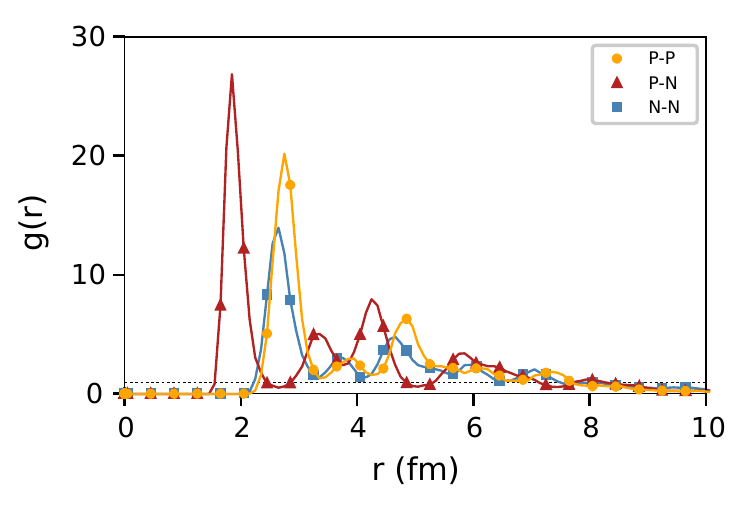}
}
\caption{\label{fig:gr_rho01_asym_separated}  Radial distribution 
function for nucleons at $T=0.2\,$MeV, $\rho=0.01\,$fm$^{-3}$, and (a) $x=0.2$ 
and (b) $x=0.4$. ``P-P'' stands for proton-proton distances only, ``P-N'' 
corresponds to proton-neutron distances  only, and ``N-N'' for neutron-neutron 
only.}
\end{figure*}

Figure~\ref{fig:gr_rho01_asym_separated} splits Figure~\ref{fig:gr_rho01_all} 
into the corresponding $g(\mathbf{r})$ for protons and neutrons. Comparing 
Figures~\ref{fig:gr_rho01_x02_separated} and \ref{fig:gr_rho01_x04_separated} it 
is clear that the smoothing of Figure~\ref{fig:gr_rho01_all} for decreasing $x$ 
is related to the distribution of nucleons within the gnocchi structure (at, 
say, $r\leq6\,$fm). Specifically, the proton-neutron and neutron-neutron 
distributions smoothen when the fraction $x$ diminishes. \\

We may further examine the simulation cell for the intermediate density
$\rho=0.04\,$fm$^{-3}$. Figure~\ref{fig:frame_rho04} shows the corresponding 
proton structures (neutron not shown) for a sequence of fractions $x<0.5$. As 
$x$ decreases, the pasta splits into smaller pieces. The structures for nearly 
symmetric neutron star matter ($x\simeq0.5$) are lasagna-like structures, while 
the low proton fractions ($x<0.3$) form gnocchi-like structures. The 
spaghetti-like structures appear in between ($x\simeq0.4$).  \\

Comparing Figure~\ref{fig:eos_x} and Figure~\ref{fig:frame_rho04}
confirm that the changing morphology is due to the $\cup$-shape
pattern for the internal energy as a function of $x$. The spaghetti-like 
structure achieves the minimum energy, while the lasagnas and the gnocchis 
correspond to higher energies on either branch of the $\cup$-shape.\\

In summary, two major effects appear when departing from symmetry. A
morphological re-arrangement of the pasta structures occurs, attaining some kind 
of fragmentation as the proton fraction diminishes. Structures with the lowest 
$x$ attain the gnocchi structure, but, the gnocchis themselves experience 
(inner) topological changes during its formation process. The resulting energy 
level surpasses the one at $x\sim0.3$ in agreement with the observations made in 
reference to the energy 
$\cup$-shape dependence on $x$.

\subsection{The symmetry energy} \label{subsec:sym_energy_3}

In Sections~\ref{subsec:asym_energy} and \ref{subsec:asym_topology} we learned 
that morphological changes taking place for decreasing $x$ are related to the 
$\cup$-shape dependence of the energy on $x$ shown in Figure~\ref{fig:eos_x}. 
These changes are bound to modify the symmetry energy too.  In this section we 
calculate $E_{Sym}$ according to the procedure detailed in 
Section~\ref{subsec:sym_energy_3} and Appendix~\ref{esymmNSM}.\\

\begin{figure*}[!htbp]
\centering
\subfloat[$\rho=0.04$ fm$^{-3}$\label{fig:fitting_rho04}]{
\includegraphics[width=0.45\columnwidth]
{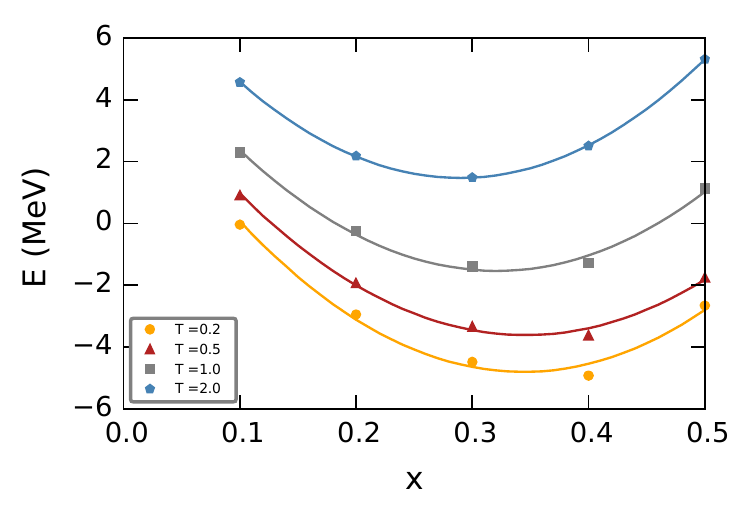}
}
\subfloat[$\rho=0.085$ fm$^{-3}$\label{fig:fitting_rho085}]{
\includegraphics[width=0.45\columnwidth]
{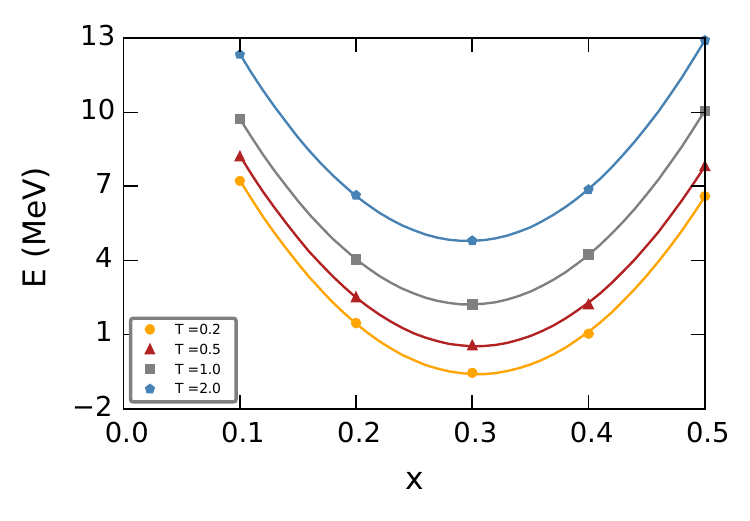}
}
 \caption{\label{fig:fitting_rhos}  Internal energy per nucleon for neutron star 
matter at the indicated densities. The points correspond to CMD data, and the 
lines correspond to the two-step fitting procedure mentioned in 
Section~\ref{subsec:sym_energy_3} and Appendix~\ref{esymmNSM}.}
\end{figure*}

Figure~\ref{fig:fitting_rhos} shows the fittings of the internal energy, 
Equation~(\ref{fitting_2}), as a function of $x$; in this case the term of 
$\mathcal{O}(\alpha^4)$ was neglected for simplicity, see 
Appendix~\ref{esymmNSM} for details. Data appears to be fitted appropriately by 
a quadratic profile, however, the low temperature curves shown in 
Figure~\ref{fig:fitting_rho04} depart from $x\sim0.4$ data. This is somewhat an 
$\mathcal{O}(\alpha^4)$ discrepancy since a noticeable improvement can be 
obtained if an $\alpha^4$ term is added into the fitting procedure (not 
shown).\\

Figure~\ref{fig:fitting_rho085} shows that the fittings for $\rho=0.085$ 
fm$^{-3}$ match better the CMD data than in the case of $\rho=0.04$ fm$^{-3}$ 
(Figure~\ref{fig:fitting_rho04}). Recall that the higher density corresponds to 
more compact proton structures (Figure~\ref{fig:frame_rho085_asym}) than at 
$\rho=0.04$ fm$^{-3}$ (Figure~\ref{fig:frame_rho04}). Thus, as already noticed 
in Section~\ref{subsec:asym_energy} and \ref{subsec:asym_topology}, more
``fragmented'' structures (say, spaghettis instead of lasagnas) flatten the 
right-branch of the U-shape pattern in Figure~\ref{fig:fitting_rho04}, departing 
from a seemingly quadratic profile.\\

\begin{figure*}[!htbp]
\centering
\subfloat[nuclear matter\label{fig:esym_nsm_nm}]{
\includegraphics[width=0.45\columnwidth]
{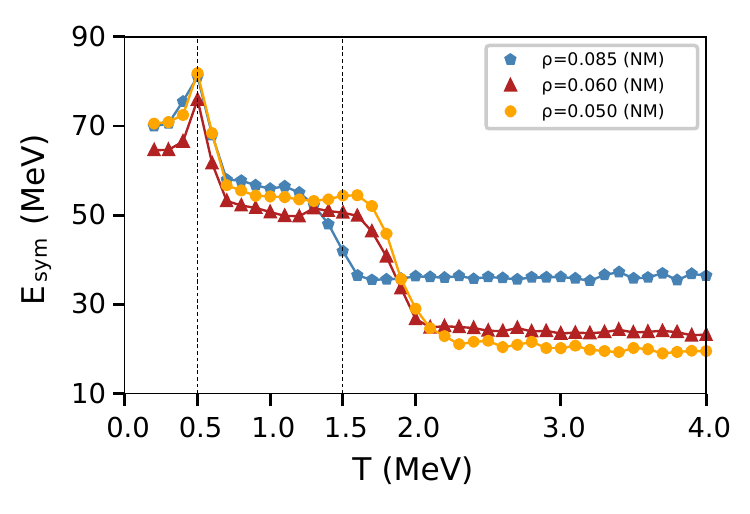}
}
\subfloat[neutron star matter\label{fig:esym_non_shifted}]{
\includegraphics[width=0.45\columnwidth]
{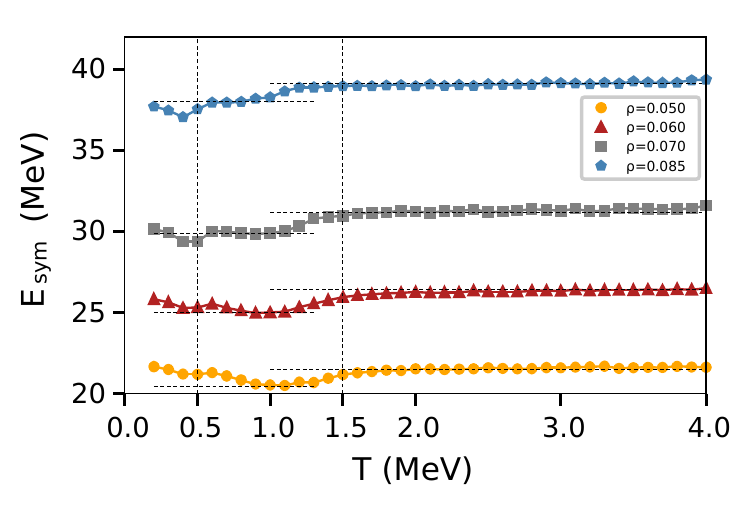}
}
 \caption{Symmetry energy as a function of the temperature for (a) NM  and (b) 
NSM, calculated for several densities (in fm$^{-3}$). Plot (a) is similar to 
Figure~\ref{fig:esym}. The horizontal and vertical lines are a guide to the 
eye.}
\label{fig:esym_all_densities}
\end{figure*}

We now proceed to compute $E_{Sym}$ for neutron star matter. For comparison, 
Figure~\ref{fig:esym_all_densities} shows $E_{Sym}$ for nuclear matter (similar 
to in Figure~\ref{fig:esym}) and for neutron star matter. It is reassuring that 
the values of $E_{Sym}$ for NM and NSM are approximately equal at $T \gtrsim 2$ 
MeV.\\

Figure~\ref{fig:esym_non_shifted} shows a change in the behavior of $E_{Sym}$ 
for NSM as compared to NM at temperatures below $1.5$ MeV. An important 
observation is that for $T < 1.5$ MeV, $E_{Sym}$ changes slope at the pasta 
regime, and decreases with lower values of $T$; this is the opposite behavior of 
the NM case. In broad terms, this is due to the changes in pasta structure and 
increase of gnocchi multiplicity that occur at low densities.  
See~\cite{dor2018} for complete details.\\

More recently, analytical expression oft he symmetry energy, as well as the 
connection of the pseudo-pastas with the Minkoswki functionals were studied 
using machine learning technology~\cite{munoz}.

\subsection{Neutrino transport properties}\label{transport}

A property of the NSM pasta of interest for neutron stars is the so-called 
neutrino opacity, which refers to the neutrino-pasta scattering. Neutrino 
opacity is of crucial importance for the evolution of neutron stars as they cool 
down by means of
neutrino emission.  Here we study the dynamics of the neutrino opacity of the 
heterogeneous matter at different thermodynamic conditions. For different 
densities, proton fractions and temperature, we calculate the very long range 
opacity and the cluster distribution. \\

\subsubsection*{Structure function}

As can be expected, the absorption of neutrinos by the neutron star crust 
depends on the structure of the nuclear pasta existing in the crust. The cross 
section for neutrino scattering by a structured medium is related to the cross 
section for scattering by a uniform medium by
\begin{equation*}
 \frac{\text{d}\sigma}{\text{d}\Omega}
 = \left( \frac{\text{d}\sigma}{\text{d}\Omega} \right)_{\text{uniform}} \times
 S(q).
\end{equation*}

\noindent where $S(q)$ is the static structure factor of nuclear 
pasta~\cite{horo_lambda}, and $q$ is the relative momentum. The structure factor 
of the system is directly related to the radial correlation function 
$g(\mathbf{r})$ via a Fourier transform. Here we calculate $S(q)$ according the 
prescription developed for systems in periodic cells in~\cite{dorso2017, 
alcain}.\\

\begin{figure}
  \includegraphics[width=0.5\columnwidth]{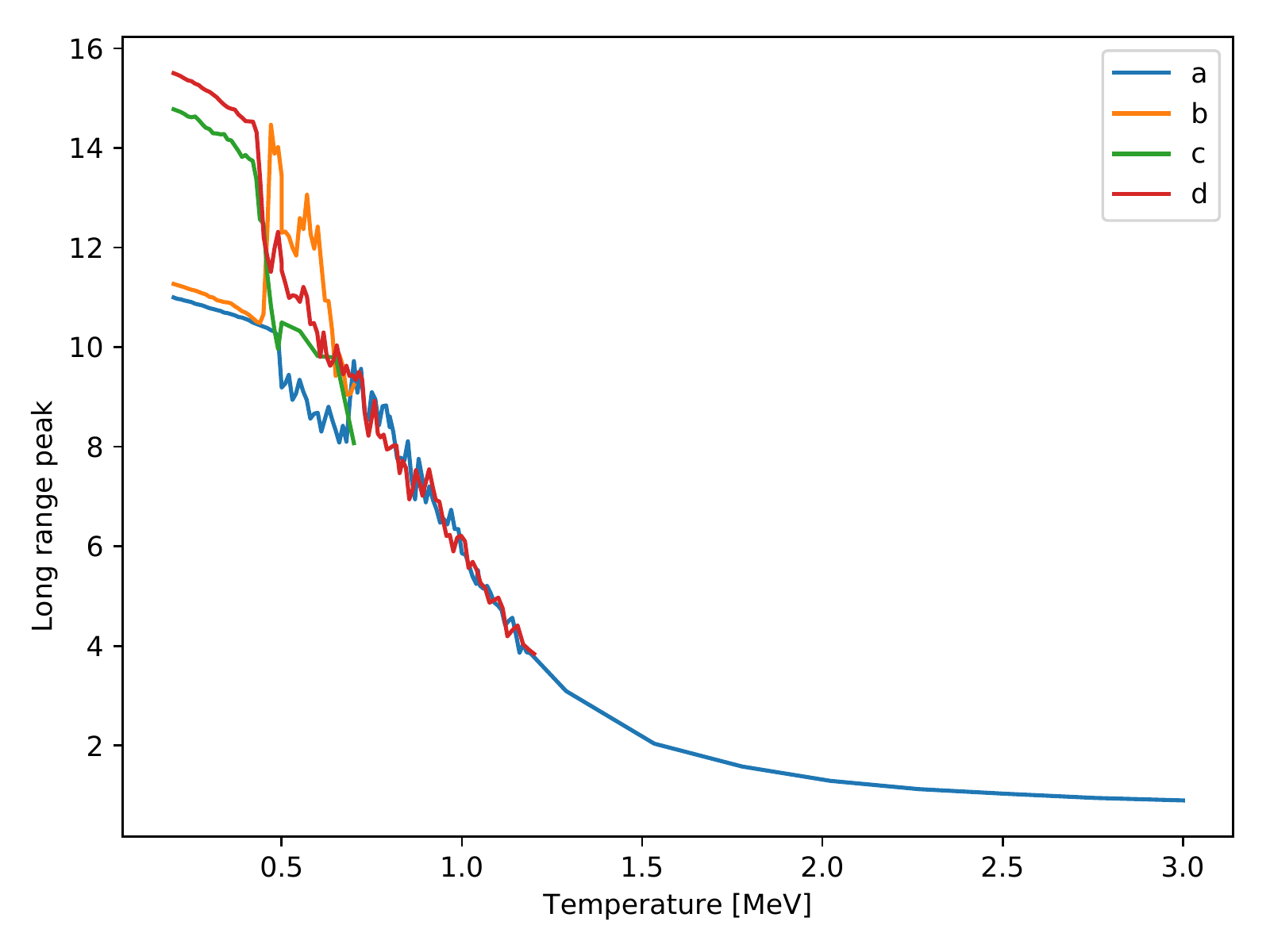}
  \caption{Peak of $S(k)$ for low momenta as a function of the
    temperature, for $\rho=0.05\,\text{fm}^{-3}$. The four curves correspond
    to those of Figure~\ref{fig:cool_morph}.}
  \label{fig:sk_peak_0-05}
\end{figure}

In Figure~\ref{fig:sk_peak_0-05} we plot the height of the low momenta peak 
$S(q<0.5\,\text{fm}^{-1})$ ($\lambda\gtrsim13\,\text{fm}$) as a function of the 
temperature for the four structures of Figure~\ref{fig:rdf}, all at 
$\rho=0.05\,\text{fm}^{-3}$.  Different behaviors can be seen at $T>0.5$ MeV in 
which all curves are practically identical, and at $T<0.5$ MeV where the 
structures have different absorption peaks for low momenta. This is in agreement 
with Section~\ref{subsec:rdf}, where Figure~\ref{fig:rdf} showed that 
$g(\mathbf{r})$ can develop a very-long range ordering characteristic of some 
pasta phases. This very long-range order is responsible for a peak at very low 
momentum $k$ ($\sim 10\,\text{fm}$ wave-length) in $S(k)$. \\

At high temperatures the nucleons are rather uniformly distributed and no 
structure is evidenced by $S(k)$: the height of the peaks tend to 1, the value 
for homogeneous systems. As the temperature is decreased, a peak at low momentum 
develops. The
transition described before manifests in Figure~\ref{fig:sk_peak_0-05} as the 
vanishing of fluctuations below the transition temperature $T \lesssim 
0.5\,\text{MeV}$. Even at temperatures as high as $T=1.0\,\text{MeV}$ there is 
still a recognizable low momentum absorption peak (with height well over 1), but 
it does not always 
correspond to a usual pasta (gnocchi, spaghetti or lasagna) in our simulations. 
At such high temperatures and for most densities, the system is in a 
``sponge-like'' structure 
which is, nevertheless, ordered enough to produce a recognizable 
peak in $S(k)$.\\

\begin{figure}
  \includegraphics[width=0.5\columnwidth]{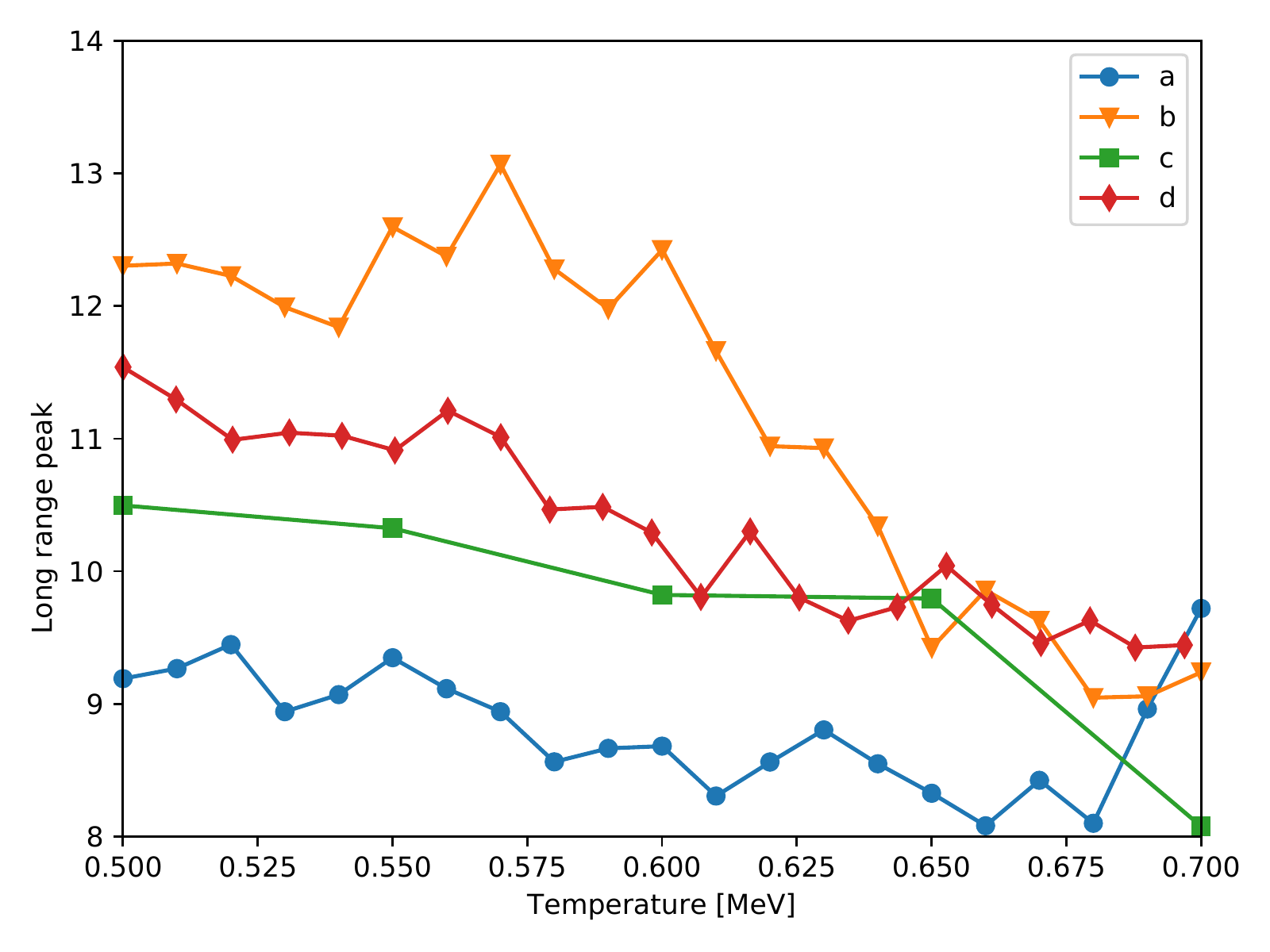}
  \caption{Peak of $S(k)$ for low momenta: zoom into
    the temperature region between $T=0.5\,\text{MeV}$ and
    $T=0.7\,\text{MeV}$. The labels of the different curves correspond
    to those of the Figure~\ref{fig:cool_morph}, where we can see that
    the structures obtained for these runs are different among each
    other and they yield different absorption peaks for low
    momenta.}
  \label{fig:sk_peak_zoom}
\end{figure}

Similarly, Figure~\ref{fig:sk_peak_zoom} plots the height of the low momenta 
peak as a function of the temperature for the four structures of 
Figure~\ref{fig:cool_morph}. These structures correspond to unusual 
\emph{intertwined lasagna} and other irregular pasta which tend to appear when 
the cooling procedure drives the system at
temperatures below $T\sim 0.7\,\text{MeV}$. In such conditions the system may 
collapse into several distinct structures, in addition to the usual lasagna 
which is the ground state at this density.  Figure~\ref{fig:sk_peak_zoom} shows 
the peaks of $S(k)$ in the low-momentum region corresponding to the structures 
in Figure~\ref{fig:cool_morph}. When we compare the absorption of each 
structure, we see that not only the usual lasagna has a peak on the low momenta 
region, but also unusual pasta phases: two that resemble an intertwined lasagna 
and another one that does not look like any other pasta (see captions for 
details).\\

Despite being different from the usual pasta phases, these shapes have a peak 
for low momentum in the structure factor. In Figure~\ref{fig:sk_peak_zoom} we 
see the corresponding absorption peaks for each structure. The unusual pasta 
phases show a larger absorption than the usual lasagna in the range of 
temperatures shown in the figure mentioned above.

\subsection{Properties of non-traditional pasta}
  \label{unusual_pasta}
  
Usual pasta shapes are ground states (potential energy minima). The 
nontraditional structures described in the previous section are likely to be 
local potential energy minima, which abound in frustrated systems like this. The 
complexity of the energy landscape (many local minima separated by energy 
barriers) makes it difficult to reach the actual ground state by simple cooling 
in molecular dynamics simulations. However, since we are working at fixed number 
of particles, volume and temperature ($(N,V,T)$ ensemble implemented through a 
thermostat, like Nos\`e-Hoover), the equilibrium state of the system at finite 
temperatures is not that which minimizes the internal energy but that which 
minimizes the Helmholtz free energy, $A = E - T\,S$. All of these structures may 
then be actual equilibrium solutions, as long as they are free energy minima.\\

\begin{figure*}[!htbp]
\centering
\subfloat[Distribution of energies for $T=0.6$ MeV \label{subfig:histo_T_0-06}]{
\includegraphics[width=0.45\columnwidth]
{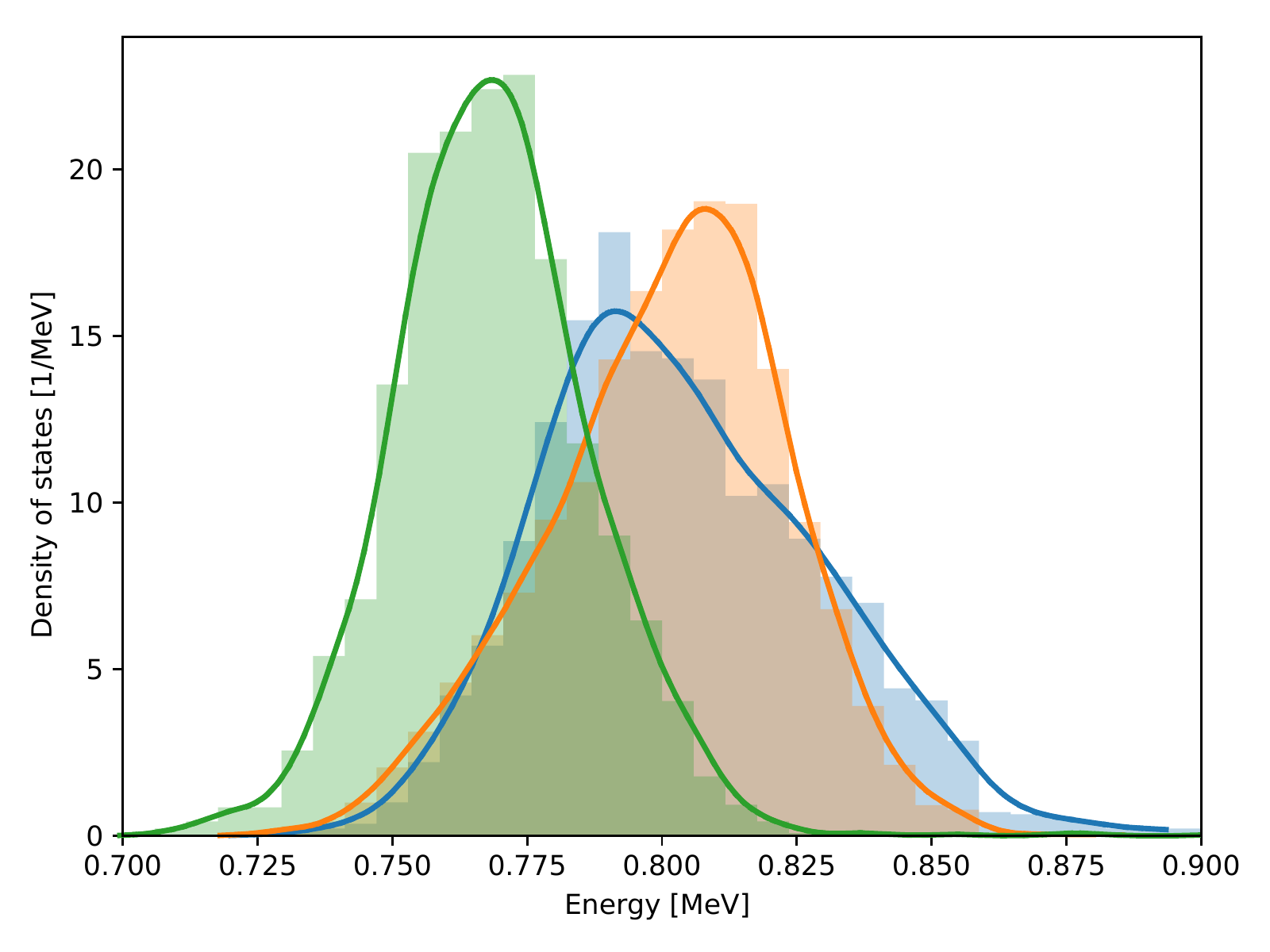}
}
\subfloat[Distribution of energies for $T=0.8$ MeV \label{subfig:histo_T_0-08}]{
\includegraphics[width=0.45\columnwidth]
{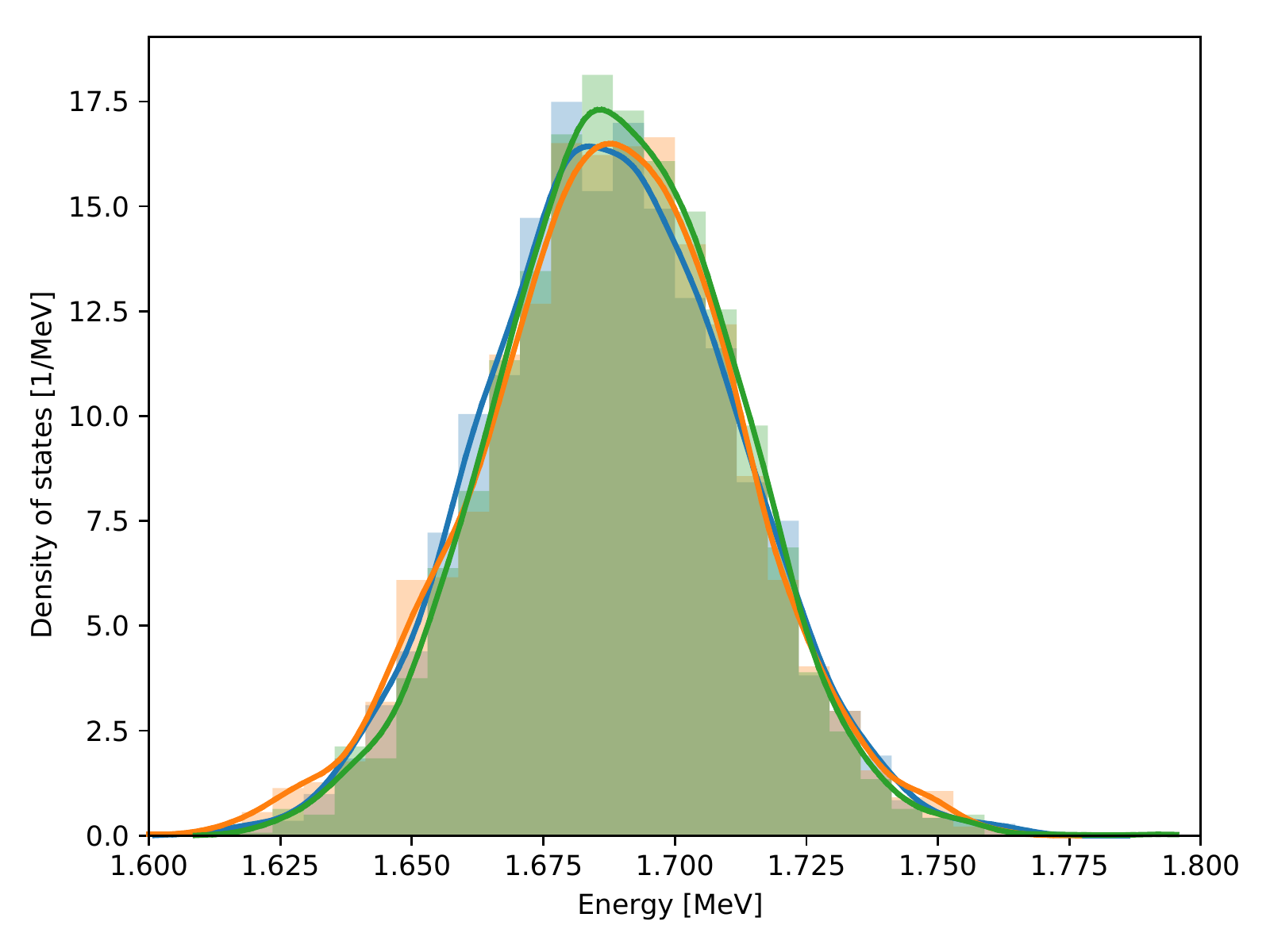}
}
\caption{Energy distribution for a canonical ensemble. It can 
    be seen that, for $T=0.8\,\text{MeV}$, all three distributions
    overlap completely. However, in $T=0.6\,\text{MeV}$, the histograms, 
    albeit split, still overlap significatively.}\label{fig:histo}
\end{figure*}

An accurate calculation of free energies from MD simulations is computationally 
very expensive~\cite{frenkel}, specially at low temperatures when overcoming 
energy barriers 
become very improbable events. However, we can easily compute the internal 
energy distributions over a long evolution at constant temperature. In 
Figure~\ref{fig:histo} we show internal energy histograms constructed from very 
long thermalized evolutions at $T=0.6\,\text{MeV}$ using three of the systems 
shown in~\ref{fig:cool_morph} as initial conditions.  We see that, although the 
histograms clearly differ, they overlap significantly. This fact indicates that 
the full ensemble of equilibrium configurations at $T=0.6\,\text{MeV}$ contains 
all of these structures, not only lasagna. In light of this we propose that at 
low but finite temperatures, the state of the system should be described as an 
ensemble of both traditional and nontraditional structures rather than by a 
single one.\\

When we heat up the system to $T=0.8\,\text{MeV}$, these three histograms become 
indistinguishable, hinting that, for this temperature, the free energy barriers 
can be surpassed, and the system is more likely to be ergodic.

These observations are relevant because all of these structures show
peaks in $S(k)$ at the same wavelenght (within the uncertainty),
although of different heights. And more importantly, we find from our
calculations that the seemingly amorphous, sponge-like structures can
be more efficient in scattering neutrinos of the same momentum that
any usual pasta (i.e. have higher peaks), usually invoked as a
necessity for coherent neutrino scattering. This result shows that
unusual pasta shapes should also be considered when studying the
structure of a neutron star's crust.


\subsection{The nucleon thermal conductivity} 

Neutron stars are expected to reach temperatures as high as $T\sim 9\,$MeV, and 
be cooled soon after birth. The late-time cooling of neutron stars (e.g. MXB 
1659-29) was shown to be consistent with low thermal conductivities 
\cite{reddy2016,brown}.  Thermal conductivity is believed to be due to electron 
flow, and not to energy flux in collisions between nuclear species 
\cite{horowitz2008,horowitz2016,nandi2018a}. \\

As the pasta is expected to dominate the inner crust at sub-saturation 
densities~\cite{dor12A,dor2018}, the pasta structures may enhance or hinder the 
energy transport~\cite{dunn}.  Models based on this scenario have attempted to 
explain the cooling of neutron stars, finding that the conductivity can vary 
with the alignment of the pasta, reducing it by 37\% for randomly oriented pasta 
slabs~\cite{horowitz2016}. \\

In this section the thermal conductivity of NSM will be studied.  We will not 
consider the energy transport due to electrons, regardless that a screening 
potential is always present, our concern is with the nucleons' energy transport. 
 In summary, the phononic thermal conductivity of nucleons embedded in an 
electron gas will be studied as it undergoes the transition to the pasta regime 
and to the solid-liquid phase transition. We first present the theoretical 
background for the thermal conductivity $\kappa$ in the context of the CMD, 
followed by the measurements of $\kappa$ within the pasta scenario divided into 
isospin symmetric and non-symmetric cases. The procedure used to calculate the 
thermal conductivity is described in Appendix~\ref{conductivity}.\\ 

\subsubsection*{Procedure}\label{simulations}

The MD simulations were performed as specified in Appendix~\ref{CMD-LT}. We 
restricted the study to elementary shapes, namely lasagnas or spaghettis, 
created with 4,000 nucleons at $x=0.5$. The procedure was to cool the system 
from $T=4$ MeV down to $T=0.1$ MeV where a solid pasta state was formed. The 
density ranged from $0.03$ fm$^{-3}$ to $0.05$ fm$^{-3}$. \\

The lasagnas and spaghettis were produced and aligned with the cell coordinates 
by means of a transformation. For the thermal conductivity measurements, the 
temperature was increased from $0.1\,$MeV to $2.1\,$MeV, and the configurations 
were set as the initial conditions for the measurements; the data were collected 
after steady states were reached.\\

The nucleonic thermal conductivities were obtained following the M\"uller-Plathe 
procedure~\cite{muller97} which requires the binning of the primary cell to 
compute the temperature gradient across the bins; the number of bins was set to 
20 across which a linear temperature profile was established. For each pasta 
shape the ``parallel'' (along the pasta) and ``transverse'' (across the pasta) 
thermal conductivities were computed. Furthermore, the thermal conductivity was 
calculated for protons and neutrons separately and for all the nucleons. 
Computing the conductivity for each specie means 
that only such specie contributes to the velocity exchange in 
Eq.~(\ref{eqn:J_transported}). \\  

\subsubsection*{$\kappa$ of symmetric NSM} \label{thermal_conductivity} 

\begin{figure*}[htbp!]
\subfloat[wide view]{\includegraphics[width=0.5\columnwidth]
{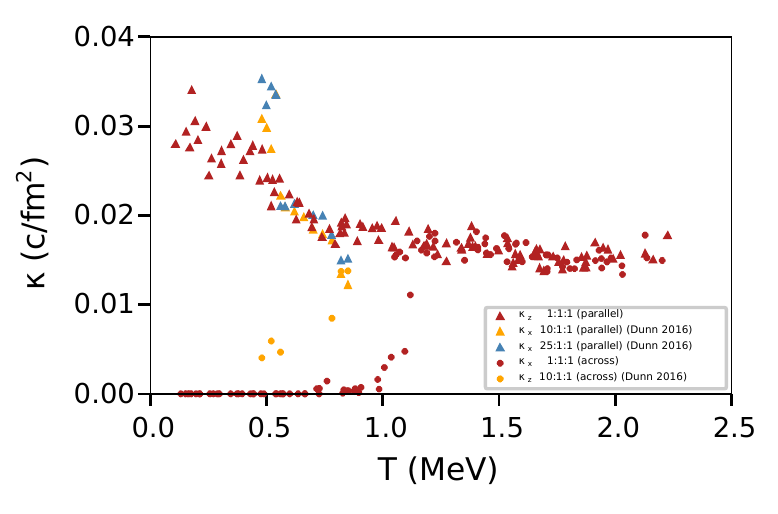}
\label{kappa_all_raw_data}}
\subfloat[detailed view]{\includegraphics[width=0.5\columnwidth]
{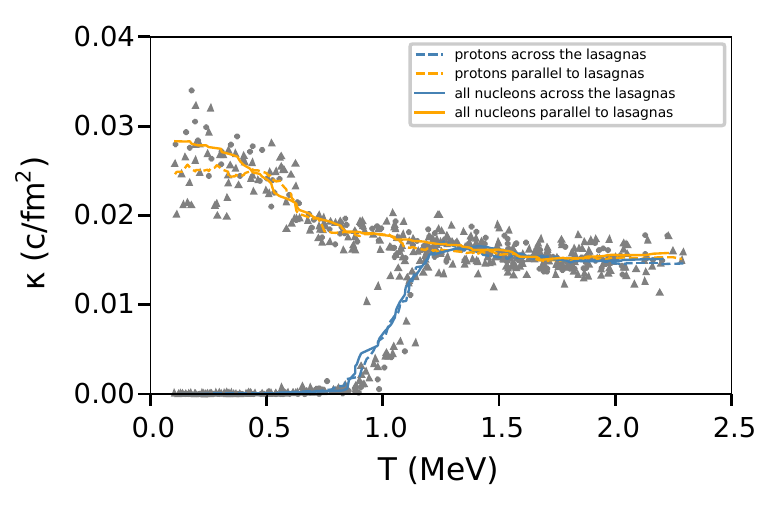}
\label{kappa_all_vs_protons_2}}
\caption[width=0.47\columnwidth]{Nucleonic thermal conductivities versus 
temperature obtained during the heating evolution of lasagnas with $\rho=0.05$ 
and $x=0.5$ system; see Fig.~\ref{snaphots_breakdown} for the corresponding 
structures. (a) $\kappa$ obtained using all the nucleons, the orange and blue 
symbols correspond to literature data. (b) $\kappa$ obtained using all the 
nucleons (circles), and using protons only (triangles). The curve is a moving 
average of $\pm10$ points. The aspect ratio of simulation cells is indicated in 
the inset of panel (a).} 
\label{kappa_all_vs_protons} 
\end{figure*}

Figure~\ref{kappa_all_vs_protons_2} shows the nucleonic thermal conductivities 
of symmetric neutron star matter at $\rho=0.05$ fm$^{-3}$ obtained during 
heating from $0.1$ to $2.1$ MeV. Clearly seen are changes at the solid-liquid 
transition found before at $T\sim 0.5$ MeV, and the pasta formation at $T\sim 1$ 
MeV. The computation was performed for all the nucleons and for the protons 
only; the figure also includes data reported in literature~\cite{dunn}. \\

Fig.~\ref{kappa_all_vs_protons_2} shows that $\kappa$ goes from a smooth curve 
above $T\simeq 1.25$ MeV to a ``decoupling'' between the thermal conductivity 
parallel to the lasagna ($\kappa_z$) and the one orthogonal to this direction 
($\kappa_x$). The ``decoupling'' pattern is essentially the same whether all the 
nucleons are considered or only the protons. \\

\begin{figure*}[!htbp]
\subfloat[well-formed]{\includegraphics[width=0.30\columnwidth]
{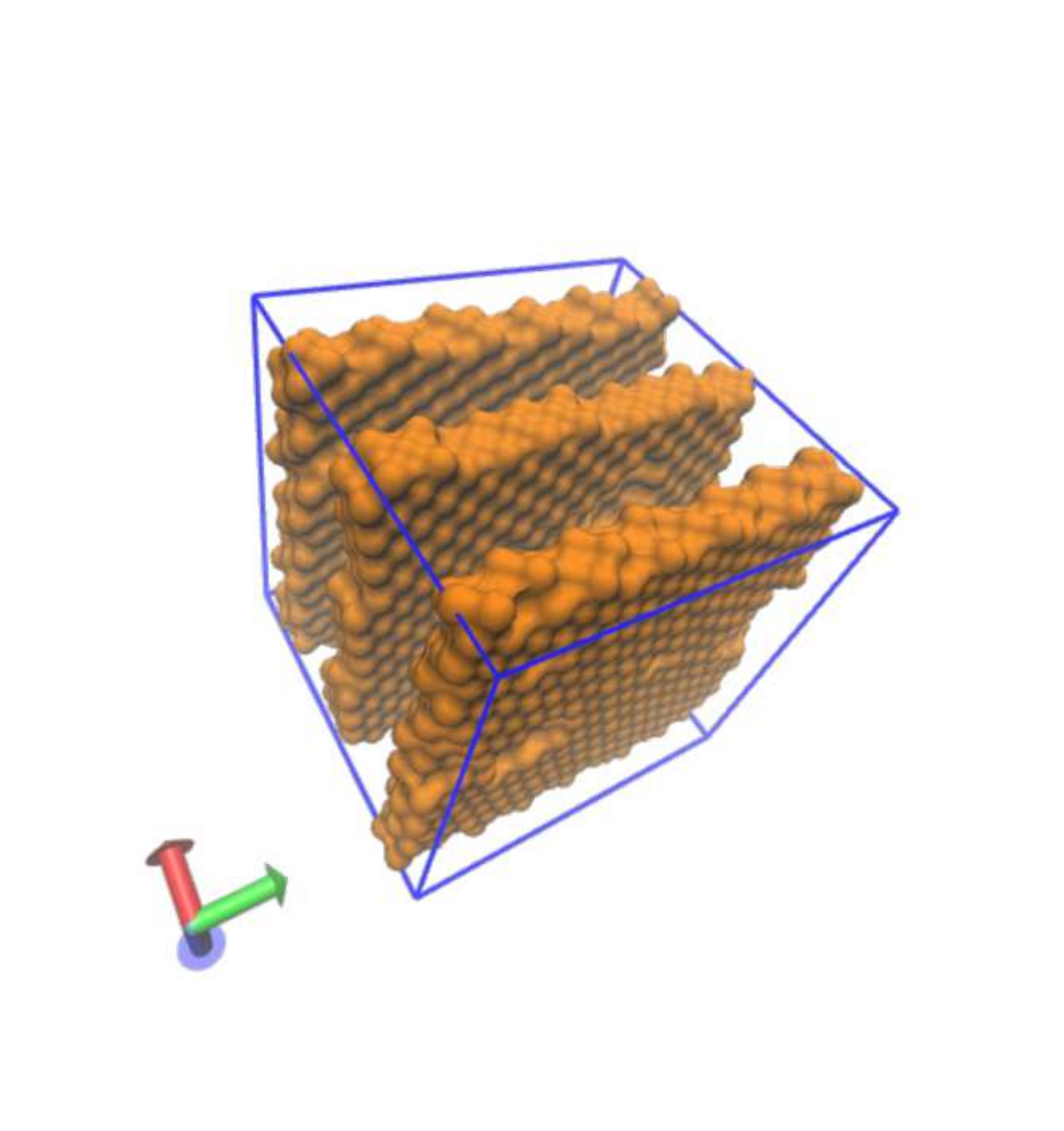}
\label{snaphots_breakdown_1}}
\subfloat[weak connections]{\includegraphics[width=0.30\columnwidth]
{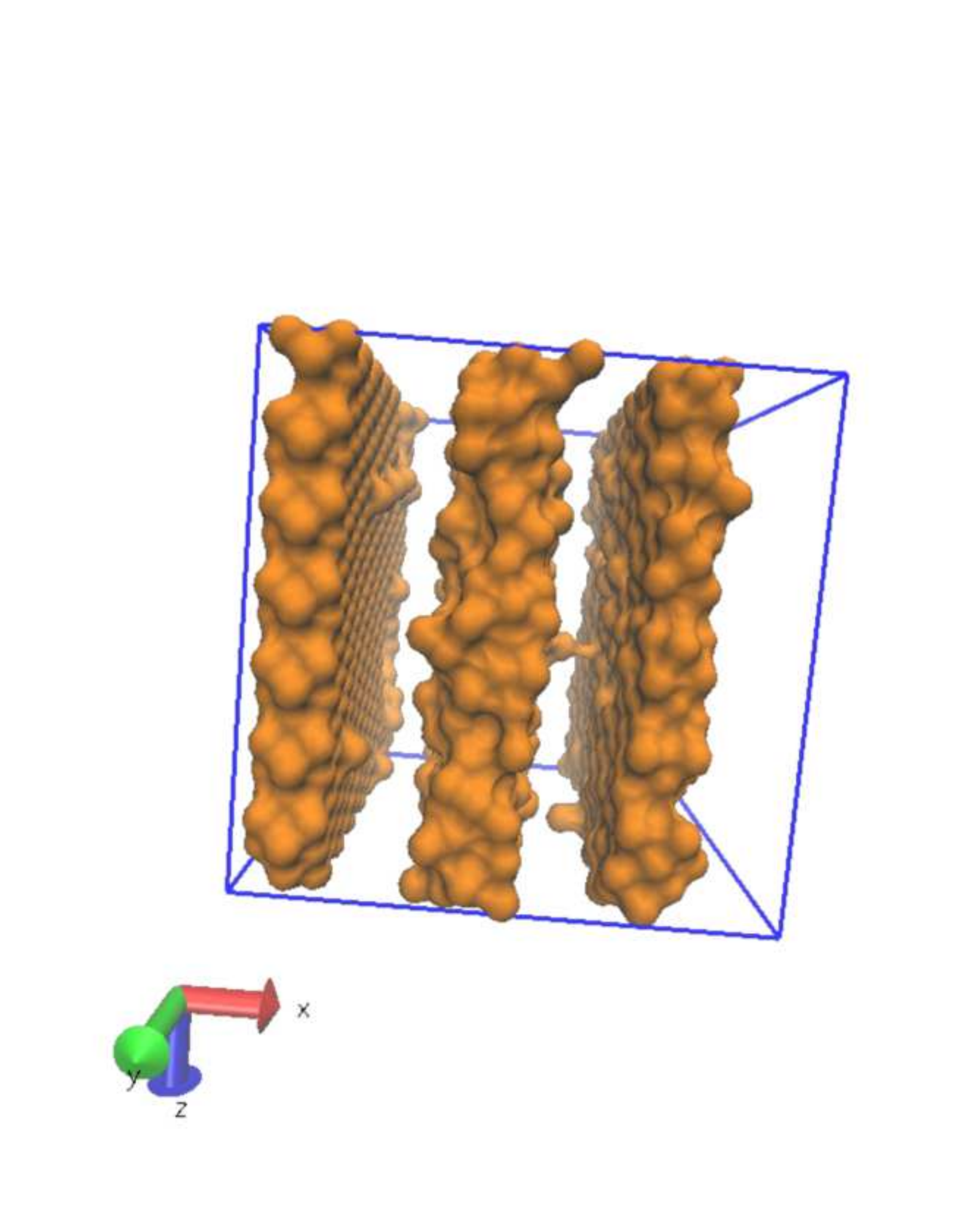}
\label{snaphots_breakdown_2}}
\subfloat[breakdown of two slabs]{\includegraphics[width=0.30\columnwidth]
{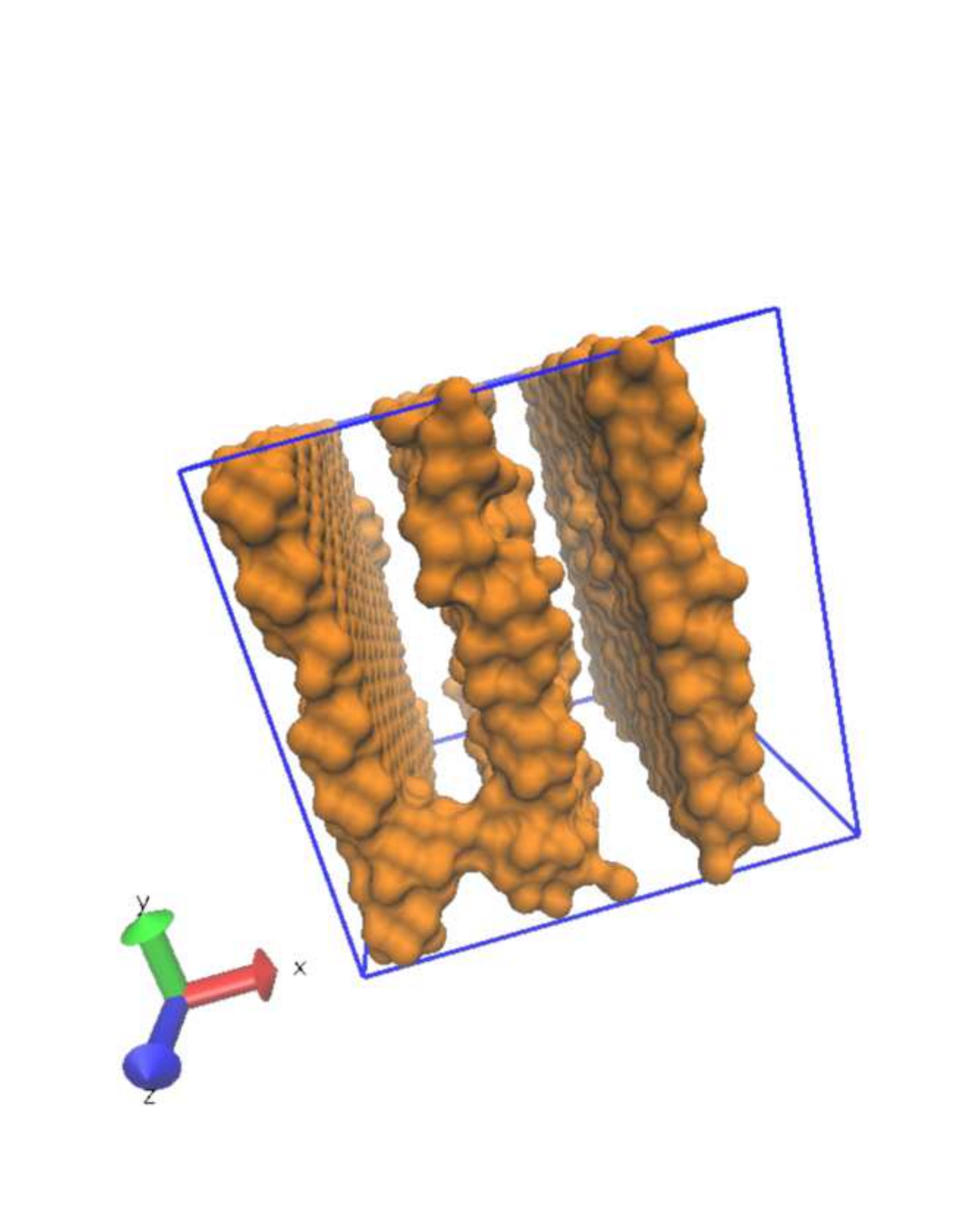}
\label{snaphots_breakdown_3}}
\caption[width=\columnwidth]{Surface plots for protons at $\rho=0.05$ and 
$x=0.5$. (a) Well-formed lasagnas at $T=0.1$ MeV. (b-c) The lasagna breakdown at 
$T\approx 1.25$ MeV.} 
\label{snaphots_breakdown}
\end{figure*}

To understand the decoupling we measured the number of clusters within the 
simulation cell. The cut-off distance between neighbors belonging to the same 
cluster was set to $r_c=4$ fm, in order to get exactly three clusters at $T=0.1$ 
MeV (see Figure~\ref{snaphots_breakdown_1}).  Weak connections between slabs 
could be observed at an early stage of the breakdown (say, $1-1.25$ MeV), as can 
be seen in Figure~\ref{snaphots_breakdown_2}. But the coupling of the parallel 
and transverse conductivities (see Fig.~\ref{kappa_all_vs_protons_2}) occurred 
when the connectivity between slabs was established; 
Figure~\ref{snaphots_breakdown_3} illustrates this phenomenon.  \\   

\begin{figure*}[!htbp]
\subfloat[$\rho=0.03$]{\includegraphics[width=0.33\columnwidth]
{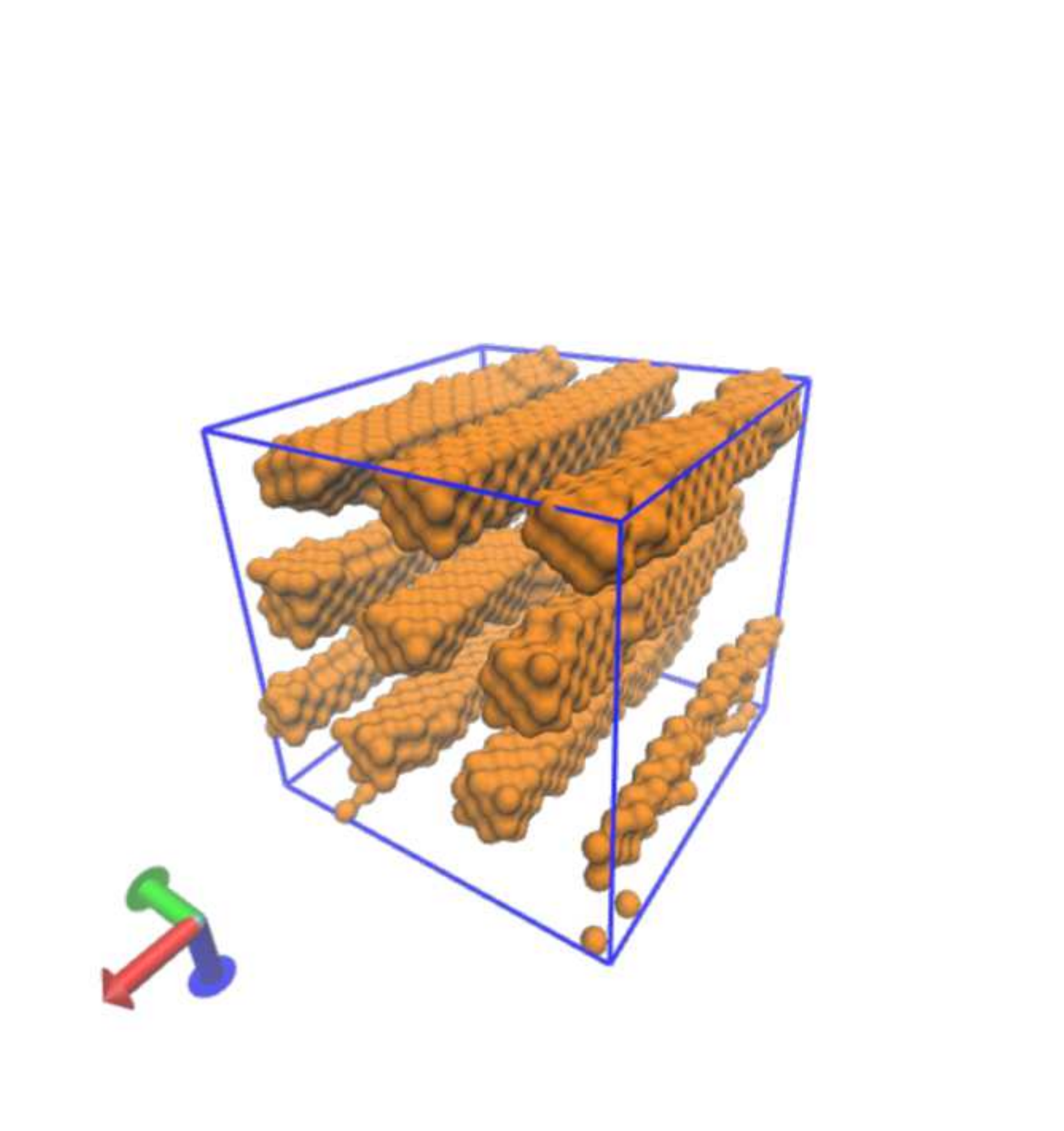}
\label{output_rho03_05}}
\subfloat[$\rho=0.04$]{\includegraphics[width=0.33\columnwidth]
{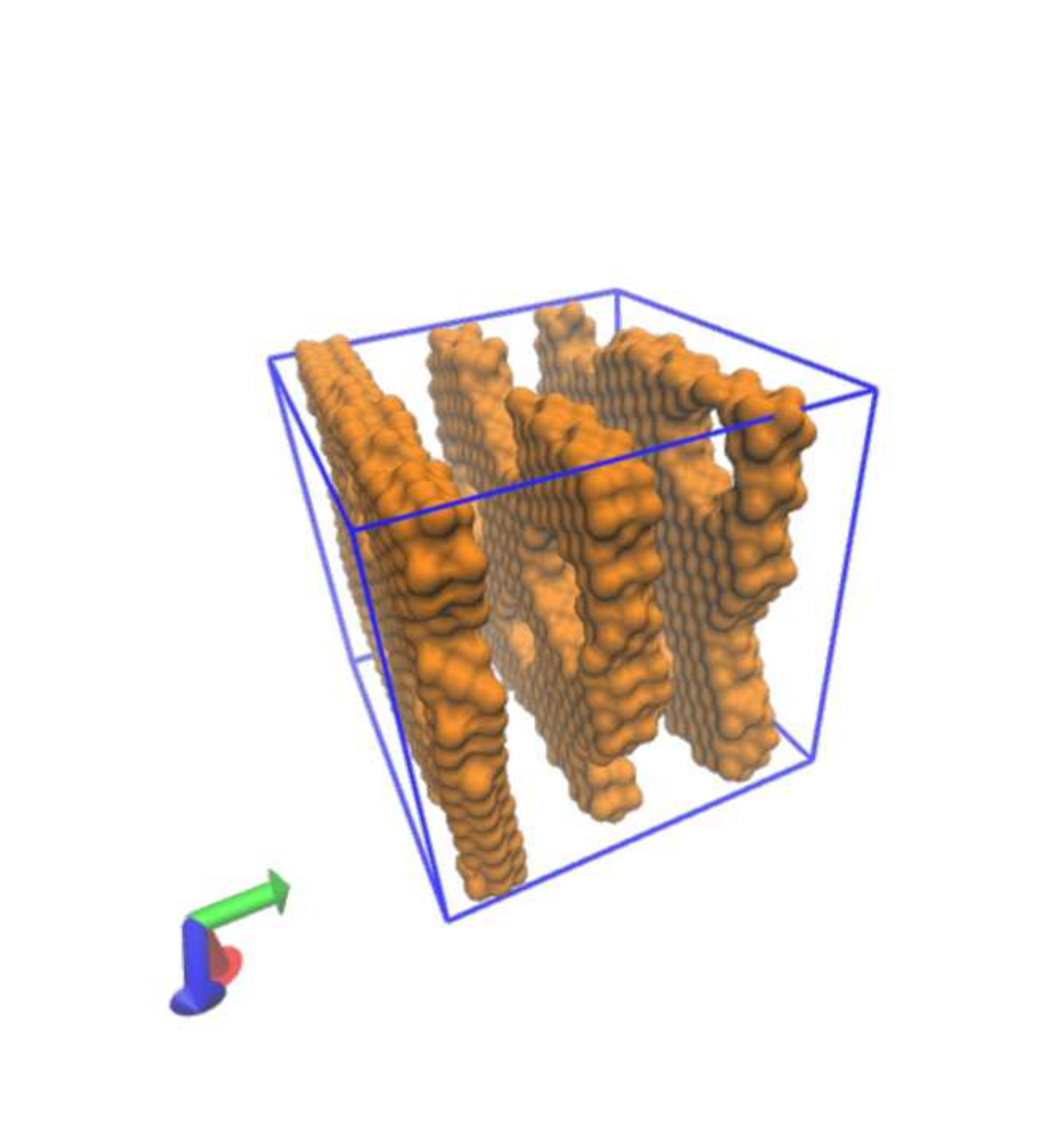}
\label{output_rho04_05}}
\subfloat[$\rho=0.05$]{\includegraphics[width=0.33\columnwidth]
{output_rho05_x05_nsm_cooling_heating_T08_new.pdf}
\label{output_rho045_05}}
\caption[width=0.47\columnwidth]{Surface plots for protons at $T=0.1$ MeV and 
$x=0.5$.} 
\label{snaphots_densities}
\end{figure*}

\begin{figure}[htbp!]
\includegraphics[width=0.5\columnwidth]
{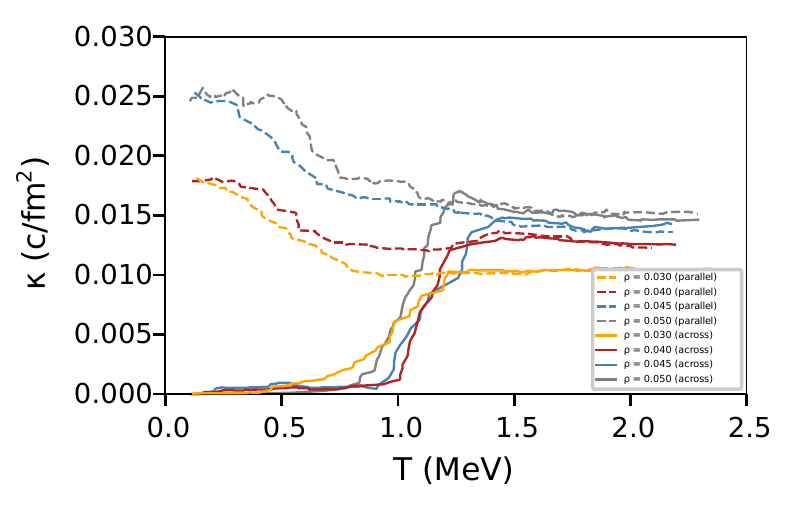}
\caption{\label{kappa_protons_densities} Proton thermal conductivity versus 
temperature for densities in the range 0.03 to 0.05 fm$^{-3}$ and $x=0.5$ 
produced by a moving average procedure of $\pm10$ points. The dashed lines 
correspond to the thermal conductivity parallel to the pasta structure, and the 
continuous lines to $\kappa$ across the pasta structure.} 
\end{figure}

The reduction of the thermal conductivity across a lasagna (say, for $T<1$ MeV) 
is due to the existence of voids between the slabs. The negative slope for the 
parallel $\kappa$ (i.e. along the lasagna) is due to the larger conductivity in 
the solid pasta with respect to the liquid one. Figure~\ref{snaphots_densities} 
shows how these openings spread over the slabs until the lasagna becomes more of 
an spaghetti-like structure (see Figure~\ref{output_rho03_05}). 
Fig.~\ref{kappa_protons_densities} exhibits the corresponding proton thermal 
conductivity. \\

In summary, cold pastas can only conduct heat along the pasta structure, but 
warming the 
pastas above $T\simeq 1.25$ MeV connects regions that were separated allowing 
heat transfer on any direction, and thus thermal conductivity becomes an 
(isotropic) value, that may depend on the system density. \\ 

\subsubsection*{$\kappa$ of non-symmetric NSM} 
\label{thermal_conductivity_asymmetric} 

\begin{figure}[htbp!]
\includegraphics[width=0.5\columnwidth]
{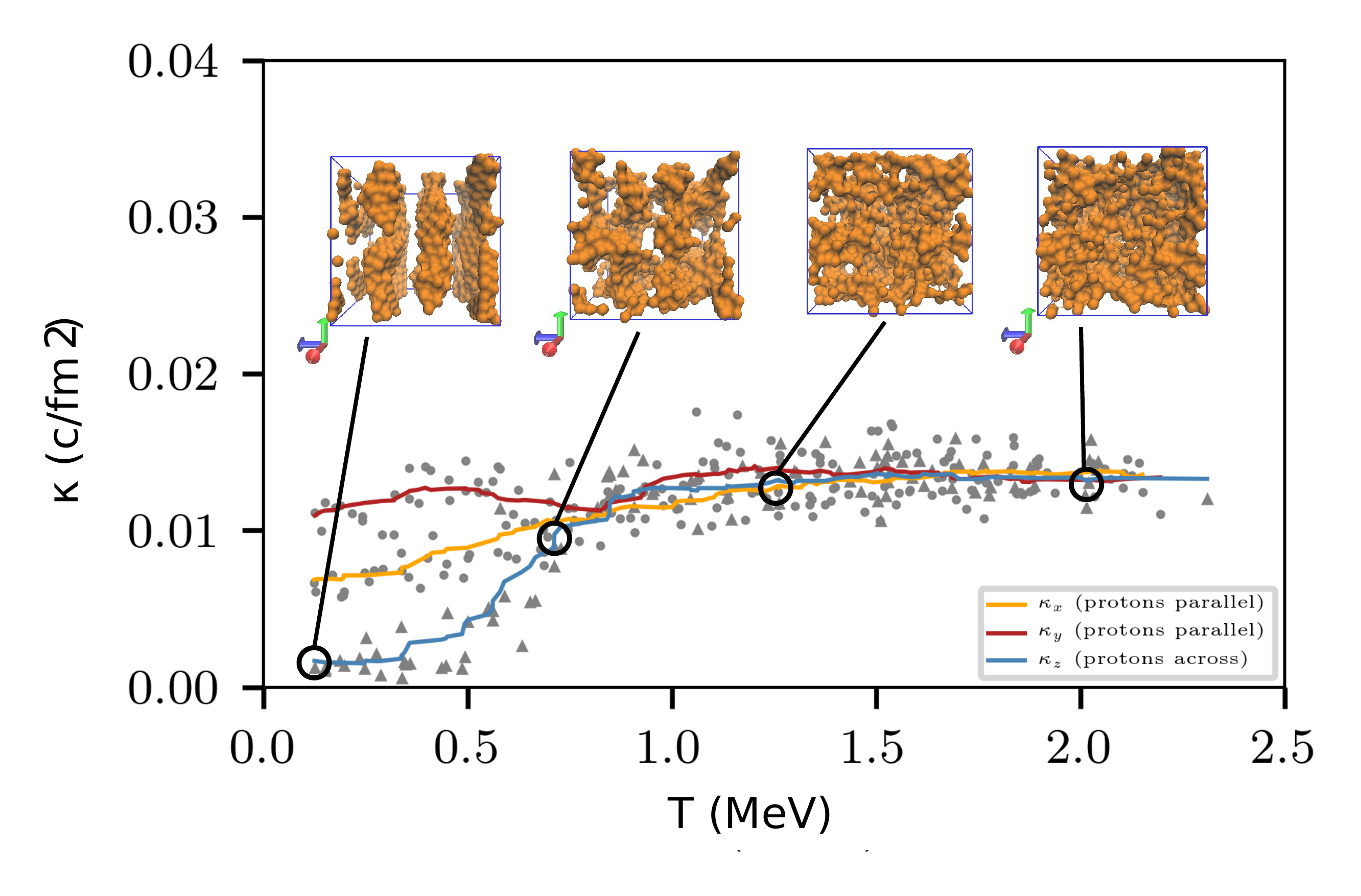}
\caption{\label{kappa_protons_asymmetric_snapshots} Proton 
thermal conductivity versus temperature for $\rho=0.05$ fm$^{-3}$ and proton 
ratio 
$x=0.3$. Shown are the data obtained along the lasagna direction (circles), 
across the 
lasagna structure (triangles), and moving averages of $10$ points.} 
\end{figure}

The thermal conductivity for non-symmetric neutron star matter was computed in 
the same way as in the previous Section. 
Figure~\ref{kappa_protons_asymmetric_snapshots} shows the proton thermal 
conductivity behavior for systems with $x=0.3$, evolving from ``cold'' (solid) 
temperatures to ``warm'' ones, along with the corresponding structures (protons 
only). \\

Figure~\ref{kappa_protons_asymmetric_snapshots} is for $x=0.3$ as 
Figure~\ref{kappa_all_vs_protons} is for the symmetric case, and they both show 
qualitatively similar trends. The proton conductivity across slabs vanishes, and 
the decoupling also occurs at around the same temperature. \\

\begin{figure}[htbp!]
\includegraphics[width=0.5\columnwidth]
{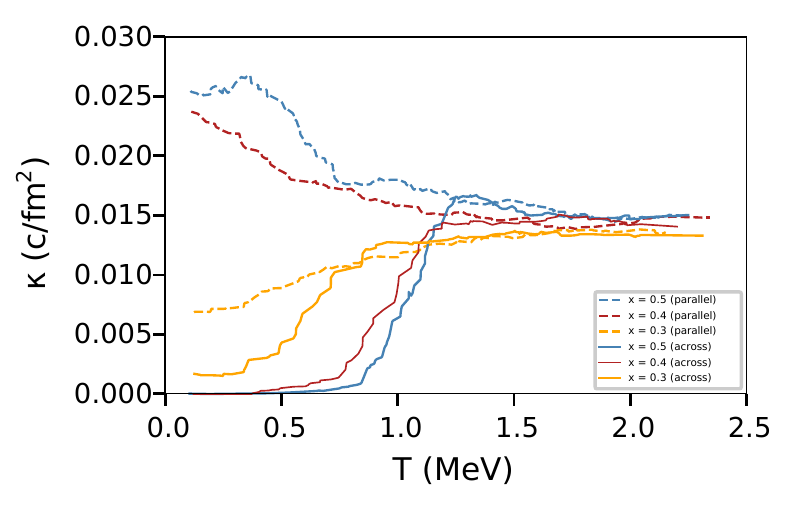}
\caption{\label{kappa_protons_asymmetric} Average proton thermal conductivity 
versus temperature for $\rho=0.05$ fm$^{-3}$ and isospin content $x=0.3$, 0.4 
and 0.5. The averaging used a moving average of $10$ points. The dashed lines 
correspond to $\kappa$ along (parallel) the pasta, while the continuous lines to 
the direction across the pasta structure.  } 
\end{figure}

The dependence of $\kappa$ on $x$ is shown in 
Fig.~\ref{kappa_protons_asymmetric} for the proton thermal conductivity. It is 
not surprising that on the pasta phase ($T \lesssim 1$ MeV) the proton $\kappa$ 
along the lasagna increases with the $x$ as more protons are available for 
transmitting heat. The proton $\kappa$ appears to be independent of $x$ except 
at the temperature at which the structures melt. \\

\begin{figure}[htbp!]
\includegraphics[width=0.5\columnwidth]
{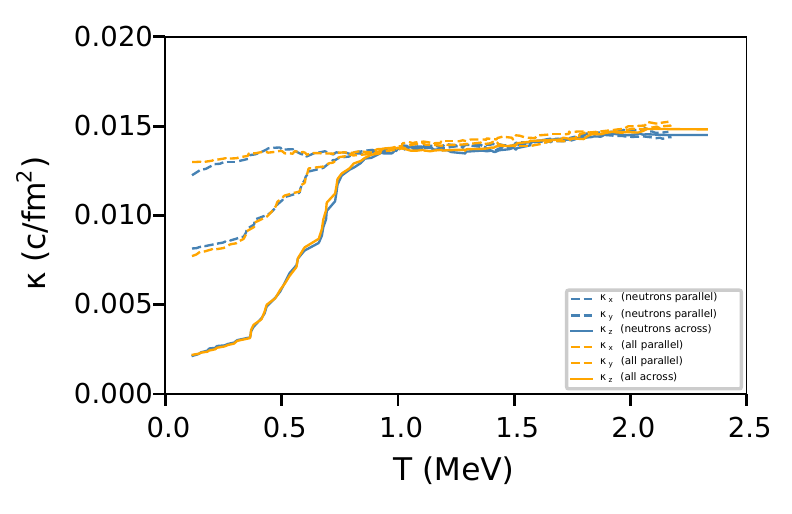}
\caption{\label{kappa_neutrons_all_asymmetric} Nucleon thermal conductivity 
versus temperature for $\rho=0.05$ fm$^{-3}$ and proton ratio $x=0.3$. The blue 
lines correspond to the neutron conductivity, and the orange ones are for all 
nucleons (see insert for details). The dashed lines are for the conductivity 
along (parallel) the spaghetti structure, and the continuous ones correspond to 
perpendicular directions, the lines are moving averages with $10$ points.}
\end{figure}

Figure~\ref{kappa_neutrons_all_asymmetric} shows the thermal conductivities, 
$\kappa_x$, $\kappa_y$ and $\kappa_z$, of all nucleons and of neutrons only for 
systems with $x=0.3$. The curves are very similar due to the fact that at 
$x=0.3$ most of the conduction is due to the neutrons. \\

\subsubsection*{Summary of findings}

Studying pasta structures in the form of lasagnas and spaghettis made out of 
4000 nucleons at $x=0.3$, 0.4 and 0.5, it was determined that the nucleonic 
thermal conductivity changes whenever the structure of neutron star matter 
switches from pastas to more isotropic structures at around  $T\approx 1$ MeV as 
a fraction of nucleons begin to bridge the slabs and rods. \\

In the pasta phase the nucleonic thermal conductivity in directions parallel and 
transverse to the pasta axis separate at $T\lesssim 1$ MeV due to the creation 
of void regions.  This effect occurs both for isospin symmetric and 
non-symmetric NSM. The thermal conductivity for non-symmetric matter shares the 
same qualitative behavior as the symmetric matter, with the differences being 
explained by the structure of the pasta and the neutrons in exces at different 
values of $x$. \\


\subsection{Summary of NSM properties}\label{concluding}

In this Section we have studied neutron star matter, the formation of pastas, 
phase transitions, symmetry energy, structure function and nucleonic thermal 
conductivity. The study was performed for isospin symmetric and for asymmetric 
NSM.\\

This was done with molecular dynamics studies of systems with 4000 nucleons in a 
cell with replicas in periodic boundary conditions. The systems were cooled and 
heated between temperatures in the range of 0.2 MeV $\leq T\leq$ 4 MeV and 
densities from 0.02 fm$^{-3}$ to 0.085 fm$^{-3}$ for isospin content symmetric 
$x=0.1$, 0.2, 0.3, 0.4 and 0.5. \\

The caloric curve presented changes in its slope at $T\approx 0.5$ MeV and at 
$T\approx 1.5$ MeV. These changes are identified as the onset of the pasta (1.5 
MeV) and a liquid-to-solid phase transitions of the nucleons inside the pasta 
(0.5 MeV). These transitions were characterized with the Lindemann coefficient 
and with the Minkowski functionals as a function of the temperature. The 
$T\approx 0.5$ MeV phase transition doesn't alter the typical pasta shape 
(lasagna and spaghetti).\\

The morphology of the pastas at different conditions of $T$, $\rho$ and $x$ was 
investigated by means of the Euler characteristic $\chi$. For temperatures above 
$T\simeq 1$ MeV, $\chi$ went from negative values (tunnel dominated 
configurations) to positive ones (configurations dominated by cavities) at 
around $\rho=0.05$ fm$^{-3}$.  \\

The structure function (and the neutrino absorption) are also connected to the 
morphology of the pasta. The maxima of the structure function maintained high 
values at low temperatures ($0 \lesssim T\lesssim 0.7$ MeV), as the pasta shapes 
are relatively well ordered in this temperature range.\\

The symmetry energy and the nucleonic thermal conductivity also presented 
changes as the phase changes. For temperatures above 2 MeV, $E_{Sym}$ remained 
basically constant, but it jumped in magnitude at temperatures around which the 
slope of the caloric curves changed. On the other hand, $\kappa$ showed two 
distinct regions above and below $T\approx 1.5$ MeV showing again a change of 
the energy transport due to a change of the pasta morphology.\\


\newpage

\section{Conclusion}\label{Concl}

In this Review we studied the nuclear pastas with the ultimate goal of 
understanding properties of neutron stars.  With masses between 1 and 3 solar 
masses and a radius of about 10 km, neutron stars have a crust of about 1 km 
where $\beta$ decays yield a neutron-rich environment immersed in a sea of 
electrons. Densities of neutron stars range from normal nuclear density down to 
practically zero in the neutron star envelope, and initial temperatures of up to 
9 MeV cooling down rapidly by thermal radiation and neutrino emission. Nuclear 
systems at such conditions form non-uniform structures, known as pastas, which 
influence the cooling of the star by modifying the neutrino-crust scattering.\\

Up until recently it was believed that pastas were due solely to the interaction 
between nuclear and Coulomb forces which constitute a frustrated system, but it 
is now known that pastas can also form due entirely to the interplay of the 
attractive-repulsive nuclear forces. We divided the complete study of pastas 
into one focused on those formed in nuclear matter, in 
Section~\ref{Nuclear-Matter}, followed by a study of the role of the electron 
gas on the formation of pastas, in Section~\ref{Electron-gas}, and finally one 
on the pastas created in neutron star matter, in Section~\ref{nsm}.

\subsection{Nuclear matter}

The properties of nuclear matter, i.e. in a system of neutrons and protons 
without a gas of electrons, were studied in Section~\ref{Nuclear-Matter}. 
Because of the different properties NM has at different temperatures, the study 
was divided into intermediate and low temperatures.

\subsubsection*{NM at intermediate temperatures}

In Section~\ref{intT} the properties of NM at intermediate temperatures ($1$ MeV 
$\lesssim T \lesssim 15$ MeV) were studied. Using molecular dynamics with 
systems with a ratio of protons of $x=Z/A=0.3$, 0.4 and 0.5, temperatures from 
$T = 1$ to 15 MeV, and $\rho \lesssim 0.2$ fm$^{-3}$, the energy per nucleon, 
pressure saturation densities, compressibility, phases, phase diagram and 
symmetry energy were investigated.\\

The density dependence of the internal energy showed characteristic $\cup$ 
shapes around saturation densities, which signaled the existence of regions 
where NM is bound, unbound, and in homogeneous (crystal) or non-homogeneous 
structures (liquid-gas mixed phase). Although the general shape of the $E-\rho$ 
curves was maintained for different values of $x$, the saturation densities and 
the compressibility varied. The radial distribution function and the mass 
distribution helped confirm the existence of a liquid phase and a liquid-gas 
mixture density at sub-saturation densities. The region were such phases exists, 
i.e. the phase diagram of NM,  was determined from the pressure-density 
isotherms by means of the Maxwell construction; a byproduct was the 
determination of the critical points for different values of $x$.\\

The symmetry energy of NM was obtained at sub-saturation densities and compared 
to experimental data and other theories. The CMD-based symmetry energies was 
able to explain the non-zero values of $E_{Sym}$ in the limit of low density, as 
demanded by experiments.\\

\subsubsection*{NM at low temperatures}

In Section~\ref{nm-LT} the properties of NM at low temperatures ($T \lesssim 1$ 
MeV) were studied. At such temperatures NM was found to produce crystal-like 
structures around saturation densities, and pasta-like structures at 
sub-saturation densities. Properties of these phases were studied using 
molecular dynamics with systems with a ratio of protons from $x=Z/A=0.1$, to 
0.5, temperatures from $T = 0.01$ to 4 MeV, and $\rho \lesssim 0.2$ fm$^{-3}$. 
\\

The energy per nucleon as a function of the density also indicated the existence 
of non-homogeneous structures at lower densities. $E(\rho,T)$ showed three 
distinct behaviors: pasta for $\rho < 0.085$ fm$^{-3}$, the crystalline 
structure for $\rho > 0.14$ fm$^{-3}$, and an intermediate region in between the 
first two.  NM at zero temperature develops a simple-cubic crystalline structure 
at densities around saturation nuclear density. The zero-temperature binding 
energy found was $-16$ MeV at a density of $0.16$ fm$^{-3}$, and with a 
compressibility of $283$ MeV. At higher temperatures ($0.001$ MeV $\le T \le 
1.0$ MeV), the $\cup$ shape is maintained still showing crystalline phases at 
densities $\rho \gtrsim 0.13$ fm$^{−3}$.\\

Spherical bubbles, gnocchi, tunnels, lasagnas, and perfect crystals were found 
at $\rho < 0.085$ fm$^{-3}$. The various shapes were studied with 
cluster-recognition algorithms, caloric curve, the radial distribution function, 
the Lindemann coefficient, Kolmogorov statistics, Minkowski functionals. \\

The caloric curve had a change of slope at $T=0.5$ MeV and a less-conspicuous 
one at $T\approx 2.0$ MeV; the changes were observed at all values of $x$. The 
Lindemann coefficient, the radial correlation function and $p-T$ curves helped 
identify the slope-changes as a solid-liquid phase transition \textit{within the 
pasta regime} at the lower temperature, and as the pasta onset at the larger 
one. These results from the caloric curves,  $g(\mathbf{r})$, Lindemann 
coefficient, and the $p-T$ curves, together indicate that the nucleons inside 
the pasta structures exist in different phases; these results were obtained both 
in isospin symmetric and non-symmetric nuclear matter. \\

The onset of the creation of the pasta was further studied with the Kolmogorov 
statistics. Major discrepancies to homogeneous distributions were detected at 
$T\approx 2\,$MeV indicating an early stage of the pasta formation. The changes 
in pasta morphology were detected at lower temperatures as bubbles widens into 
tunnels, and into slabs.  Apparently, the excess of neutrons inhibits the pasta 
formation until lower temperatures where protons manage to form the pasta. The 
released neutrons get distributed along the cell disrupting the pasta structure. 
Likewise, the Euler characteristic $\chi$ helped to understand the changes in 
morphology. Unfortunately, the $\chi$-curvature classification that worked well 
for symmetric systems becomes meaningless for the non-symmetric case. In spite 
of this, a general observation is that the system departs from homogeneity  at 
$T\sim 1.5$ MeV. \\

Finally, the symmetry energy of nuclear matter was studied in the pasta region. 
The $E_{sym}$, computed through a numerical procedure, showed a connection 
between the symmetry energy and the morphology of the system. A low value of 
$E_{sym}$ at $T>2$ MeV, a larger value in the region $1.5$ MeV $<T<2$ MeV while 
the liquid pasta is formed, an intermediate value in the range of $0.5$ 
MeV$<T<1.5$ MeV where the liquid pasta exists, and the highest value at $T<0.5$ 
MeV when the liquid-to-solid phase transition happens within the pasta. \\

In summary, NM at intermediate temperatures can exist in liquid form, gaseous 
form, and in a mixture between these phases, whereas at lower temperatures 
crystals develop around normal saturation nuclear density and the pasta is 
created at sub-saturation densities. All of the properties of the medium reflect 
these different phases.\\

\subsection{The electron gas}

The role the Coulomb interaction strength and screening length of the electron 
gas have on the formation of the pastas in NM was studied in 
Section~\ref{Electron-gas}.  An important observation is the existence of the 
pasta even without the presence of the electron gas, due to the competition 
between the attractive and the repulsive nuclear interactions. The strength of 
the Coulomb interaction emphasizes the pasta structures, makes less-compact 
objects, decreases the $x$ content of fragments, and increases of nucleon 
mobility.\\

Our study determined that there is a minimal screening length of the Coulomb 
interaction of the electron gas for the CMD to produce structures independent of 
the cell size. For the Pandharipande potential $\lambda_c$ lies between 
$10\,\text{fm}$ and $15\,\text{fm}$ depending on the density. \\

\subsection{The pasta in neutron star matter}

In Section~\ref{nsm} we studied the formation of pastas in neutron star matter. 
In particular, phases, phase transitions, symmetry energy, structure function 
and thermal conductivity were investigated for isospin symmetric and asymmetric 
NSM. Using molecular dynamics systems of 4000 nucleons per cell in periodic 
boundary conditions were cooled and heated between 0.2 MeV $\leq T\leq$ 4 MeV, 
densities from 0.02 fm$^{-3}$ to 0.085 fm$^{-3}$ for isospin content symmetric 
from $x=0.1$ to 0.5. \\

Again, using the caloric curve, the Lindemann coefficient and the Minkowski 
functionals, the pasta was found to form at around $T\approx 1.5$ MeV, and a 
liquid-to-solid phase change was detected on the nucleons inside the pasta at 
$T\approx 0.5$ MeV without altering the pasta shape.  Similarly, the Euler 
characteristic $\chi$ helped to determine that for $T\simeq 1$ MeV, the pasta 
went from being tunnel-dominated to becoming configurations dominated by 
cavities.  \\

The structure function, related to neutrino absorption, was also found to be 
dependent on the morphology of the pasta. with the largest neutrino-crust 
scattering taking place at $0 \lesssim T\lesssim 0.7$ MeV, where the pastas are 
well ordered. Likewise, the symmetry energy and the thermal conductivity present 
values according to the pasta structures. In particular, $\kappa$ had only two 
main values, above and below $T\approx 1.5$ MeV. \\


\begin{acknowledgments}
The participation of J.A.L. was partly financed by the National Science Foundation grant NSF-PHY 1066031, USA DOE's Visiting Faculty Program, and by the China-US Theory Institute for Physics with Exotic Nuclei (CUSTIPEN). C.O.D. received support from the Carrera de Investigador CONICET, by CONICET grants PIP0871, PIP 2015-2017 GI, founding D4247(12-22-2016), and Inter-American Development Bank (IDB),  Grant Number PICT 1692.

The three-dimensional figures were prepared using Visual Molecular Dynamics~\cite{QS}. Part of the calculations were carried out in the High Performance Computing Center of the University of Texas at El Paso which has a beowulf class of linux clusters with 285 processors, and some with Graphic Processing Units~\cite{lammps} at the University of Buenos Aires.
\end{acknowledgments}

\section{Appendices}
\appendix{}


\section{Classical Molecular Dynamics}\label{cmd}

In this work we use the Classical Molecular Dynamics (CMD) model, which 
represents nucleons as classical particles interacting through pair potentials 
and calculates their dynamics by solving their equations of motion numerically.  
A definite advantage of CMD over other methods is that its dynamics includes all 
particle
correlations at all levels, i.e. 2-body, 3-body, etc. Indeed the
method can describe nuclear systems ranging from highly correlated
cold nuclei (such as two approaching heavy ions in their ground
state), to hot and dense nuclear matter (nuclei fused into an
excited blob), to phase transitions (fragment and light particle
production), to hydrodynamics flow (after-breakup expansion) and
secondary decays (nucleon and light particle emission).  \\

For the nuclear case CMD is used with the Pandharipande potentials,
which were designed by the Urbana group to reproduce experimental
cross sections in nucleon-nucleon collisions of up to 600
MeV~\cite{pandha}. Such potential mimics infinite systems with realistic
binding energy, density and compressibility and to produce
heavy-ion dynamics comparable to those predicted by the
Vlasov-Nordheim equation.  This parameter-free model has been
successfully used to study nuclear reactions obtaining mass
multiplicities, momenta, excitation energies, secondary decay
yields, critical phenomena and isoscaling behavior that have been
compared to experimental
data~\cite{14a,Che02,16a,Bar07,CritExp-1,CritExp-2,TCalCur,
EntropyCalCur,8a,Dor11}.  More recently, and of interest to the
present work, the model was used to study infinite nuclear systems
at low temperatures~\cite{2013} and in neutron star crust
environments, including the pasta structures that form in NM 
and NSM~\cite{dor12, dor12A, lopram2015, dorso2014, lopez2014, dorso2017}.  \\

\subsection*{Potentials}\label{potentials}

Neutron star matter is composed of protons, neutrons and electrons, CMD uses 
$pp$, $nn$ and $np$ potentials, as well as an screening potential to mimic the 
effect 
of the electron gas; these potentials are now described in turn.\\

\subsubsection*{Nucleon-nucleon potentials}\label{cmd_star-1}

\begin{figure}
\begin{center}
   \includegraphics[width=0.5\columnwidth]{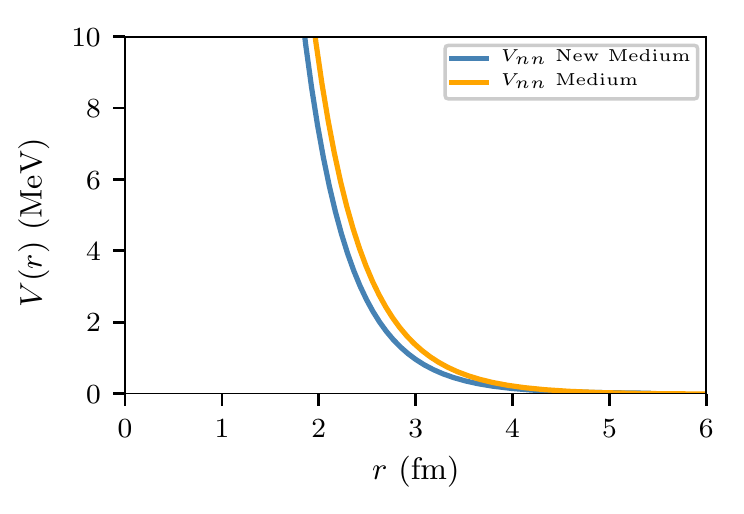}
\caption{Potential profiles for interacting nucleons nn and pp (in MeV). The 
orange curves correspond to the Pandharipande Medium model. The blue curves 
correspond to the improved New Medium model (see text for 
details).}\label{fig:nn}
\end{center}
\end{figure}

\begin{figure}
\begin{center}
   \includegraphics[width=0.5\columnwidth]{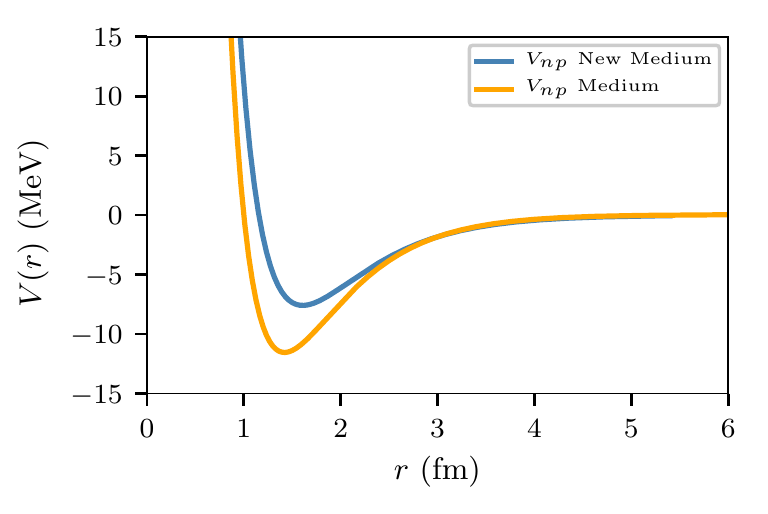}
\caption{Potential profiles for interacting neutrons and protons (in MeV). The 
orange curves correspond to the Pandharipande Medium model. The blue curves 
correspond to the improved New Medium model (see text for 
details).}\label{fig:np}
\end{center}
\end{figure}

In the CMD model nucleons interact through the Pandharipande (Medium) 
potentials. These potentials attain a binding energy $E(\rho_0)=-16$ 
MeV/nucleon and a compressibility of about $250\,$MeV. The corresponding 
mathematical expressions are\\

\begin{equation}
\begin{array}{rcl}
        V_{np}(r) & = &
\displaystyle\frac{V_{r}}{r}e^{-\mu_{r}r}-\displaystyle\frac{V_{r}}{r_c}e^{-\mu_
{ r } r_ { c } } -\displaystyle\frac{V_ { a }}{r}
e^{-\mu_{a}r}+\displaystyle\frac{V_{a}}{r_{c}}e^{-\mu_{a}r_{c}}\\
       & & \\
       V_{nn}(r) & = &
\displaystyle\frac{V_{0}}{r}e^{-\mu_{0}r}-\displaystyle\frac{V_{0}}{r_{c}}e^{
-\mu
_{0}r_{c}}
       \end{array}
\end{equation}

\noindent where $r_c$ is the cutoff radius after which the
potentials are set to zero. Although the parameters $\mu_r$,
$\mu_a$, $\mu_0$ and $V_r$, $V_a$, $V_0$ were first set by
Pandharipande for cold nuclear matter~\cite{pandha}, a recent 
improvement~\cite{dor2018}, here named \textit{New Medium}, 
reproduces the cold nuclear matter binding energies more accurately 
and, thus, is used in this work. The corresponding values are 
summarized in Table~\ref{table_parameter}. Figs.~\ref{fig:nn} and \ref{fig:np}
contrasts these potentials with those of Pandharipande Medium potentials.\\

\begin{table}
{\begin{tabular}{l @{\hspace{15mm}}@{\hspace{6mm}} r
@{\hspace{23mm}} r @{\hspace{10mm}} l}
\hline
      Parameter &  \multicolumn{1}{l}{Pandharipande} & $\ $New Medium  & 
\multicolumn{1}{c}{Units}
\\
\hline
$V_r$ &  3088.118 & 3097.0  & MeV \\
$V_a$ &  2666.647 & 2696.0  & MeV\\
$V_0$ &  373. 118 & 379.5  & MeV\\
$\mu_r$ & 1.7468 &  1.648 & fm$^{-1}$ \\
$\mu_a$ & 1.6000 &  1.528 & fm$^{-1}$ \\
$\mu_0$ & 1.5000 &  1.628 & fm$^{-1}$ \\
$r_c$   & 5.4    &  5.4/20 & fm \\
\hline
\end{tabular}
}
\caption{Parameter set for the CMD computations. The values used in 
this work correspond to the New Medium Model. }
\label{table_parameter}
\end{table}

It is worth mentioning that these potentials were crafted by 
Pandharipande~\cite{pandha} to reproduce experimental cross sections in 
nucleon-nucleon collisions without the use of an explicit Coulomb potential for 
the proton-proton interactions; the Coulomb interactions among protons can be 
assumed to be embedded in the Pandharipande potentials.


\subsubsection*{The Coulomb potential}\label{cmd_star-2}

As in neutron star crusts electrons filled the space between protons and 
neutrons, it is necessary to include the Coulomb effect of an electron gas. The 
main effect of the electron cloud is to introduce an screening effect on the 
Coulomb potential of the protons. 
Given the infinite range of Coulomb interaction some approximation is needed 
when simulating infinite systems. \\

In the CMD model protons and neutrons are then assumed to be immersed in a 
uniform gas of non-interacting electrons, and the effect of the electron gas is 
implemented either by the Thomas-Fermi Screening method, or by the Ewald 
Summations. In the present work we use first one and refer the reader to 
\cite{frenkel,nymand_linse,dor12} for details on the second method.\\

The implementation of the Thomas-Fermi Screening considers the electron gas as 
an ideal Fermi gas at the same number density as protons. This electron gas, 
being uniform, does not exert any force on protons but becomes polarized in 
their presence effectively ``Screening'' the proton's charge. After solving the 
corresponding Poisson equation, the field generated by a screened proton takes 
the form:


\begin{equation}
V_{C}=\displaystyle\frac{q^2}{r}e^{-r/\lambda},\label{eq:coulomb}
\end{equation}

where $\lambda$ is a screening length. Thus, the effective Coulomb potential 
becomes finite ranged.\\

The relativistic Thomas-Fermi screening length is given by 

\[\lambda=\frac{\pi^2}{2e}\left(k_F \sqrt{(k_F^2+m_e^2)}\right)^{-\frac{1}{2} 
}\]

where $m_e$ is the electron mass and $k_F$ is the electron Fermi momentum 
defined as $k_F=\left( 3\pi^2\rho_e\right)^{1/3}$, where $\rho_e$ is the 
electron gas number density (taken equal to that of the proton's).\\

To avoid finite size effects, $\lambda$ should be significantly smaller than the 
size $L=\left( A/\rho \right)^{ \frac{1}{3} }$ of the simulation cell. Since the 
screening length $\lambda$ depends on the density of the system, it is always 
possible to satisfy this condition by increasing the simulation box size along 
with the number of particles.
However, this can lead to prohibitively large systems for our current 
computation capabilities. Following a prescription given in \cite{horo_lambda} 
we set $\lambda=10$ fm, and 
set $V_C^{(Scr)} = 0$ at a cutoff distance of 20 fm; these values are long 
enough to reproduce the density fluctuations in the cell size used and, in the 
low temperature case, they ensure that the properties of the resulting pasta 
remain essentially constant; see~\cite{dor14}. This implementation has been used 
before~\cite{dor12,Horo2004,Maruyama}.\\



\begin{figure}
\begin{center}
   \includegraphics[width=0.5\columnwidth]{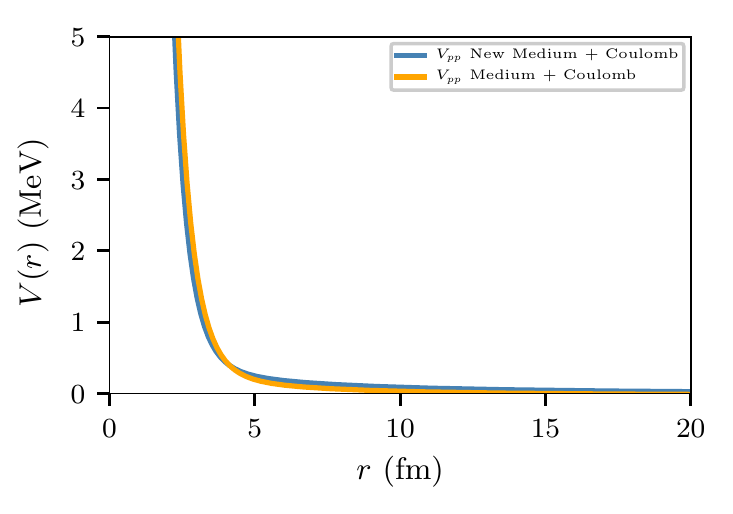}
\caption{  Potential profiles for interacting protons (in MeV). The orange curve 
corresponds to the Pandharipande Medium model embedded in the Thomas-Fermi 
potential 
(see Eq.~(\ref{eq:coulomb})). The blue curve corresponds to the improved New 
Medium model embedded in the Thomas-Fermi potential (see 
Eq.~(\ref{eq:coulomb})).}\label{fig:coulomb}

\end{center}
\end{figure}

Figs.~\ref{fig:nn} and~\ref{fig:np} shows the interaction potentials between 
nucleons
without the existence of the surrounding electron gas, while 
Fig.~\ref{fig:coulomb} 
shows proton-proton complete potential (including the Coulomb screening). The 
exponential cut-off renders the Coulomb effective interaction short ranged. In 
this way the energy and the entropy is additive (\textit{i.e.} energy scales 
with the number of particles).\\

\begin{figure*}[!htbp]
\centering
\captionsetup[subfigure]{justification=centering}
\subfloat[$\rho=0.05$\label{fig:rho05}]{
\includegraphics[width=0.5\columnwidth]
{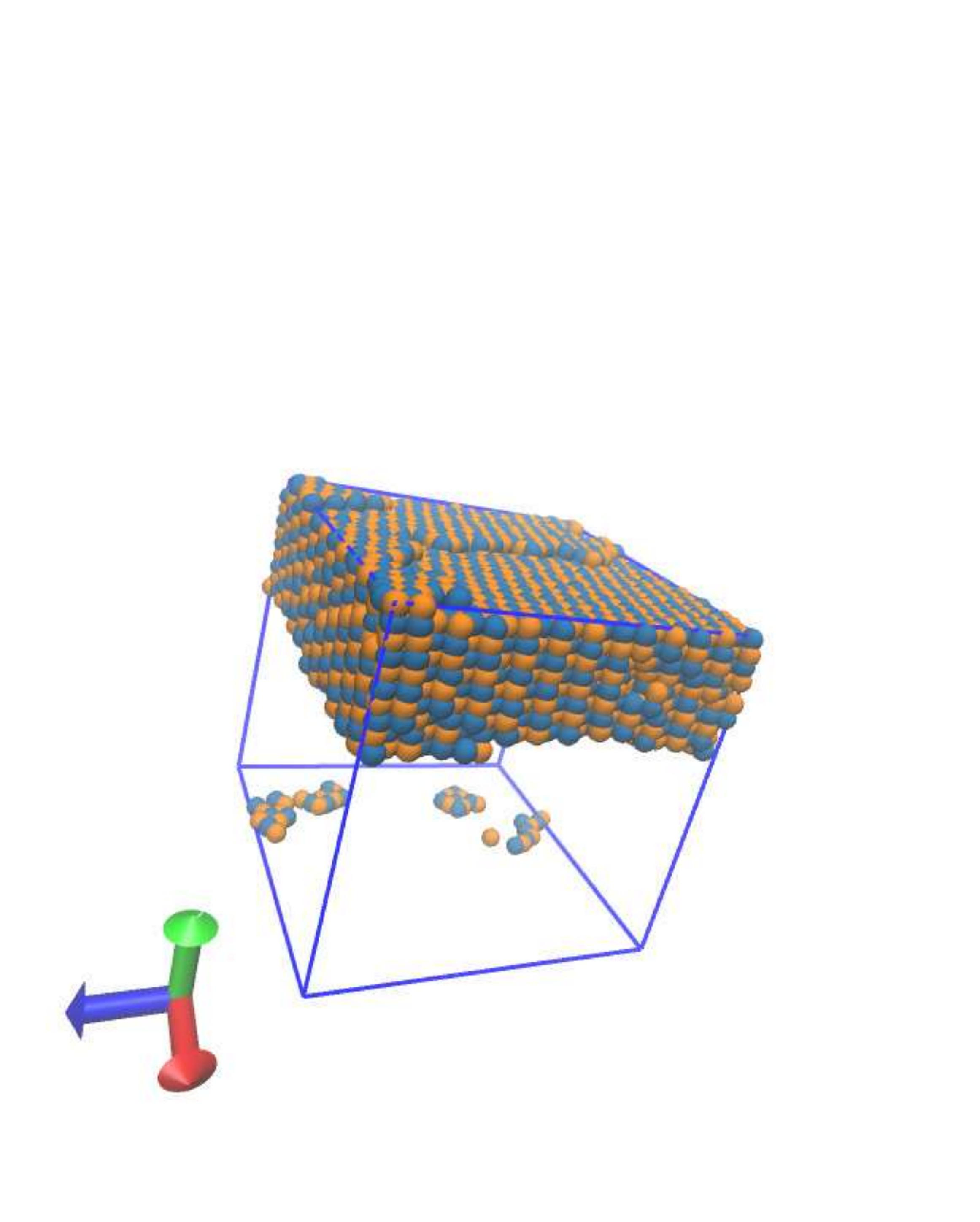}
}
\subfloat[$\rho=0.06$\label{fig:rho06}]{
\includegraphics[width=0.4\columnwidth]
{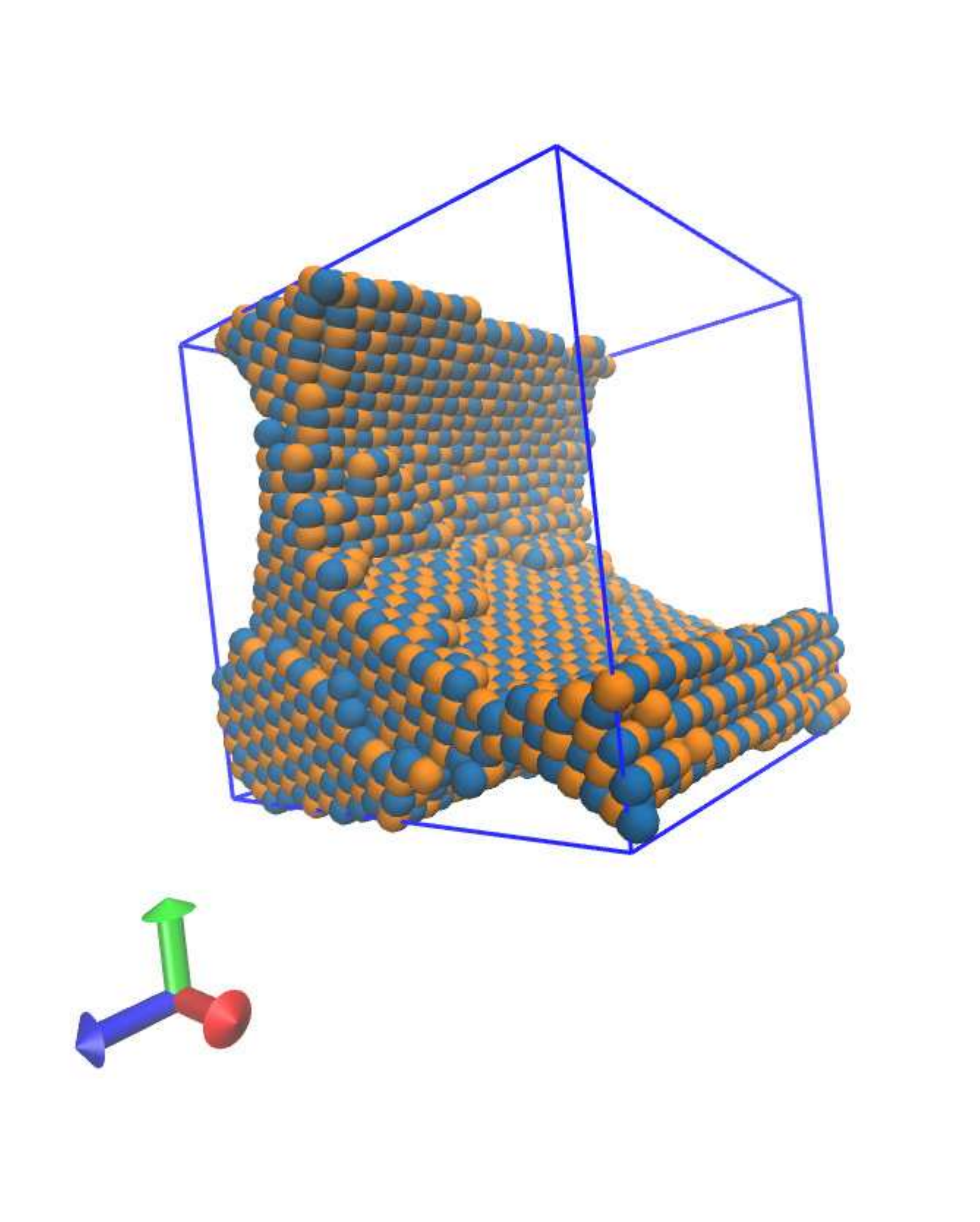}
}
\\
\subfloat[$\rho=0.07$\label{fig:rho07}]{
\includegraphics[width=0.5\columnwidth]
{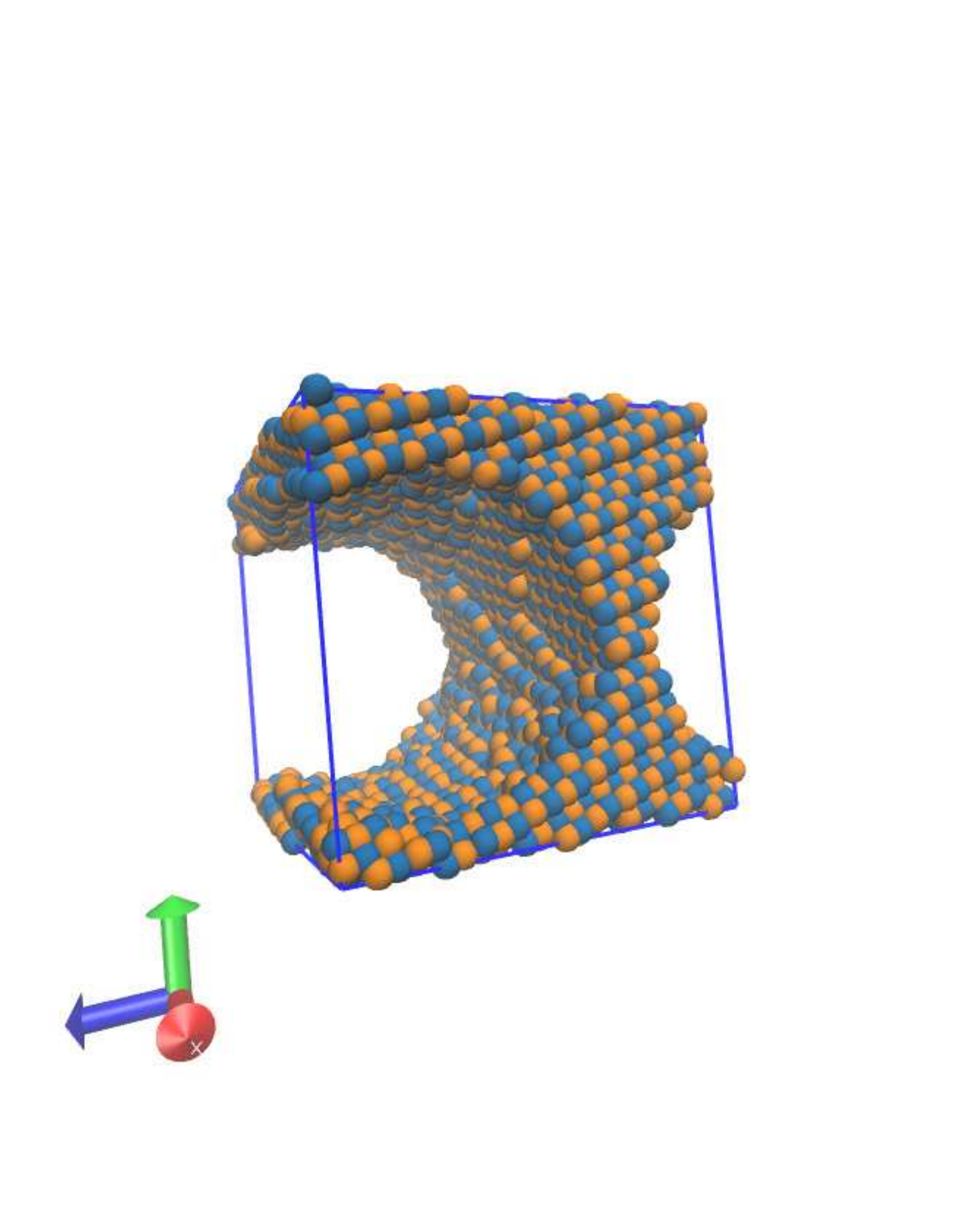}
}
\subfloat[$\rho=0.085$\label{fig:rho085}]{
\includegraphics[width=0.4\columnwidth]
{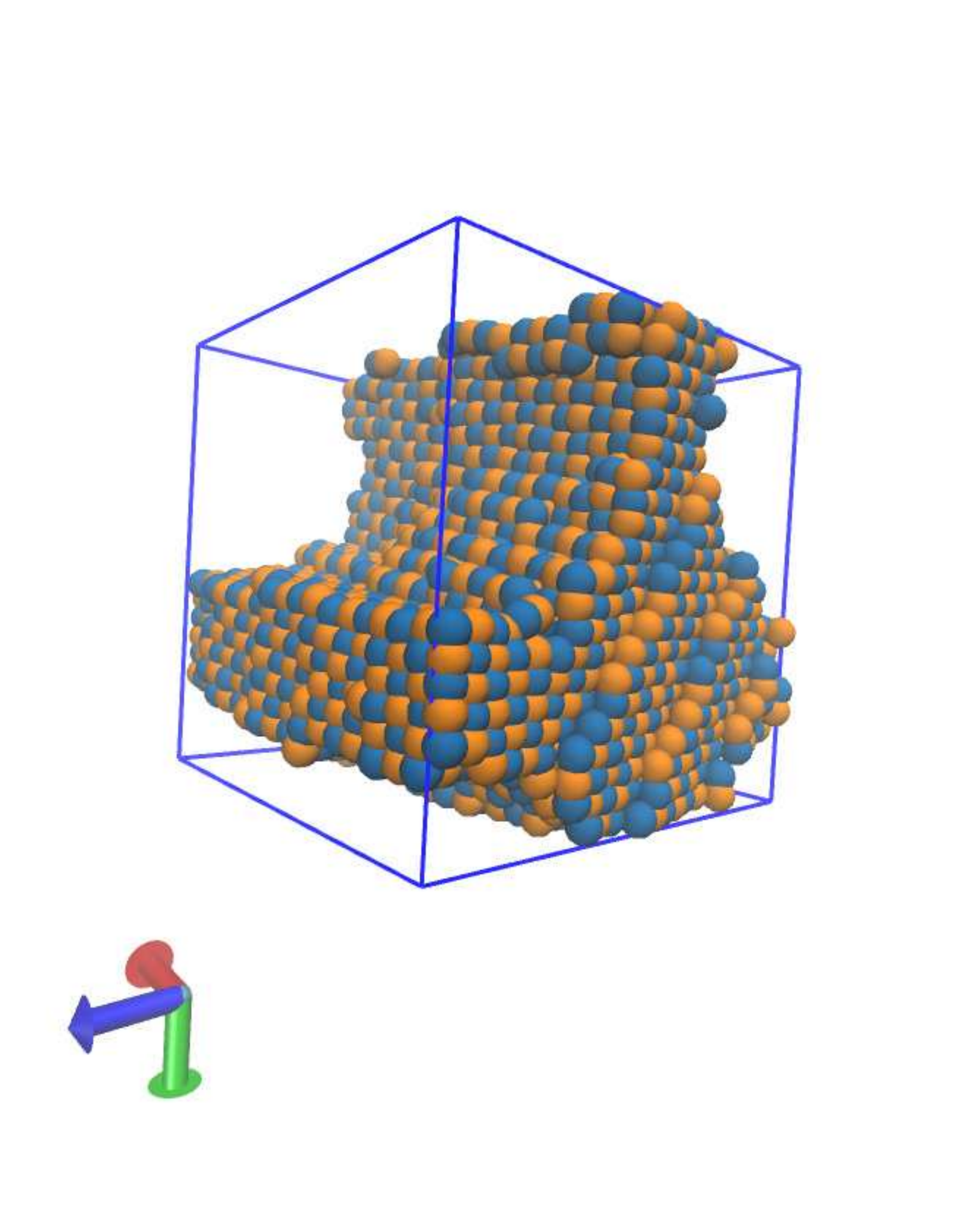}
}
\caption{Pasta structures for nuclear matter systems with 6000 nucleons,
$x=0.5$ at $T=0.2\,$MeV and densities $\rho$=0.05, 0.06, 0.07 and 
0.085$\,$fm${}^{-3}$. Protons are represented in light color (orange), while  
neutrons are represented in darker color (blue). The red arrow (out of the page) 
corresponds to the $x$ coordinate, the green 
arrow (vertical) to the $y$ coordinate and the  blue one (horizontal) to the $z$ 
coordinate.}
\label{fig:pasta}
\end{figure*}

Figure~\ref{fig:pasta} shows an example of the pasta structures for
nuclear matter with 6000 nucleons in the simulating cell (and periodic 
boundary conditions), $x=0.5$ at $T = 0.2$ MeV, and
densities $\rho$=0.05, 0.06, 0.07 and 0.085 fm${}^{-3}$, respectively.\\

\subsection*{Implementation}

\subsubsection*{Intermediate temperature}\label{CMD-IT}

To study the bulk properties of nuclear matter systems must be created with 
specific values of density and temperature. For this purpose it is convenient to 
accelerate the reaching of equilibrium by adding a heath 
reservoir~\cite{ravelo95}.  The addition of the heat flow variable to the 
classical equations of motion results in the Nos\'e-Hoover equations of motion 
which
can be integrated by St{\o}rmer finite differences.  In principle
this approach corresponds to a canonical ensemble and does not
conserve energy which is added or removed by the heat reservoir;
configurations in thermal equilibrium, however, can be achieved
faster than with the usual microcanonical formalism of newtonian
mechanics and an Andersen's thermostat~\cite{andersen}.\\

To mimic an infinite system $A=2000$ nucleons were placed in cubic
cells under periodic boundary conditions. We focus on systems with
isospin content of $x=Z/A=0.3$, 0.4 and 0.5, where $Z$ is the
number of protons. The number densities were enforced by placing
the nucleons in cubical boxes with sizes selected to adjust the
density. The temperatures of the systems studied are T = 1, 2, 3,
4, and 5 MeV, and their densities were selected to be around and
below the corresponding saturation densities values, which vary
with isospin content and temperature. \\

The procedure followed is straightforward: the nucleons are placed at random 
within the cell
avoiding overlaps (i.e. interparticle distances smaller than 0.1 fm) and endowed 
with a Maxwell-Boltzmann velocity distribution corresponding to the desired 
temperature.  The system then is
rapidly evolved until the temperature is maintained within 1$\%$. After reaching 
thermal equilibrium, the system continues evolving and its information at 
selected time steps (nucleon positions and momenta, energy per nucleon, 
pressure, temperature, density, etc.) is stored for future analysis. Each data 
point represents the average of 200 thermodynamically independent 
configurations, the average of the standard deviations is 0.036 MeV. \\

For the radial potentials used the pressure and the energy per nucleon
can be calculated through

\begin{equation}
p = \left< \rho T \right> + \left< {{1}\over{3V}}\sum_i
\sum_{j<i}\mathbf{r} \cdot \mathbf{F} \right>, \ \ \ \ \left<
\varepsilon \right> = {{1}\over{N}} \left< \sum_iV_i  + \sum_{i=i}^N
{{m_i}\over{2}} \mathbf{v_i} \cdot \mathbf{v_i} \right> \ ,
\label{PandE}
\end{equation}

\noindent where the angular brackets indicate an average over the appropriate 
ensemble,
the number density is $\rho=N/V$, N is the number of nucleons, T is
the temperature (in MeV), and $\mathbf{r}$ and $\mathbf{F}$ are the
position and force vectors between pairs of nucleons, respectively,
$V_i$ is the potential energy of each configuration included in the
ensemble average, and $\mathbf{v_i}$ and $m_i$ are the velocity and
masses of each of the nucleons. The first term of the pressure is
the ideal gas contribution and the second one is the contribution
arising from the inter-nucleon potentials.\\

\subsubsection*{Low temperature}\label{CMD-LT}
To study infinite nuclear matter systems we resort to the LAMMPS 
code~\cite{lammps}. We study the properties of a system of 6000 particles (with 
periodic boundary conditions) using the Pandharipande and screened Coulomb 
potentials. The total number of particles is divided into
protons (P) and neutrons (N) according to values of $x= P/(N+P) =$
0.3, 0.4 and 0.5. The nuclear system is cooled down from a
relatively high temperature ($T\geq 4.0\,$MeV) to a desired cool
temperature in small temperature steps ($\Delta T = 0.01\,$MeV) with the
Nos\'e Hoover thermostat~\cite{nose}, and assuring that the energy,
temperature, and their fluctuations are stable.\\

\begin{figure}
\begin{center}
   \includegraphics[width=0.5\columnwidth]{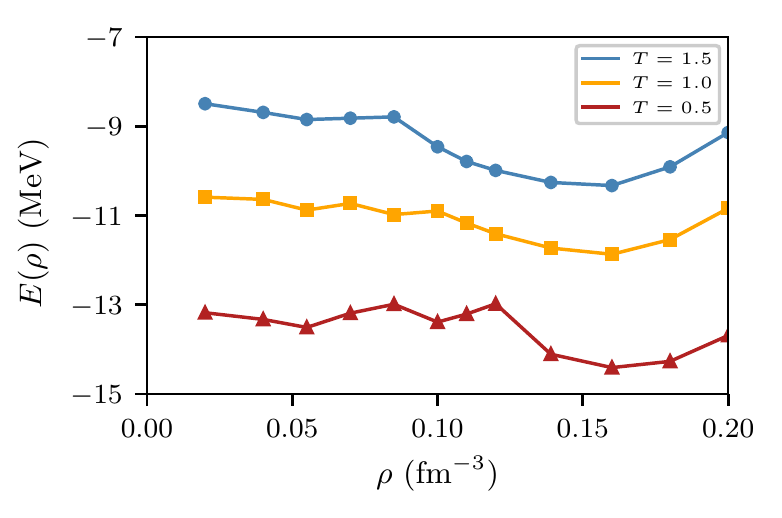}
\caption{Energy per nucleon versus density for nuclear matter with $x=0.5$ at
$T=1.5$ (circles), $1.0$ (squares) and $0.5$~MeV (triangles). The total number
of nucleons in the simulating cell was $N=2000$. The system was cooled from
2~MeV to 0.5~MeV.}\label{E-dens} 
\end{center}
\end{figure}

Figure~\ref{E-dens} shows an example of the energy per nucleon
versus the density for systems with 2000 particles at $x=0.5$ and $T =
1.5$, $1.0$ and $0.5\,$MeV. Clearly visible are the homogeneous phase 
(\textit{i.e.}
those under the ``$\cup$'' part of the energy-density curve), and the loss
of homogeneity at lower densities. \\

From Fig.~\ref{fig:pasta} and Fig.~\ref{E-dens} it is possible to distinguish 
three distinct regions to be analyzed. The first one, that goes from very low 
densities up to approximately $0.08\,$fm$^{-3}$ in which the system displays 
pasta structures, a crystal-like region at densities above approximately 
$0.14\,$fm$^{-3}$, and a transition region between these two~\cite{dor12A}.\\

\subsubsection*{CMD for neutron star matter}\label{CMD-NSM}

In order to mimic (asymmetric) neutron star matter, the LAMMPS CMD 
code~\cite{lammps} was fitted with the potentials mentioned in 
Sections~\ref{cmd_star-1} and 
\ref{cmd_star-2}. We tracked the evolution of systems with $A=$ 4000 nucleons 
situated in a cubic cell under periodic boundary conditions. The simulated 
isospin content was $x = z/A= 0.1$, $0.2$, $0.3$, $0.4$, and $0.5$. The 
densities were varied between $0.02\,$fm$^{-3}$ to $0.085\,$fm$^{-3}$. The 
temperature was controlled with a Nos\'e-Hoover thermostat slowly varying from 
$T$ = 4 MeV down to 0.2 MeV ($\Delta T < 0.1\%$). After placing the nucleons at 
random, but with a minimum inter-particle distance 
of $0.01\,$fm, the nucleons were endowed with velocities according to a 
Maxwell-Boltzmann distribution to correspond to a desired temperature, and the 
equations of motion were solved to mimic the evolution of the system. The 
nucleon positions, momenta, energy per nucleon, pressure, temperature, and 
density, were stored at fixed time-steps.\\

\section{The Maxwell construction}\label{App-Max}

Within the coexistence region, the liquid and gaseous phases are in
thermal and chemical equilibrium with one another and the energy
needed to add or subtract a nucleon from either phase is zero.
As such energy is the Gibbs free energy per particle, $g=\epsilon-Ts+pV$,
its infinitesimal changes are given by
$\mathrm{d}g=-s\mathrm{d}T+V\mathrm{d}p$ and, for isothermic
processes (where $\mathrm{d}T=0$), it reduces to
$\mathrm{d}g=V\mathrm{d}p$. The determination of the boundary of the
coexistence region is then reduced to finding the points where
$g=\int \! V \, \mathrm{d}p=0$, which can be done through an
interesting geometrical technique known as the {\it Maxwell
construction}.\\

[Notice that the present calculation does not consider the isospin chemical 
potential in its conditions for equilibrium.  That is, the Maxwell construction 
calculated only satisfies that the two phases have the same baryon chemical 
potential, but not the same isospin chemical potential. This is a limitation of 
the model used and its effect on the results has yet to be bounded.] \\

To implement the Maxwell construction, the isothermal
pressure-density curves must be first turned into pressure volume
curves (simply through $V=1/\rho$), and then inverted to yield
isothermal $V(p)$ curves; the edges of the coexistence region for
such temperature will be the limits of the integral, $V_{Gas}$ and $V_{Liq}$, at 
which
$g=\int_{V_{Liq}}^{V_{Gas}} \! V \, \mathrm{d}p=0$.\\

To carry out this method one can construct a complete pressure-density isotherms 
by using the curves obtained from CMD (such as those in Figure~\ref{pressure}), 
connected at very low densities with the $p-\rho$ isotherms of a free nucleon 
gas by means of an interpolation.\\

At low densities, the pressure of a mildly-interacting gas of
nucleons approaches that of a free Fermi gas, $p(\rho,T)=
{{2}\over{3}} \rho \, \varepsilon_{F}(\rho,T)$ which, using the
parametrization of the Fermi energy introduced in
Ref.~\cite{lopezlibro}, can be approximated by
\begin{equation}\label{eosp}
p(\rho,T) =    {{\rho}\over 3}\sum_{i=2}^5
i a_i {(\rho/\rho_0)^{i/3+1}}  + {{2\rho}\over 3}\sum_{i=0}^2
\varepsilon_{i}(T) \rho^i .
\end{equation}
where $\varepsilon_i(T) = \sum_{j=1}^2 \varepsilon_{ij}T^j$, and
the coefficients $\varepsilon_{ij}$ and $a_i$ are listed in
table~\ref{tabfit}.\\

The first term of Equation (\ref{eosp}) is a simple approximation
to $2\rho/3$ times the Fermi energy of a cold nucleon gas, and
the second term is an approximation to $2\rho/3$ times the
$T$-dependent Fermi gas energy. Notice that, as expected at low
densities, expression~(\ref{eosp}) does not depend on the isospin
content $x$, as the system is highly noninteracting.\\


\begin{table}
{\begin{tabular}{@{}c@{\quad}c@{\qquad\quad}c@{\quad}c@{}}\toprule
Coefficient       &   Value    &   Coefficient & Value \\ 
$\varepsilon_{01}$ & 0.693  & $\varepsilon_{02}$ & 0.037 MeV$^{-1}$
\\
$\varepsilon_{11}$ & $-5.420$ & $\varepsilon_{12}$ & 0.082 MeV$^{-1}$
\\
$\varepsilon_{21}$ & $11.447$ & $\varepsilon_{22}$ & -0.312 MeV$^{-1}$
\\
$a_2$ & 21.1 MeV & $a_3$ & -38.3 MeV\\
$a_4$ & -26.7 MeV & $a_5$ & 35.9 MeV\\
\bottomrule
\end{tabular}
}
\caption{Coefficients $\varepsilon_{ij}$ and $a_i$.}
\label{tabfit}
\end{table}

\begin{figure}  
\begin{center}
\includegraphics[width=3.7in]{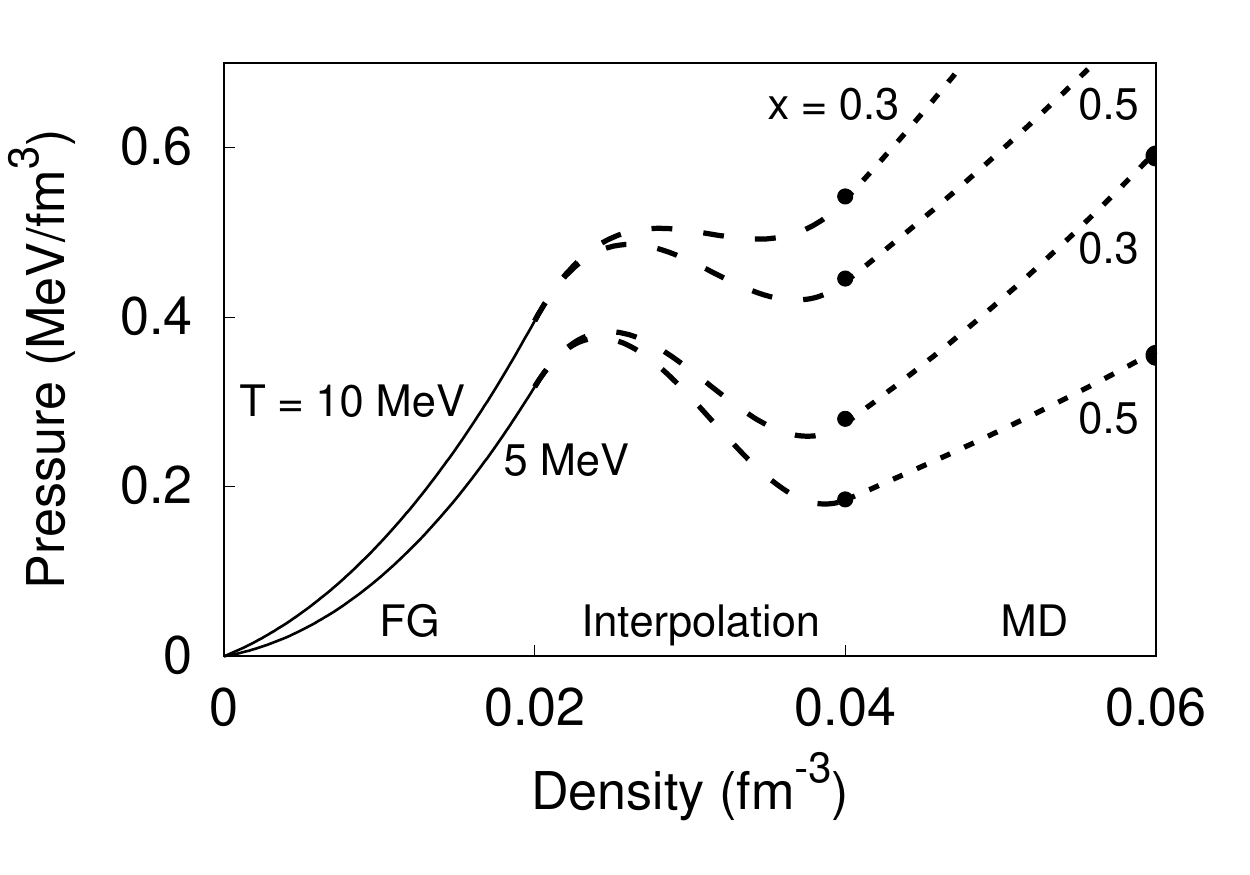}
\end{center}
\caption{Example of pressure density interpolated curves for the cases of $T$ = 
5 and 10 MeV, and for $x$ = 0.3 and 0.5. ``FG'' stands for Fermi gas and signals 
the left-most segments of the curves (continuous line) obtained with 
Eq.~(\ref{eosp}), ``MD'' indicates the right-most segments of the pressure 
isentropes (short dashes) and corresponds to the fit to the molecular dynamics 
results, and ``Interpolation'' refers to the middle segment obtained by a 
least-squares cubic interpolation to match the FG and MD segments. The points 
indicate MD-calculated points.}
\label{Fig-3}
\end{figure}

Using this approximation for the low density region ($\rho
\lesssim\rho_0/6$), and the CMD pressure density curves for the
liquid phase (from $\rho \gtrsim\rho_{L}$), it is possible to obtain a
complete pressure-density curve by least-squares cubic interpolation
constrained to match the three segments and the first derivatives of
the curves at the two matching points. Figure~\ref{Fig-3} shows one
example of a resulting pressure-density curve; the values
used for the matching liquid density $\rho_L$ vary depending on the
values of $T$ and $x$.\\

\begin{figure}  
\begin{center}
\includegraphics[width=3.7in]{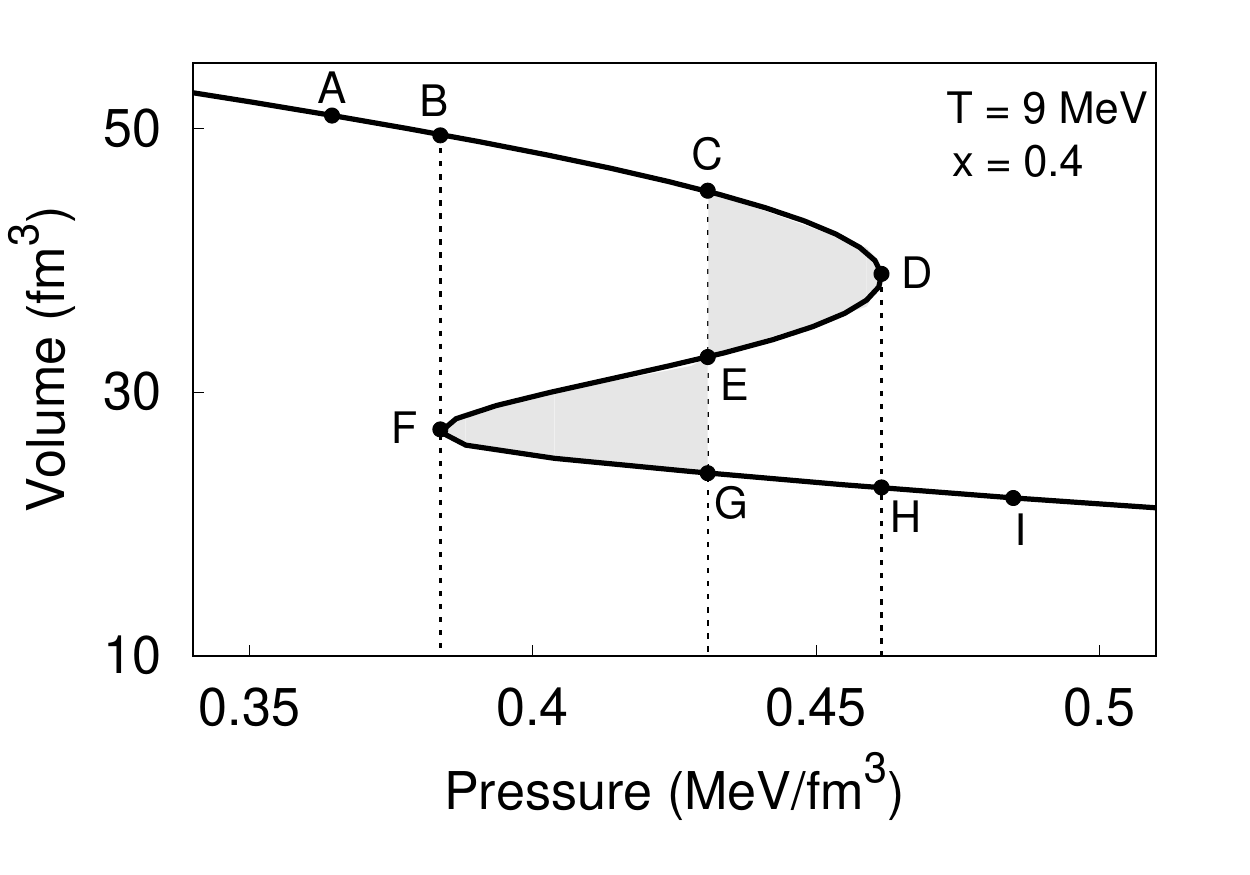}
\end{center}
\caption{\label{eos}Pressure Isotherm for $T$ = 9 MeV and $x$ = 0.4 plotted 
sideways versus the volume. 
The shaded area to the right of the dotted line $C$-$G$ and the curve 
$C$-$D$-$E$ equals the area to the 
left of the dotted line and the curve $E$-$F$-$G$. The liquid-gas coexistence 
region is limited by volumes $V_C$ and $V_G$, and the unstable region lies 
between volumes $V_D$ and $V_F$.} \label{Fig-4}
\end{figure}

To implement the Maxwell construction, the pressure-density curves,
$p(\rho,T)$, must be inverted to yield $V(p,T)$ as shown in
Figure~\ref{Fig-4} for the $T=9$ MeV, $x$ = 0.4 pressure isotherm.
In this curve the gaseous phase is the region that includes the
points from $A$ to $D$, the unstable (negative compressibility) region 
goes from point $D$ to point $F$, and the liquid phase is from $F$ to $I$ 
and beyond. The boundary of the coexistence region are the points $C$ 
and $G$, such that the area to the right of the dotted line $C$-$G$ and 
the curve $C$-$D$-$E$ equals that to the left of the dotted line and the curve
$E$-$F$-$G$, i.e. $g=\int_{V_{G}}^{V_{C}} \! V \, \mathrm{d}p=0$.\\

\begin{figure}  
\begin{center}
\includegraphics[width=3.7in]{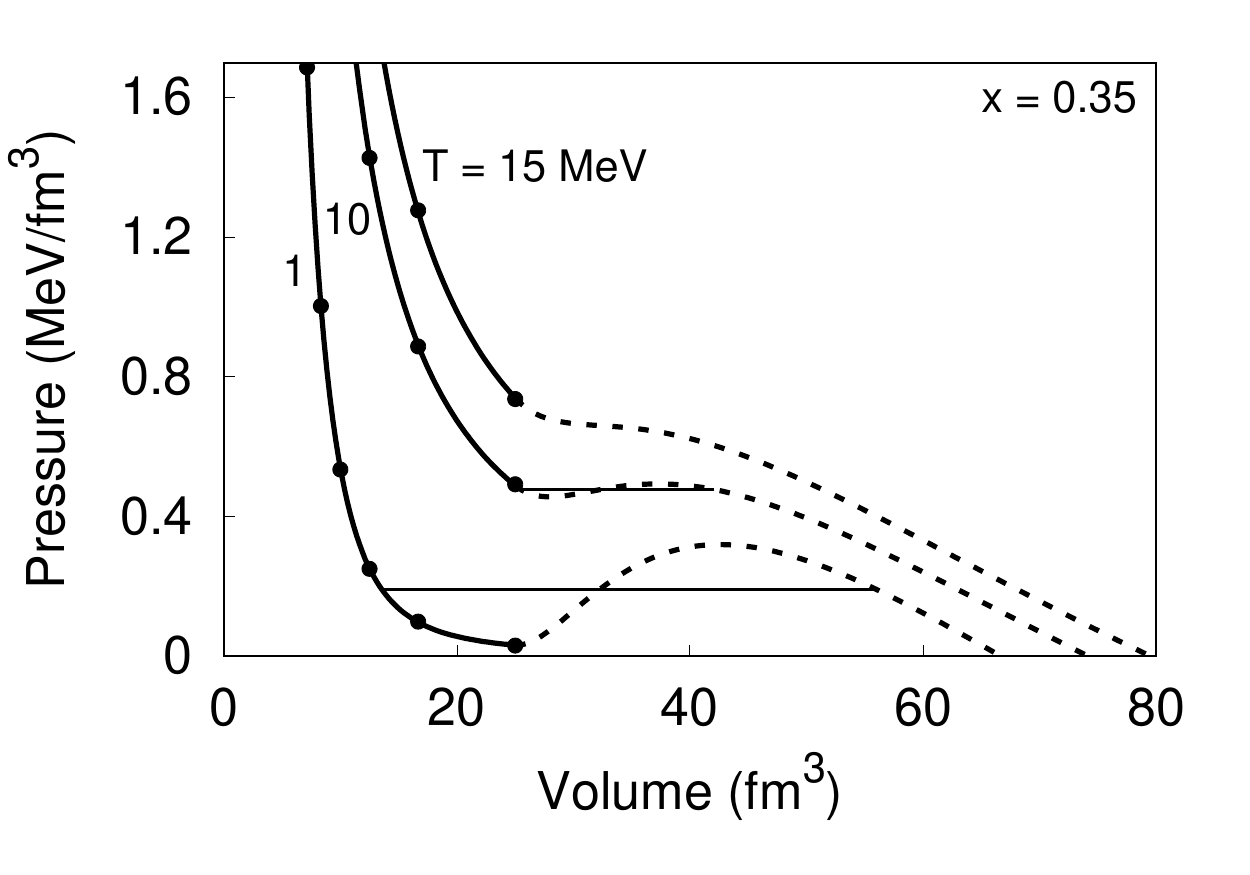}
\end{center}
\caption{\label{Fig-5}Pressure isotherms for $T$ = 1, 10 and 15 MeV and $x$ = 
0.35. The points are the CMD results, the continuous curves are fits to such 
results, and the dashed curves are the cubic least square fits. The pressures at 
which $g=\int_{V_{Liq}}^{V_{Gas}} \! V \, \mathrm{d}p=0$ are shown (horizontal 
lines) for the cases $T$ = 1 and 10 MeV; at $T$ = 15 MeV nuclear matter at $x$ = 
0.35 never reaches a liquid-gas mix phase.}
\end{figure}

To determine the boundary points of the coexistence region the method outlined 
in the previous paragraph was implemented for all the cases studies with CMD. 
Figure~\ref{Fig-5} shows three pressure isotherms plotted versus the volume for 
the cases $T$ = 1, 10 and 15 MeV and for $x$ = 0.35. The continuous curves of 
the left show a fit to the CMD results (points), and the dashed curves of the 
right are the least square fits inverted to appear as a function of the volume. 
The coexistence region is obtained by repeatedly integrating the area between a 
given pressure value and the pressure isotherm, until the pressure at which the 
integral is zero is found. The resulting pressures for the cases $T$ = 1 and 10 
MeV are indicated with the horizontal lines; for the case of $T$ = 15 MeV no 
pressure satisfied the condition of zero integral indicating that for $x$ = 0.35 
the coexistence region ends at a lower temperature. For further details 
see~\cite{lopezAIP2016}.\\

\section{Nuclear symmetry energy from CMD at intermediate 
temperatures}\label{nse}

The effect of the excess of neutrons to protons in the nuclear equation of
state (EOS) is characterized by the symmetry energy, $E_{sym}(T,\rho)$, and its 
importance in phenomena ranging from nuclear structure to astrophysical 
processes has prompted intense investigations~\cite{li,EPJA}. \\

Some of the latest experimental and theoretical studies of the symmetry energy 
have been at subsaturation densities and warm 
temperatures~\cite{hagel,lopez2017}. Experimental 
reactions~\cite{Kowalski,wada,hagel}, for example, have shown that $E_{sym}$ is
affected by the formation of clusters. A recent calculation of the symmetry
energy at clustering densities and temperatures~\cite{lopez2017} obtained
good agreement with experimental data, corroborating the Natowitz
conjecture~\cite{hagel}, namely that the asymptotic limit of
$E_{sym}$ would not tend to zero at small densities as predicted by mean-field
theories.\\

\begin{figure}  
\begin{center}
\includegraphics[width=3.4in]{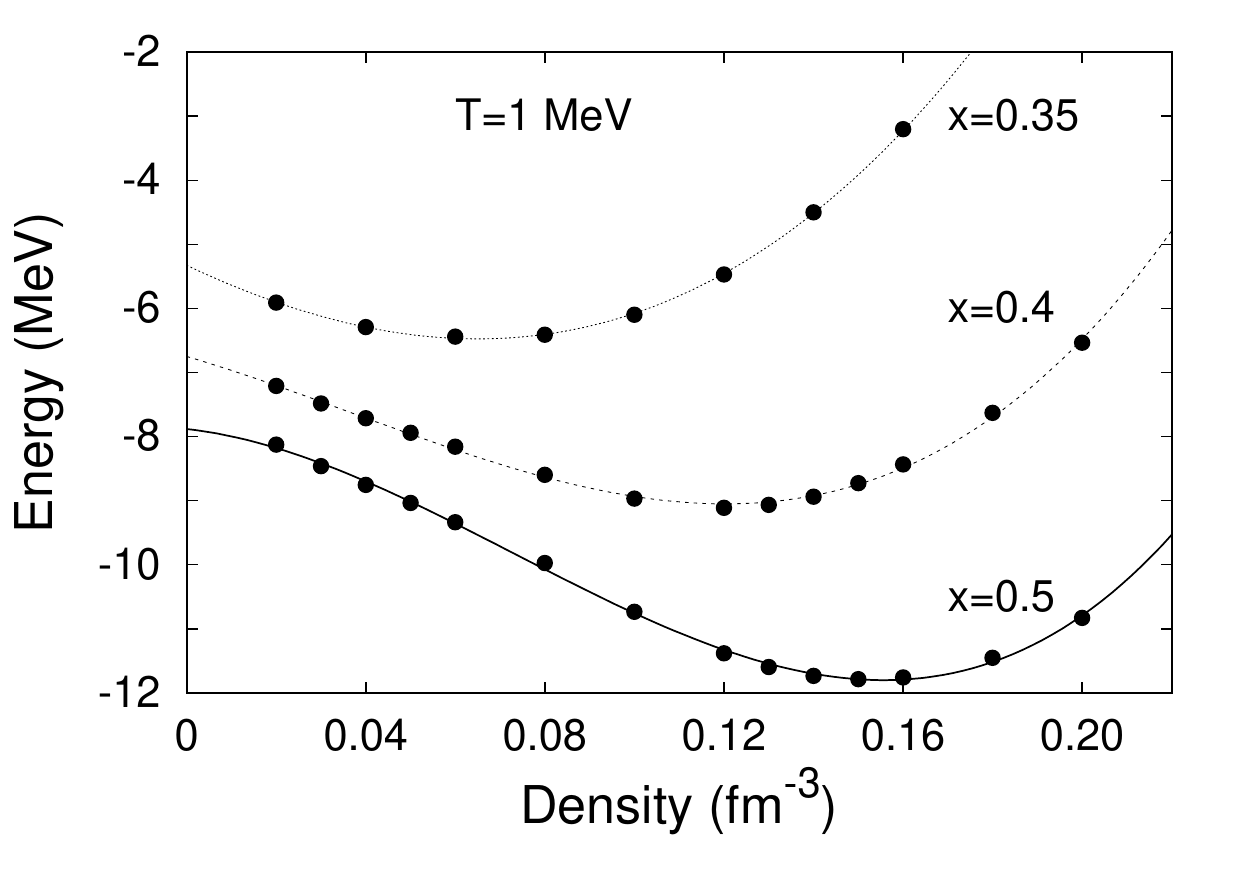}
\end{center}
\caption{Energy per nucleon versus density for systems with
$x=0.35$, 0.4 and 0.5 at $T$ = 1 MeV. Also shown are the cubic fits
used in the determination of the symmetry energy.} \label{ED-Fit}
\end{figure}



Formally, the symmetry energy is the correction that must be added or subtracted 
to the binding energy in systems with different
numbers of protons and neutrons. Defining the symmetry energy as
\begin{equation}
E_{Sym}(\rho,T)={{1}\over{2!}} \left[{{\partial^2
E(\rho,T,\alpha)}/{\partial \alpha^2}}\right]_{\alpha=0},
\label{eqesym}
\end{equation}
with $\alpha=(N-Z)/(N+Z)=1-2x$. The evaluation of the symmetry energy follows 
the procedure introduced before~\cite{lopez2014,lopez2017}. \\

It is possible to use the CMD results of the internal energy $E(\rho,T,x)$ to 
construct a continuous function by means of fitting techniques, and use it to 
obtain the internal symmetry energy. This can be done by fitting the CMD results 
of $E(T,\rho,\alpha)$ for each $T$ and $x$ with an expression of the type
\begin{eqnarray}
E(T,\rho,\alpha)&=& E_0(T,\alpha)+E_1(T,\alpha)\rho \nonumber \\ &+&
E_2(T,\alpha)\rho^2+E_3(T,\alpha)\rho^3. \label{eq2}
\end{eqnarray}
Figure~\ref{ED-Fit} shows the behavior of the energy as a function of the 
density at $T=1$ MeV and $x=0.5$, 0.4 and 0.35 (or $\alpha=0$, 0.2 and 0.3, 
respectively). The liquid phase of the system is easily identifiable by its 
``$\cup$'' shape, as well as the liquid saturation density by the minima of the 
energy curves. Furthermore, the liquid-gas coexistence region can be identified 
as the lower density region where the CMD data separates from the ``$\cup$'' 
curve; this is clearly noticeable for $x=0.4$ and 0.5, but not so much for 
$x=0.35$, which appears to be in the continuous phase up to very low densities, 
perhaps up to 0.02 $fm^{-3}$.\\

\begin{figure}  
\begin{center}
\includegraphics[width=3.4in]{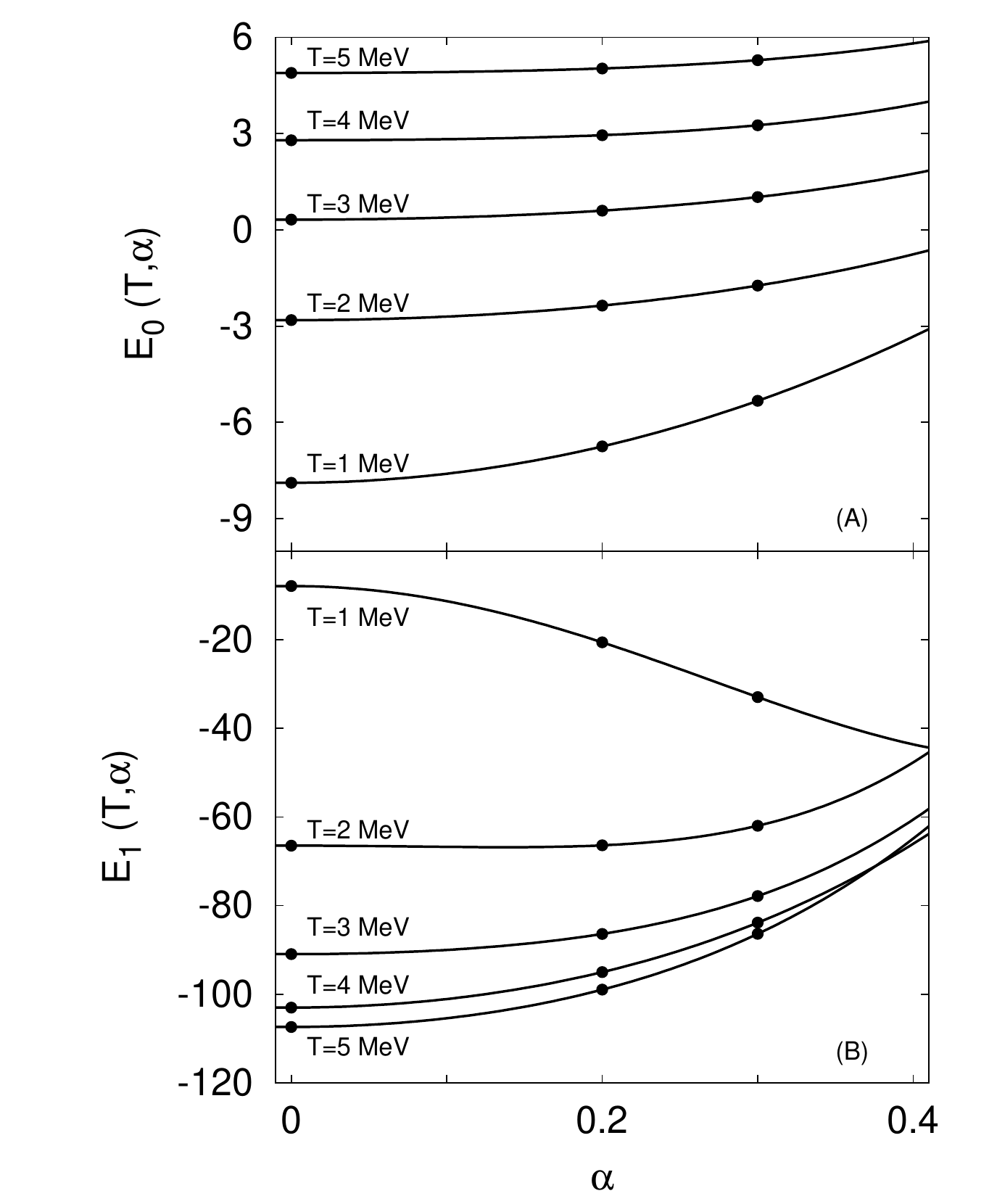}
\end{center}
\caption{Coefficients $E_0(T,\alpha)$ and $E_1(T,\alpha)$ as a function of 
$\alpha$ at various temperatures.} \label{E0E1}
\end{figure}

\begin{figure}  
\begin{center}
\includegraphics[width=3.4in]{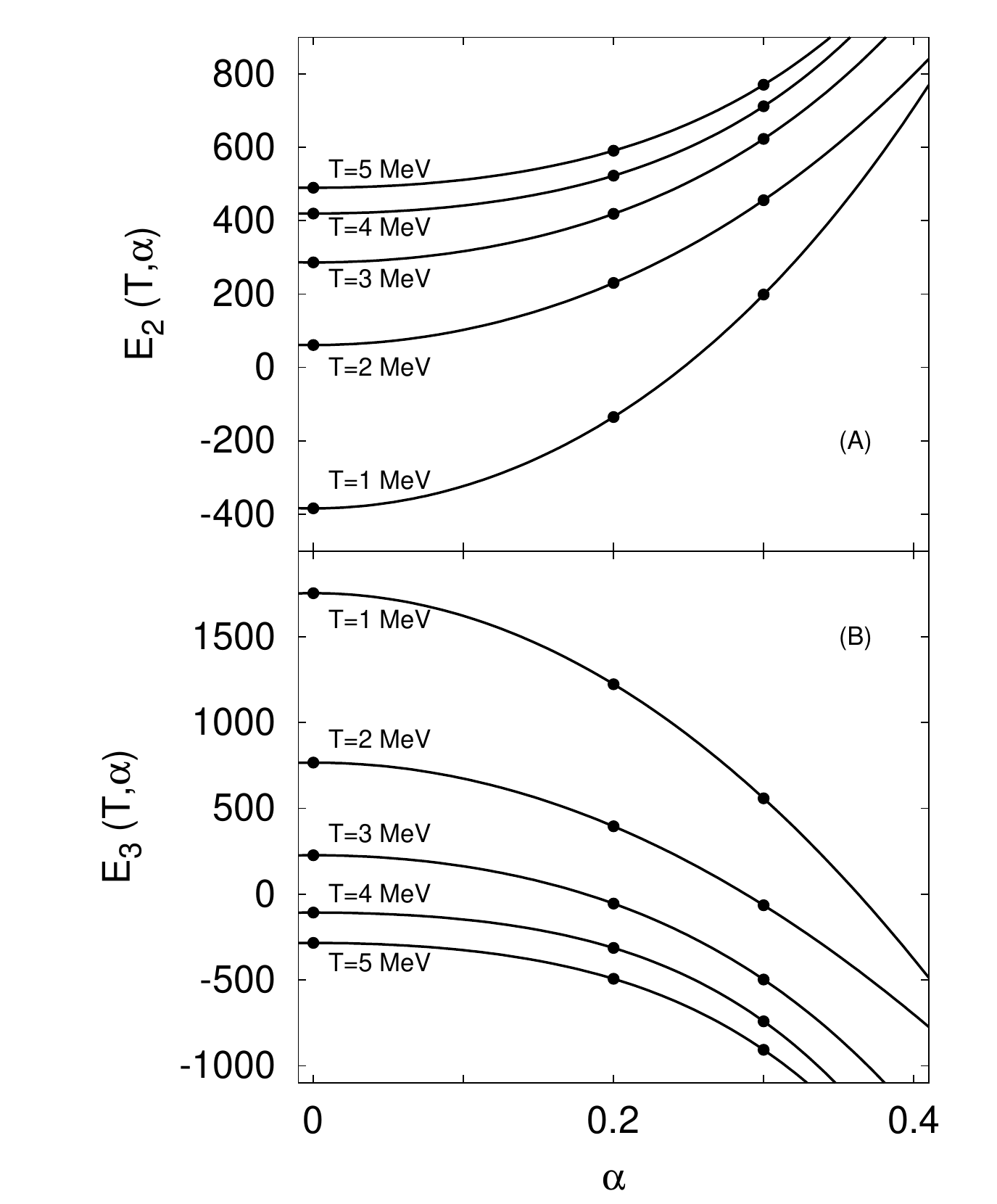}
\end{center}
\caption{Coefficients $E_2(T,\alpha)$ and $E_3(T,\alpha)$.}
\label{E2E3}
\end{figure}

The $\alpha$ dependence of the coefficients $E_0(T,\alpha)$, $E_1(T,\alpha)$, 
$E_2(T,\alpha)$, $E_3(T,\alpha)$ can be easily extracted from the CMD data 
assuming an $\alpha$ dependence of the type
\begin{equation}
E_i(T,\alpha)=E_{i0}(T)+E_{i2}(T)\alpha^2+E_{i4}(T)\alpha^4
\end{equation}
for $i=0$, 1, 2 and 3; odd terms in $\alpha$ are not included to respect the 
isospin symmetry of the strong force (without the Coulomb potential).  The 
smooth behavior of these coefficients with respect to $\alpha$ can be inspected 
in Figures~\ref{E0E1} and \ref{E2E3}; the curves on these figures are the 
results of least squares fits.\\

With this, the symmetry energy is then given by
\begin{eqnarray}
E_{Sym}(T,\rho)&=&E_{02}(T)+ E_{12}(T)\rho \nonumber
\\ &+&E_{22}(T)\rho^2+E_{32}(T)\rho^3,
\end{eqnarray}
with the coefficients $E_{i2}(T)$ given in Table~\ref{tab}.
Figure~\ref{ESym-Dens} shows the symmetry energy in the low density
(i.e. in the liquid-gas mixture) region and its comparison to experimental 
data~\cite{Kowalski, wada}.\\

\begin{table}
    \begin{tabular}{| c || c | c | c | c |}
    \hline
    $T$ (MeV) & $E_{02}$ & $E_{12}$ & $E_{22}$ & $E_{32}$ \\ \hline \hline
    1 & 28.2803 & -349.243 & 6029.93 & -13241.5 \\ \hline
    2 & 10.8479 & -36.8188 & 4101.04 & -9321.43 \\ \hline
    3 & 6.37652 & 88.9445 & 2966.63 & -6241.85 \\ \hline
    4 & 2.92684  & 189.568 & 2040.28 & -3654.5 \\ \hline
    5 & 2.73406 & 192.057 & 2055.3 & -3875.61 \\ \hline
    \end{tabular}\caption{Symmetry energy coefficients.}\label{tab}
\end{table}

Notice that the $E_{Sym}(T,\rho)$ obtained from the CMD values of 
$E(T,\rho,\alpha)$ shows a smooth dependence on the density and temperature, and 
lie on the appropriate range of temperatures for the experimental data, which 
are known to vary between $T=3$ to 11 MeV~\cite{Kowalski,wada, nato2010, hagel}; 
thus corroborating the conjecture of Natowitz, et al. namely that 
$E_{Sym}(T,\rho)$ should approach a constant value as the density approaches 
zero.\\

It is worth mentioning that the mechanism to extract $E_{Sym}$ from CMD data is 
sensitive to the type of data used. For instance, the study of 
Ref.~\cite{lopez2014} was performed solely using data from the liquid phase, and 
produced results comparable to other theories at saturation densities, but 
incorrect results at low densities.  Indeed, since such calculation did not use 
the $E-\rho$ information about the liquid-gas mixed phase, it yielded the 
incorrect result of $E_{Sym}\rightarrow 0$ as $\rho\rightarrow 0$, as all other 
theories that do not include any clusterization; the validity of the results 
of~\cite{lopez2014} is thus limited to near saturation densities. On the other 
hand, the results presented here for the liquid-gas mixed phase contain 
clustarization data and, thus, yield the correct non-zero value for $E_{Sym}$ as 
$\rho\rightarrow 0$~\cite{lopez2017} . \\

In summary, the applications of the method must be adapted to the type of data 
used. The fits used for the ``$\cup$'' data of $E(T,\rho,\alpha)$ in the liquid 
phase, for instance, should be different than those used for the data of the 
liquid-gas mixed phase; the latter needing an extra term in the 
$\rho$-expansion, Eq.~(\ref{eq2}). As we will see in Section~\ref{nse-LT}, the 
computation of $E_{Sym}$ at low temperatures will require further adaption for 
the method to work with frozen nuclear matter.\\

\subsubsection*{Comparison to other theories}\label{theories}

Comparison of symmetry energies obtained with different models is
difficult due to a series of factors. As explained in detail
elsewhere~\cite{typel}, there are complications in the various
methodologies used both in experimental and theoretical works to
extract $E_{Sym}$. This becomes more pronounced when dealing with
systems in the liquid phase as opposed to those in the liquid-gas
coexistence region, and in dealing with warm systems instead of cold
ones. The seriousness of this problem was quantified in the
comparison of different theories through the symmetry energy
parameters, which managed to approve but a handful of cases out of
hundreds of models~\cite{dutra12,dutra14}. We start this section
with a brief outline of some of these complications.\\

To begin with, there is no unique procedure used to obtain the
symmetry energy. Although most experimental studies use the method
of finite differences instead of that of derivatives, as
Eq.~(\ref{eqesym}), there are some others that use
isoscaling~\cite{nato2010}. These methods have different ranges of
validity, finite differences, for instance, assumes a quadratic
behavior for the dependence on $\alpha$ thus limiting its validity
to near symmetric homogeneous matter, and excluding cases with
larger isotopic asymmetry. Indeed substantial differences in
symmetry energies calculated with the derivative definition and with
finite differences have been reported in recent
studies~\cite{typel}.\\

Secondly, these methodological differences become more pronounced at
subsaturation densities. At low densities phase transitions produce
coexisting regions of low and high densities which can have
significant effects on the symmetry energy. This is important as it
is known that the liquid and gaseous phases have different isotopic
asymmetry: the high-density phase is more isospin symmetric and the
low-density phase more asymmetric; a phenomenon called isospin
distillation or fractionation~\cite{colonna}. [In fact the present
results (finite values of $E_{Sym}$ at low densities) are believed
to be due more to the separation of phases than to the process of
cluster formation which is not taken into account in our
calculations in thermal equilibrium~\cite{typel}.]\\

Another source of discrepancies surfaces when the concept of
symmetry energy is applied to finite systems (as in experiments) or
to infinite matter, as well as at zero or finite temperature.
Experimental constraints obtained from laboratory experiments, which
should apply to finite cold-to-warm systems, are fundamentally
different than those coming from stellar systems which obey
considerably distinct equations of state.\\

Yet another complication in comparing finite temperature theories to
zero temperature calculations is that their thermodynamics are
different. At finite temperatures, and depending on the
thermodynamic conditions, the appropriate quantity that needs to be
studied may be the free symmetry energy instead of the internal
symmetry energy; although the two energies are equal to each other
at $T=0$ MeV, they can be substantially different at large
temperatures, as exemplified in a recent calculation~\cite{typel}.\\

In summary, the symmetry energy is sensitive both to the methodology
used for its calculation, and to the specific thermodynamic
conditions of the system. A comparison of symmetry energies from
different theories and experiments demands a careful consideration
of possible discrepancies, especially if comparing cluster-forming
calculations with theories that are unable to undergo phase
transitions.\\

\section{Nuclear symmetry energy from CMD at low temperatures}\label{nse-LT}

The fitting procedure outlined in Section~\ref{nse} corresponds to a two
step-fitting method. The coefficients $E_i$ ($i=0...3$) from
Eq.~(\ref{eq2}) are computed at the first stage of the procedure. These
coefficients, however, depend on the fraction $\alpha$, although not directly
on the fraction $x$ (recall that $\alpha=1-2x$). A more suitable (low
order) expression for the $E_i$'s instead of Eq.~(\ref{eq2}) is

\begin{equation}
\begin{array}{lcl}
         E_i(T,x) & = &
E_{i0}+E_{i1}\overbrace{(1-2x)}^{\alpha}+E_{i2}\,\overbrace{(1-4x+4x^2)}^
{\alpha^2}+\mathcal{O}(\alpha^4)\\
        & & \\
        & \simeq &
(E_{i0}+E_{i1}+E_{i2})-2(E_{i1}+2E_{i2})\,x+4E_{i2}\,x^2\\
        \end{array}\label{fitting_3}
\end{equation}

The above expression corresponds to the lowest non-trivial expansion
$E_i(T,x)\simeq \tilde{E}_{i0}+\tilde{E}_{i1}\,x+\tilde{E}_{i2}\,x^2$. Thus,
the $\tilde{E}_{ij}$'s ($j=0,1,2$) are related to the $E_{i0}$, $E_{i2}$
coefficients according to the following matrix relation\\

\begin{equation}
 \left(\begin{array}{rrr}
        1 & 1 & 1 \\
           & &  \\
         0 & -2 & -4  \\
          &  & \\
         0  & 0 & 4  \\
        \end{array}\right)\,\left(\begin{array}{c}
                                  E_{i0} \\
                                  \\
                                  E_{i1} \\
                                  \\
                                      E_{i2} \\
                                   \end{array}\right)=\left(\begin{array}{c}
                                     \tilde{E}_{i0}\\
                                     \\
                                     \tilde{E}_{i1} \\
                                     \\
                                     \tilde{E}_{i2} \\
                                   \end{array}\right)\label{fitting_4}
\end{equation}

Notice that $\tilde{E}_{i0}$, $\tilde{E}_{i1}$, $\tilde{E}_{i2}$
actually correspond to the fitting parameters at the second stage of the
procedure outlined in Section~\ref{nse} (that is, after the $E_i$'s were
obtained). The  $E_{i0}$, $E_{i1}$ and $E_{i2}$ parameters, though, are now 
computed indirectly from the linear system (\ref{fitting_4}). It is 
straightforward that $E_{i2}=\tilde{E_{i2}}/4$ at this (low order) 
approach. \\

The linear system (\ref{fitting_4}) attains a single solutions for the three 
unknowns $E_{i0}$, $E_{i1}$ and $E_{i2}$. However, any additional constrain 
(say, $E_{i1}=0$ for nuclear matter) would drive the system to an
overdetermined state. The $E_{i2}$ values might still be computed (or 
estimated) in that case, although $E_{i0}$ or $E_{i1}$ would become 
meaningless.  \\

Our numerical computations for $E_{i2}$ at the examined temperatures and
densities indicated in Section~\ref{subsec:sym_energy_3} are detailed in
Table~\ref{table_esym}.\\

\begin{table}[!htbp]
{
\begin{tabular}{c@{\hspace{29mm}}r@{\hspace{15mm}}r@{\hspace{15mm}}r@{\hspace{
15mm}}r}
\hline
$T$ & $E_{02}$ & $E_{12}$ & $E_{22}$ & $E_{32}$ \\
\hline
0.2   & -18.101   &  1441  & -15373 & 87273\\
0.5   &  -53.816  &  3184 & -43548 & 235478\\
1.0   &  -2.303   &  612  & -2398 & 25368\\
2.0   &   3.292   &  344  &  3389 & -13235\\
\hline
\end{tabular}
}
\caption{The computed values $E_{i2}=\tilde{E_{i2}}/4$, after the two-steps
fitting procedure, according to Eq.~(\ref{eq2}) and the approach 
(\ref{fitting_3}). The density ranged from $\rho = 0.04$ to 0.085 fm$^{-3}$. }
\label{table_esym}
\end{table}

In closing, formally speaking the symmetry energy is approximated from a 
second order expansion of the energy with respect to the parameter $\alpha$, and 
evaluated at $\alpha=0$ and at saturation density. Our $E_{sym}$, however, is a 
generalization of such expansion that makes it dependent on the density and 
temperature. Since the use of the liquid drop mass formula that inspired such 
approximation of the symmetry energy is limited to finite systems at $T=0$ and 
near saturation density, the $E_{sym}$ here obtained cannot be simply inserted 
into such formula. A more complete equation of state of nuclear matter, 
\textit{i.e.} one for infinite systems dependent on density, temperature and 
isospin content, could benefit from the analytical fit obtained here for 
non-zero temperature infinite systems at sub-saturation densities.\\

\section{Symmetry energy for neutron star matter}\label{esymmNSM}

The evaluation of the symmetry energy for neutron star matter follows the 
procedure
introduced in Sections~\ref{nse} and~\ref{nse-LT}. The symmetry energy
is defined as in Equation~\ref{eqesym}, again using an $E(T,\rho,\alpha)$ as 
given by Equation~\ref{eq2}, but the $\alpha$ dependence modified as follows.\\


The $\alpha$ dependence of the coefficients $E_i(T,\alpha)$ can be extracted
from the CMD data calculated at various values of $\alpha$, and assuming an
$\alpha$ dependence of the type\\

\begin{equation}
E_i(T,\alpha)=E_{i0}(T)+E_{i1}(T)\,\alpha+E_{i2}(T)\,\alpha^2+E_{i3}(T)\,
\alpha^3+E_{i4}(T)\, \alpha^4\label{fitting_2}
\end{equation}

Notice that no constrains on the powers of $\alpha$ are introduced in 
Eq.~\ref{fitting_2}, 
as opposed for nuclear matter. This is a major difference between NM and NSM, 
since the latter disregards the isospin symmetry of the strong force due to the 
presence the Coulomb screening potential. The symmetry energy is then given by

\begin{eqnarray}
E_{sym}(T,\rho)&=&E_{02}(T)+ E_{12}(T)\rho+ \nonumber
\\ &+&E_{22}(T)\rho^2+E_{32}(T)\rho^3
\end{eqnarray}

\noindent with the coefficients $E_{ij}(T)$ obtained from the fit of the CMD
data.\\


\section{Thermal conductivity}\label{conductivity}

The thermal conductivity $\kappa$ relates the energy flux $\mathbf{J}$ to the 
temperature gradient $\nabla T$, through the Fourier law
\begin{equation}
 \mathbf{J}(t)=-\mathbf { \kappa }\,\nabla T\label{eqn:def_kappa}.
\end{equation}
\noindent The energy flux $\mathbf{J}$ is obtained from the mean flux density 
$\langle\mathbf{j}\rangle$ transported across a small volume $\mathcal{V}$, that 
is $\mathbf{J}=\langle\mathbf{j}\rangle.\mathcal{V}$. The calculation of thermal 
transport properties  from atomistic simulations is well 
established~\cite{muller97,zhou2007phonon,dunn2016role,lin2013thermal}. For the 
case of $\mathbf{J}$ and $\nabla T$ being collinear (say, for example, along the 
$\hat{z}$ axis) a non-equilibrium method for computing the thermal conductivity 
$\kappa_z$ is proposed by M\"uller-Plathe~\cite{muller97} from the average heat 
flux and temperature gradient.
\begin{equation}
 \kappa_z=-\lim_{t\rightarrow\infty}\displaystyle\frac{\langle 
J_z\rangle}{\langle\partial T/\partial z\rangle}\label{eqn:kappa_lim}
\end{equation}

\noindent If the medium is isotropic, common practice sets the mean thermal 
conductivity as $(\kappa_x+\kappa_y+\kappa_z)/3$. Notice that the linear nature 
of Eqs.~(\ref{eqn:def_kappa}) and (\ref{eqn:kappa_lim}) requires relatively 
small temperature gradients.  \\

\begin{figure}[htbp!]
\includegraphics[width=0.5\columnwidth]
{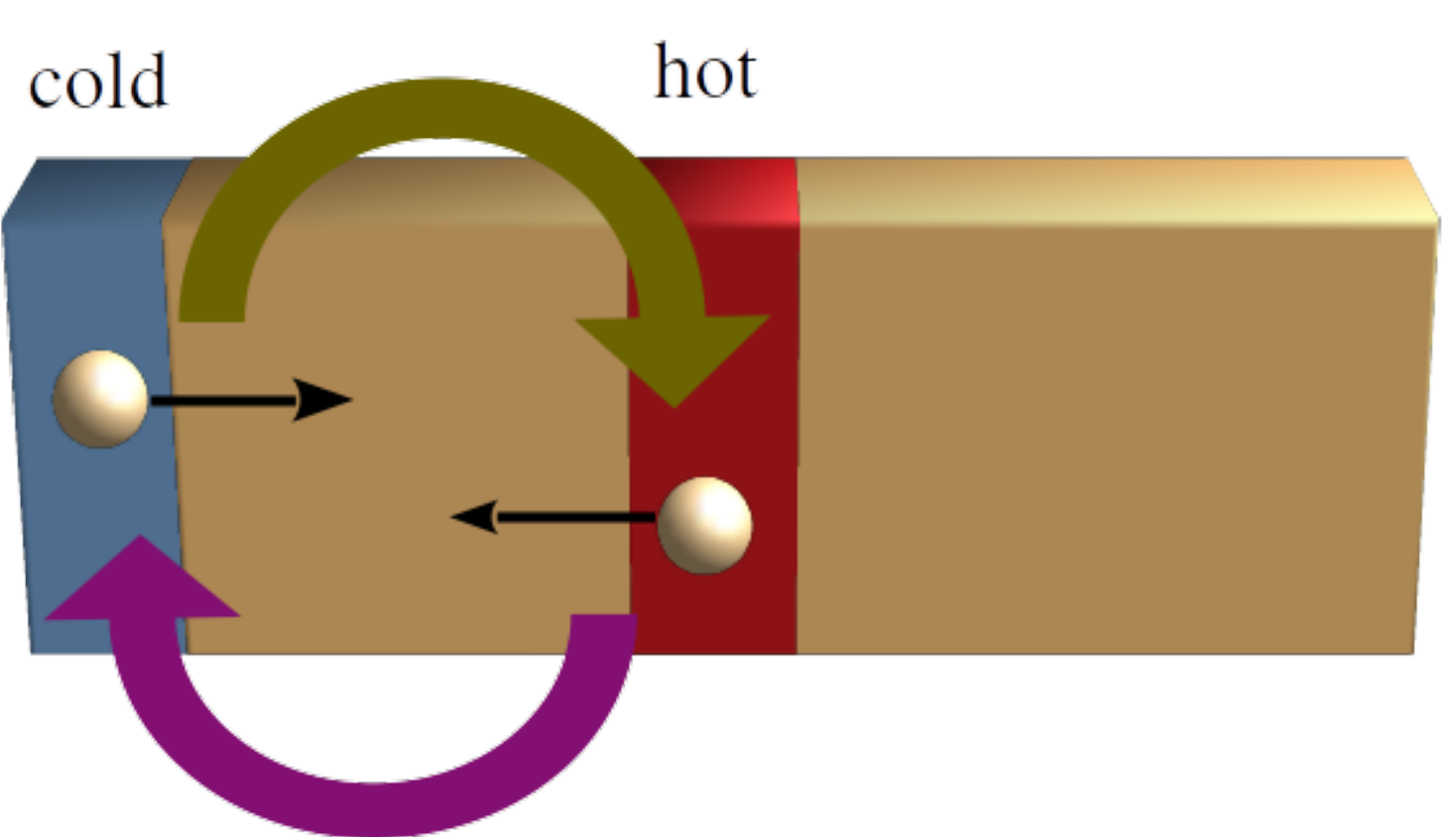}
\caption{\label{scheme} Schematic representation of the 
M\"uller-Plathe procedure. The blue and red bins correspond the the 
``cold'' and ``hot'' slabs, respectively. The horizontal flat arrows stand for 
the particles velocity. The curved arrows (green and violet, respectively) 
represent the velocity exchange process. } 
\end{figure}

In a nutshell, the M\"uller-Plathe procedure \cite{muller97} generates a heat
flux of known magnitude and the temperature gradient is obtained as local
averages of the kinetic energy. The system is divided in thin bins along the 
the  heat flux direction (see Fig.~\ref{scheme} for  details); the first slab 
is 
labeled as the ``cold'' slab, while the slab in the middle is labeled as 
``hot''. A heat flux is generated by exchanging the velocities of two particles 
(with the same mass), the hottest particle in the ``cold'' bin and the coldest 
one in the ``hot'' bin (see Fig.~\ref{scheme}). Thus, the system is 
(artificially) driven out of equilibrium, and a heat flux $\mathbf{J}$ develops 
through the system of interest in the opposite direction for the equilibrium 
restoration. This flux is expected to reach the stationary state if the 
exchanging rate is held regularly for a long time. \\

In order to generate an external heat flux, particle velocity exchanges are
performed periodically during the MD simulation.  Recall that the species 
themselves are not exchanged, but only the velocities. Thus, the ``pumping'' 
process only transports kinetic energy (for particles with the same mass). This 
procedure conserves total energy and linear momentum. \\ 

The heat flux introduced by the velocity exchange is hard to compute from 
dynamical magnitudes. The computation from the net transported (kinetic) energy 
is somewhat easier since

\begin{equation}
 \langle J_z\rangle+\displaystyle\frac{1}{2A_{xy}}\,\bigg[ 
\displaystyle\frac{1}{\tau}\sum_{n=1}^T\displaystyle\frac{1}{2}m(v_{h}^{2}-
v_{c}^{2})\bigg] =0
\label{eqn:J_transported}
\end{equation}

\noindent where the expression between the square brackets represents the mean 
(kinetic) energy exchanged during the time period $\tau$. $v_h$ and $v_c$ 
refers to the velocities of the hot particle and cold particle, respectively.  
The factor $2A_{xy}$ corresponds to the cross section of the slabs (two faces). 
 \\

The temperature profile is obtained by computing the local (kinetic) temperature 
for each slab. Once the steady state is reached, the temperature profile is 
expected to be linear away from the cold and hot bins where velocities are 
exchanged, provided the heat flux remains small. Further details can be found in 
Ref.~\cite{muller97}. 
 \\

We stress the fact that the balance condition (\ref{eqn:J_transported}) 
links the heat flux $\mathbf{J}$ to the (artificial) kinetic energy 
transportation introduced by the M\"uller-Plathe procedure. The velocity 
exchange is not restricted to pairs of similar particles, but also across 
species (with the same mass). Therefore, the procedure enables the computation 
of the thermal conductivity for the set of \textit{all} the nucleons, or for 
the set of protons and neutrons separately. The meaning of either coefficients, 
though, will be quite different.  \\


\section{Analytical tools}\label{tools}

\subsection{The Radial Distribution Function}\label{RadialDist}

A rigorous definition for the radial distribution function $g(\mathbf{r})$ 
starts from
the following distance distribution

\begin{equation}
g(\mathbf{r})=\displaystyle\frac{1}{\rho_0}\bigg\langle\displaystyle\frac{1}{N}
\displaystyle\sum_{i=1}^N\displaystyle\sum_{j\neq i}^N
\delta(\mathbf{r}-\mathbf{r}_{ij} )\bigg\rangle\label{eqn:gr_def_rigorous}
\end{equation}

\noindent where $\rho_0=N/V$ is the (mean) density in the simulation cell of 
volume $V$ (or equivalently $L^3$). 
$\mathbf{r}_{ij}=\mathbf{r}_{j}-\mathbf{r}_{i}$ is the distance vector between 
the particle $i$ and the particle $j$. The $\delta(\cdot)$ function corresponds 
to the Dirac
delta. The mean value indicated by $\langle\cdot\rangle$ corresponds to the 
average operation over successive time-steps. \\

\subsubsection*{$g(\mathbf{r})$ for intermediate temperatures}\label{g-for-IT}

For cases where the existence of phases is suspected, it is instructive to 
compare the local density to the average one, $g(\mathbf{r})=\rho(r)/\rho$; 
peaks in $g(\mathbf{r})$ will indicate the existence (or lack of) neighbors, 
thus signaling transitions from one phase to another, say from liquid to gas. \\

For intermediate temperatures, say $T > 1$ MeV, 
expression~\ref{eqn:gr_def_rigorous} is applied cumulatively to a large number 
of systems (say, 200) at the same temperature and densities. The strengths of 
the nearest-neighbor peaks will indicate higher correlation at higher densities 
(continuous liquid phase) than at lower densities (gaseous phase composed of a 
mixture of free nucleons and small clusters). The difference in correlation is 
due to the larger nucleon mobility in the liquid-gas mixture compared to the 
homogeneous liquid phase. Similar effect is also observed in the second-neighbor 
peaks which show stronger correlations in homogeneous phases than in mixed ones. 
Figure~\ref{rdf} shows the radial distribution functions of a system at $T = 1$ 
MeV. \\

\subsubsection*{$g(\mathbf{r})$ for pasta stuctures}\label{g-for-pasta}
To calculate $g(\mathbf{r})$ for a pasta structure, we assume that particles 
form a homogeneous slab (that is, a \textit{lasagna}-like structure) bounded 
between the $\pm z_0$ (horizontal) planes. The slab is quasi static, meaning 
that no averaging over successive time-steps is required. \\

We first consider a small region $\Omega(\mathbf{r})$ with volume $\delta
x\,\delta y\,\delta z$ in the $\mathbf{r}$ domain. Thus, we may evaluate the 
expression (\ref{eqn:gr_def_rigorous}) as follows

\begin{equation}
\displaystyle\int_{\Omega(\mathbf{r})}g(\mathbf{r})\,d^3x=
\displaystyle\frac{1}{\rho_0}\,
\displaystyle\frac{1}{N}
\displaystyle\sum_{i=1}^N \sum_{j\neq i}^N \int_{\Omega(\mathbf{r})}
\displaystyle\delta(\mathbf{r}-\mathbf{r}_{ij})\,d^3x
\label{eqn:gr_def_rigorous_2}
\end{equation}

\noindent where the integral on the right-hand side equals unity at the
positions $\mathbf{r}=\mathbf{r}_{ij}$ within $\Omega(\mathbf{r})$. \\

We now proceed to evaluate the sum over the $j\neq i$ neighbors of the
particle $i$. The counting of neighbors is proportional to $\delta x\,\delta
y\,\delta z$ since the volume $\Omega(\mathbf{r})$ is very small. Thus, the
tally is

\begin{equation}
\sum_{j\neq i}^N \int_{\Omega(\mathbf{r})}
\displaystyle\delta(\mathbf{r}+\mathbf{r}_{i}-\mathbf{r}_{j})\,d^3x\approx
\rho(\mathbf{r}+\mathbf{r}_i)\,\delta x\,\delta y\,\delta z
\label{eqn:gr_def_rigorous_3}
\end{equation}

The magnitude $\rho(\mathbf{r}+\mathbf{r}_i)$ refers to the density within the 
small region $\Omega(\mathbf{r}+\mathbf{r}_i)$, its value remains constant 
inside the slab and vanishes outside. In fact, $\rho$ can be expressed as 
$\rho_s\,\Theta(z_0-|z+z_i|)$, with $\Theta(\cdot)$ representing the Heaviside 
function, and $\rho_s$ being the slab density ($\rho_s>\rho_0$).   \\

The sum over the $i$ particles is evaluated through the integral of the 
infinitesimal pieces $\rho_s\,d^3x'$. Thus,

\begin{equation}
\displaystyle\sum_{i=1}^N
\rho(\mathbf{r}+\mathbf{r}_i)\approx\displaystyle\int_\mathrm{slab}\!\!\rho_s^2
\,\Theta(z_0-|z+z'|)\,d^3x'
\label{eqn:gr_def_rigorous_4b}
\end{equation}

\noindent where the slab domain corresponds to the bounded region $|z'|\leq
z_0$. Notice that this integral vanishes for $|z|\geq 2z_0$ and equals
$\rho_s^2L^2\,(2z_0-|z|)$ otherwise. The $g(\mathbf{r})$ function is
then

\begin{equation}
g(\mathbf{r})\approx \displaystyle\frac{\rho_s}{\rho_0}\,\bigg(1-
\displaystyle\frac{|z|}{2z_0}\bigg)\ \ \ , \ \ \ |z|<2z_0
\label{eqn:gr_def_rigorous_4}
\end{equation}

\noindent assuming that $g(\mathbf{r})$ is somewhat fixed inside the small 
domain $\Omega(\mathbf{r})$ in (\ref{eqn:gr_def_rigorous_2}). We further 
replaced $N$ by $\rho_s\,2z_0L^2$ in (\ref{eqn:gr_def_rigorous_2}). \\

The expression (\ref{eqn:gr_def_rigorous_4}) is not exactly the radial 
distribution function yet because of the angular dependency of $g(\mathbf{r})$. 
This dependency may be eliminated by integrating $g(\mathbf{r})$ over a 
spherical shell of radius $r$, which introduces the normalization factor $4\pi 
r^2$. The expression for the radial distribution then reads

\begin{equation}
g(\mathbf{r})=\displaystyle\frac{1}{4\pi r^2}\,
\displaystyle\int_S \displaystyle\frac{\rho_s}{\rho_0} \, \bigg(1-
\displaystyle\frac{r|\cos\theta|}{2z_0}\bigg)\,r^2\sin\theta\,d\theta\,d\varphi
\label{eqn:gr_def_rigorous_5}
\end{equation}

\noindent where $z=r\,\cos\theta$. Notice that this expression is valid along
the interval $|\cos\theta|<2z_0/r$ whenever $2z_0<r$, but it is constrained
to the natural bound $|\cos\theta|\leq 1$ if $2z_0\geq r$. The integration 
of (\ref{eqn:gr_def_rigorous_5}) finally yields

\begin{equation}
g(\mathbf{r})=\left\{\begin{array}{lcl}
             \displaystyle\frac{\rho_s}{\rho_0} \,
\bigg(1-\displaystyle\frac{r}{4z_0}\bigg) & \mathrm{if} & r<2z_0\\
             & & \\
             \displaystyle\frac{\rho_s}{\rho_0}\,\displaystyle\frac{z_0}{r}  &
\mathrm{if}  &  r>2z_0 \\
            \end{array}\right.\label{eqn:gr_def_rigorous_6}
\end{equation}

Notice that as $r\rightarrow0$, $g(\mathbf{r})$ correctly goes to the limit of 
$g(\mathbf{r}) \rightarrow \rho_s/\rho_0$. Similarly, for larger values of $r$, 
and up 
to $r<2z_0$, $g(\mathbf{r})$ decreases linearly, as observed in Fig.~\ref{rad2} 
for different values of $\rho_0$. \\

Eq. (\ref{eqn:gr_def_rigorous_6}) also indicates that $g(2z_0)\approx 1$ in the 
case that $\rho_s/\rho_0\approx 2$, in agreement with Fig.~\ref{rad2} where 
$g(\mathbf{r})$ 
tends to unity at $20\,$fm (for simulation cells of $L\sim 45\,$fm); this 
figure, however, does not show the behavior beyond $20\,$fm as the statistics 
that can be collected for distances above $L/2$ are very poor. \\

From literature references, $g(\mathbf{r})$ is supposed to converge to unity as
$r\rightarrow\infty$. But, according to (\ref{eqn:gr_def_rigorous_6}), the
radial distribution vanishes at this limit. The disagreement corresponds to
the fact that the condition $g(\infty)=1$ is only valid for homogeneous
systems. The expression (\ref{eqn:gr_def_rigorous_6}) can actually meet this
condition if the slab occupies all the simulation cell, since
$\rho_s\rightarrow\rho_0$ and $z_0\rightarrow\infty$ (for periodic boundary
conditions). \\

\subsection{Lindemann coefficient}\label{lind}

The Lindemann coefficient~\cite{lindemann} provides an estimation of
the root mean square displacement of the particles respect to their
equilibrium position in a crystal state, and it serves as an
indicator of the phase where the particles are in, as well as of transitions
from one phase to another. Formally,

\begin{equation}
\Delta_L =
\displaystyle\frac{1}{a}\displaystyle\sqrt{\displaystyle\sum_{i=1}
^N\bigg\langle\displaystyle\frac{\Delta r^2_i}{N}\bigg\rangle}
\end{equation}

\noindent where $\Delta r^2_i=(r_i-\langle r_i\rangle)^2$, $N$  is the number
of particles, and $a$ is the crystal lattice constant; for the nuclear case we
use the volume per particle to set the length scale through $a = (V/N)^{1/3}$.
\\

\subsection{Kolmogorov statistic}\label{kolm}
The Kolmogorov statistic measures the difference between a sampled (cumulative)
distribution $F_n$ and a theoretical distribution $F$. The statistic, as
defined by Kolmogorov \cite{birnbaum}, applies to univariate distributions (1D)
as follows

\begin{equation}
D_N = \mathrm{sup}_{\{x\}}|F_N(x) - F(x)|
\end{equation}

\noindent where ``sup'' means the supremum value of the argument along the $x$
domain, and $N$ is the total number of samples. This definition is proven to
represent univocally the greatest absolute discrepancy between both
distributions. \\

An extension of the Kolmogorov statistic to multivariate distributions,
however, is not straight forward and researchers moved in different directions
for introducing an achievable statistic \cite{gosset}. The Franceschini's
version seems to be ``well-behaved enough'' to ensure that the computed
supremum varies in the same way as the ``true'' supremum. It also appears to be
sufficiently distribution-free for practical purposes \cite{franceschini}. \\

The three dimensional (3D) Franceschini's extension of the Kolmogorov statistic
computes the supremum for the octants

\begin{eqnarray}
 (x<X_i,y<Y_i,z&<&Z_i),   \nonumber
\\ (x<X_i,y&<&Y_i,z>Z_i),  \nonumber
\\ ...\ (x&>&X_i,y>Y_i,z>Z_i)
\end{eqnarray}

\noindent for any sample $(X_i,Y_i,Z_i)$, $i$ denoting each of the $N$ 
particles, 
and chooses the supremum from this set of eight values. The method assumes 
that the variables $X_i$, $Y_i$ and $Z_i$ are not highly correlated. \\

In the nuclear case, the Kolmogorov 3D (that is, the Franceschini's version)
quantifies the discrepancy in the nucleons positions compared to the
homogeneous case. \\

It is worth mentioning that the reliability of the 3D Kolmogorov statistic has
been questioned in recent years \cite{babu}. The arguments, however, focus on
the correct confidence intervals when applying the 3D Kolmogorov statistic to
the null-hypothesis tests. Our investigation does not require computing these
intervals, and thus, the questionings are irrelevant to the matter.  \\

\subsection{Minkowski functionals}\label{mink}

Complex geometry may be decomposed into simple structures 
connected each other side by side. The basic 3D structure is the 
\textit{voxel}. Thus, the complexity of any body may be associated to 
the way the voxels are connected between each other. \\

The connectivity pattern across voxels (required to represent a 
complex body) may be characterized by counting the number of unique vertices 
($n_v$), edges ($n_e$), faces ($n_f$) and volume ($n_c$). However, a more 
practical set of indices corresponds to the set of the Minkowski functionals, 
defined as follows~\cite{michielsen}  \\

\begin{equation}
 V=n_c\ \ \ , \ \ \ S=-6n_c+2n_f\ \ \ , \ \ \ 2B=3n_c-2n_f+n_e\ \ \ , \ \ \ 
\chi=-n_c+n_f-n_e+n_v\label{eq:indices}
\end{equation}

$V$ and $S$ correspond to the body volume and surface, 
respectively. The $B$ functional represents the mean breadth (or integral mean 
curvature) of the body. The positive values of $B$ have been associated to 
spherical or cylindrical bodies (with no holes inside), while negative ones 
have been associated to bubble-like bodies or hollow cylinders 
\cite{schuetrumpf2,sonoda}. The $B=0$ value (with $\chi=0$) appears to 
be associated to slab-like bodies. \\

An alternative expression for the Euler characteristic 
$\chi$ reads as follows 

\begin{equation}
\chi = \mathrm{isolated\ regions} + \mathrm{cavities} -
\mathrm{tunnels}\label{eq:chi}
\end{equation}

This means that solid (including isolated voids inside) may be 
associated to positive values of $\chi$. But sponge-like or jungle-gym 
structures may be associted to negative values $\chi$ values, since the number 
of tunnels exceeds the number of isolated regions. \\

A more detailed set of pastas associated to positive or 
negative signs of $B,\chi$ can be seen in Table~\ref{tab1} (more 
information in Ref.~\cite {dor12A}). As mentioned above, the \textit{lasagnas} 
(that is, slab-like structures) correspond to $B\sim\chi\sim 0$. The 
\textit{gnocchi} (spherical-like) or \textit{spaghetti} (cylinder-like) 
attain positive values of $B$ and $\chi\geq0$. The \textit{jungle-gym} 
(sponge-like) structures attain $\chi<0$. The corresponding opposite 
structures in Table~\ref{tab1} are named with the prefix \textit{anti-}.\\

\begin{table}
{
\begin{center}
\begin{tabular}{c @{\hspace{9mm}}|@{\hspace{9mm}} c
@{\hspace{9mm}}|@{\hspace{9mm}} c @{\hspace{9mm}}|@{\hspace{9mm}} c}
\hline
           &   B $<$ 0  & B $\sim$ 0  &   B $>$ 0 \\
\hline
$\chi > 0$ & anti-gnocchi &  & Gnocchi\\
$\chi \sim 0$ & anti-spaghetti  & lasagna & spaghetti\\
$\chi < 0$ & anti-jungle gym &  & jungle gym \\
\hline
\end{tabular}
\end{center}
}
\caption{Integral mean curvature and Euler characteristic values for
pasta shapes. The ``anti'' prefix means the inverted situation between
occupied and empty regions. The ``jungle gym'' stands for a 3D rectangular
trellis.}
\label{tab1}
\end{table}

\subsection{\label{sec:voxel}The Minkowski voxels}

The Minkowski functionals require the binning of space into ``voxels''
(that is, tridimensional ``pixels''). Each voxel is supposed to include
(approximately) a single nucleon. But this is somehow difficult to achieve if
the system is not completely homogeneous (and regular). \\

\begin{figure*}[!htbp]
\centering
\captionsetup[subfigure]{justification=centering}
\subfloat[]{
\includegraphics[width=3.2in]
{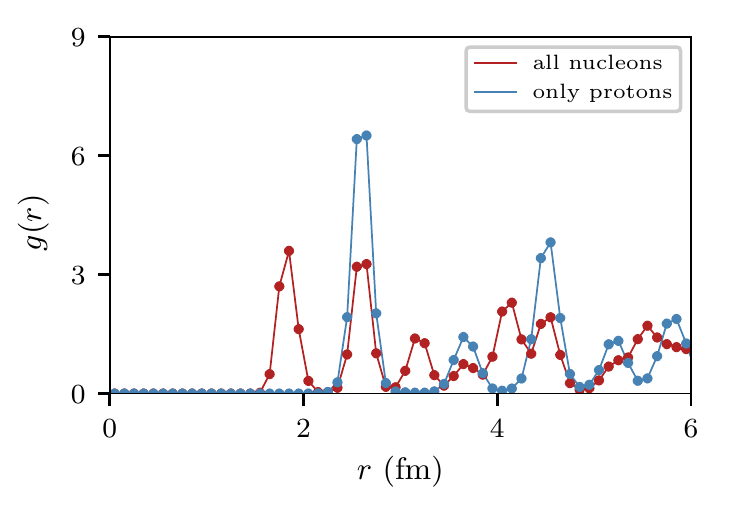}
}
\subfloat[]{
\includegraphics[width=0.33\columnwidth]
{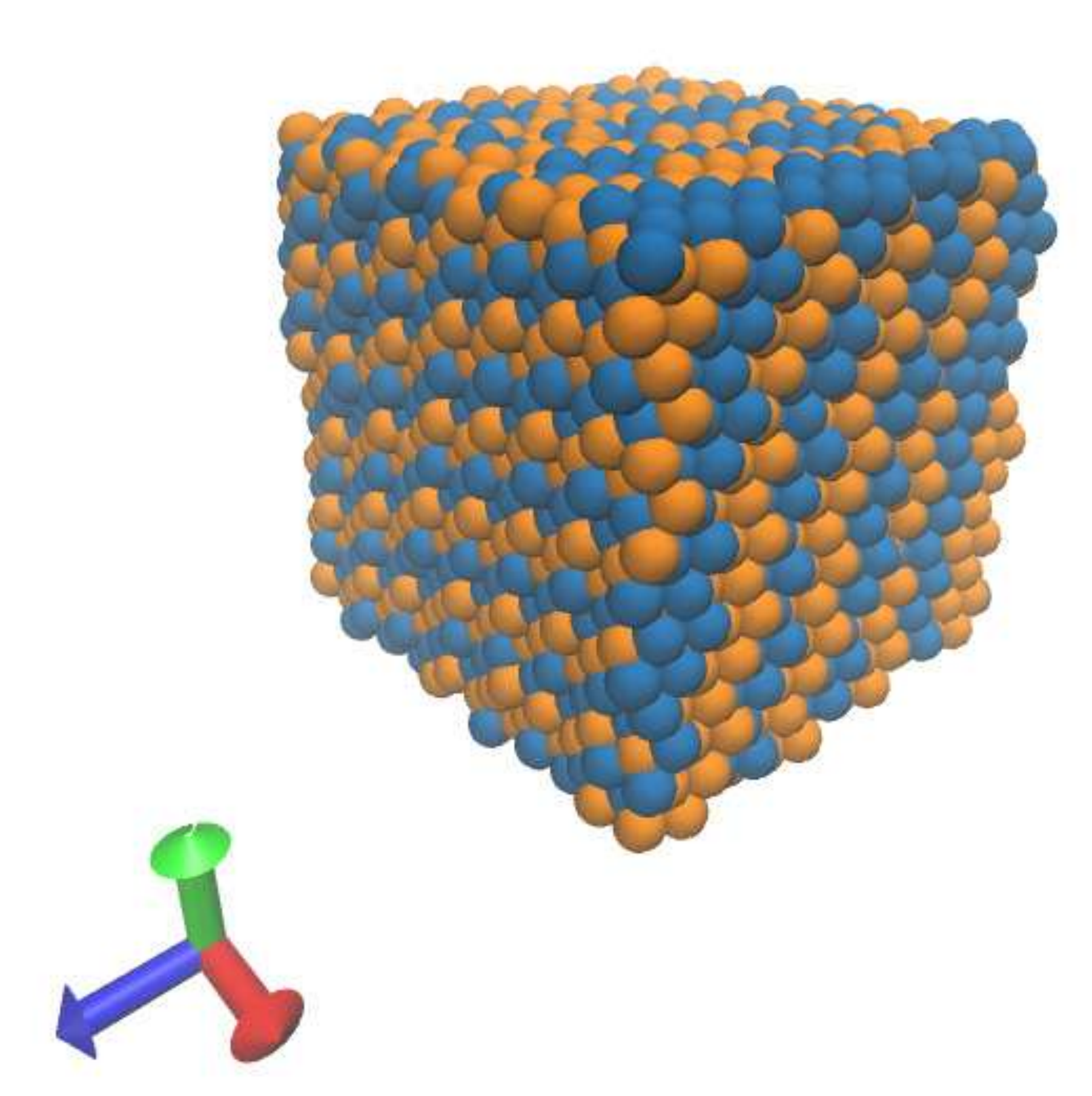}
}
\caption{\label{fig:rho016_x05} (Color online) (a) Radial
distribution function $g(\mathbf{r})$ for the case of 5832 nucleons ($x=0.5$) at
$\rho=0.16\,$fm$^{-3}$ and $T=0.1\,$MeV. The binning is $0.05\,$fm. The symbols
in blue correspond to the $g(\mathbf{r})$ for protons only. The symbols in red 
correspond
to the $g(\mathbf{r})$ computed over all the nucleons. The first peak for the 
blue
symbols occurs at $2.65\,$fm. The first peak for the red symbols occurs at
$1.85\,$fm. (b) A snapshot of the system at $T=0.1\,$MeV.  }
\end{figure*}

We start with a simple cubic arrangement of 50\% protons and 50\% neutrons as
shown in Fig.~\ref{fig:rho016_x05}. The system is at the saturation density
$\rho_0=0.16\,$fm$^{-3}$ (see caption for details).  \\

Notice from Fig.~\ref{fig:rho016_x05} that the nearest distance between protons
and neutrons is $1.85\,$fm. Likewise, the nearest distance between nucleons of
the same species is approximately $2.65\,$fm (that is, the position of the
second peak). It can be verified that the latter is approximately
$\sqrt{2}\times 1.85\,$fm, as expected for the simple cubic arrangement (within
the $g(\mathbf{r})$ binning errors).  \\

\begin{figure*}[!htbp]
\centering
\captionsetup[subfigure]{justification=centering}
\subfloat[\label{fig:rho016_x05_chi_euler}]{
\includegraphics[width=0.5\columnwidth]
{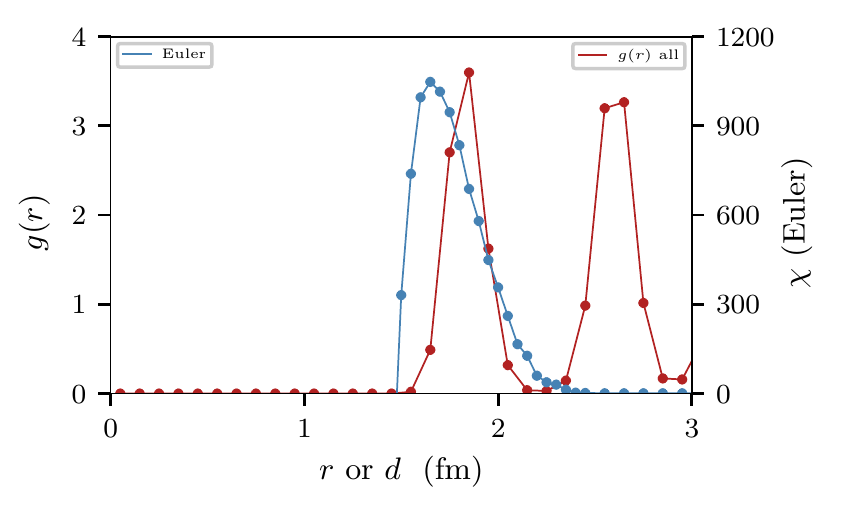}
} 
\subfloat[\label{fig:rho016_x05_chi_vol}]{
\includegraphics[width=0.5\columnwidth]
{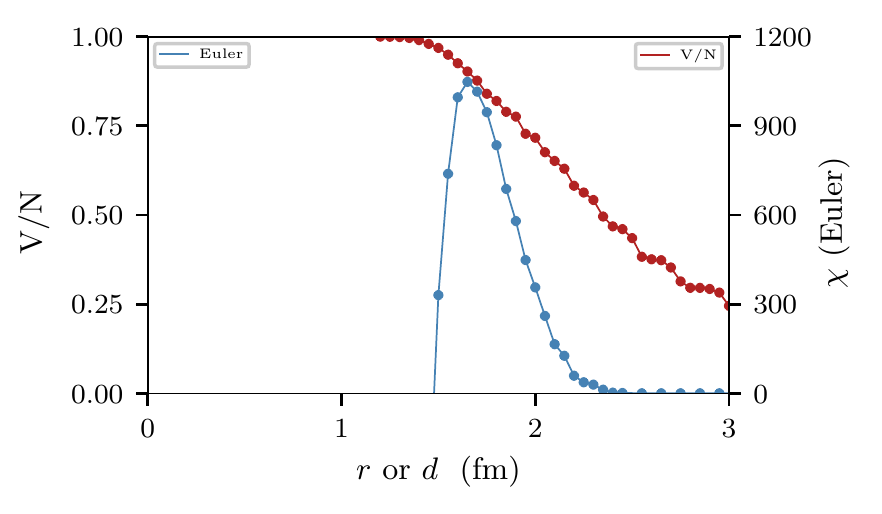}
}
\caption{\label{fig:rho016_x05_chi} (Color online) Analysis of the
same system as in Fig.~\ref{fig:rho016_x05} (with $N=5832$ nucleons). (a) On
the left scale (red symbols), the radial distribution function $g(\mathbf{r})$. 
On the
right scale, the Euler functional $\chi$ as a function of the voxel's edge
length $d$. (b) On the left scale (blue symbols), the first Minkowski
functional (volume) as a function of the voxel's edge length $d$. The volume is
normalized by $N$. On the right scale, the same the Euler functional $\chi$ as
in (a) for comparison reasons. }
\end{figure*}

Fig.~\ref{fig:rho016_x05_chi_euler} reproduces the same pattern for the
distribution $g(\mathbf{r})$ over all the nucleons. It further shows the Euler
functional $\chi$ as a function of the voxel's width (see caption for details).
Both curves share the same abscissa for comparison reasons. For small values of
$d$ the functional $\chi$ is negative (not shown), meaning that the voxels are
so small that tunnels prevail in the (discretized) system. At $d=1.65\,$fm this
functional arrives to a maximum where cavities or isolated regions prevail.
Some cavities may be ``true'' empty regions but other may simply be fake voids.
The most probable ones, however, correspond to fake voids since the 
$g(\mathbf{r})$
pattern actually presents a maximum at $r>1.65\,$fm. Thus, increasing the
voxel's size will most probably cancel the fake cavities.  \\

The particles located at the first maximum of $g(\mathbf{r})$ (at the saturation
density) may be envisaged as touching each other in a regular (simple cubic)
array. Therefore, the mean radius for a nucleon should be $1.85/2\,$fm. This
means, as a first thought, that binning the space into voxels of width
$1.85\,$fm will include a single nucleon per voxels. This is, however, not
completely true since approximately half of the first neighbors exceeds the
$1.85\,$fm (see first peak in Fig.~\ref{fig:rho016_x05}). Many voxels will be
empty, and thus, a relevant probability of finding fake voids exists.
Fig.~\ref{fig:rho016_x05_chi_euler} illustrates this situation. \\

\begin{figure*}[!htbp]
\centering
\captionsetup[subfigure]{justification=centering}
\subfloat[$d=1.85$\label{fig:voxels_width_a}]{
\includegraphics[width=0.5\columnwidth]
{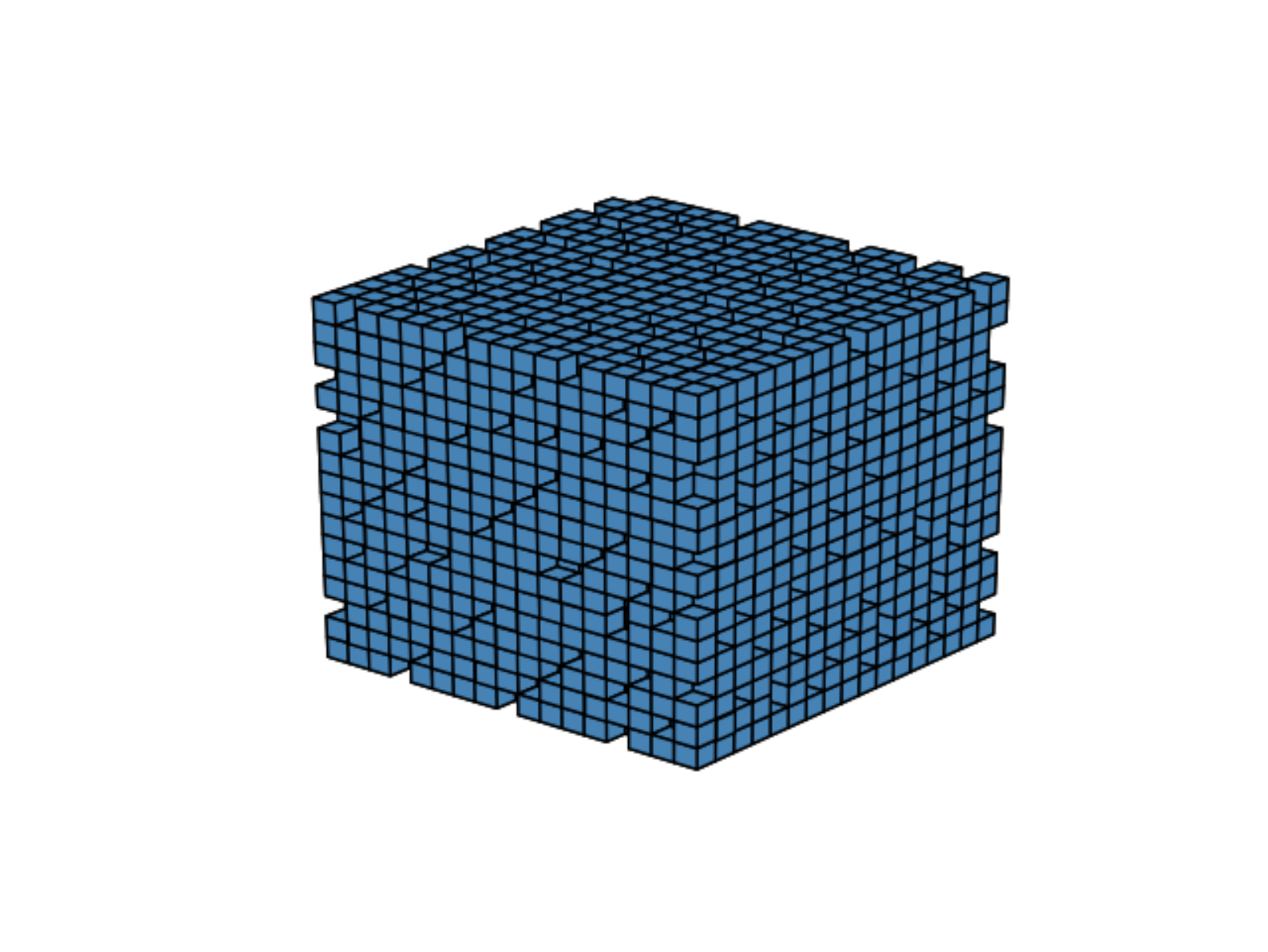}
} 
\subfloat[$d=2.35$\label{fig:voxels_width_b}]{
\includegraphics[width=0.5\columnwidth]
{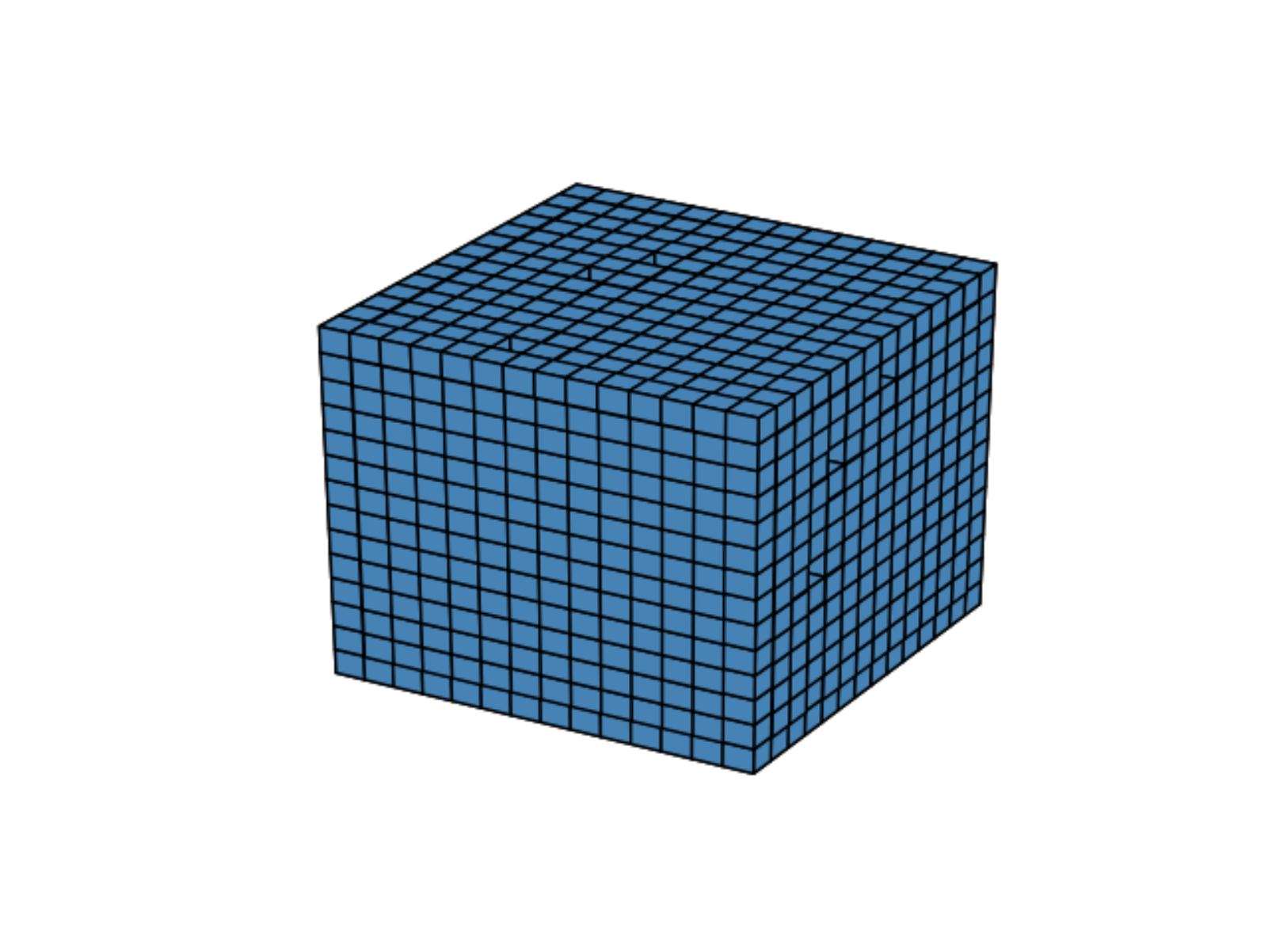}
} 

\caption{\label{fig:voxels_width} Example of the space binning into voxels
corresponding to the situation shown in Fig.~\ref{fig:rho016_x05}. (a) The edge
length of the cubic voxel is $d=1.85\,$fm. (b) The edge length of the cubic
voxel is $d=2.35\,$fm. }
\end{figure*}

Notice that whenever a fake empty voxel exists, the contiguous one will perhaps
host two nucleons. This is because the nucleon that exceeds the $1.85\,$fm
distance to the neighbor, say on the left, may have shorten the distance to the
neighbor on the right. Thus, the number of occupied voxels will probably not
match the number of nucleons. Fig.~\ref{fig:rho016_x05_chi_vol} shows a decrease
in the number of occupied voxels (\textit{i.e.} the volume) for $r>1.6\,$fm. \\

The space binning should be done wider in order to avoid fake empty voxels.
But, too wide voxels may include second neighbors. The most reasonable binning
distance appears to be around the first minimum of $g(\mathbf{r})$ (see
Fig.~\ref{fig:rho016_x05}). That is, at some point between $2.15\,$fm and
$2.35\,$fm. \\

A reasonable \textit{criterion} for the space binning may raise from the
Euler functional: the right binning distance should drive the $\chi$ functional 
to unity, that is, to a single compact region. This occurs at $d=2.35\,$fm for 
sure, as can be seen in Fig.~\ref{fig:voxels_width_b}. It can be further checked 
from Fig.~\ref{fig:rho016_x05_chi_vol} that this binning allows the hosting of 
approximately two nucleons per voxels, meaning that the number of fake voids is 
negligible. \\

The binning width $d=2.35\,$fm may be further compared with 
other literature values. The nucleon radius used in Ref.~\cite{horo13} is 
$1.5\,$fm, although these authors realize that this value seems rather 
large. Indeed, we were able to associated a nucleon radius of $0.925\,$fm at the 
saturation density for a simple cubic arrangement. As mentioned above (and 
exhibited in Fig.~\ref{fig:rho016_x05_chi_vol}) the distance $1.5\,$fm 
corresponds (roughly) to the maximum distance that can host 
one nucleon per voxel (at the saturation density). Therefore, this binning 
width is right if maximum ``contrast'' is required, as in Ref.~\cite{horo13}. 
Our criterion, however, is somewhat more conservative since it requires 
$\chi\simeq 1$ (at the saturation density), as explained above.  \\

\subsection{Cluster recognition}\label{cluster}
The nucleon positions and momenta are used to identify the fragment structure of 
the system by means of the ``Minimum Spanning Tree'' ($MST$) cluster-detection 
algorithm of~\cite{Str97}. In summary, $MST$ looks for correlations in 
configuration space: a particle $i$ belongs to a cluster $C$ if there is another 
particle $j$ that belongs to $C$ and $|r_i-r_j| \leq r_{cl}$, where $r_{cl}$ is 
a clusterization radius which, for the present study, was set to $r_{cl} = 3.0$ 
fm.

The main drawback of $MST$ is that, since only correlations in
$r$-space are used, it neglects completely the effect of momentum
giving incorrect information for dense systems and for highly
dynamical systems such as those formed in colliding nuclei.
Although more robust algorithms which look at relative momenta
between nucleons or pair-binding energies have been devised for
such systems (e.g. as the "Early Cluster Recognition Algorithm",
$ECRA$~\cite{dor-ran}), in the case of relatively cold systems,
such as nuclear crusts, the $MST$ is sufficient. In our case of
periodic boundary conditions, the $MST$ detection of fragments has
been modified to take into account the image cells and recognize
fragments that extend into adjacent cells.

\begin{figure}[h]
\begin{center}
\includegraphics[width=4.5in]{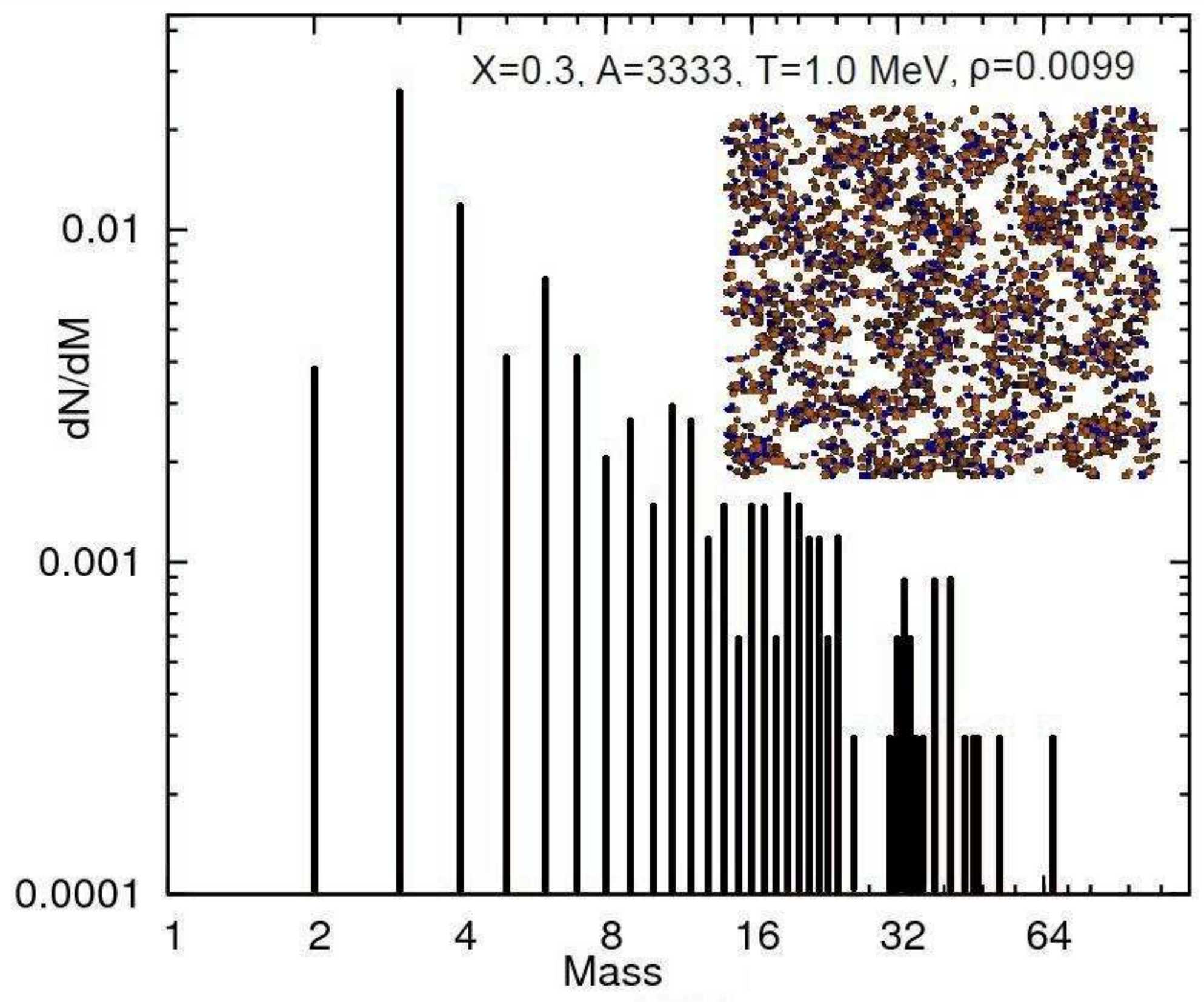}
\end{center}
\caption{Typical size distribution of clusters as obtained with
$MST$.  The inset shows a projection of the particle spatial
distribution.}\label{histo}
\end{figure}

The Figure~\ref{histo} shows an example of the size distribution
of the clusters obtained for a case with $3,333$ nucleons,
$x=0.3$, and at $T=0.3 \ MeV$ and $\rho=0.009 \ fm^{-3}$.  The
inset shows a projection of the position of the nucleons within
the cell.  The shown structure was obtained with the ``screened
Coulomb'' treatment that will be described in the next section.


\end{document}